\documentclass[12pt]{article}
\usepackage[body={17.5cm, 21cm},right=2cm]{geometry}

\usepackage{color}
\usepackage{graphicx}
\usepackage{epsf}
\usepackage{graphicx,epsfig}
\begin{document}

\vspace{5mm}
\vspace{0.5cm}
\begin{center}

\def\thefootnote{\fnsymbol{footnote}}

{\Large \bf Non-Gaussianity and the 
\\ [0.3cm]
Cosmic Microwave Background Anisotropies}
\\[0.5cm]
{\large  N. Bartolo$^{a,b}$, S. Matarrese$^{a,b}$ and A. Riotto$^{b,c}$}
\\[0.5cm]

\vspace{.3cm}
{\normalsize { \sl $^{a}$ Dipartimento di Fisica ``G.\ Galilei'', 
        Universit\`{a} di Padova, via Marzolo 8, I-131 Padova, Italy}}\\

\vspace{.3cm}
{\normalsize { \sl $^{b}$ INFN, Sezione di Padova, via Marzolo 8, I-35131 Padova, Italy}}\\

\vspace{.3cm}
{\normalsize {\sl $^{\rm c}$  CERN, Theory Division, CH-1211 Geneva 23, Switzerland}}\\

\vspace{.3cm}
{\normalsize {\sl E-mail: nicola.bartolo@pd.infn.it, sabino.matarrese@pd.infn.it and riotto@mail.cern.ch}}



\end{center}

\vspace{.8cm}

\hrule \vspace{0.3cm}
{\small  \noindent \textbf{Abstract} \\[0.3cm]
\noindent We review in a pedagogical way the present status of the impact  of
non-Gaussianity (NG) on  the Cosmic Microwave Background (CMB)
anisotropies. We first show how to set the initial conditions at
second-order for the (gauge invariant) CMB anisotropies when some
primordial NG is present. However, there are many sources of NG in CMB
anisotropies, beyond the primordial one, which can contaminate the
primordial signal. We mainly focus on the NG generated from the post-inflationary evolution  
of the CMB anisotropies at second-order in perturbation theory at large and small angular scales, such as the ones 
generated at the recombination epoch. We show how to derive
the equations to study the second-order CMB anisotropies and provide
analytical computations to evaluate their contamination to primordial
NG (complemented with numerical examples). We also offer a brief summary
of other secondary effects.
This review requires basic knowledge of the theory of cosmological
perturbations at the linear level.

\vspace{0.5cm}  \hrule
\def\thefootnote{\arabic{footnote}}
\setcounter{footnote}{0}



\newpage 
\tableofcontents

\newcommand{\fix}{\Phi(\mathbf{x})}
\newcommand{\fiLx}{\Phi_{\rm L}(\mathbf{x})}
\newcommand{\fiNLx}{\Phi_{\rm NL}(\mathbf{x})}
\newcommand{\fik}{\Phi(\mathbf{k})}
\newcommand{\fiLk}{\Phi_{\rm L}(\mathbf{k})}
\newcommand{\fiLkone}{\Phi_{\rm L}(\mathbf{k_1})}
\newcommand{\fiLktwo}{\Phi_{\rm L}(\mathbf{k_2})}
\newcommand{\fiLkthree}{\Phi_{\rm L}(\mathbf{k_3})}
\newcommand{\fiLkfour}{\Phi_{\rm L}(\mathbf{k_4})}
\newcommand{\fiNLk}{\Phi_{\rm NL}(\mathbf{k})}
\newcommand{\fiNLkone}{\Phi_{\rm NL}(\mathbf{k_1})}
\newcommand{\fiNLktwo}{\Phi_{\rm NL}(\mathbf{k_2})}
\newcommand{\fiNLkthree}{\Phi_{\rm NL}(\mathbf{k_3})}

\newcommand{\kernel}{f_{\rm NL} (\mathbf{k_1},\mathbf{k_2},\mathbf{k_3})}
\newcommand{\dirac}{\delta^{(3)}\,(\mathbf{k_1+k_2-k})}
\newcommand{\dirackonektwokthree}{\delta^{(3)}\,(\mathbf{k_1+k_2+k_3})}

\newcommand{\beq}{\begin{equation}}
\newcommand{\eeq}{\end{equation}}
\newcommand{\beqarr}{\begin{eqnarray}}
\newcommand{\eeqarr}{\end{eqnarray}}

\newcommand{\angk}{\hat{k}}
\newcommand{\angn}{\hat{n}}

\newcommand{\tfnow}{\Delta_\ell(k,\eta_0)}
\newcommand{\tf}{\Delta_\ell(k)}
\newcommand{\tfone}{\Delta_{\el\ell_1}(k_1)}
\newcommand{\tftwo}{\Delta_{\el\ell_2}(k_2)}
\newcommand{\tfthree}{\Delta_{\el\ell_3}(k_3)}
\newcommand{\tffour}{\Delta_{\el\ell_1^\prime}(k)}
\newcommand{\deltatilde}{\tilde{\Delta}_{\el\ell_3}(k_3)}

\newcommand{\alm}{a_{\ell m}}
\newcommand{\almL}{a_{\ell m}^{\rm L}}
\newcommand{\almNL}{a_{\ell m}^{\rm NL}}
\newcommand{\almone}{a_{\el\ell_1 m_1}}
\newcommand{\almLone}{a_{\el\ell_1 m_1}^{\rm L}}
\newcommand{\almNLone}{a_{\el\ell_1 m_1}^{\rm NL}}
\newcommand{\almtwo}{a_{\el\ell_2 m_2}}
\newcommand{\almLtwo}{a_{\el\ell_2 m_2}^{\rm L}}
\newcommand{\almNLtwo}{a_{\el\ell_2 m_2}^{\rm NL}}
\newcommand{\almthree}{a_{\el\ell_3 m_3}}
\newcommand{\almLthree}{a_{\el\ell_3 m_3}^{\rm L}}
\newcommand{\almNLthree}{a_{\el\ell_3 m_3}^{\rm NL}}

\newcommand{\YLMstar}{Y_{L M}^*}
\newcommand{\Ylmstar}{Y_{\ell m}^*}
\newcommand{\Ylmstarone}{Y_{\el\ell_1 m_1}^*}
\newcommand{\Ylmstartwo}{Y_{\el\ell_2 m_2}^*}
\newcommand{\Ylmstarthree}{Y_{\el\ell_3 m_3}^*}
\newcommand{\Ylmstarfour}{Y_{\el\ell_1^\prime m_1^\prime}^*}
\newcommand{\Ylmstarfive}{Y_{\el\ell_2^\prime m_2^\prime}^*}
\newcommand{\Ylmstarsix}{Y_{\el\ell_3^\prime m_3^\prime}^*}

\newcommand{\YLM}{Y_{L M}}
\newcommand{\Ylm}{Y_{\ell m}}
\newcommand{\Ylmone}{Y_{\el\ell_1 m_1}}
\newcommand{\Ylmtwo}{Y_{\el\ell_2 m_2}}
\newcommand{\Ylmthree}{Y_{\el\ell_3 m_3}}
\newcommand{\Ylmfour}{Y_{\el\ell_1^\prime m_1^\prime}}
\newcommand{\Ylmfive}{Y_{\el\ell_2^\prime m_2^\prime}}
\newcommand{\Ylmsix}{Y_{\el\ell_3^\prime m_3^\prime}}

\newcommand{\jl}{j_\ell(k r)}
\newcommand{\jlfourone}{j_{\el\ell_1^\prime}(k_1 r)}
\newcommand{\jlfivetwo}{j_{\el\ell_2^\prime}(k_2 r)}
\newcommand{\jlsixthree}{j_{\el\ell_3^\prime}(k_3 r)}
\newcommand{\jlsix}{j_{\el\ell_3^\prime}(k r)}
\newcommand{\jlthree}{j_{\el\ell_3}(k_3 r)}
\newcommand{\jlthreetau}{j_{\el\ell_3}(k r)}

\newcommand{\Gaunt}{\mathcal{G}_{\el\ell_1^\prime \, \el\ell_2^\prime \, 
\el\ell_3^\prime}^{m_1^\prime m_2^\prime m_3^\prime}}
\newcommand{\Gaunttwo}{\mathcal{G}_{\el\ell_1^\prime \, \el\ell_2^\prime \, 
\el\ell_3}^{m_1^\prime m_2^\prime m_3}}
\newcommand{\Gauntstardef}{\mathcal{H}_{\el\ell_1 \, \el\ell_2 \, \el\ell_3}^{m_1 m_2 m_3}}
\newcommand{\Gauntstarone}{\mathcal{G}_{\el\ell_1 \, L \,\, \el\ell_1^\prime}
^{-m_1 M m_1^\prime}}
\newcommand{\Gauntstartwo}{\mathcal{G}_{\el\ell_2^\prime \, \el\ell_2 \, L}
^{-m_2^\prime m_2 M}}

\newcommand{\dangn}{d \angn}
\newcommand{\dangk}{d \angk}
\newcommand{\dangkone}{d \angk_1}
\newcommand{\dangktwo}{d \angk_2}
\newcommand{\dangkthree}{d \angk_3}
\newcommand{\dk}{d^3 k}
\newcommand{\dkone}{d^3 k_1}
\newcommand{\dktwo}{d^3 k_2}
\newcommand{\dkthree}{d^3 k_3}
\newcommand{\dkfour}{d^3 k_4}
\newcommand{\dallk}{\dkone \dktwo \dk}

\newcommand{\FT}{ \int  \! \frac{d^3k}{(2\pi)^3} 
e^{i\mathbf{k} \cdot \angn \eta_0}}
\newcommand{\planewave}{e^{i\mathbf{k \cdot x}}}
\newcommand{\dallkfourier}{\frac{\dkone}{(2\pi)^3}\frac{\dktwo}{(2\pi)^3}
\frac{\dkthree}{(2\pi)^3}}

\newcommand{\Bis}{B_{\el\ell_1 \el\ell_2 \el\ell_3}^{m_1 m_2 m_3}}
\newcommand{\Avbis}{B_{\el\ell_1 \el\ell_2 \el\ell_3}}

\newcommand{\los}{\mathcal{L}_{\el\ell_3 \el\ell_1 \el\ell_2}^{L \, 
\el\ell_1^\prime \el\ell_2^\prime}(r)}
\newcommand{\loszero}{\mathcal{L}_{\el\ell_3 \el\ell_1 \el\ell_2}^{0 \, 
\el\ell_1^\prime \el\ell_2^\prime}(r)}
\newcommand{\losone}{\mathcal{L}_{\el\ell_3 \el\ell_1 \el\ell_2}^{1 \, 
\el\ell_1^\prime \el\ell_2^\prime}(r)}
\newcommand{\lostwo}{\mathcal{L}_{\el\ell_3 \el\ell_1 \el\ell_2}^{2 \, 
\el\ell_1^\prime \el\ell_2^\prime}(r)}
\newcommand{\losfNL}{\mathcal{L}_{\el\ell_3 \el\ell_1 \el\ell_2}^{0 \, 
\el\ell_1 \el\ell_2}(r)}
\newcommand\ee{\end{equation}}

\newcommand\be{\begin{equation}}

\newcommand\eea{\end{eqnarray}}

\newcommand\bea{\begin{eqnarray}}
\def\d{d}
\def\C{{\rm CDM}}
\def\me{m_e}
\def\te{T_e}
\def\ti{\tau_{\rm initial}}
\def\tci#1{n_e(#1) \sigma_T a(#1)}
\def\tr{\eta_r}
\def\dtr{\delta\eta_r}
\def\dd{\tilde\Delta^{\rm Doppler}}
\def\dsw{\Delta^{\rm Sachs-Wolfe}}
\def\clsw{C_\ell^{\rm Sachs-Wolfe}}
\def\cldop{C_\ell^{\rm Doppler}}
\def\Dt{\tilde{\Delta}}
\def\mut{\mu}
\def\vt{\tilde v}
\def\hp{ {\bf \hat p}}
\def\sdv{S_{\delta v}}
\def\svv{S_{vv}}
\def\bvt{\tilde{\bv}}
\def\delt{\tilde{\delta_e}}
\def\cos{{\rm cos}}
\def\nn{\nonumber \\}
\def\bq{ {\bf q} }
\def\ba{ {\bf p} }
\def\bap{ {\bf p'} }
\def\bqp{ {\bf q'} }
\def\bp{ {\bf p} }
\def\bpp{ {\bf p'} }
\def\bk{ {\bf k} }
\def\bx{ {\bf x} }
\def\bv{ {\bf v} }
\def\qp{ p^{\mu}k_{\mu} }
\def\qpp { p^{\mu} k'_{\mu} }
\def\bgm{ {\bf \gamma} }
\def\bkp{ {\bf k'} }
\def\gq{ g(\bq)}
\def\gqp{ g(\bqp)}
\def\fp{ f(\bp)}
\def\h#1{ {\bf \hat #1}}
\def\fpp{ f(\bpp)}
\def\fz{f^{(0)}(p)}
\def\fpz{f^{(0)}(p')}
\def\f#1{f^{(#1)}(\bp)}
\def\fps#1{f^{(#1)}(\bpp)}
\def\dq{ {d^3\bq \over (2\pi)^32E(\bq)} }
\def\dqp{ {d^3\bqp \over (2\pi)^32E(\bqp)} }
\def\dpp{ {d^3\bpp \over (2\pi)^32E(\bpp)} }
\def\dtq{ {d^3\bq \over (2\pi)^3} }
\def\dtqp{ {d^3\bqp \over (2\pi)^3} }
\def\dtpp{ {d^3\bpp \over (2\pi)^3} }
\def\part#1;#2 {\partial#1 \over \partial#2}
\def\deriv#1;#2 {d#1 \over d#2}
\def\Done{\Delta^{(1)}}
\def\Dtwo{\tilde\Delta^{(2)}}
\def\fone{f^{(1)}}
\def\ftwo{f^{(2)}}
\def\tg{T_\gamma}
\def\delpp{\delta(p-p')}
\def\delb{\delta_B}
\def\tc{\eta_0}
\def\DD{\langle|\Delta(k,\mu,\eta_0)|^2\rangle}
\def\DDL{\langle|\Delta(k=l/\tc,\mu)|^2\rangle}
\def\bkpp{{\bf k''}}
\def\kmkp{|\bk-\bkp|}
\def\kmkpsq{k^2+k'^2-2kk'x}
\def\tt{ \left({\tau' \over \tau_c}\right)}
\def\kt{ k\mu \tau_c}

%
%
%

\section{Introduction}
Cosmic Microwave Background (CMB) anisotropies play a special role 
in cosmology, as they allow an accurate 
determination of cosmological parameters and may provide 
a unique probe of the physics of the early universe and in particular of
the processes that gave origin to the primordial 
perturbations. 

Cosmological inflation \cite{lrreview} is nowadays considered 
the dominant paradigm for the generation of the initial seeds for 
structure formation. 
In the inflationary picture, 
the primordial cosmological perturbations are created from quantum 
fluctuations ``redshifted'' out of the horizon during an early period of 
accelerated expansion of the universe, where they remain ``frozen''.  
They are observable through CMB temperature anisotropies (and polarization) and
the large-scale clustering properties of the matter 
distribution in the Universe. 

This picture has recently received further spectacular confirmations 
from the  results of the Wilkinson Microwave Anisotropy Probe (WMAP) 
five year set of data \cite{wmap5}.
Since the observed cosmological perturbations are of the order
of $10^{-5}$, one might think that first-order perturbation theory
will be adequate for all comparisons with observations. This might not be
the case, though.  Present \cite{wmap5,planck} and future experiments \cite{Baumann}  
may be sensitive to the non-linearities of the cosmological
perturbations at the level of second- or higher-order perturbation theory.
The detection of these non-linearities through the non-Gaussianity
(NG) in the CMB \cite{review} has become one of the
primary experimental targets. 

There is one fundamental  reason why a positive detection of NG 
is so relevant: it might help in discriminating among the various mechanisms
for the generation of the cosmological perturbations. Indeed,
various models of inflation, firmly rooted in modern 
particle physics theory,   predict a significant amount of primordial
NG generated either during or immediately after inflation when the
comoving curvature perturbation becomes constant on super-horizon scales
\cite{review}. While standard single-field models of slow-roll inflation \cite{noi}
and -- in general -- two (multi)-field \cite{two} models of inflation predict
a tiny level of NG, ``curvaton''-type models,  in which
a  significant contribution to the curvature perturbation is generated after
the end of slow-roll inflation by the perturbation in a field which has
a negligible effect on inflation, may predict a high level of NG \cite{luw}.
Alternatives to the curvaton model are models 
where a curvature perturbation mode is  
generated by an inhomogeneity in the decay rate \cite{hk,varcoupling},
the mass  \cite{varmass} or the interaction rate \cite{thermalization}
of the particles responsible for the reheating after inflation. 
Other opportunities for generating the curvature perturbations occur
at the end of inflation \cite{endinflation}, during
preheating \cite{preheating}, and 
at a phase-transition producing cosmic strings \cite{matsuda}. Also, within single-field models of inflation, 
a high level of NG can be generated breaking 
the standard conditions of canonical kinetic terms and initially vaccum states:  e.g. this is the case of Dirac-Born-Infeld 
(DBI) models of inflation~\cite{DBI}, and initially excited states respectively~\cite{HT}.  
For every scenario there exists a well defined prediction for the \emph{strength} of NG and its 
\emph{shape}~\cite{Shapes,FS} as a function of the parameters.

Statistics like the bispectrum and the trispectrum of the 
CMB can then be used to assess the level of primordial NG (and possibly its shape) on 
various cosmological scales 
and to discriminate it from the one induced by 
secondary anisotropies and systematic effects \cite{review,hu,dt,jul}. 
A positive detection of a primordial NG in the CMB at some level 
might therefore 
confirm and/or rule out a whole class of mechanisms by which the cosmological
perturbations have been generated.

Despite the importance of evaluating the impact of primordial 
NG in a crucial observable like the  CMB anisotropy, the vast majority of the
literature has been devoted to the computation of the bispectrum
of either the comovig curvature perturbation or the gravitational 
potential on large scales within given inflationary models. 
These, however, are not the  physical
quantities which are observed. One should instead provide a full prediction
for the second-order radiation transfer function. 
A preliminary step towards this goal has been taken in Ref.~\cite{fulllarge}
(see also~\cite{twoT,enh,prl,tom,beyond,cre})
where the full second-order radiation transfer function 
for the CMB  anisotropies on large angular scales in a flat universe 
filled with matter and cosmological constant was computed, including 
the second-order generalization of the Sachs-Wolfe effect,
both the early and late Integrated Sachs-Wolfe (ISW) effects and the 
contribution of the second-order tensor modes.\footnote{A similar computation of the CMB anisotropies up to third-order from gravitational perturbations
has been performed in Ref.~\cite{Dnoi}, which is particularly relevant to provide a complete theoretical predition for cubic non-linearities 
characterizing the level of NG in the CMB through the connected four-point correlation function (trispectrum)~\cite{hu}.}.

There are many sources of NG in CMB anisotropies, beyond the primordial
one.  The most relevant  sources are the so-called  secondary anisotropies,
which arise  after the last scattering epoch. These anisotropies can be 
divided into two categories: scattering secondaries, when the CMB photons 
scatter with electrons along the line of sight, 
and gravitational secondaries when effects are mediated by 
gravity \cite{Hu:2001bc}.  Among the  scattering secondaries 
we may list the thermal Sunyaev-Zeldovich effect, where 
hot electrons in clusters transfer energy to the CMB photons, 
the kinetic Sunyaev-Zeldovich effect produced by the bulk motion of the 
electrons in clusters, the Ostriker-Vishniac effect, produced by bulk 
motions modulated by linear density perturbations, and effects due to  
reionization processes. The scattering secondaries 
are most significant on small angular scales
as density inhomogeneities, bulk and thermal motions  grow and become 
sizeable on small length-scales when structure formation proceeds.

Gravitational secondaries arise from the change 
in energy of photons when the gravitational potential is time-dependent, the 
ISW effect, and  gravitational lensing.  At late times, when the 
Universe becomes dominated by the dark energy, 
the gravitational potential on linear scales starts to decay, causing 
the ISW effect mainly on large angular scales. Other secondaries 
that result from a 
time dependent potential are the Rees-Sciama effect, produced during
the matter-dominated epoch by the time 
evolution of the potential on non-linear scales. 

The fact that the potential never grows appreciably means that most 
second order effects created by gravitational secondaries are generically small
compared to those created by 
scattering ones. However, when a photon propagates from the 
last scattering to us, its path may be deflected because of 
the gravitational lensing. This effect 
does not create anisotropies, but only modifies existing ones. Since
photons with large wavenumbers $k$  are lensed over many regions ($\sim k/H$, 
where $H$ is the Hubble rate) along the
line of sight, the corresponding second-order effect may be sizeable.
The  
three-point function arising from the correlation of the gravitational lensing 
and ISW effects generated by  
the matter distribution along the line of sight 
\cite{Seljak:1998nu,Goldberg:xm} and the Sunyaev-Zeldovich effect \cite{sk}
are large and detectable by Planck~\cite{ks,SZ}. A crucial issue is the level of \emph{contamination} 
to the extraction of the primordial NG the seconndary effects can produce. In Sec.~\ref{cont} we briefly summarize some recent results about the level of 
CMB NG generated by some of these secondary effects.

Another relevant source of NG comes from the physics operating at
the recombination. A naive estimate would tell that these non-linearities
are tiny being suppressed by an extra power of the gravitational 
potential. However, the dynamics  at recombination is quite involved because
all the non-linearities in the evolution of the baryon-photon fluid at 
recombination and the ones coming from general relativity should be 
accounted for. This complicated dynamics might lead
to  unexpected suppressions or enhancements of the NG at recombination. 
A step towards the evaluation of the three-point correlation function
has been taken in Ref. \cite{rec} where some effects were taken into account 
in the  so-called  squeezed triangle limit, corresponding to the case 
when one wavenumber is much smaller than the other two and was outside 
the horizon at recombination. Refs.~\cite{paperI,paperII} (see also \cite{pit}), 
present the computation of the full system of Boltzmann 
equations, describing the evolution of the photon, 
baryon and Cold Dark Matter (CDM) 
fluids, at second order and neglecting polarization, 
These equations allow to follow the time evolution of the CMB anisotropies 
at second order on all angular scales
from the early epochs, when the cosmological perturbations were generated,
to the present  time, through the recombination era. These calculations 
set the stage for the computation of the full second-order
radiation transfer function at all scales and for a 
a generic set of initial conditions specifying the level of primordial 
non-Gaussianity. 
Of course, for specific effects on small angular scales like Sunyaev-Zel'dovich,
gravitational lensing, etc., fully non-linear
calculations would provide a more
accurate estimate of the resulting CMB anisotropy, however,
as long as the leading contribution to 
second-order statistics like the bispectrum  is
concerned, second-order perturbation theory suffices.

The goal of this review is to summarize in a pedagogical form the present status of the
evaluation of the impact of NG on the CMB anisotropies. This implies first of all determining how to set the initial conditions at second-order for the (gauge-invariant) CMB anisotropy when some source of primordial NG is present. The second step will be determining how primordial NG flows on small angular  scales. In this review we will focus ourselves on the
study of the second-order effects appearing at the recombination era when the CMB anisotropy
is left imprinted. We will  show how to derive the equations to evaluate 
CMB anisotropies, 
by computing the Boltzmann equations describing the evolution of the 
baryon-photon fluid up to second order. This permits to follow the 
time evolution of CMB anisotropies (up to second order) on all scales, 
from the early epoch, when the cosmological perturbations were generated,
to the present time, through the recombination era. We will also provide the reader with some
simplified analytical computation to evaluate   the contamination of the recombination secondary effects 
onto the detection of primordial NG. The formalism for a more refined numerical analysis is also displayed and results 
for some worked examples will be also reported. The review is mainly based
on a series of papers written by the authors along the past years on the subject (with various updates) and, as such,
follows both a  logic and a chronological order.  It requires    knowledge of the theory
of cosmological perturbation at the linear level (which however we summarize in the Appendices). 
We have tried to write the different sections in a self-contained way. Nevertheless, we alert the reader that  the level of complexity increases with the number of the sections. 

The review is organized as follows. In Section 2 we provide a simple, but illuminating example to show why we do expect some NG present in the CMB anisotropy regardless if there is or not some primordial NG. In Section 3 we provide the reader with the necessary tools to study the dynamics at second-order in the gravity sector. In Section 4  we show how to set the initial conditions for the primordial NG, while in Section 5 we provide a gauge-invariant way to define the CMB temperature anisotropy at second-order on large scales. In Section 6 we go to small scales and present the full procedure to compute the Boltzmann equations necessary to follow the evolution of the non-linearities from the recombination epoch down to the present epoch. Section 7 presents some analytical solutions of the Boltzmann equations in the tight coupling limit, along the same lines of what is done at the linear level. The issue of contamination is addressed in Section 8, while in Section 9 we offer the reader with an analytical estimate of such a contamination. A more refined numerical work is presented in Section 10. Finally, in Section 11 
some conclusions are given.
This review has also some hopefully useful appendices: in Appendix A 
the reader can find the energy-momentm tensors at second-order, Appendix B gives the solutions of Eistein equations for the perturbed fluids up to second-order, 
while Appendix C offers the analytical solutions of the linearized Boltzmann equations in the tight coupling limit.
\section{Why do we expect NG in the cosmological perturbations?}
 
Before tackling the problem of interest -- the computation of the cosmological perturbations
at second-order after the inflationary era-- 
we first provide a simple, but insightful computation, derived from Ref. \cite{beyond}, 
which illustrates
why we expect that the cosmological perturbations develop some NG even if 
the latter is not present at some primordial epoch. This example
 will help the reader to understand why
the cosmological perturbations are inevitably affected by nonlinearities, beyond those
arising at some primordial epoch. The reason is clear: gravity is nonlinear and it  feeds
this property into the cosmological perturbations during the post-inflationary
evolution of the universe. As  gravity talks to all fluids, this transmission is 
inevitable. To be specific, we focus on the CMB anisotropies.   
We will adopt the Poisson gauge which eliminates 
one scalar degree of freedom from the $g_{0i}$ component of the 
metric and one scalar and two vector degrees of freedom from $g_{ij}$. 
We will use a metric of the form   
\begin{equation}
\label{metric}
ds^2=
-e^{2\Phi} dt^2+2a(t)\omega_i dx^i dt+a^2(t)(e^{-2\Psi}\delta_{ij}+\chi_{ij}) 
dx^i dx^j\, ,
\end{equation}
where $a(t)$ is the scale factor as a function of the cosmic time 
$t$, and $\omega_i$ and $\chi_{ij}$ 
are the vector and tensor peturbation modes 
respectively. Each metric perturbation can be expanded into a 
linear (first-order) and a second-order part, as for example, the 
gravitational potential $\Phi=\Phi^{(1)}+\Phi^{(2)}/2$. However 
in the metric~(\ref{metric}) the choice of the exponentials greatly 
helps in computing the relevant expressions, and thus we will 
always keep them where it is convenient. From Eq.~(\ref{metric}) one 
recovers at linear order the well-known  
longitudinal gauge while at second order, one finds 
$\Phi^{(2)}=\phi^{(2)}-2 (\phi^{(1)})^2$ and $\Psi^{(2)}=
\psi^{(2)}+2(\psi^{(1)})^2$ where $\phi^{(1)}$, $\psi^{(1)}$ 
and $\phi^{(2)}$, $\psi^{(2)}$ (with 
$\phi^{(1)}=\Phi^{(1)}$ and $\psi^{(1)}=\Psi^{(1)}$) are the first and 
second-order gravitational 
potentials in the longitudinal (Poisson) gauge adopted in 
Refs.~\cite{mol,review} as far as  scalar perturbations are concerned.

We now consider the long wavelength modes of the CMB anisotropies,
i.e. we focus on scales larger than the horizon at last-scattering. We can therefore neglect vector and tensor 
perturbation modes in the metric. For the vector perturbations the 
reason is that we are 
they   contain gradient terms being produced 
as non-linear combination of scalar-modes and thus they will be more important 
on small scales (linear vector modes 
are not generated in standard mechanisms for 
cosmological perturbations, as inflation). 
The tensor contribution 
can be neglected for two reasons. First, the tensor perturbations    
produced from inflation on large scales give a negligible contribution to 
the higher-order statistics of the Sachs-Wolfe effect
being of the order of (powers of) the slow-roll 
parameters during inflation (this holds for linear tensor modes as well as for 
tensor modes generated by the non-linear evolution of scalar perturbations 
during inflation).

Since we are interested  in   the cosmological
perturbations on large scales, that is in 
perturbations whose wavelength is  larger than the Hubble radius at last 
scattering, a local
observer would see them in  the form of  a classical -- possibly 
time-dependent -- 
(nearly zero-momentum) homogeneous and isotropic
background. Therefore,  it should be
possible to perform a change of coordinates in such a way as to
absorb  the  
super-Hubble modes and work with a metric
of an homogeneous and isotropic Universe (plus, of course, cosmological
perturbations on  scale smaller than the horizon). 
We  split the gravitational potential $\Phi$ as
\begin{equation}
\Phi=\Phi_\ell +\Phi_s\, ,
\end{equation}
where $\Phi_\ell$ stands for the part of the
gravitational potential receiving contributions only from the
super-Hubble modes; $\Phi_s$ receives contributions only
from the sub-horizon modes

\begin{eqnarray}
\Phi_\ell&=&\int\frac{d^3\!k}{(2\pi)^3}\, \theta\left(aH-k\right)
\, \Phi_{\vec{k}} \ e^{i\vec{k}\cdot\vec{x}} \, ,\nonumber\\
\Phi_s&=&\int\frac{d^3\!k}{(2\pi)^3}\, \theta\left(k-aH\right)
\, \Phi_{\vec{k}} \ e^{i\vec{k}\cdot\vec{x}} \, ,
\end{eqnarray}
where $H$ is the Hubble rate computed with respect to the cosmic time,
$H=\dot{a}/a$, and $\theta(x)$ is the step function. Analogous definitions
hold for the other gravitational potential $\Psi$. 

By construction $\Phi_\ell$ and $\Psi_\ell$  are a collection of 
Fourier modes whose wavelengths are larger than the horizon
length and we may safely neglect  their spatial gradients.
Therefore $\Phi_\ell$ and 
$\Psi_\ell$ are only  functions of time.
This amounts to saying that  
we can absorb the large-scale perturbations in the metric (\ref{metric})
by the following redefinitions
\begin{eqnarray}
\label{tbar}
d\overline{t}&=&e^{\Phi_\ell} dt\, , \\
\label{abar}
\overline{a}&=&a\, e^{-\Psi_\ell}\, .
\end{eqnarray}
The new metric  describes a  homogeneous and 
isotropic Universe
\begin{equation}
\label{newmetric}
ds^2=-d\overline{t}^2 + \overline{a}^{2} \delta_{ij} \,dx^i\,dx^j\, ,
\end{equation} 
where for simplicity we have not included  the sub-horizon modes.
On super-horizon scales one can regard  the Universe as a collection
of regions of size of the Hubble radius 
evolving like  unperturbed patches with metric (\ref{newmetric})
~{\cite{Salopek1}.

Let us now go back to  the quantity we are interested in, namely the 
anisotropies of the CMB as measured today by an observer ${\mathcal O}$.
If she/he is interested in the CMB anisotropies at large scales, the effect
of super-Hubble modes is encoded in the metric (\ref{newmetric}). 
During their travel    
from the last scattering surface -- to be considered as the  
emitter point ${\mathcal E}$ --  to the observer,  
the CMB photons suffer a redshift 
determined by the ratio of the emitted 
frequency $\overline{\omega}_{\mathcal E}$ to the observed one 
$\overline{\omega}_{\mathcal O}$ 
\begin{equation}
\label{Texact}
\overline{T}_{\mathcal O}=\overline{T}_{\mathcal E}\,
\frac{\overline{\omega}_{\mathcal O}}{\overline{\omega}_{\mathcal E}}\, , 
\end{equation}
where $\overline{T}_{\mathcal O}$ and 
$\overline{T}_{\mathcal E}$ are  the temperatures at the
observer point and  at the
last scattering surface, respectively.

What is then the temperature anisotropy measured by the  observer?
The expression (\ref{Texact}) shows that 
the measured large-scale anisotropies  
are made of two contributions:  the intrinsic 
inhomogeneities in the temperature at the last scattering
surface  and the inhomogeneities in the scaling factor
provided by the ratio of the frequencies of the photons
at the departure and arrival points. 
Let us first consider the second contribution.
As the 
frequency of the photon  is the inverse of a time period, we get
immediately the fully non-linear relation
\begin{equation}
\label{1result}
\frac{\overline{\omega}_{\mathcal E}}{\overline{\omega}_{\mathcal O}}
=\frac{\omega_{\mathcal E}}{\omega_{\mathcal O}}e^{-\Phi_{\ell
{\mathcal E}}+\Phi_{\ell
{\mathcal O}}}\, .
\end{equation}             
As for the   temperature anisotropies coming from the intrinsic 
temperature fluctuation at the emission point, 
it maybe worth to recall   how to obtain this quantity 
in the longitudinal gauge at first order. By expanding the photon energy
density $\rho_\gamma \propto T_\gamma^4$, the intrinsic temperature 
anisotropies at last scattering are given by  $\delta_1 T_{\mathcal E}/
T_{\mathcal E}=(1/4)\delta_1 
\rho_\gamma/\rho_\gamma$. One relates the photon energy density 
fluctuation to the gravitational perturbation first by implementing the 
adiabaticity condition $\delta_1 
\rho_\gamma/\rho_\gamma=(4/3)\delta_1 
\rho_m/\rho_m$, 
where $\delta_1\rho_m/\rho_m$ 
is the relative fluctuation in the matter component, 
and then using the energy constraint of Einstein equations 
$\phi^{(1)}=-(1/2)\delta_1\rho_m/\rho_m$. The result is $\delta_1 T_{\mathcal E}/
T_{\mathcal E}=-2 \Phi_{1\mathcal E}/3$. Summing this contribution
to the anisotropies coming from 
the redshift factor (\ref{1result}) expanded at first order provides the  
standard (linear) Sachs-Wolfe effect 
$\delta_1 T_{\mathcal O}/T_{\mathcal O}
=\Phi_{1\mathcal E}/3$. Following the same steps,  
we may easily obtain its full non-linear generalization.   

Let us first relate the photon energy density 
$\overline{\rho}_\gamma$ to the energy density of the non-relativistic matter 
$\overline{\rho}_m$ by using the adiabaticity conditon. 
Again here  a bar indicates  that we are considering  
quantities in the locally homogeneous Universe described by the metric
(\ref{newmetric}). Using the 
energy continuity equation on large scales $\partial \overline{\rho}/  
\partial\overline{t} =-3 \overline{H} (\overline{\rho}+\overline{P})$, where 
$\overline{H}=d \ln \overline{a}/d \overline{t}$ 
and $\overline{P}$ is the pressure of the fluid, one can easily show that 
there exists a conserved 
quantity in time at any order in perturbation theory ~\cite{KMNR}
\begin{equation}
{\cal F} \equiv \ln \overline{a} +\frac{1}{3}\int^{\overline{\rho}}\, 
\frac{d \overline{\rho'}}{\left(\overline{\rho'}+\overline{P'}\right)}\, .
\label{fun}
\end{equation}
The perturbation $\delta {\cal F}$ is a gauge-invariant quantity representing 
the non-linear extension of the curvature perturbation $\zeta$ 
on uniform energy density hypersurfaces on superhorizon scales 
for adiabatic fluids~\cite{KMNR}. Indeed, expanding it at 
first and second order one gets the corresponding definition 
$\zeta_1=-\psi_1-\delta_1\rho/ \dot{\rho}$ and the quantity 
$\zeta_2$ introduced in Ref.~\cite{MW}. We will come back to these definitions later. 
At first order the adiabaticity condition 
corresponds to set $\zeta_{1\gamma}=\zeta_{1 m}$ for the curvature 
perturbations relative to each component. At the non-linear level
the adiabaticity condition generalizes to 

\begin{equation}
\frac{1}{3} \int \frac{d \overline{\rho}_m}{\overline{\rho}_m} =
\frac{1}{4} \int  
\frac{d \overline{\rho}_\gamma}{\overline{\rho}_\gamma}\, ,
\end{equation}
or
\begin{equation}
\ln \overline{\rho}_m = \ln \overline{\rho}_\gamma^{3/4}\, .
\label{2result}
\end{equation}
To make contact with the standard second-order result, we may 
expand  in Eq.~(\ref{2result}) the photon energy density 
perturbations as 
$\delta \overline{\rho}_\gamma/\rho_\gamma=
\delta_1 {\rho}_\gamma/ \rho_\gamma+\frac{1}{2} 
\delta_2 {\rho}_\gamma/\rho_\gamma $, and similarly for the matter 
component. We   immediately recover 
the adiabaticity condition  

\begin{equation}
\frac{\delta_2 {\rho}_\gamma}{\rho_\gamma}=
\frac{4}{3} \frac{\delta_2 {\rho}_m}{\rho_m}+
\frac{4}{9} \left(\frac{\delta_1 {\rho}_m}{\rho_m}\right)^2
\end{equation}
given in
Ref. \cite{review}.
 
Next we need to relate the photon energy density to the gravitational 
potentials at the non-linear level. The energy 
constraint inferred from the (0-0) component of Einstein equations in the 
matter-dominated era with the  
 ``barred'' metric~(\ref{newmetric}) is
\begin{equation}
\label{0-0}
\overline{H}^2=\frac{8\pi G_N}{3} \overline{\rho}_m\, .
\end{equation}     
Using Eqs.~(\ref{tbar}) and~(\ref{abar}) the 
Hubble parameter $\overline{H}$ reads
\begin{equation}
\overline{H}=\frac{1}{\overline{a}}\frac{d\overline{a}}{d \overline{t}}=
e^{-\Phi_\ell}
(H-\dot{\Psi}_\ell)\, ,
\end{equation} 
where $H=d \ln a/dt$ is the  Hubble parameter in the
``unbarred'' metric.   
Eq.~(\ref{0-0}) thus yields an expression for the energy density of the 
non-relativistic matter which is fully nonlinear, being expressed in terms of 
the gravitational potential $\Phi_\ell$
\begin{equation}
\label{3result}
\overline{\rho}_m=\rho_m e^{-2 \Phi_\ell}\, ,
\end{equation}
where we have dropped $\dot{\Psi}_\ell$ which is negligible on 
large scales. 
By perturbing the expression ~(\ref{3result}) we are able to 
recover in a straightforward way the solutions
of  the (0-0) component of Einstein equations for 
a matter-dominated 
Universe in the large-scale limit obtained at 
second-order in perturbation theory. Indeed,  recalling that 
$\Phi$ is perturbatively related 
to the quantity $\phi=\phi^{(1)}+\phi^{(2)}/2$ used in Ref.~\cite{review} by 
$\phi^{(1)}=\phi^{(1)}$ and $\phi^{(2)}=\phi^{(2)}-2 (\phi^{(1)})^2$,  one immediately  
obtains 

\begin{eqnarray}
\frac{\delta_1\rho_m}{\rho_m} &=& -2 \phi^{(1)}\, ,\nonumber\\  
\frac{1}{2}\frac{\delta_2 \rho_m}{\rho_m}&=&- \phi^{(2)}+4 (\phi^{(1)}) ^2\, .
\end{eqnarray}
The expression for the intrinsic temperature of the photons at 
the last scattering surface $\overline{T}_{\mathcal E} 
\propto \overline{\rho}^{1/4}_\gamma$ follows from  
Eqs.~({\ref{2result}) and (\ref{3result})  
\begin{equation}
\label{4result}
\overline{T}_{\mathcal E}=T_{\mathcal E}\, e^{-2 \Phi_\ell/3}\, .
\end{equation}
Plugging Eqs.~(\ref{1result}) and (\ref{4result}) into 
the expression (\ref{Texact}) we are finally able 
to provide the expression for the 
CMB temperature which is fully nonlinear and takes 
into account both the gravitational redshift of the photons due to the 
metric perturbations at last scattering and the intrinsic temperature 
anisotropies 
\begin{equation}
\label{final}
\overline{T}_{\mathcal O}=\left(
\frac{\omega_{\mathcal O}}{\omega_{\mathcal E}}\right) 
T_{\mathcal E}\, e^{\Phi_\ell/3}\, .
\end{equation}
From Eq.~(\ref{final}) we read the {\it non-perturbative} 
anisotropy corresponding to the Sachs-Wolfe effect
\begin{equation}
\label{final2}
\frac{\delta_{np} \overline{T}_{\mathcal O}}{T_{\mathcal O}}=
e^{\Phi_\ell/3}-1\, .
\end{equation}        
Eq.~(\ref{final2}) is  one of the main results of this paper and 
represents
{\it at any order in perturbation theory} the extension of the 
linear Sachs-Wolfe effect. At first order one gets
\begin{equation}
\frac{\delta_1 T_{\mathcal O}}{T_{\mathcal O}}=\frac{1}{3} \phi^{(1)}\, ,
\end{equation} 
and at  second order 
\begin{equation}
\label{SW2nd}
\frac{1}{2} \frac{\delta_2 T_{\mathcal O}}{T_{\mathcal O}}=
\frac{1}{6} \phi^{(2)}+\frac{1}{18} \left( \phi^{(1)} \right)^2\, .
\end{equation}
This result shows that the CMB anisotropies is nonlinear on large scales and that a source of NG is inevitably
sourced by gravity.

\section{Perturbing gravity}
\noindent 
In this Section  we  provide the
necessary tools to deal with perturbed gravity, giving 
 the expressions for the Einstein tensor 
perturbed up to second-order around  a flat 
Friedmann-Robertson-Walker background, and the relevant Einstein equations. 
In the following we will adopt the Poisson gauge which 
eliminates one scalar degree of freedom from the $g_{0i}$ component of the 
metric and one scalar and two vector degrees of freedom from $g_{ij}$. We 
rewrite the metric (\ref{metric}) as 
\begin{equation}
\label{metricconf}
ds^2=a^2(\eta)\left[
-e^{2\Phi} d\eta^2+2\omega_i dx^i d\eta+(e^{-2\Psi}\delta_{ij}+\chi_{ij}) dx^i dx^j
\right]\, ,
\end{equation}
where $a(\eta)$ is the scale factor as a 
function of the conformal time $\eta$. As we previously mentioned, 
for the vector and tensor perturbations,  we will neglect linear vector modes since they are not produced in standard 
mechanisms for the generation of cosmological perturbations (as inflation), 
and we also neglect tensor modes at linear order, since they give a negligible contribution to second-order 
perturbations. Therefore we take $\omega_i$ and $\chi_{ij}$ to be 
second-order vector and tensor perturbations of the metric. 

Let us now give our definitions for the connection coefficients and their expressions for the metric~(\ref{metric}). 
The number of spatial dimensions is $n=3$.
Greek indices ($\alpha, \beta, ..., \mu, \nu, ....$)
 run from 0 to 3, while latin indices ($a,b,...,i,j,k,....
m,n...$) run from 1 to 3. 
The total spacetime metric $g_{\mu \nu}$ has signature ($-,+,+,+$). 
The connection coefficients are defined as
\begin{equation}
\label{conness} \Gamma^\alpha_{\beta\gamma}\,=\,
\frac{1}{2}\,g^{\alpha\rho}\left( \frac{\partial
g_{\rho\gamma}}{\partial x^{\beta}} \,+\, \frac{\partial
g_{\beta\rho}}{\partial x^{\gamma}} \,-\, \frac{\partial
g_{\beta\gamma}}{\partial x^{\rho}}\right)\, .
\end{equation}
The Riemann tensor is defined as
\begin{equation}
R^{\alpha}_{~\beta\mu\nu}=
\Gamma^{\alpha}_{\beta\nu,\mu}-\Gamma^{\alpha}_{\beta\mu,\nu}+
\Gamma^{\alpha}_{\lambda\mu}\Gamma^{\lambda}_{\beta\nu}-
\Gamma^{\alpha}_{\lambda\nu}\Gamma^{\lambda}_{\beta\mu} \,.
\end{equation}

The Ricci tensor is a contraction of the Riemann tensor
\begin{equation}
R_{\mu\nu}=R^{\alpha}_{~\mu\alpha\nu} \,,
\end{equation}
and in terms of the connection coefficient it is given by
\begin{equation}
R_{\mu\nu}\,=\, \partial_\alpha\,\Gamma^\alpha_{\mu\nu} \,-\,
\partial_{\mu}\,\Gamma^\alpha_{\nu\alpha} \,+\,
\Gamma^\alpha_{\sigma\alpha}\,\Gamma^\sigma_{\mu\nu} \,-\,
\Gamma^\alpha_{\sigma\nu} \,\Gamma^\sigma_{\mu\alpha}\,.
\end{equation}
The Ricci scalar is given by contracting the Ricci tensor
\begin{equation}
R=R^{\mu}_{~\mu} \,.
\end{equation}
The Einstein tensor is defined as
\begin{equation}
G_{\mu\nu}=R_{\mu\nu}-\frac{1}{2}g_{\mu\nu}R \,.
\end{equation}

\subsection{The connection coefficients}
\noindent 
For the connection coefficients we find
\begin{eqnarray}
\Gamma^0_{00}&=& {\mathcal H}+\Phi'\, ,\nonumber\\
\Gamma^0_{0i}&=& \frac{\partial\Phi}{\partial x^i}+
{\mathcal H}\omega_i\, ,\nonumber\\
\Gamma^i_{00}&=& \omega^{i'}+{\mathcal H}\omega^i+e^{2\Psi+2\Phi}
\frac{\partial\Phi}{\partial x_i}
\, ,\nonumber\\
\Gamma^0_{ij}&=& -\frac{1}{2}\left(\frac{\partial \omega_j}{\partial x^i}+
\frac{\partial \omega_i}{\partial x^j}\right)+e^{-2\Psi-2\Phi}
\left({\mathcal H}-\Psi'\right)\delta_{ij}+\frac{1}{2}\chi_{ij}'+
{\mathcal H}\chi_{ij}
\, ,\nonumber\\
\Gamma^i_{0j}&=&\left({\mathcal H}-\Psi'\right)\delta_{ij}+
\frac{1}{2}\chi_{ij}'+\frac{1}{2}\left(\frac{\partial \omega_i}{\partial x^j}-
\frac{\partial \omega_j}{\partial x^i}\right)\, ,\nonumber\\
\Gamma^i_{jk}&=&-{\cal H}\omega^i\delta_{jk}-
\frac{\partial\Psi}{\partial x^k}\delta^i_{~j}-
\frac{\partial\Psi}{\partial x^j}\delta^i_{~k}+
\frac{\partial\Psi}{\partial x_i}\delta_{jk}+ \frac{1}{2} \left(\frac{\partial\chi^i_{~j}}{\partial x^k}+
\frac{\partial\chi^i_{~k}}{\partial x^j}+\frac{\partial\chi_{jk}}{\partial x_i}
\right)\, .\nonumber\\
&&
\end{eqnarray}
\subsection{Einstein equations}
\noindent 
The Einstein equations are written as $G_{\mu\nu}=\kappa^2 T_{\mu\nu}$, so that $\kappa^2=8\pi G_{\rm N}$, 
where $G_{\rm N}$ is the usual Newtonian gravitational constant. They read
\begin{eqnarray}
\label{00}
G^0_{~0}&=&-\frac{e^{-2\Phi}}{a^2}\left[3{\mathcal H}^2-6{\mathcal H}\Psi'
+3(\Psi')^2-e^{2\Phi+2\Psi}\left(\partial_i\Psi\partial^i\Psi-2\nabla^2
\Psi\right)\right]= \kappa^2 T^0_{~0}\, ,\\
\label{i0}
G^i_{~0}&=&2\frac{e^{2\Psi}}{a^2}\left[\partial^i\Psi'+\left({\mathcal H}-
\Psi'\right)\partial^i\Phi\right]-\frac{1}{2a^2}\nabla^2\omega^i
+\left(4{\mathcal H}^2-2\frac{a''}{a}\right)\frac{\omega^i}{a^2}=\kappa^2 T^i_{~0}\, ,\\
\label{ij}
G^i_{~j}&=&\frac{1}{a^2}\left[e^{-2\Phi}\left({\mathcal H}^2-2\frac{a''}{a}-
2\Psi'\Phi'-3(\Psi')^2+2{\mathcal H}\left(\Phi'+2\Psi'\right)
+2\Psi''\right)\right.\nonumber\\
&+&\left. e^{2\Psi}\left(\partial_k\Phi\partial^k\Phi+\nabla^2\Phi-\nabla^2\Psi
\right)\right]\delta^i_j\nonumber\\
&+&\frac{e^{2\Psi}}{a^2}
\left(-\partial^i\Phi\partial_j\Phi-
\partial^i\partial_j\Phi+\partial^i\partial_j\Psi-\partial^i\Phi\partial_j\Psi
+\partial^i\Psi\partial_j\Psi-\partial^i\Psi\partial_j\Phi\right)\nonumber\\
&-&\frac{{\mathcal H}}{a^2}\left(\partial^i\omega_j+\partial_j\omega^i\right)
-\frac{1}{2a^2}\left(\partial^{i}\omega_j'+\partial_j\omega^{i'}\right)
+\frac{1}{a^2}\left(
{\mathcal H}\chi^{i'}_j+\frac{1}{2}\chi_j^{i''}-\frac{1}{2}\nabla^2
\chi^i_j\right)=\kappa^2 T^i_{~j}\, .\nonumber\\
&&
\end{eqnarray}
Taking the traceless part of Eq.~(\ref{ij}), we find
\begin{equation}
\label{Q}
\Psi-\Phi={\cal Q}\,,
\end{equation} 
where ${\cal Q}$ is defined by 
$\nabla^2{\cal Q}=-P+3N$, 
with $P\equiv P{^i}_{~i}$, 
\begin{equation}
P^i_{~ j} = \partial^i\Phi\partial_j\Psi+\frac{1}{2} \left( 
\partial^i \Phi\partial_j\Phi- \partial^i \Psi\partial_j\Psi\right) 
+4\pi G_{\rm N}a^2e^{-2\Psi}T^i_{~j}
\end{equation}
and $\nabla^2 N=\partial_i\partial^j P^i_{~ j}$. 
The trace of Eq.~(\ref{ij}) gives
\begin{eqnarray}
\label{LH2trace}
& & e^{-2\Phi}\left({\mathcal H}^2-2\frac{a''}{a}-2\Phi'\Psi'-3(\Psi')^2+
2{\mathcal H}\left(3\Psi'-{\cal Q}'\right)+2\Psi''\right) \nonumber \\
&&
+\frac{e^{2\Psi}}{3}\left(2\partial_k\Phi\partial^k\Phi+
\partial_k\Psi\partial^k\Psi-
2\partial_k\Phi\partial^k\Psi+2(P-3N)\right) \nonumber \\
& & = \frac{8\pi G_{\rm N}}{3} a^2 T^k_{~k}\, .
\end{eqnarray}
From Eq.~(\ref{i0}), we may deduce an equation for $\omega^i$
\begin{eqnarray}
\label{LH2omegai}
&-&\frac{1}{2}\nabla^2\omega^i
+\left(4{\mathcal H}^2-2\frac{a''}{a}\right)\omega^i \\
&=&
-\left(\delta^i_j-\frac{\partial^i\partial_j}{\nabla^2}\right)
\left(2\left(\partial^j\Psi'+\left({\mathcal H}-
\Psi'\right)\partial^j\Phi\right)-8\pi G_{\rm N}a^2 e^{-2\Psi} 
T^j_{~ 0}\right)\, . \nonumber
\end{eqnarray}
Here $T^\mu_{~\nu}$ is the energy momentum tensor accounting for different components, photons, baryons, dark matter. 
We will give the expressions later for each component in terms
 of the distribution functions.

\section{Setting the initial conditions from the primordial NG}
\label{PNG}
\noindent
In this review we are 
 concerned with the second-order evolution of the cosmological perturbations. This requires
that we well define the initial conditions of the cosmological perturbations at second order. These boundary conditions
may or may not contain already some level of NG. It they do, we say that there exist some primordial NG. The latter
is usually defined in the epoch in which the comoving curvature perturbation remains constant on large superhorizon scales. 
In the standard single-field inflationary model, 
the first seeds of density fluctuations are generated on super-horizon scales 
from the fluctuations of a scalar field, the inflaton \cite{lrreview}. 

In order to follow the evolution on super-horizon scales of the  
density fluctuations coming from the various  mechanisms, we 
use the curvature perturbation of uniform density hypersurfaces $\zeta=\zeta^{(1)}+\zeta^{(2)}/2+\cdots$, where 
$\zeta^{(1)}=-\psi^{(1)}-{\mathcal H} {\delta \rho}^{(1)}/{\bar{\rho}'}$ and
the expression for $ \zeta^{(2)}$ is given by~\cite{MW}
\begin{equation}
\label{defz2}
\zeta^{(2)}=-\psi^{(2)}-{\cal H} \frac{\delta^{(2)} \rho}{{\rho}'}+\Delta \zeta^{(2)}\, ,
\end{equation}
with
\begin{eqnarray}
\label{deltaz2}
\Delta \zeta^{(2)} &=& 
2 {\cal H} \frac{\delta^{(1)} \rho'}{{\rho}'} \frac{\delta^{(1)} \rho}{\rho'}
\nonumber\\
&+&2
\frac{\delta^{(1)} \rho}{\rho'} (\psi^{(1)\prime}+2{\cal H} \psi^{(1)})
\nonumber\\
&-& 
\left( \frac{\delta^{(1)} \rho}{\rho'} \right)^2 \left({\cal H} \frac{\rho''}{\rho} -{\cal H}' 
-2{\cal H}^2\right)+2\psi^{(1)2} \, .
\end{eqnarray}
The crucial point is that the  gauge-invariant curvature perturbation
$\zeta$ remains  {\it constant} on super-horizon scales after it 
has been generated during a primordial epoch and possible isocurvature 
perturbations are no longer present. Therefore, we may set
the initial conditions at the time when $\zeta$  becomes
constant. In particular,  $\zeta^{(2)}$ 
provides  the necessary information about the
``primordial'' level of non-Gaussianity generated either during inflation, 
as in the 
standard scenario, or immediately after it, as in the curvaton scenario. 
Different scenarios are  characterized by different values of 
$\zeta^{(2)}$. For example, in    
the standard single-field inflationary model
 $\zeta^{(2)}=2\left( 
\zeta^{(1)}\right)^2+{\cal O}\left(\epsilon,\eta\right)$~\cite{noi}, 
where 
$\epsilon$ and $\eta$ are the standard slow-roll parameters~\cite{lrreview}. 
In general, we may  parametrize the primordial non-Gaussianity level 
in terms of the conserved curvature perturbation as in Ref. \cite{prl}  
\begin{equation}
\label{param}
\zeta^{(2)}=2 a_{\rm NL}\left(\zeta^{(1)} \right)^2\, ,
\end{equation}
where the parameter $a_{\rm NL}$ depends on the physics of a given scenario.
For example in the standard scenario $a_{\rm NL}\simeq 1$, while in the  
curvaton case $a_{\rm NL}=(3/4r)-r/2$, where 
$r \approx (\rho_\sigma/\rho)_{\rm D}$ is the relative   
curvaton contribution to the total energy density at curvaton 
decay~\cite{ngcurv,review}. 
Alternatives to the curvaton model are those models 
characterized by the curvature perturbation being 
generated by an inhomogeneity in 
the decay rate \cite{hk,varcoupling} or 
  the mass   \cite{varmass} or 
of the particles responsible for the reheating after inflation. 
Other opportunities for generating the curvature perturbation occur
 at the end of inflation \cite{endinflation} and  during
preheating \cite{preheating}.
All these models generate a level of NG which is local as the NG part of the primordial curvature
perturbation is a local function of the Gaussian part, being generated on superhorizon scales. 
In momentum space, the three point function, or bispectrum, arising from the local NG is dominated by the
so-called ``squeezed'' configuration, where one of the momenta is much smaller than the other two. 
Other models, such as DBI inflation
\cite{DBI} and ghost inflation \cite{ghost}, predict a different kind of primordial
NG, called ``equilateral'', because the three-point function for this kind of NG is peaked on equilateral configurations, 
in which the lenghts 
of the three wavevectors forming a triangle in Fourier
space are equal \cite{Shapes}. 

One of the best tools 
to detect or constrain the primordial large-scale non-Gaussianity is 
through the analysis 
of the CMB anisotropies, for example by studying the bispectrum
~\cite{review}. In that case the standard procedure is to
introduce  the primordial non-linearity 
parameter $f_{\rm NL}$ characterizing the primordial non-Gaussianity via the curvature perturbation~\cite{review}\footnote{The sign covention is such that it follows the WMAP covention for $f_{\rm NL}$, 
see Sec.~\ref{Sfive}.}

\begin{equation}
\label{fnl}
\zeta=\zeta^{(1)}+\frac{3}{5}f_{\rm NL}\left(\zeta^{(1)}\star \zeta^{(1)}\right)\, ,
\end{equation}
where the coefficient 3/5 arises from the first-order relation connecting
the comoving curvature perturbation and the gravitational potential,
$\zeta^{(1)}=-5\phi^{(1)}/3$, and the $\star$-product reminds us that one has to perform a convolution
product in momentum space and that $f_{\rm NL}$ is indeed momentum-dependent. 
 To give the feeling
of the resulting size of $f_{\rm NL}$ when $|a_{\rm NL}| \gg 1$, 
$f_{\rm NL} 
\simeq 5 a_{\rm NL}/3$~(see Refs.~\cite{review,prl}). Present limits
on NG are summarized by $-4<f^{\rm loc}_{\rm NL}<80$ \cite{leo} 
and   $-125<f^{\rm equil}_{\rm NL}<435$ \cite{equil} at $95\%$ CL,
where $f^{\rm loc}_{\rm NL}$ and $f^{\rm equil}_{\rm NL}$ stand for the non-linear parameter in the case in 
which the squeezed and the  equilateral configurations dominate, respectively.  

\section{CMB anisotropies at second-order on large scales}
\label{Sfive}
\noindent 
In this section,  we provide the exact 
expression for large-scale CMB temperature fluctuations at second order 
in perturbation theory. What this Section contains, therefore, should be considered as a more technical
elaboration of what the reader can find in Section 2.  
The final  expression we will find has various virtues. First, it is gauge-invariant. Second, 
from it one can unambiguously extract the exact definition of the 
nonlinearity parameter $f_{\rm NL}$ which is used by the experimental
collaborations to pin down the level of NG 
in the temperature fluctuations. Third, it contains 
a ``primordial'' term encoding all the information
about the NG generated in primordial epochs, namely
during or immediately after inflation, and depends upon the various 
fluctuation generation mechanisms. As such, the expression 
neatly disentangles the primordial contribution to the NG
from that arising after inflation. Finally, 
the expression applies to all scenarios for the generation of cosmological
perturbations. 

In order to obtain our gauge-independent formula for the temperature 
anisotropies,  
we again  perturb the  spatially flat Robertson-Walker background. We  expand 
the metric perturbations (\ref{metricconf}) in a first and a second-order part as     
\begin{eqnarray} \label{metric1}
g_{00}&=&-a^2 \left( 1+2 \phi^{(1)}+\phi^{(2)}+2 \left(\phi^{(1)}\right)^2\right)\, ,
g_{0i}=a^2 \left( \omega_i^{(1)}+\frac{1}{2} 
\omega_i^{(2)} \right)\, ,
 \nonumber \\
g_{ij}&=&a^2\left[
(1 -2 \psi^{(1)} - \psi^{(2)}+2\left(\psi^{(1)}\right)^2)\delta_{ij}+
\left( \chi^{(1)}_{ij}+\frac{1}{2}\chi^{(2)}_{ij} \right)\right] 
\, .\nonumber\\
&&
\end{eqnarray}
Again, the functions $\phi^{(r)}, \omega_i^{(r)}, 
\psi^{(r)}$ and $\chi^{(r)}_{ij}$, where $(r)=(1,2)$, stand for the 
$r$th-order perturbations of the metric. 
We can split 
$\omega_i^{(r)}=\partial_i\omega^{(r)}+\omega_i^{(r)}$, 
where $\omega^{(r)}$ is the scalar part and $\omega^{(r)}_i$ 
is a transverse vector, {\it i.e.} $\partial^i\omega^{(r)}_i=0$.
The metric perturbations will transform according to an infinitesimal 
change of coordinates. From now on we limit ourselves to a second-order time 
shift 

\begin{equation}
\eta\rightarrow 
\eta-\alpha_{(1)}+\frac{1}{2} ({\alpha'_{(1)}}\alpha_{(1)}
-\alpha_{(2)})\, , 
\end{equation}
where a prime denotes differentiation w.r.t. conformal 
time.
In general a gauge corresponds to a choice of coordinates 
defining a slicing of spacetime into hypersurfaces (at fixed time $\eta$) 
and a threading into lines (corresponding to fixed spatial coordinates 
${\bf x}$), but in this Section only the former is relevant so that
gauge-invariant can be taken to mean independent of the slicing.
For example, under the time shift, 
the first-order spatial curvature perturbation 
$\psi^{(1)}$ transforms as $
\psi^{(1)}\rightarrow\psi^{(1)}-{\mathcal H} \,\alpha_{(1)}$ 
(here ${\mathcal H}=a'/a$), 
while $\phi^{(1)} \rightarrow \phi^{(1)}+\alpha'_{(1)}
+{\mathcal H} \alpha^{(1)}$, 
$\omega_i^{(1)} \rightarrow
\omega_i^{(1)}-\partial_i \alpha^{(1)}$, and the traceless 
part of the spatial metric $\chi^{(1)}_{ij}$ turns out 
to be gauge-invariant.
At second order in the perturbations 
we just give some useful examples like the 
transformation of the energy density and the curvature perturbation~\cite{mol}. 

\begin{equation}
\delta^{(2)} \rho\rightarrow\delta^{(2)} \rho +\rho^\prime\alpha_{(2)} +
\alpha_{(1)}\left(\rho^{\prime\prime}\alpha_{(1)}
+\rho^\prime\alpha_{(1)}^\prime+2\delta^{(1)} \rho^\prime\right)
\end{equation} 
and 

\begin{eqnarray}
\psi^{(2)}&\rightarrow&\psi^{(2)}
+2\alpha_{(1)}\psi^{(1)\prime}
-{\mathcal H}\alpha^2_{(1)}
-{\mathcal H} \alpha_{(1)} \alpha_{(1)}^{\prime}\nonumber\\
&-& \frac{1}{3}\left(2\omega^{i}_{(1)}-
\alpha^{,i}_{(1)}\right)\alpha^{(1)}_{,i} -{\mathcal H} \alpha_{(2)}\, .
\end{eqnarray}
We now construct in a gauge-invariant way temperature anisotropies 
at second order. Temperature anisotropies beyond the linear regime   
have been calculated in Refs.~\cite{twoT,mol},  
following the photons path from last-scattering 
to the observer in terms of perturbed geodesics. 
The linear temperature anisotropies read~\cite{twoT,mol}
\begin{equation} 
\label{T1}
\frac{\Delta T^{(1)}}{T}=\phi^{(1)}_{\mathcal E} -v^{(1)i}_{\mathcal E}e_i
+ \tau^{(1)}_{\mathcal E}
-\int_{\lambda_{\mathcal O}}^{\lambda_{\mathcal E}} d\lambda A^{(1) \prime} \;,
\end{equation} 
where $A^{(1)}\equiv \psi^{(1)}+\phi^{(1)}+\omega^{(1)}_i e^i-
\frac{1}{2}\chi^{(1)}_{ij}e^i e^j$, the subscript 
${\mathcal E}$ indicates that quantities are evaluated at 
last-scattering, $e^i$ is a spatial unit vector 
specifying the direction of observation and the integral 
is evaluated along the line-of-sight parametrized by the affine parameter 
$\lambda$. Eq.~(\ref{T1}) includes the intrinsic fractional 
temperature fluctuation at emission  
$\tau_{\mathcal E}$, the Doppler effect due to emitter's velocity 
$v^{(1)i}_{\mathcal E}$ and the
gravitational redshift of photons, including the Integrated 
Sachs-Wolfe (ISW) effect. We omitted monopoles 
due to the observer ${\mathcal O}$ ({\it e.g.} the gravitational 
potential $\psi^{(1)}_{\mathcal O}$ evaluated at the event of observation), 
which, being independent of the angular coordinate, can be always recast  
into the definition of temperature anisotropies. 
Notice however that the physical meaning of 
each contribution in Eq.~(\ref{T1}) is not gauge-invariant, as 
the different terms are gauge-dependent. However, 
it is easy to show that the whole expression~(\ref{T1}) is 
gauge-invariant. 
Since the temperature $T$ is a scalar,     
the intrinsic temperature fluctuation transforms as 
$\tau_{\mathcal E}^{(1)}\rightarrow\tau_{\mathcal E}^{(1)}
+(T'/T)\alpha_{(1)}=\tau_{\mathcal E}^{(1)}
-{\mathcal H}\alpha_{(1)}$, 
having used the fact that the temperature scales as 
$T \propto a^{-1}$. Notice, instead, 
that the velocity $v^{(1)i}_{\mathcal E}$ 
does not change. Therefore, using the transformations of metric 
perturbations we find  
\begin{eqnarray}
\label{proof2}
\frac{\Delta T^{(1)}}{T} &\rightarrow &\frac{\Delta T^{(1)}}{T}+\alpha'_{(1)}
-\int_{\eta_{\mathcal O}}^{\eta_{\mathcal E}} 
d\eta \frac{d \alpha'_{(1)}}{d \eta}
=\frac{\Delta T^{(1)}}{T} + {\mathcal O}\, , \nonumber \\
&& 
\end{eqnarray}
where we have used the fact that the integral is evaluated along 
the line-of-sight which can be parametrized by the background 
geodesics $x^{(0) \mu}=
\left( \lambda, (\lambda_{\mathcal O}-\lambda_{\mathcal E}) e^i \right)$ 
(with $d \lambda/d \eta=1$), and the decomposition for the total 
derivative along the path 
for a generic function $f(\lambda,x^i(\lambda))$, 
$f'=\frac{\partial f}{\partial \lambda}=
\frac{d f}{d \lambda} + \partial_i f e^i$. 
Eq.~(\ref{proof2}) shows that the expression~(\ref{T1}) for 
first-order temperature 
anisotropies is indeed gauge-invariant 
(up to monopole terms related to the observer ${\mathcal O}$). 
Temperature anisotropies can be easily written in 
terms of particular combinations of perturbations which are manifestly 
gauge-invariant. For the gravitational potentials we consider 
the gauge-invariant 
definitions $\psi^{(1)}_{\rm GI}=\psi^{(1)}-{\mathcal H} \omega^{(1)}$ and 
$\phi^{(1)}_{\rm GI}=\phi^{(1)}+{\mathcal H} \omega^{(1)}+\omega^{(1)'}$.
For the $(0-i)$ component of the metric and the traceless part of 
the spatial metric we define $\omega_i^{(1) \rm GI}=\omega_i^{(1)}$ and $
\chi^{(1) \rm GI}_{ij}=\chi^{(1)}_{ij}$.
For the matter variables we use a 
gauge-invariant intrinsic temperature fluctuation
$\tau^{(1)}_{\rm GI}=\tau^{(1)}- {\mathcal H}\omega^{(1)}$, 
while the velocity itself is gauge-invariant 
$v^{(1)i}_{\rm GI}=v^{(1)i}$ under time shifts.
Following the same steps 
leading to Eq.~(\ref{proof2}) one gets the linear 
temperature anisotropies in Eq.~(\ref{T1}) in terms of 
these gauge-invariant quantities
\begin{eqnarray}
\label{T1GI}
& &\frac{\Delta T^{(1)}_{\rm GI}}{T}=\phi^{(1)}_{\rm GI} -v^{(1)i}_{\rm GI}e_i
+ \tau^{(1)}_{\rm GI} -\int_{\lambda_{\mathcal O}}^{\lambda_{\mathcal E}} 
d \lambda\,  A^{(1) \prime}_{\rm GI}\, ,
\end{eqnarray}  
where $A^{(1)}_{\rm GI}=
\phi^{(1)}_{\rm GI}+\psi^{(1)}_{\rm GI}+\omega_i^{(1) \rm GI} 
e_i-\frac{1}{2}\chi^{(1) \rm GI}_{ij} e^i 
e^j$ and we omitted the subscript ${\mathcal E}$. 
For the primordial fluctuations we are interested in the large-scale 
modes set by the curvature perturbation $\zeta^{(1)}$. Defining a 
gauge-invariant density perturbation $\delta^{(1)}\rho_{\rm GI}=
\delta^{(1)} \rho+\rho' \omega^{(1)}$, we write the curvature 
perturbation as $\zeta^{(1)}_{\rm GI}=-\psi^{(1)}_{\rm GI}-
{\mathcal H} (\delta^{(1)} \rho_{\rm GI}/\rho')$. 
Since for adiabatic perturbations in the radiation ($\gamma$) and 
matter ($m$) eras
$(1/4)(\delta^{(1)} \rho_\gamma/\rho_{\gamma}) = (1/3)
(\delta^{(1)} \rho_m/\rho_{m})$, one can write the 
intrinsic temperature fluctuation as 
$\tau^{(1)}=(1/4)(\delta^{(1)} \rho_\gamma/\rho_{\gamma})=
-{\mathcal H} (\delta^{(1)} \rho/\rho')$ and a gauge-invariant
definition is $\tau^{(1)}_{\rm GI}=-{\mathcal H} (\delta^{(1)} 
\rho_{\rm GI}/\rho')$. 
In the large-scale limit, from Einstein equations, in the 
matter era $\phi^{(1)}_{\rm GI}=\psi^{(1)}_{\rm GI}=-\frac{3}{5}
\zeta^{(1)}_{\rm GI}$. 
Thus we obtain the large-scale limit of temperature anisotropies~(\ref{T1GI}) 
$\frac{\Delta T^{(1)}_{\rm GI}}{T}=
2\psi^{(1)}_{\rm GI} +\zeta^{(1)}_{\rm GI}= 
\psi^{(1)}_{\rm GI}/3$, i.e. the usual Sachs-Wolfe effect. 

At second order, the procedure is similar to the one described 
so long, though more lengthy and cumbersome. 
We only provide the reader with the  
main steps to get the final expression. The 
second-order temperature fluctuations 
in terms of metric perturbations read~\cite{twoT,mol}
\begin{eqnarray}
\label{T2}
& &\frac{\Delta T^{(2)}}{T}=\frac{1}{2}\phi^{(2)}_{\mathcal{E}}
-\frac{1}{2}v^{(2)i}_{\mathcal E}e_i
+ \frac{1}{2}\tau^{(2)}_{\mathcal E}
-I_2(\lambda_{\mathcal{E}}) \nonumber\\
&+&\left(I_1(\lambda_{\mathcal{E}})+v^{(1)i}_{\mathcal{E}} e_i\right)
\left(-\phi^{(1)}_{\mathcal{E}}-\tau^{(1)}_{\mathcal E}
+v^{(1)i}_{\mathcal{E}} e_i
+I_1(\lambda_{\mathcal{E}})\right)\nonumber\\
&+&x^{(1)0}_{\mathcal{E}} A ^{(1)'}_{\mathcal{E}}+(x^{(1)j}_{\mathcal{E}}
+x^{(1)0}_{\mathcal{E}} e^j)\left(\phi^{(1)}_{,j}-v^{(1)}_{i,j} e^i+
\tau^{(1)}_{,j}\right)_{\mathcal{E}} \nonumber\\
&-&\frac{1}{2}v^{(1)}_{{\mathcal{E}}i} v^{(1)i}_{\mathcal{E}}
+\phi^{(1)}_{\mathcal{E}}\tau^{(1)}_{\mathcal E}+
\frac{\partial \tau^{(1)}}{\partial d^i}d^{(1)i} 
-v^{(1)i}_{\mathcal{E}} e_i \phi^{(1)}_{\mathcal{E}}
\nonumber\\
&+&v^{(1)}_{{\mathcal{E}}i}\left(-\omega^{(1)i}_{\mathcal{E}}
- I_1^i(\lambda_{\mathcal {E}})\right)\, . 
\end{eqnarray}
Here $I_2$ is the second-order ISW
\begin{eqnarray} 
I_2(\lambda_{\mathcal{E}})&=&\int_{\lambda_{\mathcal{O}}}
^{\lambda_{\mathcal{E}}} d\lambda [\frac{1}{2}A^{(2)'}
-(\omega^{(1)'}_i-\chi^{(1)'}_{ij} e^j)
(k^{(1)i}+e^i k^{(1)0})+2 k^{(1)0} A^{(1)'}\nonumber\\
&+&2 \psi^{(1)'} A^{(1)}
+x^{(1)0} A^{(1)''}+x^{(1)i} A^{(1)'}_{,i}]\, ,
\end{eqnarray}
where $A^{(2)}\equiv \psi^{(2)}+\phi^{(2)}+\omega^{(2)}_i e^i-
\frac{1}{2}\chi^{(2)}_{ij}e^i e^j$,
while $k^{(1)0}(\lambda)=-2 \phi^{(1)}
-\omega^{(1)i} e_i +I_1(\lambda)$ and 
$k^{(1)i}(\lambda)=-2 \phi^{(1)} e^i-
\omega^{(1)i}+\chi^{(1)ij} e_j
- I_1^i(\lambda)$ are the photon wave vectors, with 
$I_1(\lambda)$ given by the integral in Eq.~(\ref{T1}) and $I_1^i(\lambda)$ 
is obtained from the same integral replacing the time derivative with a
spatial gradient. 
Finally in Eq.~(\ref{T2})

\begin{equation} 
x^{(1)0}(\lambda)=
\int_{\lambda_{\mathcal{O}}}^{\lambda}d\lambda'
\left[-2 \phi^{(1)}-\omega^{(1)}_i e^i+(\lambda-\lambda')
A^{(1)'}\right]\, 
\end{equation} 
and 

\begin{equation}
x^{(1)i}(\lambda)=
-\int_{\lambda_{\mathcal{O}}}^{\lambda}d\lambda'
\left[2 \psi^{(1)} e^i+\omega^{(1)i}-\chi^{(1)ij} e_j
+(\lambda-\lambda')A^{(1),i}\right]
\end{equation}
are the geodesics at first 
order, and $d^{(1)i}=e^i-\frac{e^i-k^{(1)i}}{|e^i-k^{(1)i}|}$ is the 
direction of the photon emission. As usual we have omitted the monopole 
terms due to the observer. Using the transformation rules of 
Ref.~\cite{mol}, it is possible to check that the expression~(\ref{T2}) 
is gauge-invariant. We can express the 
second-order anisotropies in terms of explicitly gauge-invariant 
quantities, whose definition proceeds as for the linear case, 
by choosing the shifts $\alpha^{(r)}$ such that 
$\omega^{(r)}=0$. For example, we consider the gauge-invariant gravitational 
potential~\cite{enh} 
\begin{eqnarray}
\label{PhiGI}
&\phi^{(2)}_{\rm GI}&=\phi^{(2)}+\omega^{(1)}\left[2\left(
\psi^{(1)'}+2\frac{a'}{a}\psi^{(1)}\right)+\omega^{(1) \prime \prime} 
+ 5 \frac{a'}{a}\omega^{(1) '}\right.\nonumber \\
&+&\left. \left( {\mathcal H}'+2 {\mathcal H}^2 \right)
\omega^{(1)}\right]
+2\omega^{(1)'}\left(2\psi^{(1)}+\omega^{(1)'}\right)+
\frac{1}{a} \left( a\alpha^{(2)} \right)' \, , \nonumber \\
\end{eqnarray}
where $\alpha^{(2)}=\omega^{(2)}+\omega^{(1)}\omega^{(1)'}
+\nabla^{-2}\partial^i[-4\psi^{(1)}\partial_i\omega^{(1)}-2
\omega^{(1)'}\partial_i\omega^{(1)}]$. 
Expressing the second-order temperature anisotropies~(\ref{T2}) 
in terms of our gauge-invariant quantities and taking the large-scale limit 
we find

\begin{equation}
\Delta T^{(2)}_{\rm GI}/T=(1/2)\phi^{(2)}_{\rm GI}
-(1/2)\left( \phi_{\rm GI}^{(1)} \right)^2
+ (1/2)\tau^{(2)}_{\rm GI}+\phi^{(1)}_{\rm GI} \tau^{(1)}_{\rm GI}\, ,
\end{equation}
(having dropped the subscript ${\mathcal E}$), and the 
gauge-invariant intrinsic temperature fluctuation at emission is 

\begin{equation}
\tau^{(2)}_{\rm GI}=(1/4) 
(\delta^{(2)} \rho^{\rm GI}_\gamma/\rho_\gamma)
-3( \tau^{(1)}_{\rm GI})^2\, .
\end{equation} 
We have dropped those terms which represent integrated 
contributions and other second-order 
small-scale effects that can be distinguished from the large-scale part 
through their peculiar scale dependence.
At this point we make use of Einstein's equations. We take the  
expression for $\zeta^{(2)}$ in Eqs.~(\ref{defz2}) and (\ref{deltaz2}), and we use 
the $(0-0)$ component and the traceless part of the $(i-j)$ Einstein's 
equations (\ref{00}) and (\ref{ij}) after having appropriately expanded the
exponentials. Thus, on large scales we find that the temperature 
anisotropies are given by
\begin{equation}
\label{main}
\frac{\Delta T^{(2)}_{\rm GI}}{T}=
\frac{1}{18} \left( \phi^{(1)}_{\rm GI} \right)^2 
-\frac{{\mathcal K}}{10}-\frac{1}{10} \left[ \zeta^{(2)}_{\rm GI}-
2 \left( \zeta^{(1)}_{\rm GI} \right)^2 \right]\, , 
\end{equation} 
where we have defined a kernel  

\begin{equation}
{\mathcal K}=
10 \nabla^{-4} \partial_i \partial^j 
 (\partial^i \psi^{(1)} \partial_j
\psi^{(1)}) -\nabla^{-2} 
( \frac{10}{3} \partial^i \psi^{(1)} \partial_i \psi^{(1)} )\, .
\end{equation} 
Eq.~(\ref{main}) is the main result of this Section. It clearly shows that 
there are two contributions to the final nonlinearity in the 
large-scale temperature anisotropies. 
The  contribution, $[\zeta^{(2)}_{\rm GI}-
2 ( \zeta^{(1)}_{\rm GI} )^2]$, 
comes from the ``primordial'' conditions set during or after inflation. 
They are encoded in 
the curvature perturbation $\zeta$ which remains constant 
once it has been generated. 
The remaining part of Eq.~(\ref{main}) describes the post-inflation
processing of the primordial non-Gaussian signal due to the nonlinear 
gravitational dynamics, including also second-order corrections at last 
scattering to the Sachs-Wolfe effect~\cite{twoT,mol}. 
Thus, the expression in 
Eq.~(\ref{main}) allows to neatly disentangle the 
primordial contribution to NG from that coming from that arising 
after inflation and from it 
 we can extract the nonlinearity parameter 
$f_{\rm NL}$, see the expression (\ref{fnl}) which is usually adopted to phenomenologically 
parametrize the NG level of cosmological perturbations and 
has become the standard quantity to be observationally
The definition of $f_{\rm NL}$ adopted in the analyses performed in 
Refs.~\cite{ks} goes through the conventional Sachs-Wolfe formula 
$\Delta T/T= - \Phi/3$ 
where $\Phi$ is Bardeen's potential, which is conventionally expanded as 
(up to a constant offset, which only affects the temperature monopole)
$\Phi = \Phi_{\rm L} + f_{\rm NL} \left(\Phi_{\rm L}\right)\star \left(\Phi_{\rm L}\right)$.
Therefore, using  $\zeta^{(1)}=-\frac{5}{3} 
\psi^{(1)}_{\rm GI}$ during matter domination, 
from Eq.~(\ref{main}) we read the nonlinearity parameter in 
momentum space  
\begin{equation}
\label{f_NL}
f_{\rm NL}({\bf k}_1,{\bf k}_2)
=-\left[ \frac{5}{3} \left(1-a_{\rm NL} \right) 
+\frac{1}{6}-\frac{3}{10} {\mathcal K} 
\right]\, ,
\end{equation}
where 
\begin{equation}
{\mathcal K}=10\, ({\bf k}_1 \cdot {\bf k}_3) 
({\bf k}_2 \cdot {\bf k}_3)/k^4 -\frac{10}{3}
{\bf k}_1 \cdot {\bf k}_2/k^2 \, , 
\end{equation}
with ${\bf k}_3+{\bf k}_1+{\bf k_2}=0$
and $k=\left|
{\bf k}_3\right|$. To obtain Eq.~(\ref{f_NL}) we have made use of the expression (\ref{param}) to set the initial conditions.
In particular within the standard scenario 
where cosmological perturbations are due to the inflaton the 
primordial contribution to NG is given by 
$a_{\rm NL}=1-\frac{1}{4} (n_{\zeta}-1)$~\cite{noi}, where the 
spectral index is expressed in terms of the usual slow-roll parameters as
$n_{\zeta}-1=-6 \epsilon +2 \eta$~\cite{lrreview}. The 
nonlinearity parameter from inflation now reads
\begin{equation}
f^{\rm inf}_{\rm NL}=-\frac{5}{12} (n_{\zeta}-1)  
-\frac{1}{6}+\frac{3}{10} {\mathcal K}\, . 
\end{equation} 
Therefore the main NG contribution comes from the 
post-inflation evolution of the second-order perturbations 
which give rise to order-one coefficients, while the primordial 
contribution is proportional to $|n_{\zeta}-1|\ll 1$. 
This is true even in   
the ``squeezed'' limit first discussed by Maldacena in ~\cite{noi}, 
where one of the wavenumbers is much smaller than the other two, 
\emph{e.g.} $k_1 \ll k_{2,3}$ and ${\cal K}\rightarrow 0$.

\section{CMB anisotropies at second-order at all scales}
\label{allscales}
As we already mentioned in the Introduction,  
despite the importance of NG in CMB anisotropies, little effort
has ben made so far to provide accurate
theoretical predictions of it. On the contrary, the vast majority of the
literature has been devoted to the computation of the bispectrum
of either the comovig curvature perturbation or the gravitational 
potential on large scales within given inflationary models. 
These, however, are not the  physical
quantities which are observed. One should instead provide a full prediction
for the second-order radiation transfer function. 
A preliminary step towards this goal has been taken in Refs.~
\cite{fulllarge,twoT,beyond} (see also \cite{cre})
where the full second-order radiation transfer function 
for the CMB  anisotropies on large angular scales in a flat universe 
filled with matter and cosmological constant was computed, including 
the second-order generalization of the Sachs-Wolfe effect,
both the early and late Integrated Sachs-Wolfe (ISW) effects and the 
contribution of the second-order tensor modes.\footnote{For some recent papers focusing on the generation and evolution of 
tensor perturbations at second-order see, e.g., Refs.~\cite{WandsGW,BS,Mnoi}.} We have partly reported about these
works in the previous Sections. 

In this Section we wish to offer a summary of some of the second-order effects in the
CMB anisotropies on small scales. 
There are many sources of NG in CMB anisotropies, beyond the primordial
one.  The most relevant  sources are the so-called  secondary anisotropies,
which arise  after the last scattering epoch. These anisotropies can be 
divided into two categories: scattering secondaries, when the CMB photons 
scatter with electrons along the line of sight, 
and gravitational secondaries when effects are mediated by 
gravity \cite{Hu:2001bc}.  Among the  scattering secondaries 
we may list the thermal Sunyaev-Zeldovich effect, where 
hot electrons in clusters transfer energy to the CMB photons, 
the kinetic Sunyaev-Zeldovich effect produced by the bulk motion of the 
electrons in clusters, the Ostriker-Vishniac effect, produced by bulk 
motions modulated by linear density perturbations, and effects due to  
reionization processes. The scattering secondaries 
are most significant on small angular scales
as density inhomogeneities, bulk and thermal motions  grow and become 
sizeable on small length-scales when structure formation proceeds.

Gravitational secondaries arise from the change 
in energy of photons when the gravitational potential is time-dependent, the 
ISW effect, and  gravitational lensing.  At late times, when the 
Universe becomes dominated by the dark energy, 
the gravitational potential on linear scales starts to decay, causing 
the ISW effect mainly on large angular scales. Other secondaries 
that result from a 
time dependent potential are the Rees-Sciama effect, produced during
the matter-dominated epoch by the time 
evolution of the potential on non-linear scales. 

The fact that the potential never grows appreciably means that most 
second order effects created by gravitational secondaries are generically small
compared to those created by 
scattering ones. However, when a photon propagates from the 
last scattering to us, its path may be deflected because of 
the gravitational lensing. This effect 
does not create anisotropies, but only modifies existing ones. Since
photons with large wavenumbers $k$  are lensed over many regions ($\sim k/H$, 
where $H$ is the Hubble rate) along the
line of sight, the corresponding second-order effect may be sizeable.
The  
three-point function arising from the correlation of the gravitational lensing 
and ISW effects generated by  
the matter distribution along the line of sight 
\cite{Seljak:1998nu,Goldberg:xm} and the Sunyaev-Zeldovich effect \cite{sk}
are large and detectable by Planck~\cite{ks}.

Another relevant source of NG comes from the physics operating at
the recombination. A naive estimate would tell that these non-linearities
are tiny being suppressed by an extra power of the gravitational 
potential. However, the dynamics  at recombination is quite involved because
all the non-linearities in the evolution of the baryon-photon fluid at 
recombination and the ones coming from general relativity should be 
accounted for. This complicated dynamics might lead
to  unexpected suppressions or enhancements of the NG at recombination. 
A step towards the evaluation of the three-point correlation function
has been taken in Ref. \cite{rec} where some effects were taken into account 
in the in so-called  squeezed triangle limit, corresponding to the case 
when one wavenumber is much smaller than the other two and was outside 
the horizon at recombination. 

This Section, which is based on Refs.~\cite{paperI,paperII}, 
present the computation of the full system of Boltzmann 
equations, describing the evolution of the photon, 
baryon and Cold Dark Matter (CDM) 
fluids, at second order and neglecting polarization, 
These equations allow to follow the time evolution of the CMB anisotropies 
at second order on all angular scales
from the early epochs, when the cosmological perturbations were generated,
to the present  time, through the recombination era. These calculations 
set the stage for the computation of the full second-order
radiation transfer function at all scales and for a 
a generic set of initial conditions specifying the level of primordial 
non-Gaussianity. 
Of course on small angular scales, fully non-linear
calculations of specific effects like Sunyaev-Zel'dovich,
gravitational lensing, etc.  would provide a more
accurate estimate of the resulting CMB anisotropy, however,
as long as the leading contribution to 
second-order statistics like the bispectrum  is
concerned, second-order perturbation theory suffices.

\subsection{The collisionless Boltzmann equation for photons}
\noindent
We are interested in the anisotropies in the cosmic
distribution of photons and inhomogeneities in the matter. Photons
are affected by gravity and by Compton scattering
with free electrons. The latter are tightly coupled to protons. Both
are, of course, affected by gravity. The metric which
determines the gravitational forces is influenced by all these components
plus CDM (and neutrinos). Our plan is to write down Boltzmann equations for 
the phase-space distributions of each species in the Universe.

The phase-space distribution of particles $g(x^i,P^\mu,\eta)$ 
is a function of spatial coordinates $x^i$, conformal time $\eta$, and 
momentum of the particle 
$P^\mu=dx^\mu/d\lambda$ where $\lambda$ parametrizes the particle path. 
Through the constraint $P^2 \equiv g_{\mu\nu} P^\mu P^\nu 
=- m^2$, where $m$ is 
the mass of the particle one can eliminate $P^0$ and $g(x^i,P^j,\eta)$ 
gives the number of particles in the differential phase-space volume 
$dx^1 dx^2 dx^3 dP^1 dP^2 dP^3$. 
In the following we will denote the 
distribution function for photons with $f$. 

The photons' distribution evolves according to the Boltzmann equation 
\begin{equation}
\label{Boltzgeneric}
\frac{df}{d\eta}= {\overline C}[f]\, ,
\end{equation}  
where the collision term is due to the scattering of photons off free 
electrons. In the following we will derive the left-hand side of 
Eq.~(\ref{Boltzgeneric}) while in the next section we will compute 
the collision term.

For photons we can impose $P^2 \equiv g_{\mu\nu} P^\mu P^\nu =0$ 
and using the metric~(\ref{metric}) in the conformal time $\eta$ we find
\begin{equation}
\label{P=0}
P^2=a^2\left[ - e^{2\Phi} (P^0)^2+\frac{p^2}{a^2} + 
2 \omega_i P^0 P^i\right]=0\, ,
\end{equation} 
where we define
\begin{equation}
\label{defp}
p^2= g_{ij}  P^iP^j\, .
\end{equation}
From the constraint~(\ref{P=0})
\begin{equation}
\label{P0}
P^0=e^{-\Phi}\left( \frac{p^2}{a^2}+2 \omega_i P^0P^i\right )^{1/2}\, .
\end{equation}
Notice that we immediately recover the usual zero and first-order relations
$P^0=p/a$ and $P^0=p(1-\Phi^{(1)})/a$. 

The components $P^i$ are proportional to $p n^i$, where $n^i$ is a unit 
vector with $n^in_i = \delta_{ij} n^in^j=1$.  
We can write $P^i=C n^i$, where $C$ is determined by
\begin{equation}
g_{ij} P^iP^j=C^2 \, a^2 (e^{-2\Psi}+\chi_{ij}n^in^j)=p^2\, ,
\end{equation}
so that 
\begin{equation}
\label{Pi}
P^i=\frac{p}{a} n^i\left(e^{-2\Psi}+\chi_{km}n^kn^m \right)^{-1/2} = 
\frac{p}{a} n^i e^{\Psi} \left( 1-\frac{1}{2} \chi_{km}n^kn^m \right)\, , 
\end{equation}
where the last equality holds up to second order in the perturbations. Again 
we recover the zero and first-order relations $P^i=p n^i/a$ and 
$P^i=p n^i (1+\Psi^{(1)})/a$ respectively. Thus up to second order we can write
\begin{equation}
\label{P0bis}
P^0=e^{-\Phi} \frac{p}{a} \left( 1+\omega_i\, n^i \right)\, .
\end{equation} 
Eq.~(\ref{Pi}) and~(\ref{P0bis}) allow us to replace $P^0$ and $P^i$ in terms 
of the variables $p$ and $n^i$. Therefore, as 
it is standard in the literature, from now on we will consider the phase-space 
distribution $f$ as a function of the momentum ${\bf p}=p n^i$ with magnitude 
$p$ and angular direction $n^i$, $f\equiv f(x^i, p, n^i,\eta)$. 

Thus, in terms of these variables, the total time derivative of the 
distribution function reads
\begin{equation}
\label{Df}
\frac{d f}{d \eta} = \frac{\partial f}{\partial \eta}+
\frac{\partial f}{\partial x^i} \frac{d x^i}{d \eta}+
\frac{\partial f}{\partial p} \frac{d p}{d \eta}+
\frac{\partial f}{\partial n^i} \frac{d n^i} {d \eta} \, . 
\end{equation}
In the following we will compute $d x^i/d \eta$, $d p/d \eta$ and 
$d n^i/d \eta$.
\\

\noindent
a) $d x^i/d \eta$:

From 
\begin{equation}
P^i=\frac{d x^i}{d \lambda}=\frac{d x^i}{d \eta} 
\frac{d\eta}{d\lambda}=\frac{d x^i}{d \eta} P^0
\end{equation}
and from Eq.~(\ref{Pi}) and~(\ref{P0bis}) 
\begin{equation}
\label{dxi}
\frac{d x^i}{d \eta}= n^i e^{\Phi+\Psi}
\left( 1- \omega_{j}\, n^j - \frac{1}{2} 
\chi_{km} n^k n^m \right)\, .
\end{equation}   

\noindent
b) $d p/d \eta$:

For $dp/d \eta$ we make use of the time component of the geodesic equation 
$d P^0/d \lambda= - \Gamma^0_{\alpha \beta} 
P^{\alpha} P^{\beta}$, where   
$d/d\lambda = (d\eta/d\lambda) 
\, d/d\eta=P^0 \, d/d\eta$, and  
\begin{equation}
\label{0geod}
\frac{d P^0}{ d \eta}= - \Gamma^0_{\alpha \beta} 
\frac{P^{\alpha} P^{\beta}}{P^0}\, ,
\end{equation}
Using the metric~(\ref{metric}) we find 
\begin{eqnarray}
\label{GPP}
2 \Gamma^0_{\alpha \beta} P^{\alpha} P^{\beta}&=& g^{0\nu} 
\left[ 2 \frac{\partial g_{\nu \alpha}}{\partial x^\beta} 
-\frac{\partial_{\alpha \beta}}{\partial x^\nu}
\right]P^\alpha P^\beta \nonumber \\
&=&2 ({\mathcal H}+\Phi')\left( P^0 \right)^2+4 \Phi_{,i} P^0P^i 
+4 {\mathcal H} \omega_i P^0 P^i \nonumber \\ 
&+&
2 e^{-2\Phi}\left[ ({\cal H}-\Psi') e^{-2 \Psi} \delta_{ij}  
-\omega_{i,j} + \frac{1}{2} \chi'_{ij}+{\cal H}\chi_{ij} \right] P^iP^j \, .
\nonumber \\
\end{eqnarray}
On the other hand the expression~(\ref{P0bis}) 
of $P^0$ in terms of $p$ and $n^i$ gives
\begin{eqnarray}
\frac{dP^0}{d \eta}&=&-\frac{p}{a} \frac{d \Phi}{d \eta} e^{-\Phi} 
\left( 1+ \omega_i n^i
\right)+e^{-\Phi} \left( 1+\omega_i \, n^i \right) \frac{d(p/a)}{d\eta}
\nonumber \\
&+&
\frac{p}{a} e^{-\Phi} \frac{d(\omega_i \, n^i)}{d\eta}\, .
\end{eqnarray} 
Thus Eq.~(\ref{0geod}) allows us express $dp/d\eta$ as
\begin{eqnarray}
\label{dp1}
\frac{1}{p} \frac{dp}{d\eta}=- {\cal H} +\Psi'-\Phi_{,i} \,n^i e^{\Phi+\Psi} 
- \omega'_{i}\,n^i-\frac{1}{2} \chi'_{ij}n^in^j\, ,
\end{eqnarray} 
where in Eq.~(\ref{GPP}) we have replaced $P^0$ and $P^i$ by 
Eqs.~(\ref{P0bis}) and 
(\ref{Pi}). Notice that in order to obtain Eq.~(\ref{dp1}) 
we have used the following 
expressions for the total time derivatives of the metric perturbations
\begin{eqnarray}
\label{dPhi}
\frac{d \Phi}{d\eta}&=&\frac{\partial \Phi}{\partial \eta}+
\frac{\partial \Phi}{\partial x^i} \frac{d x^i}{d \eta} 
\nonumber \\
&= &
\frac{\partial \Phi}{\partial \eta}+\frac{\partial \Phi}{\partial x^i}n^i 
e^{\Phi+\Psi}\left( 1-\omega_j\, n^j -\frac{1}{2} \chi_{km}n^k n^m \right)
\end{eqnarray}
and 
\begin{eqnarray}
\frac{d (\omega_{i} n^i)}{d \eta}=n^i\left( 
\frac{\partial \omega_i}{\partial \eta}+
\frac{\partial \omega_i}{\partial x^j} \frac{dx^j}{d\eta}\right)=
\frac{\partial \omega_i}{\partial \eta}n^i+
\frac{\partial \omega_i}{\partial x^j} n^i n^j \, ,
\end{eqnarray}
where we have taken into account that $\omega_i$ is already a second-order 
perturbation 
so that we can neglect $dn^i/d\eta$ which is at least a first order quantity, 
and we can 
take the zero-order expression in Eq.~(\ref{dxi}), $dx^i/d\eta=n^i$.     
In fact there is also an alternative expression for 
$dp/d\eta$ which turns out to be useful later and which can be obtained by 
applying once more Eq.~(\ref{dPhi}) 
\begin{eqnarray}
\label{dp2}
\frac{1}{p}\frac{dp}{d\eta}=-{\cal H} -\frac{d\Phi}{d\eta}
+\Phi'+\Psi'-\omega'_{i}\,n^i
-\frac{1}{2}\chi'_{ij} n^in^j\, .
\end{eqnarray}
\\ 
\\
c) $d n^i/d \eta$: 
\\
\\
We can proceed in a similar way to compute $dn^i/d\eta$. 
Notice that since in Eq.~(\ref{Df}) it multiplies 
$\partial f/\partial n^i$ which is first order, we need only 
the first order perturbation of $dn^i/d\eta$.  
We use the spatial components of the geodesic equations $dP^i/d\lambda=- 
\Gamma^i_{\alpha \beta}P^{\alpha}P^{\beta}$ written as
\begin{eqnarray}
\label{geoi}
\frac{dP^i}{d\eta}=-\Gamma^i_{\alpha \beta}\frac{P^{\alpha}P^{\beta}}{P^0}\, .
\end{eqnarray}
For the right-hand side we find, up to second order,
\begin{eqnarray}
\label{RHSgeoi}
&&2\Gamma^i_{\alpha \beta} P^{\alpha}P^{\beta}=g^{i\nu} \left[ 
\frac{\partial g_{\alpha\nu}}{\partial x^\beta} +
\frac{\partial g_{\beta \nu}}{\partial x^{\alpha}}
-\frac{\partial g_{\alpha \beta}}{\partial x^\nu}\right] 
P^\alpha P^\beta \\
&&= 4 ({\cal H}-\Psi')P^i P^0 +2\left( \chi^{i\prime}_{~k}+
\omega^i_{,k}-\omega_k^{~,i} \right) P^0P^k \nonumber \\
&&+ 
\left(2 \frac{\partial \Phi}{\partial x^i} e^{2\Phi+2\Psi} 
+2 \omega^{i\prime} +2 {\cal H} \omega^i \right)
\left( P^0 \right)^2 \nonumber - 4 \frac{\partial 
\Psi}{\partial x^k} P^i P^k \nonumber \\
&&+ 2 \frac{\partial \Psi}{\partial x^i}\delta_{km} P^kP^m
- \left[ 2{\cal H} \omega^i \delta_{jk}-
\left(\frac{\partial\chi^i_{~j}}{\partial x^k} +
\frac{\partial\chi^i_{~k}}{\partial x^j} - 
\frac{\partial\chi_{jk}}{\partial x_i }\right) \right]P^jP^k
\, \nonumber, 
\end{eqnarray}
while the expression~(\ref{Pi}) of $P^i$ in terms of our variables $p$ 
and $n^i$ in the left-hand side of Eq.~(\ref{geoi}) brings
\begin{eqnarray}
\frac{dP^i}{d\eta} &=&\frac{p}{a} e^{\Psi}\left[ \frac{d\Psi}{d\eta} n^i+
\frac{a}{p} \frac{d(p/a)}{d\eta} n^i+\frac{dn^i}{d\eta} \right]
\left(1-\frac{1}{2} \chi_{km} n^kn^m \right) \nonumber \\
&-&\frac{p}{a}n^i e^{\Psi} \frac{1}{2} 
\frac{d \left(\chi_{km}n^kn^m\right)}{d\eta} \, . 
\end{eqnarray}
Thus, using the expression~(\ref{Pi}) for $P^i$ and~(\ref{P0}) for $P^0$ in 
Eq.~(\ref{RHSgeoi}), together with the previous result~(\ref{dp1}), 
the geodesic equation~(\ref{geoi}) gives the following 
expression $dn^i/d\eta$ (valid up to first order)
\begin{eqnarray}
\label{dni}
\frac{d n^i}{d\eta}=\left( \Phi_{,k}+
\Psi_{,k} \right) n^k n^i-\Phi^{,i}-\Psi^{,i}\, .
\end{eqnarray}

To proceed further we now expand the distribution function for 
photons around the zero-order value $f^{(0)}$ which is that of a Bose-Einstein 
distribution
\begin{equation}
\label{BEd}
f^{(0)}(p,\eta)=2\,\, \frac{1}{\exp\left\{\frac{p}{T(\eta)}\right\}-1}\, ,
\end{equation}
where $T(\eta)$ is the average (zero-order) temperature and the factor $2$ 
comes from the spin degrees of photons. The perturbed 
distribution of photons will depend also on $x^i$ and on the propagation 
direction $n^i$ so as to account for inhomogeneities and anisotropies
\begin{equation}
\label{expf}
f(x^i,p,n^i,\eta)=f^{(0)}(p,\eta)+f^{(1)}(x^i,p,n^i,\eta)+\frac{1}{2} 
f^{(2)}(x^i,p,n^i,\eta)\, ,
\end{equation} 
where we split the perturbation of the distribution function into a 
first and a second-order part. The Boltzmann equation 
up to second order can be written in a straightforward way by 
recalling that the total time derivative of a given $i$-th perturbation, as 
{\it e.g.} $df^{(i)}/d\eta$ is {\it at least} a quantity of the $i$-th order. 
Thus it is easy to realize, looking at Eq.~(\ref{Df}), that 
 the left-hand side of Boltzmann equation can be written 
up to second order as 
\begin{eqnarray}
\label{LHSBoltz}
\frac{df}{d\eta}&=&
\frac{d f^{(1)}}{d\eta}+\frac{1}{2} \frac{df^{(2)}}{d\eta}
-p\frac{\partial f^{(0)}}{\partial p}
\frac{d}{d\eta}\left(\Phi^{(1)}+\frac{1}{2} 
\Phi^{(2)}\right) \nonumber \\
&+&
p \frac{\partial f^{(0)}}{\partial p} \frac{\partial }{\partial \eta}
\left( \Phi^{(1)}+\Psi^{(1)}+\frac{1}{2} \Phi^{(2)}+
\frac{1}{2} \Psi^{(2)} \right) \nonumber \\
&-& p \frac{\partial f^{(0)}}{\partial p} 
\frac{\partial \omega_i}{\partial \eta}n^i
- \frac{1}{2} p \frac{\partial f^{(0)}}{\partial p} \frac{\partial \chi_{ij}} 
{\partial \eta} n^in^j\, ,
\end{eqnarray}
where for simplicity in Eq.~(\ref{LHSBoltz}) we have already used the 
background Boltzmann equation $(df/d\eta)|^{(0)}=0$. 
In Eq.~(\ref{LHSBoltz}) there are all the terms which will give rise to 
the integrated Sachs-Wolfe 
effects (corresponding to the terms which explicitly 
depend on the gravitational perturbations), 
while other effects, such as the gravitational lensing, are still 
hidden in the (second-order part) of the first term. In fact in order to 
obtain Eq.~(\ref{LHSBoltz}) we just need for the time being  
to know the expression for $dp/d\eta$, Eq.~(\ref{dp2}). 
\subsection{Collision term}
 
\subsubsection{The Collision Integral}
\label{CI}

In this section we focus on the collision term due to Compton scattering 
\begin{equation}
e({\bf q}) \gamma({\bf p}) \longleftrightarrow e({\bf q}') 
\gamma({\bf p}')\, ,
\end{equation}
where we have indicated the momentum of the photons  and electrons involved in 
the collisions. The collision term will be important for 
small scale anisotropies and 
spectral distortions. The important point to compute the 
collision term is that for 
the epoch of interest very little energy is transferred. Therefore one can 
proceed by expanding the right hand side of Eq.~(\ref{Boltzgeneric}) 
both in the small perturbation, Eq.~(\ref{expf}), and in the small energy 
transfer. 

The collision term is given (up to second order) by
\begin{eqnarray}
\label{collisionterm0}
{\overline C}({\bf p})=C({\bf p})a e^{\Phi}\, ,
\end{eqnarray}
where $a$ is the scale factor and
\footnote{The reason why we write the collision term as 
in Eq.~(\ref{collisionterm0}) is that the starting point of the 
Boltzmann equation requires differentiation with respect to an 
affine parameter $\lambda$, $df/d\lambda=C'$. 
In moving to the conformal time $\eta$ one rewrites the Boltzmann 
equation as $df/d\eta=C'(P^{0})^{-1}$, 
with $P^0=d\eta/d\lambda$ given by Eq.~(\ref{P0bis}). Taking 
into account that the collision term is at least of first order, 
Eq.~(\ref{collisionterm0}) then follows.}
\begin{eqnarray}
\label{collisionterm}
C({\bf p})&=&\frac{1}{E({\bf p})} \int \frac{d{\bf q}}{(2 \pi)^3 2E({\bf q})} 
\frac{d{\bf q}'}{(2 \pi)^3 2E({\bf q}')} 
\frac{d{\bf p}'}{(2 \pi)^3 2E({\bf p}')} \nonumber \\
&\times & (2\pi)^4 \delta^4(q+p-q'-p') \left| M \right|^2  \nonumber \\
& \times & \{ g({\bf q}')f({\bf p}')[ 1+f({\bf p})]-
g({\bf q})f({\bf p})[ 1+f({\bf p}')]\} 
\end{eqnarray}
where $E({\bf q})=(q^2+m_e^2)^{1/2}$, $M$ is the amplitude of 
the scattering process,
$\delta^4(q+p-q'-p')=\delta^3({\bf q}+{\bf p}-{\bf q}'-{\bf p}') 
\delta(E({\bf q})+p-E({\bf q}') -p')$ 
ensures the energy-momentum conservation and $g$ is 
the distribution function for electrons. The Pauli suppression factors $(1-g)$
have been dropped since for the epoch of interest the density of 
electrons $n_e$ is low. 
The electrons are kept in thermal equilibrium by Coulomb 
interactions with protons and 
they are non-relativistic, thus we can take a Maxwell-Boltzmann 
distribution around some bulk velocity ${\bf v}$
\begin{eqnarray}
\label{gel}
g({\bf q})=n_e \left( \frac{2 \pi}{m_e T_e}\right)^{3/2} 
\exp\left\{-\frac{({\bf q}-m_e{\bf v})^2}{2m_e T_e} \right\} 
\end{eqnarray}      
By using the three dimensional delta function the energy transfer is given by 
$E({\bf q})-E({\bf q}+{\bf p}-{\bf p}')$ and it turns out to be 
small compared to the typical thermal energies
\begin{equation}
\label{par}
E({\bf q})-E({\bf q}+{\bf p}-{\bf p}')\simeq \frac{({\bf p}-{\bf p}') 
\cdot {\bf q}}{m_e}
={\cal O}(Tq/m_e)\, . 
\end{equation}  
In Eq.~(\ref{par}) we have used $E({\bf q})=m_e+q^2/2m_e$ and the fact 
that, since the scattering 
is almost elastic ($p\simeq p'$), $({\bf p}-{\bf p'})$ is of order 
$p\sim T$, with $q$ much bigger 
than $({\bf p}-{\bf p'})$. In general, 
the electron momentum has two contributions, 
the bulk velocity ($q=m_e v$)  and the thermal 
motion ($q \sim (m_e T)^{1/2}$) and thus the parameter expansion 
$q/m_e$ includes 
the small bulk velocity ${\bf v}$ and the ratio 
$(T/m_e)^{1/2}$ which is small because 
the electrons are non-relativistic. 

The expansion of all the quantities entering the collision term 
in the energy transfer parameter and the integration over the momenta 
${\bf q}$ and 
${\bf q}'$ is described in details in Ref.~\cite{DJ}. 
It is easy to realize that we just need the scattering amplitude 
up to first order since at zero order 
$g({\bf q}')=g({\bf q}+{\bf p}-{\bf p}')=g({\bf q})$ 
and $\delta(E({\bf q})+p-E({\bf q}')-p')=\delta(p-p')$ so that 
all the zero-order quantities 
multiplying $\left| M \right|^2$ vanish. To first order 
\begin{equation}
\left| M \right|^2=6\pi\sigma_T 
m_e^2[(1+\cos^2\theta)-2\cos\theta(1-\cos\theta){\bf q}\cdot 
({\bf \hat{p}}+{\bf \hat{p}}')/m_e]\, ,
\end{equation}
where $\cos\theta={\bf n} \cdot {\bf n'}$ is the 
scattering angle and $\sigma_T$ the 
Thompson cross-section. The resulting collision term up to second order 
is given by~\cite{DJ} 
\begin{eqnarray}
\label{Integralcolli}
C(\bp) &=&  
{3n_e\sigma_T\over 4p} \int dp' p' {d\Omega' \over 4 \pi}
\bigg[ c^{(1)}(\bp,\bpp) + c^{(2)}_\Delta(\bp,\bpp)+c^{(2)}_v(\bp,\bpp) 
\nonumber \\
& +& c^{(2)}_{\Delta v}(\bp,\bpp) + 
c^{(2)}_{vv}(\bp,\bpp)+c^{(2)}_K(\bp,\bpp)   \bigg]\, ,
\end{eqnarray}
where we arrange the different contributions following Ref.~\cite{DJ}.
The first order term reads 
\begin{eqnarray}
c^{(1)}(\bp,\bpp) &=& (1+\cos^2\theta)\Bigg[
\delta(p-p') (\fps1-\f1) \nonumber \\
&+&(\fpz-\fz) (\bp-\bpp)\cdot\bv {\partial \delta(p-p')
			\over \partial p'} \Bigg]\, ,
\end{eqnarray}
while the second-order terms  have been separated into four parts. 
There is the so-called anisotropy suppression term 
\begin{eqnarray}
c^{(2)}_\Delta(\bp,\bpp) =\frac{1}{2} \left(1+\cos^2\theta\right)
		 \delta(p-p')(\fps2 - \f2)\, ;
\end{eqnarray}
a term which depends on the second-order velocity 
perturbation defined by the expansion 
of the bulk flow as ${\bf v}={\bf v}^{(1)}+{\bf v}^{(2)}/2$
\begin{equation}
c^{(2)}_v(\bp,\bpp)=\frac{1}{2}(1+\cos^2 \theta)  
(\fpz-\fz) (\bp-\bpp)\cdot\bv^{(2)}\, {\partial \delta(p-p')
			\over \partial p'}\, ;
\end{equation}
a set of terms coupling the photon perturbation to the velocity
\begin{eqnarray}
c^{(2)}_{\Delta v}(\bp,\bpp)&=&
\left(\fps1 - \f1\right)
\Bigg[  \left(1+\cos^2\theta\right)(\bp-\bpp)\cdot\bv \nonumber \\
&\times& {\partial \delta(p-p') \over \partial p'} 
- 2\cos\theta(1-\cos\theta) \delta(p-p')
		({\bf n}+ {\bf n}')\cdot \bv
\Bigg]\, , \nonumber
\end{eqnarray}
and a set of source terms quadratic in the velocity
\begin{eqnarray}
c^{(2)}_{vv}(\bp,\bpp) & = &
	\left(\fpz-\fz\right)\ (\bp-\bpp)\cdot\bv
\Bigg[ \left(1+\cos^2\theta\right) \nonumber \\
&\times & {(\bp-\bpp)\cdot\bv\over2}
		 {\partial^2 \delta(p-p')
			\over \partial p'^2} \nonumber \\
&-& 2\cos\theta(1-\cos\theta)
({\bf n}+ {\bf n}')\cdot \bv{\partial \delta(p-p')
			\over \partial p'} \Bigg]\,. \nonumber \\
\end{eqnarray}
The last contribution are the Kompaneets terms 
describing spectral distortions to the CMB
\begin{eqnarray}
c^{(2)}_K(\bp,\bpp)& = & \left(1+\cos^2\theta\right) {(\bp-\bpp)^2\over2\me}
	\Bigg[  \left( \fpz-\fz\right) \te \\
&\times & {\partial^2 \delta(p-p')
			\over \partial p'^2} 
- \left(\fpz+\fz+2\fpz\fz\right) \nonumber \\
& \times & {\partial \delta(p-p')
			\over \partial p'} \Bigg] 
+ {2(p-p')\cos\theta(1-\cos^2\theta)\over\me}
	\Bigg[ \delta(p-p') \nonumber \\
& \times & \fpz (1+\fz) \left(\fpz-\fz\right){\partial \delta(p-p')
			\over \partial p'} \Bigg]\, . \nonumber 
\end{eqnarray}
Let us make a couple of comments about the various contributions to the 
collision term. First, 
notice the term $c^{(2)}_v(\bp,\bpp)$ due to second-order 
perturbations in the velocity of electrons which is absent in Ref.~\cite{DJ}. 
In standard cosmological scenarios (like inflation) vector perturbations 
are not generated at 
linear order, so that linear velocities are irrotational $v^{(1)i}=\partial^i 
v^{(1)}$. However at 
second order vector perturbations are generated after horizon crossing as 
non-linear combinations 
of primordial scalar modes. Thus we must take into account
also a transverse (divergence-free) component, $v^{(2)i}=\partial^i v^{(2)}+ 
v^{(2)i}_T$ with 
$\partial_i v^{(2)i}_{T}=0$. As we will see such vector perturbations will 
break azimuthal symmetry 
of the collision term with respect to a given mode ${\bf k}$, which instead  
usually holds at linear order. Secondly, notice that the 
number density of electrons appearing in Eq.~(\ref{Integralcolli}) must be 
expanded as 
$n_e = \bar{n}_e(1+\delta_e)$ and then 
\begin{equation}
\label{deltac1}
\delta^{(1)}_e \,c^{(1)}(\bp,\bpp)
\end{equation}
gives rise to second-order contributions in addition to the list above, 
where we split $\delta_e=\delta^{(1)}_e+\delta^{(2)}_e/2$ into a 
first- and second-order part. In particular the 
combination with the term proportional to ${\bf v}$ in $c^{(1)}(\bp,\bpp)$ 
gives rise to the so-called Vishniac effect, as discussed in Ref.~\cite{DJ}.  

\subsubsection{Computation of different contributions to the collision term}

In the integral~(\ref{Integralcolli}) over the momentum ${\bf p}'$ the 
first-order term gives the usual collision term
\begin{equation}
\label{C1}
C^{(1)}({\bf p})=n_e \sigma_T \left[ f^{(1)}_0(p)+\frac{1}{2}f^{(1)}_2 
P_2({\bf \hat v} \cdot 
{\bf n})-f^{(1)}-p\frac{\partial f^{(0)}}{\partial p} {\bf v} \cdot {\bf n} 
\right]\, ,
\end{equation}
where one uses the decomposition in Legendre polynomials
\begin{equation} 
\label{dec1}
f^{(1)}({\bf x},p,{\bf n})=\sum_\ell (2\ell +1) f^{(1)}_\ell(p) 
P_\ell(\cos \vartheta)\, ,
\end{equation}
where $\vartheta$ is the polar angle of ${\bf n}$, $\cos \vartheta ={\bf n} 
\cdot {\bf \hat{v}}$.

In the following we compute the second-order collision term separately for 
the different 
contributions, using the notation $C(\bp)=C^{(1)}(\bp)+C^{(2)}(\bp)/2$. 
We have not reported the details of the calculation of the first-order 
term because 
for its second-order analog, $c^{(2)}_{\Delta}(\bp,\bpp)+c^{(2)}_v(\bp,\bpp)$, 
the procedure is the same. The important 
difference is that the second-order velocity term includes a vector part, 
and this leads to 
a generic angular decomposition of the distribution function (for simplicity 
drop the time 
dependence) 
\begin{equation}
\label{fangdeco}
f^{(i)}({\bf x},p,{\bf n})=\sum_{\ell} \sum_{m=-\ell}^{\ell} 
f^{(i)}_{\ell m}({\bf x},p)  
(-i)^{\ell}\ \sqrt{\frac{4\pi}{2\ell+1}} Y_{\ell m}({\bf n})\, ,
\end{equation} 
such that 
\begin{equation}
\label{angular}
f^{(i)}_{\ell m}=(-i)^{- \ell}\sqrt{\frac{2\ell+1}{4\pi}} \int d\Omega  
f^{(i)} 
Y^{*}_{\ell m}({\bf n}) \, .
\end{equation}
Such a decomposition holds also in Fourier space.
The notation at this stage is a bit confusing, so let us restate it:
superscripts 
denote the order of the perturbation; the subscripts refer to the moments
of the distribution.  Indeed at first order one can drop the dependence on $m$ 
setting $m=0$ using the fact that the distribution function does not depend 
on the azimuthal angle 
$\phi$. In this case the relation with $f^{(1)}_l$ is 
\begin{equation}
\label{rel}
f^{(1)}_{\ell m}=(-i)^{-\ell} (2\ell +1)  \delta_{m0} \, f^{(1)}_{\ell}\, .
\end{equation} 

\noindent
a) $c^{(2)}_\Delta(\bp, \bpp)$:

The integral over $\bpp$ yields
\begin{eqnarray}
C^{(2)}_\Delta(\bp) &=&\frac{3n_e \sigma_T}{4p}\int dp' p' 
\frac{d \Omega'}{4 \pi} c^{(2)}_\Delta(\bp, \bpp) =
\frac{3n_e \sigma_T}{4p}\int dp' p'\delta(p-p') \nonumber \\
&\times & \int \frac{d \Omega'}{4 \pi} 
[1+({\bf n} \cdot {\bf n}')^2] [f^{(2)}(\bpp)-f^{(2)}(\bp)]\, . 
\end{eqnarray}
To perform the angular integral we write 
the angular dependence on the scattering angle 
$\cos \theta= {\bf n} \cdot {\bf n}'$ in terms of the Legendre polynomials  
\begin{eqnarray}
\label{DL}
[1+({\bf n} \cdot {\bf n}')^2]&=&\frac{4}{3}\left[1+\frac{1}{2} 
P_2({\bf n} \cdot {\bf n}') \right] \nonumber \\
&=& 
\left[1+\frac{1}{2}\sum_{m=-2}^{2} Y_{2m}({\bf n})  Y^{*}_{2m}({\bf n}') 
\frac{4 \pi}{2\ell +1}
\right] \, ,
\end{eqnarray}
where in the last step we used the addition theorem for spherical harmonics
\begin{equation}
P_\ell=\frac{4 \pi}{2\ell +1} \sum_{m=-2}^{2} Y_{\ell m}({\bf n})  
Y^{*}_{\ell m}({\bf n}')\, . 
\end{equation}
Using the decomposition~(\ref{angular}) and the orthonormality of the 
spherical harmonics we find
\begin{eqnarray}
C^{(2)}_\Delta(\bp)=n_e \sigma_T \left[ 
f^{(2)}_{0 0}(p)-f^{(2)}(\bp)-\frac{1}{2} \sum_{m=-2}^{2} 
\frac{\sqrt{4 \pi}}{5^{3/2}}\, f^{(2)}_{2m}(p) \, Y_{2m}({\bf n}) 
\right]. \nonumber \\ 
\end{eqnarray} 
It is easy to recover the result for the corresponding first-order 
contribution in Eq.~(\ref{C1})
by using Eq.~(\ref{rel}).
\\
\noindent
b) $c^{(2)}_v(\bp,\bpp)$:

Let us fix for simplicity our coordinates such that the polar angle 
of ${\bf n}'$ is defined by $\mu'=
{\bf \hat{v}}^{(2)} \cdot {\bf n}'$ with $\phi'$ the corresponding 
azimuthal angle. The contribution 
of  $c^{(2)}_v(\bp,\bpp)$ to the collision term is then 
\begin{eqnarray}
C^{(2)}_v(\bp) &= &\frac{3 n_e \sigma_T}{4 p} v^{(2)} \int dp' p' 
[f^{(0)}(p')-f^{(0)}(p)]\frac{\partial \delta(p-p')}{\partial p'} 
\nonumber \\
&\times &\int_{-1}^{1} \frac{d \mu'}{2} (p \mu-p'\mu') \int_0^{2\pi} 
\frac{d \phi'}{2\pi} 
[1+({\bf p}\cdot {\bf p'})^2]\, .
\end{eqnarray} 
We can use Eq.~(\ref{DL}) which in our coordinate system reads
\begin{equation}
\label{DL2}
\frac{4}{3}\left[1+\frac{1}{2}  
\sum_{m=-2}^m \frac{(2-m)!}{(2+m)!} P_2^m({\bf n}\cdot{\bf 
\hat{ v}}^{(2)}) P_2^m({\bf n}' \cdot{\bf\hat{ v}}^{(2)})
e^{im(\phi'-\phi)} \right] \, ,
\end{equation}
so that 
\begin{equation}
\label{intphi}
\int \frac{d \phi'}{2 \pi} P_2({\bf n} \cdot{\bf n}') = 
P_2({\bf n} \cdot{\bf {\hat v}}^{(2)}) 
 P_2({\bf n}' \cdot{\bf {\hat v}}^{(2)}) =P_2(\mu)P_2(\mu')\, .
\end{equation}
By using the orthonormality of the Legendre polynomials and 
integrating by parts over $p'$ we find 
\begin{equation}
C^{(2)}_v(\bp)= - n_e\,  \sigma_T\,
p \frac{\partial f^{(0)}}{\partial p} {\bf v}^{(2)} \cdot {\bf n}\, . 
\end{equation}
As it is clear by the presence of the scalar product ${\bf v}^{(2)} 
\cdot {\bp}$ the final result is 
independent of the coordinates chosen.
\\
\noindent
c) $c^{(2)}_{\Delta v}(\bp,\bpp)$:

Let us consider the contribution from the first term
$$
c^{(2)}_{\Delta v(I)}(\bp,\bpp)=\left(1+\cos^2\theta\right) 
\left(\fps1 - \f1\right)
(\bp-\bpp)\cdot\bv {\partial \delta(p-p')
			\over \partial p'}\, ,  
$$
where the velocity has to be considered at first order. 
In the integral~(\ref{Integralcolli}) it brings
\begin{eqnarray}
\frac{1}{2}C^{(2)}_{\Delta v (I)} & = & \frac{3 n_e \sigma_T v}{4 p} 
\int dp' p' 
\frac{\partial \delta(p-p')}{\partial p'} 
\int_{-1}^1 \frac{d \mu'}{2} [f^{(1)}(\bpp)-f^{(1)}(\bp)] \nonumber \\
&\times &(p\mu - p'\mu') 
\int_0^{2\pi} \frac{d \phi'}{2 \pi} (1+\cos^2 \theta) \, ,
\end{eqnarray} 
The procedure to do the integral is the same as above. We use the 
same relations as in
Eqs.~(\ref{DL2}) and~(\ref{intphi}) where now the angles are those taken 
with respect to 
the first-order velocity. This eliminates the integral over $\phi'$, and 
integrating by parts over $p'$ yields
\begin{eqnarray}
\label{interm}
&&\frac{1}{2} 
C^{(2)}_{\Delta v (I)}(\bp)=-\frac{3 n_e \sigma_T v}{4p} \int_{-1}^1 
\frac{d \mu'}{2} 
\left[
\frac{4}{3}+\frac{2}{3} P_2(\mu) P_2(\mu') 
\right] \\
&&\times 
\biggl[ 
p(\mu-2\mu')(f^{(1)}(p,\mu')-f^{(1)}(p,\mu))
+p^2(\mu-\mu') \frac{\partial f^{(1)}(p,\mu')}{\partial p} 
\biggr]\, . \nonumber
\end{eqnarray} 
We now  use the decomposition~(\ref{dec1}) and the orthonormality of the 
Legendre polynomials to find
\begin{eqnarray}
&&\int \frac{d\mu'}{2} \mu' f^{(1)}(p,\mu') P_2(\mu')=
\sum_\ell \int \frac{d\mu'}{2} \mu ' P_2(\mu') P_l(\mu') f^{(1)}_\ell(p)
\nonumber \\
&&=
\sum_\ell \int \frac{d\mu'}{2} \left[ \frac{2}{5} P_1(\mu') +\frac{3}{5} 
P_3(\mu') \right] 
P_\ell(\mu')f^{(1)}_\ell(p) \nonumber \\
&&= \frac{2}{5} f^{(1)}_1(p)+\frac{3}{5} f^{(1)}_3(p)\, ,
\end{eqnarray}
where we have used $\mu ' P_2(\mu') P_l(\mu')=\frac{2}{5} P_1(\mu') +
\frac{3}{5} P_3(\mu')$, with 
$P_1(\mu')=\mu'$. Thus from Eq.~(\ref{interm}) we get 
\begin{eqnarray}
\frac{1}{2} C^{(2)}_{\Delta v (I)}(\bp)&=& n_e \sigma_T \Bigg\{ 
{\bf v} \cdot {\bf n} \biggl[ 
f^{(1)}(\bp) -f^{(1)}_0(p) - p\frac{\partial f^{(1)}_0(p)}{\partial p} 
\nonumber \\
&-& \frac{1}{2} P_2({\bf \hat{v}} \cdot {\bf n}) \left( f^{(1)}_2(p)+
p\frac{\partial f^{(1)}_2(p)}{\partial p}
\right) \biggr] \nonumber \\
&+& v \biggl[ 2f^{(1)}_1(p)+ p\frac{\partial f^{(1)}_1(p)}{\partial p} + 
\frac{1}{5} 
P_2({\bf \hat{v}}\cdot {\bf n}) \biggl( 2f^{(1)}_1(p) \nonumber \\
&+& p\frac{\partial 
f^{(1)}_1(p)}{\partial p}+
3 f^{(1)}_3(p)+\frac{3}{2} p\frac{\partial f^{(1)}_3(p)}{\partial p}
\biggr)
\biggr] \Bigg\} \, .
\end{eqnarray}
In $c^{(2)}(\bp,\bpp)$ there is a second term
$$c^{(2)}_{\Delta v(II)}= 
-2\cos\theta(1-\cos\theta) 
\left( f^{(1)}(\bpp)-f^{(1)}(\bp) \right)
\delta(p-p') ({\bf n}+ {\bf n}')\cdot \bv, $$
whose contribution to the collision term is 
\begin{eqnarray}
\frac{1}{2}C^{(2)}_{\Delta v (II)}(\bp) &=& -\frac{3n_e \sigma_T v}{2p} 
\int dp' p' \delta(p-p') \int_{-1}^1 
\frac{d\mu'}{2} (f^{(1)}(\bpp) \nonumber \\
&-& f^{(1)}(\bp)) (\mu+\mu')\int_0^{2\pi} 
\frac{d \phi'}{2\pi} 
\cos \theta(1-\cos\theta)\, .
\end{eqnarray}
This integration proceeds through the same steps as for 
$C^{(2)}_{\Delta v (I)}(\bp)$. In particular
by noting that $\cos \theta(1-\cos \theta)=-1/3+P_1(\cos \theta)
-2P_3(\cos\theta)/3$, 
Eqs.~(\ref{DL2}) and~(\ref{intphi}) allows to compute    
\begin{equation}
\int \frac{d \phi'}{2 \pi} \cos \theta (1-\cos \theta)=
-\frac{1}{3} +P_1(\mu) P_1(\mu')-\frac{2}{3} P_2(\mu) P_2(\mu')\, , 
\end{equation} 
and using the decomposition~(\ref{dec1}) we arrive at
\begin{eqnarray}
\label{interm2}
\frac{1}{2}C^{(2)}_{\Delta v(II)}(\bp) &=& - n_e \sigma_T 
\biggl\{ {\bf v} \cdot {\bf n}\, f^{(1)}_2(p) (1-P_2({\bf {\hat v}}
\cdot {\bf n})) \nonumber \\
&+& v \left[\frac{1}{5} \,
P_2({\bf {\hat v}}\cdot {\bf n})\,  \left( 
 3 f^{(1)}_1(p)-3 f^{(1)}_3(p) \right)
\right]
\biggr\}\, .
\end{eqnarray}
We then obtain
\begin{eqnarray}
\frac{1}{2}C^{(2)}_{\Delta v}(\bp) &=&
n_e \sigma_T \Bigg\{ {\bf v} \cdot {\bf n} \biggl[ 
f^{(1)}(\bp) -f^{(1)}_0(p) - p\frac{\partial f^{(1)}_0(p)}{\partial p} 
-f^{(1)}_2(p) \nonumber \\
&+ &\frac{1}{2} P_2({\bf \hat{v}} \cdot {\bf n}) \biggl( f^{(1)}_2(p)-
p\frac{\partial f^{(1)}_2(p)}{\partial p}
\biggr) \biggr] + v \biggl[ 2f^{(1)}_1(p) \nonumber \\ 
&+& p\frac{\partial f^{(1)}_1(p)}{\partial p} + 
\frac{1}{5} P_2({\bf \hat{v}}\cdot {\bf n}) 
\biggl( -f^{(1)}_1(p) \nonumber \\
&+& p\frac{\partial 
f^{(1)}_1(p)}{\partial p}+
6 f^{(1)}_3(p)+\frac{3}{2} p\frac{\partial f^{(1)}_3(p)}{\partial p}
\biggr)
\biggr]
\Bigg\} \, .
\end{eqnarray}
As far as the remaining terms, these have already 
been computed in Ref.~\cite{DJ} (see also 
Ref.~\cite{HSS})  and here we just report them
\\

\noindent
d) $c^{(2)}_{v v}(\bp,\bpp)$:

The term proportional to the velocity squared 
yield a contribution to the collision term
\begin{eqnarray}
\frac{1}{2}C^{(2)}_{v v}(\bp) &=& 
n_e \sigma_T \biggl\{ ({\bf v} \cdot 
{\bf n})^2 \left[ 
p \frac{\partial f^{(0)}}{\partial p}+\frac{11}{20} p^2 
\frac{\partial^2 f^{(0)}}{\partial p^2} \right] \nonumber \\
&+& v^2 \left[ 
p \frac{\partial f^{(0)}}{\partial p}+\frac{3}{20} p^2 
\frac{\partial^2 f^{(0)}}{\partial p^2}
\right]
\biggr\}\, .
\end{eqnarray}

\noindent
e) $c^{(2)}_K(\bp,\bp')$: 

The terms responsible for the spectral distortions give
\begin{equation}
\frac{1}{2}C_K^{(2)}(\bp)=\frac{1}{m_e^2}\frac{\partial}{\partial p}\left\{ 
p^4\left[
T_e \frac{\partial f^{(0)}}{\partial p}+f^{(0)}(1+f^{(0)})
\right]
\right\}\, .
\end{equation}
Finally, we write also the part of the collision term coming from 
Eq.~(\ref{deltac1})
\begin{eqnarray}
\delta^{(1)}_e\, c^{(1)}(\bp,\bpp) &\rightarrow &\delta^{(1)}_e C^{(1)}(\bp) =
 n_e \sigma_T\, \delta^{(1)}_e \biggl[ f^{(1)}_0(p) \nonumber \\
&+& \frac{1}{2}f^{(1)}_2 
P_2({\bf \hat v} \cdot 
{\bf n})-f^{(1)}-p\frac{\partial f^{(0)}}{\partial p} {\bf v} \cdot {\bf n} 
\biggr]\, . 
\end{eqnarray}

\subsubsection{Final expression for the collision term}

Summing all the terms we find the final expression for the collision 
term~(\ref{Integralcolli}) up to second order
\begin{eqnarray}
C(\bp)=C^{(1)}(\bp)+\frac{1}{2} C^{(2)}(\bp)
\end{eqnarray}
with 
\begin{eqnarray}
\label{C1p}
C^{(1)}(\bp)= n_e \sigma_T \left[ f^{(1)}_0(p)+\frac{1}{2}f^{(1)}_2 P_2({\bf 
\hat v} \cdot 
{\bf n})-f^{(1)}-p\frac{\partial f^{(0)}}{\partial p} {\bf v} \cdot {\bf n} 
\right]
\end{eqnarray}
and 
\begin{eqnarray}
\label{C2p}
\frac{1}{2}C^{(2)}(\bp)&=&n_e \sigma_T \Bigg\{ 
\frac{1}{2}f^{(2)}_{0 0}(p)-\frac{1}{4} \sum_{m=-2}^{2} 
\frac{\sqrt{4 \pi}}{5^{3/2}}\, f^{(2)}_{2m}(p) \, Y_{2m}({\bf n}) \nonumber \\
&-&\frac{1}{2} f^{(2)}(\bp) 
+ \delta^{(1)}_e \biggl[ f^{(1)}_0(p)+\frac{1}{2}f^{(1)}_2 P_2({\bf \hat v} 
\cdot {\bf n})-f^{(1)} \nonumber \\
&-& p\frac{\partial f^{(0)}}{\partial p} {\bf v} \cdot {\bf n} 
\biggr] 
- \frac{1}{2}p\frac{\partial f^{(0)}}{\partial p} {\bf v}^{(2)} 
\cdot {\bf n}+ {\bf v} \cdot {\bf n} \biggl[ f^{(1)}(\bp) \nonumber \\
&-& f^{(1)}_0(p) - p\frac{\partial f^{(1)}_0(p)}{\partial p} 
-f^{(1)}_2(p) 
+ \frac{1}{2} P_2({\bf \hat{v}} \cdot {\bf n}) \nonumber \\
&\times & \left( f^{(1)}_2(p)-
p\frac{\partial f^{(1)}_2(p)}{\partial p}
\right) \biggr] 
+ v \biggl[ 2f^{(1)}_1(p)+ p\frac{\partial f^{(1)}_1(p)}{\partial p} 
\nonumber \\
&+ & \frac{1}{5} 
P_2({\bf \hat{v}}\cdot {\bf n}) \biggl( - f^{(1)}_1(p)+ p\frac{\partial 
f^{(1)}_1(p)}{\partial p}+
6 f^{(1)}_3(p) \nonumber \\
&+& \frac{3}{2} p\frac{\partial f^{(1)}_3(p)}{\partial p}
\biggr) \biggr] + ({\bf v} \cdot {\bf n})^2 \biggl[ 
p \frac{\partial f^{(0)}}{\partial p}+\frac{11}{20} p^2 
\frac{\partial^2 f^{(0)}}{\partial p^2}
\biggr] \nonumber \\
&+&v^2 \biggl[ 
p \frac{\partial f^{(0)}}{\partial p}+\frac{3}{20} p^2 
\frac{\partial^2 f^{(0)}}{\partial p^2}
\biggr] \nonumber \\
&+& \frac{1}{m_e^2}\frac{\partial}{\partial p}\biggl[
p^4\biggl(
T_e \frac{\partial f^{(0)}}{\partial p}+f^{(0)}(1+f^{(0)})
\biggr) \biggr] \Bigg\}\, .
\end{eqnarray}
Notice that there is an internal hierarchy, with terms which do not depend 
on the baryon velocity 
${\bf v}$, terms proportional to ${\bf v} \cdot {\bf n}$ and then to 
$({\bf v} \cdot {\bf n})^2$, 
$v$ and $v^2$ (apart from the Kompaneets terms). In particular notice the 
term  proportional to 
$\delta^{(1)}_e {\bf v} \cdot {\bf n}$ is the one corresponding to the 
Vishniac effect. 
We point out that we have kept all the 
terms up to second order in the collision term. In Refs.~\cite{DJ,HSS} many 
terms coming from $c^{(2)}_{\Delta v}$ 
have been dropped mainly because these terms are proportional to the photon 
distribution function $f^{(1)}$ which on 
very small scales (those of interest for reionization) is suppressed by the 
diffusion damping. Here we want 
to be completely general and we have to keep them.   

\subsection{The Brightness equation}
 
\subsubsection{First order}

The Boltzmann equation for photons is obtained by combining 
Eq.~(\ref{LHSBoltz}) with Eqs.~(\ref{C1p})-(\ref{C2p}). 
At first order the left-hand side reads
\begin{eqnarray}
\label{Bdf1}
\frac{df}{d\eta}&=&\frac{d f^{(1)}}{d \eta}-p
\frac{\partial f^{(0)}}{\partial p} \frac{\partial \Phi^{(1)}}{\partial x^i} 
\frac{dx^i}{d\eta}+p\frac{\partial f^{(0)}}{\partial p} 
\frac{\partial \Psi^{(1)}}{ \partial \eta}\, .
\end{eqnarray}
At first order it is useful to characterize the perturbations to 
the Bose-Einstein distribution function~(\ref{BEd}) 
in terms of a perturbation to the temperature as 
\begin{equation}
\label{ft}
f(x^i,p,n^i,\eta)=2\, \left[ \exp\left\{ \frac{p}{T(\eta)
(1+ \Theta^{(1)})}\right\}-1 \right]^{-1}\, .
\end{equation}  
Thus it turns out that 
\begin{equation}
\label{thetaf1}
f^{(1)}=-p\frac{\partial f^{(0)}}{\partial p} \Theta^{(1)}\, ,
\end{equation} 
where we have used the fact that $\partial f/\partial \Theta|_{\Theta=0}=-p 
\partial f^{(0)}/\partial p$. In terms of this variable $\Theta^{(1)}$   
the linear collision term~(\ref{C1p}) will now become proportional to 
$- p \partial f^{(0)}/\partial p$ which contains the only explicit 
dependence on $p$, and the same happens for the left-hand side, 
Eq.~(\ref{Bdf1}). This is telling us that at first order $\Theta^{(1)}$ 
does not depend on $p$ but only on ${x^i, n^i, \eta}$,  
$\Theta^{(1)}=\Theta^{(1)}({x^i, n^i, \tau})$. This is well known 
and the physical reason is that at linear order there is no energy 
transfer in Compton collisions between photons and electrons. Therefore, 
the Boltzmann equation for $\Theta^{(1)}$ reads
\begin{eqnarray}
\label{BoltzTheta}
&&\frac{\partial \Theta^{(1)}}{\partial \eta}+n^i \frac{\partial 
\Theta^{(1)}}{\partial x^i}
+\frac{\partial \Phi^{(1)}}{\partial x^i} n^i -\frac{\partial 
\Psi^{(1)}}{ \partial \eta} \nonumber \\
&& = 
n_e\sigma_T a \left[ \Theta^{(1)}_0+\frac{1}{2} \Theta^{(1)}_2 
P_2({\bf {\hat v}} \cdot {\bf n})-\Theta^{(1)}+{\bf v} \cdot {\bf n} 
\right]\, ,
\end{eqnarray} 
where we made us of 
$f^{(1)}_{\ell}=- p \partial f^{(0)}/\partial p 
\Theta^{(1)}_\ell$, according to the decomposition of 
Eq.~(\ref{dec1}), and we have 
taken the zero-order expressions for $dx^i/d\eta$, dropping the 
contribution from $dn^i/d\eta$ in Eq.~(\ref{Df}) since it is already 
first-order. 

Notice that, since $\Theta^{(1)}$ is independent of $p$, it is 
equivalent to consider the quantity
\begin{equation}
\label{Delta1}
\Delta^{(1)}(x^i,n^i,\tau)=
\frac{\int dp p^3 f^{(1)}}{\int dp p^3 f^{(0)}}\, , 
\end{equation}
being $\Delta^{(1)}=4\Theta^{(1)}$ at this order.  
The physical meaning of $\Delta^{(1)}$ is that of a fractional 
energy perturbation (in a given direction). From Eq.~(\ref{LHSBoltz}) another 
way to write an equation for $\Delta^{(1)}$ -- the so-called 
brightness equation -- is     
\begin{eqnarray}
\label{B1}
&&\frac{d}{d\eta} \left[ \Delta^{(1)}+4 \Phi^{(1)} \right]-4 
\frac{\partial}{\partial \eta}\left( \Phi^{(1)}+\Psi^{(1)} \right) 
\nonumber \\
&& = n_e \sigma_T a \left[ \Delta^{(1)}_0+\frac{1}{2} \Delta^{(1)}_2 
P_2({\bf {\hat v}} \cdot {\bf n})-\Delta^{(1)}+4 {\bf v} \cdot {\bf n}
\right]\, .
\end{eqnarray}
\subsubsection{Second order}
 
The previous results show that at linear order the photon distribution 
function has a Planck spectrum with the temperature that at any point 
depends on the photon direction. At second order one could characterize the 
perturbed 
photon distribution function in a similar way as in Eq.~(\ref{ft})  
\begin{equation}
f(x^i,p,n^i,\eta)=2\, \left[ \exp\left\{ \frac{p}{T(\eta)\, e^{\Theta}}-1 
\right\}\right]^{-1}\, ,
\end{equation}
where by expanding $\Theta=\Theta^{(1)}+\Theta^{(2)}/2+...$ as usual one 
recovers the first-order expression. 
For example, in terms of $\Theta$, the perturbation of $f^{(1)}$ is given 
by Eq.~(\ref{thetaf1}), while at second order  
\begin{eqnarray}
\frac{f^{(2)}}{2}
=-\frac{p}{2} \frac{\partial f^{(0)}}{\partial p} \Theta^{(2)}+\frac{1}{2} 
\left(p^2 \frac{\partial^2f^{(0)}}{\partial p^2}
+ p \frac{\partial f^{(0)}}{\partial p}  \right) \left( \Theta^{(1)} 
\right)^2\, .
\end{eqnarray}
However, as discussed in details in Refs.~\cite{HSS,DJ}, now the second-order 
perturbation $\Theta^{(2)}$ will not be momentum independent 
because the collision term in the equation for $\Theta^{(2)}$ does depend 
explicitly on $p$ (defining the combination  
$- (p \partial f^{(0)}/\partial p)^{-1} f^{(2)}$ does not lead to a 
second-order momentum independent equation as above). Such dependence 
is evident, for example, in the terms of $C^{(2)}({\bf p})$, Eq.~(\ref{C2p}), 
proportional to $v$ or $v^2$, and in the Kompaneets terms. 
The physical reason is that at the non-linear level photons and electrons do 
exchange energy during Compton collisions. As a consequence 
spectral 
distorsions are generated. For example, in the isotropic limit,
 only the Kompaneets terms survive giving rise to the Sunyaev-Zeldovich 
distorsions. As discussed in Ref.~\cite{HSS}, the Sunyaev-Zeldovich 
distorsions can also be obtained with the correct coefficients by replacing 
the average over the direction electron 
$\langle v^2 \rangle$ with the mean squared thermal velocity $\langle 
v_{th}^2 \rangle=3T_e/m_e$ in Eq.~(\ref{C2p}). 
This is due simply to the fact that 
the distinction between thermal and bulk velocity of the electrons is 
just for convenience. This fact also 
shows that spectral distorsions due 
to the bulk flow (kinetic Sunyaev-Zeldovich) has the same form as the 
thermal effect. Thus 
spectral distorsions can be in general described by a global Compton 
$y$-parameter 
(see Ref.~\cite{HSS} for a full discussion of spectral distorsions). 
However in the following we will not be 
interested in the frequency dependence but only in the anisotropies of 
the radiation distribution. Therefore we can integrate over the 
momentum 
$p$ and define ~\cite{HSS,DJ} 
\begin{equation}
\label{Delta2}
\Delta^{(2)}(x^i,n^i,\tau)=\frac{\int dp p^3 f^{(2)}}{\int dp p^3 f^{(0)}}\, ,
\end{equation}
as in Eq.~(\ref{Delta1}).

Integration over $p$ of Eqs.~(\ref{LHSBoltz})-(\ref{C2p}) is straightforward 
using the following relations
\begin{eqnarray}
\label{rules}
\int dp p^3 p \frac{\partial f^{(0)}}{\partial p}&=&-4\, N\, ; \quad \quad
\int dp p^3 p^2 \frac{\partial^2 f^{(0)}}{\partial p^2}= 20\, N\, ; 
\nonumber\\
\int dp p^3 f^{(1)} &=&N \Delta^{(1)} \, ; \quad \quad
\int dp p^3 p \frac{\partial f^{(1)}}{\partial p}=-4\, N \Delta^{(1)}\, .
\end{eqnarray}
Here $N=\int dp p^3 f^{(0)}$ is the normalization factor
(it is just proportional the background energy density of 
photons ${\bar \rho}_\gamma$). At first order one recovers Eq.~(\ref{B1}). 
At second order we find
\begin{eqnarray}
\label{B2}
& & \frac{1}{2} \frac{d}{d\eta} \left[ \Delta^{(2)}+4\Phi^{(2)} \right] + 
\frac{d}{d\eta} \left[ \Delta^{(1)} +4 \Phi^{(1)} \right] 
-4 \Delta^{(1)}\left( \Psi^{(1)'}-\Phi^{(1)}_{,i}n^i \right) \nonumber \\
&&- 2 
\frac{\partial}{\partial \eta}\left( \Psi^{(2)}+\Phi^{(2)} \right) 
+4 \frac{\partial \omega_i}{\partial \eta} n^i + 2 \frac{\partial 
\chi_{ij}}{\partial \eta} n^i n^j = \nonumber \\
& &= - \frac{\tau'}{2} \Bigg[ \Delta^{(2)}_{00} -\Delta^{(2)} 
- \frac{1}{2} \sum_{m=-2}^{2} \frac{\sqrt{4 \pi}}{5^{3/2}}\, 
\Delta^{(2)}_{2m} \, Y_{2m}({\bf n}) +
2 (\delta^{(1)}_e +\Phi^{(1)}) \nonumber \\
&& \left( \Delta^{(1)}_0+\frac{1}{2} \Delta^{(1)}_2 P_2({\bf {\hat v}} 
\cdot {\bf n})-\Delta^{(1)}+4 {\bf v} \cdot {\bf n}
\right)+4{\bf v}^{(2)} \cdot {\bf n} \nonumber \\
& & + 2 ({\bf v} \cdot {\bf n}) \left[ \Delta^{(1)}+3\Delta^{(1)}_0-
\Delta^{(1)}_2 \left(1-\frac{5}{2} P_2({\bf {\hat v}} 
\cdot {\bf n}) \right)\right] \nonumber \\
&& - v\Delta^{(1)}_1 \left(4+2 P_2({\bf {\hat v}} \cdot {\bf n}) 
\right) 
+14 ({\bf v} \cdot {\bf n})^2-2 v^2  \Bigg]\, ,
\end{eqnarray}
where we have expanded the angular dependence of $\Delta$ as in 
Eq.~(\ref{fangdeco})
\begin{equation}
\label{Dlm2}
\Delta^{(i)}({\bf x}, {\bf n})=\sum_{\ell} \sum_{m=-\ell}^{\ell} 
\Delta^{(i)}_{\ell m}({\bf x})  
(-i)^{\ell}  \sqrt{\frac{4 \pi}{2\ell+1}}Y_{\ell m}({\bf n})\, ,
\end{equation} 
with  
\begin{equation}
\label{angular1}
\Delta^{(i)}_{\ell m}=(-i)^{- \ell}\sqrt{\frac{2\ell+1}{4\pi}} 
\int d\Omega  \Delta^{(i)} 
Y^{*}_{\ell m}({\bf n}) \, ,
\end{equation}
where we recall that the superscript stands by the order of the 
perturbation. When going to Fourier space some convolutions will appear and the coefficients $\Delta_\ell^{(1)}({\bf k}',\eta)$
are related to $\Delta_{\ell m}^{(1)}({\bf k}',\eta)$ as 
$ \Delta^{(1)}_{\ell m}({\bf k}')=i^\ell \sqrt{4 \pi(2\ell+1)}Y^*_{\ell m}(\hat{\bf k}') 
\break \Delta^{(1)}_\ell({\bf k}')$. In Eq.~(\ref{B2}) we have  
introduced the differential optical depth 
\begin{equation}
\label{deftau}
\tau'=-{\bar n}_e \sigma_T a \, .
\end{equation}
It is understood that
 on the left-hand side of Eq.~(\ref{B2}) one has to 
pick up for the total time derivatives only those terms which contribute to 
second order. Thus we have to take 
\begin{eqnarray}
\label{D}
&&\frac{1}{2} \frac{d}{d\eta} \left[ \Delta^{(2)}+4\Phi^{(2)} \right] + 
\frac{d}{d\eta} \left[ \Delta^{(1)} +4 \Phi^{(1)} \right]\Big|^{(2)} \\
&&= \frac{1}{2}\left( \frac{\partial}{\partial \eta}+n^i
 \frac{\partial}{\partial x^i}\right) \left( \Delta^{(2)}+4\Phi^{(2)} \right)
+n^i(\Phi^{(1)}+\Psi^{(1)}) \nonumber \\
&& \times 
\partial_i(\Delta^{(1)}+4\Phi^{(1)}) + 
\left[(\Phi^{(1)}_{,j}+\Psi^{(1)}_{,j})n^in^j 
-(\Phi^{,i}+\Psi^{,i})\right]
\frac{\partial \Delta^{(1)}}{\partial n^i}\, , \nonumber
\end{eqnarray}
where we used Eqs.~(\ref{dxi}) and ~(\ref{dni}).

\subsubsection{Hierarchy equations for multipole moments}

Let us now move to Fourier space. In the following, for a given wave-vector 
${\bf k}$ we will choose the coordinate system such that 
${\bf e}_3={\bf {\hat k}}$ and the polar angle 
of the photon momentum is $\vartheta$, with $\mu=\cos \vartheta = 
{\bf \hat{k}} \cdot {\bf n}$. Then Eq.~(\ref{B2}) can be written as 
\begin{equation}
\label{BF}
\Delta^{(2) \prime}+ik\mu \Delta^{(2)}-\tau' \Delta^{(2)}= 
S({\bf k},{\bf n},\eta)\, ,
\end{equation}
where $S({\bf k},{\bf n},\eta)$ can be easily read off Eq.~(\ref{B2}). 
We now expand the temperature anisotropy in the multipole moments 
$\Delta^{(2)}_{\ell m}$ in order to obtain a system of coupled 
differential equations. By applying the angular integral of 
Eq.~(\ref{angular1}) to Eq.~(\ref{BF}) we find
\begin{equation}
\label{H}
\Delta^{(2)\prime }_{\ell m}({\bf k},\eta)=k 
\left[\frac{{\kappa}_{\ell m}}{2\ell-1}\Delta^{(2)}_{\ell-1,m}-
\frac{{\kappa}_{\ell+1, m}}{2\ell+3}\Delta^{(2)}_{\ell+1,m} \right] + 
\tau' \Delta^{(2)}_{\ell m}+ S_{\ell m}
\end{equation}  
where the first term on the right-hand side of Eq.~(\ref{H}) has been 
obtained by using the relation 
\begin{eqnarray}
\label{gradient}
i {\bf k} \cdot {\bf n}\, \Delta^{(2)}({\bf k}) &=& \sum_{\ell m} 
\Delta^{(2)}_{\ell m}({\bf k}) \frac{k}{2\ell +1} \left[ 
{\kappa_{\ell m} {\tilde G}_{\ell-1,m}-\kappa_{\ell+1,m}{\tilde G}_{\ell+1,m}} 
\right] \nonumber \\
&=&  
k \sum_{\ell m} \left[\frac{{\kappa}_{\ell m}}{2\ell-1}
\Delta^{(2)}_{\ell-1,m}-
\frac{{\kappa}_{\ell m}}{2\ell+3}\Delta^{(2)}_{\ell+1,m} 
\right] {\tilde G}_{\ell m}\, , 
\end{eqnarray} 
where $ {\tilde G}_{\ell m}= (-i)^{\ell} \sqrt{4\pi/(2\ell+1)}
Y_{\ell m}({\bf n})$ is the 
angular mode for the decomposition~(\ref{Dlm2}) and 
${\kappa}_{\ell m}=\sqrt{l^2-m^2}$.
This relation has been discussed in Refs.~\cite{Complete,huwhite} 
and corresponds to the term $n^i \partial \Delta^{(2)}/\partial x^i$ 
in Eq.~(\ref{B2}). In Eq.~(\ref{H}) $S_{lm}$ are the multipole moments of the source term according to the decomposition~(\ref{angular1}). 
We do not show its complete expression here, since it is very long, and we prefer to make some general comments (for some specific examples about 
the terms $S_{lm}$ see Sec.~\ref{NumAn}). As it should be already clear from the calculations involving second-order perturbations performed so far, 
Eq.~(\ref{H}) has the same functional form of 
the Boltzmann equation at linear order (just replace the order of the linear perturbations $(1)$ with that of 
intrinsically second-order terms $(2)$) with the exception of the source terms $S_{lm}$ which now contain both 
intrinsic second-order perturbations and also products of 
first-order perturbations. Therefore as expected, at second order we recover some intrinsic effects which 
are characteristic of the linear regime. 
In Eq.~(\ref{H}) the relation~(\ref{gradient})  represents the free 
streaming effect: when the radiation undergoes free-streaming, 
the inhomogeneities of the photon    
distribution are seen by the obsever as angular anisotropies. At first 
order it is responsible for the hierarchy of Boltzmann 
equations coupling the different $\ell$ modes, and it represents a 
projection effect of fluctuations on a scale $k$ onto the angular scale 
$\ell \sim k\eta$. The term $\tau' \Delta^{(2)}_{\ell m}$ causes an 
exponential suppression of anisotropies in the absence of the source 
term $S_{\ell m}$. The source term 
contains additional scattering processes and gravitational effects. 
The intrinsically second-order part of the source term just 
reproduces the expression of the first order case.   
Of course the dynamics of the second-order metric and baryon-velocity 
perturbations which appear will be different and governed by the 
second-order Einstein equations and continuity equations. The remaining 
terms in the source are second-order effects generated as 
non-linear combinations of the primordial (first-order) perturbations. 
Notice in particular that they involve 
the first-order anisotropies $\Delta^{(1)}_\ell$  and as a 
consequence such terms contribute to generate the hierarchy of equations 
(apart from the free-streaming effect).
On large scales (above the horizon at recombination) 
we can say that the main effects are due to gravity, and they include 
the Sachs-Wolfe and the (late and early) Sachs-Wolfe effect due 
to the redshift photons suffer when travelling through the second-order 
gravitational potentials. These, toghether with the contribution 
due to the second-order tensor modes, have been already studied in 
details in Ref.~\cite{fulllarge} (see also~\cite{cre}). Another important gravitational effect is 
that of lensing of photons as they travel from the last scattering 
surface to us. A contribution of this type is given by the last term
of Eq.~(\ref{D}). On the other hand examples of second-order scattering effects are the terms 
proportional to the square of the baryon-velocity fluid ($v^2$), 
giving rise to the quadratic Doppler effect (like those in the last line of Eq.~(\ref{B2})), or those coupling the photon fluctuations to the baryon 
velocity (second-line from the bottom of Eq.~(\ref{B2})). The Vishniac effect corresponds to the terms proportional to $\delta^{(1)}_e$.   
Finally notice that in the Boltzmann equation~(\ref{B2})) the second-order baryon velocity appears. At linear order 
the baryon velocity is irrotational, 
meaning that it is the gradient of a potential, and thus in Fourier space 
it is parallel to 
${\bf {\hat k}}$, and following the same 
notation of Ref.~\cite{husolo}, we write 
\begin{equation}
\label{vzero}
{\bf v}^{(1)}({\bf k})=-i v^{(1)}_0({\bf k}) {\bf {\hat k}}\, .
\end{equation}
The second-order velocity perturbation will contain a transverse 
(divergence-free) part whose components are orthogonal to  
${\bf {\hat k}}={\bf e}_3$, and we can write 
\begin{equation}
\label{vdec}
{\bf v}^{(2)}({\bf k})=-i v^{(2)}_0({\bf k}) {\bf e}_3+\sum_{m=\pm1}
 v^{(2)}_m \, \frac{{\bf e}_2\mp i {\bf e}_1}{\sqrt 2}\, ,
\end{equation}
where ${\bf e}_i$ form an orhtonormal basis with  ${\bf {\hat k}}$. 
The second-order
perturbation $\omega_i$ is decomposed in a similar way, with $\omega_{\pm 1}$ 
the corresponding components 
(in this case in the Poisson gauge there is no scalar component). 
Similarly for the tensor perturbation $\chi_{ij}$ 
we have indicated its amplitudes as $\chi_{\pm 2}$ in the 
decomposition~\cite{huwhite}
\begin{equation}
\chi_{ij}=\sum_{m=\pm 2} - \sqrt{\frac{3}{8}} \, \chi_m ({\bf e}_1\pm i 
{\bf e}_2)_i ({\bf e}_1\pm i {\bf e}_2)_j\, .  
\end{equation}
In computing the source term $S_{lm}$ one has to take into account that in the gravitational part of the Boltzmann 
equation and in the collsion term there are some terms, like 
$\delta^{(1)}_e {\bf v}$, which still can be decomposed in the scalar and 
transverse parts in Fourier space as in Eq.~(\ref{vdec}). 
For a generic quantity  $f({\bf x}) {\bf v}$ one can indicate the 
corresponding scalar and vortical components with $(f {\bf v})_m$ and 
their explicit expression is easily found by projecting the Fourier 
modes of $f({\bf x}) {\bf v}$ along the ${\bf {\hat k}}={\bf e}_3$ and 
$({\bf e}_2\mp i {\bf e}_1)$ directions 
\begin{equation}
(f {\bf v})_m({\bf k})=\int \frac{d^3k_1}{(2\pi)^3} v^{(1)}_0({\bf k}_1) 
f({\bf k}_2) Y^*_{1m}({\bf \hat{k}}_1) \sqrt{\frac{4\pi}{3}}\, .
\end{equation} 
Similarly for a term like $f({\bf x}) \nabla g({\bf x})$ one can use the notation
\begin{equation}
(f\nabla g)_m({\bf k})=- \int \frac{d^3k_1}{(2\pi)^3} k_1 g({\bf k}_1) 
f({\bf k}_2) Y^*_{1m}({\bf \hat{k}}_1) 
\sqrt{\frac{4\pi}{3}}\, .
\end{equation}

\subsubsection{Integral solution of the second-order Boltzmann equation}
 
As in linear theory, one can derive an integral solution of the 
Boltzmann equation~(\ref{B2}) in terms of the source term $S$.  
Following the standard procedure (see {\it e.g.} Ref.~\cite{review}) 
for linear perturbations, 
we write the left-hand side as $\Delta^{(2) \prime}+ik\mu \Delta^{(2)}-\tau' 
\Delta^{(2)}=
e^{-ik\mu \eta + \tau} d[\Delta^{(2)} e^{ik \mu \eta-\tau}]/d\eta$ in order to 
derive the integral solution  
\begin{eqnarray}
\label{IS}
\Delta^{(2)}({\bf k},{\bf n},\eta_0)=\int_0^{\eta_0} d\eta 
S({\bf k},{\bf n},\eta) e^{ik\mu(\eta-\eta_0)-\tau}\, ,
\end{eqnarray}
where $\eta_0$ stands by the present time. The expression of the 
photon moments $\Delta^{(2)}_{\ell m}$ can be 
obtained as usual from Eq.~(\ref{angular1}). In the previous section 
we have already found the coefficients for the 
decomposition of source term $S$  
\begin{eqnarray}
\label{decS}
S({\bf k},{\bf n}, \eta)= \sum_{\ell} \sum_{m=-\ell}^{\ell} 
S_{\ell m}({\bf k},\eta) 
(-i)^{\ell}\ \sqrt{\frac{4\pi}{2\ell+1}} Y_{\ell m}({\bf n}) \, .
\end{eqnarray}
In Eq.~(\ref{IS}) there is an additional angular dependence in 
the exponential. It is easy to 
take it into account by recalling that 
\begin{equation}
\label{eikx}
e^{i{\bf k} \cdot {\bf x}}=
\sum_\ell (i)^\ell (2\ell+1) j_\ell(kx) P_\ell( {\bf {\hat k}} \cdot 
{\bf {\hat x}}) \, .
\end{equation}
Thus the angular integral~(\ref{angular1}) is computed by using the 
decomposition of the source term~(\ref{decS}) 
and Eq.~(\ref{eikx}) 
\begin{eqnarray}
\label{IS1}
\Delta^{(2)}_{\ell m}( {\bf k},\eta_0)&=&(-1)^{-m}(-i)^{-\ell} (2\ell+1) 
\int_0^{\eta_0} d\eta \, e^{-\tau(\eta)} \nonumber \\
&\times& 
\sum_{\ell_2}\sum_{m_2=-\ell_2}^{\ell_2} (-i)^{\ell_2} S_{\ell_2 m_2}  
\sum_{\ell_1} i^{\ell_1} j_{\ell_1}(k(\eta-\eta_0))
\nonumber \\  
&\times& (2\ell_1+1) 
\left(\begin{array}{ccc}\ell_1&\ell_2&\ell\\0&0&0\end{array}\right)
\left(\begin{array}{ccc}\ell_1&\ell_2&\ell\\0&m_2&- m\end{array}\right) \, ,
\end{eqnarray}
where the Wigner $3-j$ symbols appear because of the Gaunt integrals
\begin{eqnarray}
  \nonumber
  {\mathcal G}_{\ell_1\ell_2\ell_3}^{m_1m_2m_3}
  &\equiv&
  \int d^2\hat{\mathbf n}
  Y_{\ell_1m_1}(\hat{\mathbf n})
  Y_{\ell_2m_2}(\hat{\mathbf n})
  Y_{\ell_3m_3}(\hat{\mathbf n})\\
 \nonumber
  &=&\sqrt{
   \frac{\left(2\ell_1+1\right)\left(2\ell_2+1\right)\left(2\ell_3+1\right)}
        {4\pi}
        }
\nonumber \\
&\times &   \left(
  \begin{array}{ccc}
  \ell_1 & \ell_2 & \ell_3 \\ 0 & 0 & 0 
  \end{array}
  \right)
  \left(
  \begin{array}{ccc}
  \ell_1 & \ell_2 & \ell_3 \\ m_1 & m_2 & m_3 
  \end{array}
  \right),
\end{eqnarray}
Since the second of the Wigner 3-$j$ symbols in Eq.~(\ref{IS1}) 
is nonzero only if $m=m_2$, our solution can be rewritten to 
recover the corresponding expression found for linear anisotropies in 
Refs.~\cite{huwhite}
\begin{equation}
\label{comp}
\frac{\Delta^{(2)}_{\ell m}( {\bf k},\eta_0)}{2\ell+1}=
\int_0^{\eta_0} d\eta \, e^{-\tau(\eta)} \sum_{\ell_2} 
S_{\ell_2 m}\, j_{\ell}^{(\ell_2,m)}[k(\eta_0-\eta)]\, ,
\end{equation}    
where $j_{\ell}^{(\ell_2,m)}[k(\eta_0-\eta)]$ are the so called radial 
functions. Of course the main information at second order is 
included in the source term containing different effects due to the 
non-linearity of the perturbations. 
In the total angular momentum method of Ref.~\cite{huwhite} 
Eq.~(\ref{comp}) is interpreted just as the intergration over 
the radial coordinate $(\chi=\eta_0-\eta)$ of the projected source term. 
Another important comment is that, as in linear theory, the integral 
solution~(\ref{IS1}) is in fact just a formal solution, since 
the source term $S$ contains itself the second-order photon moments up
 to $l=2$. 
This means that one has anyway to resort to the 
hierarchy equations for photons, Eq.~(\ref{H}), to solve for these 
moments. Nevertheless, as in linear theory, 
one expects to need just a few moments beyond $\ell=2$ in the hierarchy 
equations, and once the moments entering in the 
source function are computed the higher moments are obtained from the 
integral solution. Thus the integral solution should in fact be 
more advantageous than solving the system of coupled 
equations~(\ref{H}).

\subsection{The Boltzmann equation for baryons and cold dark matter}
 
In this section we will derive the Boltzmann equation for massive particles, 
which is the case of interest for baryons and dark matter. These equations
 are necessary to find the time evolution of number 
densities and velocities of the baryon fluid which appear in the 
brightness equation, thus allowing to close the system of 
equations. Let us start from the baryon component. 
Electrons are tightly coupled to protons via Coulomb interactions. 
This forces the relative energy density contrasts and the 
velocities to a common value, $\delta_e=\delta_p \equiv \delta_b$ and 
${\bf v}_e={\bf v}_p \equiv {\bf v}$, so that we can identify  
electrons and protons collectively as ``baryonic'' matter. 

To derive the Boltzmann equation for baryons let us first focus 
on the collisionless equation and compute therefore $dg/d\eta$, where $g$ 
is the distribution function for a massive 
species with mass $m$. One of the differences with respect to photons is 
just that baryons are non-relativistic for the epochs of 
interest. Thus the first step is to generalize the formulae in Section 4
up to Eq.~(\ref{dni}) to the case of a massive particle. 
In this case one enforces the constraint $Q^2=g_{\mu\nu}Q^\mu Q^\nu=-m^2$ 
and it also useful to use the particle energy
$E=\sqrt{q^2+m^2}$, 
where $q$ is defined as in Eq.~(\ref{defp}). Moreover in this case it is 
very convenient to take 
the distribution function as a function of the variables $q^i=qn^i$, the 
position $x^i$ and time $\eta$, without using the explicit splitting into 
the magnitude of the momentum $q$ (or the energy E) and its direction $n^i$. 
Thus the total time derivative of the distribution functions reads
\begin{equation}
\label{DG}
\frac{d g}{d \eta}=\frac{\partial g}{\partial \eta}+
\frac{\partial g}{\partial x^i} \frac{d x^i}{d \eta}+
\frac{\partial g}{\partial q^i} \frac{d q^i}{d \eta}\, .
\end{equation}
We will not give the details of the calculation since we just need to 
replicate the 
same computation we did for the photons. For the four-momentum of the 
particle notice that $Q^i$ has the same form as Eq.~(\ref{Pi}), 
while for $Q^0$ we find
\begin{equation}
\label{P0m}
Q^0=\frac{e^{-\Phi}}{a}\, E \left(1+\omega_i \frac{q^i}{E}  \right)\, .
\end{equation}  
In the following we give the expressions for $dx^i/d\eta$ and $dq^i/d\eta$.
\\
\noindent
a) As in Eq.~(\ref{dxi}) $dx^i/d\eta=Q^i/Q^0$ and it turns out to be
\begin{equation}
\frac{d x^i}{d\eta}=\frac{q}{E} n^i e^{\Phi+\Psi}  
\left(1-\omega_i n^i \frac{q}{E} \right) 
\left(1-\frac{1}{2} \chi_{km} n^k n^m \right)\, .
\end{equation}

\noindent
b) For $dq^i/d\eta$ we need the expression of $Q^i$ which is 
the same as that of  
Eq.~(\ref{Pi}) 
\begin{equation}
\label{Qi}
Q^i= \frac{q^i}{a} e^{\Psi} \left( 1-\frac{1}{2} \chi_{km}n^kn^m \right)\, .
\end{equation}
The spatial component of the geodesic equation, up to second order, reads 
\begin{eqnarray}
\label{SPG}
\frac{dQ^i}{d\eta}&=&-2({\cal H}-\Psi')\left( 1-\frac{1}{2}\chi_{km}n^kn^m 
\right) 
\frac{q}{a}n^i e^{\Psi} + e^{\Phi+2\Psi} 
\nonumber\\
&\times & \biggl( \frac{\partial \Psi}{\partial x^k}
\frac{q^2}{aE}(2 n^in^k - \delta^{ik})
-\frac{\partial \Phi}{\partial x^i}\frac{E}{a} \biggr) 
- \frac{E}{a}\bigl[\omega^{i\prime}+{\cal H} \omega^i + 
q^k \bigl(\chi^{i\prime}_{~k} \nonumber \\
&+& \omega^{i}_{'k}-\omega_k^{,i}\bigr)\bigr]  
+\left[{\cal H}\omega^i \delta_{jk}-\frac{1}{2}(\chi^{i}_{~j,k}+\chi^i_{~k,j} -
\chi_{jk}^{~~,i})\right] \frac{q^j q^k}{Ea}\, .
\end{eqnarray}
Proceeding as in the massless case we now take the total time derivative 
of Eq.~(\ref{Qi}) and using Eq.~(\ref{SPG}) we find 
\begin{eqnarray}
\frac{dq^i}{d\eta}&=&-({\cal H}-\Psi')q^i+\Psi_{,k}\frac{q^iq^k}{E} 
e^{\Phi+\Psi}-\Phi^{,i}E e^{\Phi+\Psi} 
-\Psi_{,i}\frac{q^2}{E} e^{\Phi+\Psi} \nonumber \\
&-& E(\omega^{i\prime}+{\cal H} \omega^i)
- (\chi^{i\prime}_{~k}+\omega^{i}_{'k}-\omega_k^{,i}) E q^k \nonumber \\
&+&\left[{\cal H}\omega^i \delta_{jk}- \frac{1}{2} 
(\chi^{i}_{~j,k}+\chi^i_{~k,j} - \chi_{jk}^{~,i}) \right] \frac{q^j q^k}{E}\, .
\end{eqnarray}
We can now write the total time derivative of the distribution function as
\begin{eqnarray}
\label{Dg}
\frac{d g}{d \eta}&=& 
 \frac{\partial g}{\partial \eta}+\frac{q}{E} n^i e^{\Phi+\Psi}
\left(1-\omega_i n^i -\frac{1}{2} 
\chi_{km}n^kn^m \right) \frac{\partial g}{\partial x^i} \nonumber \\
&+& \biggl[ -({\cal H}-\Psi')q^i+\Psi_{,k}\frac{q^iq^k}{E} e^{\Phi+\Psi}-
\Phi^{,i}E e^{\Phi+\Psi} 
-\Psi_{,i}\frac{q^2}{E} e^{\Phi+\Psi} \nonumber \\
&-& E(\omega^{i\prime}+{\cal H} \omega^i)
- (\chi^{i\prime}_{~k}+\omega^{i}_{'k}-\omega_k^{,i}) E q^k \nonumber \\
&+& 
\biggl({\cal H}\omega^i \delta_{jk}-\frac{1}{2}(\chi^{i}_{~j,k}+\chi^i_{~k,j} -
\chi_{jk}^{~,i})\biggr) \frac{q^j q^k}{E} 
\biggr] \frac{\partial g}{\partial q^i}\, .
\end{eqnarray}
This equation is completely general since we have just solved for the 
kinematics of massive particles. 
As far as the collision terms are concerned, for the system of electrons 
and protons we consider the Coulomb scattering processes 
between the electrons and protons and the Compton scatterings between 
photons and electrons 
\begin{eqnarray}
\label{boltzep}
\frac{dg_e}{d\eta}({\bf x},{\bf q},\eta)&=&\langle c_{ep} \rangle_{QQ'q'} 
+\langle c_{e\gamma} \rangle_{pp'q'} \\
\label{boltzep2}
\frac{dg_p}{d\eta}({\bf x},{\bf Q},\eta)&=&\langle c_{ep} \rangle_{qq'Q'}\, ,
\end{eqnarray} 
where we have adopted the same formalism of Ref.~\cite{Dodelsonbook} 
with ${\bf p}$ and ${\bf p}'$ the initial and final momenta of the 
photons, ${\bf q}$ and ${\bf q}'$ the corresponding quantities for the 
electrons and for protons ${\bf Q}$ and ${\bf Q}'$. The 
integral  over different momenta is indicated by   
\begin{equation}
\langle \cdots \rangle_{pp'q'} \equiv \int \frac{d^3p}{(2\pi)^3}\,\int 
\frac{d^3 p'}{(2\pi)^3}\, 
\int \frac{d^3q'}{(2\pi)^3} \dots \, ,
\end{equation}
and thus one can read 
$c_{e\gamma}$ as the unintegrated part of Eq.~(\ref{collisionterm0}),
 and similarly for $c_{ep}$ (with the appropriate amplitude
$|M|^2$). In Eq.~(\ref{boltzep}) Compton scatterings between protons and 
photons can be safely neglected because the amplitude of 
this process has a much smaller amplitude than Compton scatterings with
 electrons being weighted by the inverse squared mass of the 
particles. 

At this point for the photons we considered the perturbations around the 
zero-order Bose-Einstein distribution function (which are 
the unknown quantities). For the electrons (and protons) we can take the 
thermal distribution described by Eq.~(\ref{gel}). 
Moreover we will take the moments of Eqs.~(\ref{boltzep})-(\ref{boltzep2}) 
in order to find the energy-momentum continuity equations.

\subsubsection{Energy continuity equations} 
 
We now integrate Eq.~(\ref{Dg}) over $d^3q/(2\pi)^3$. Let us recall that 
in terms of the distribution function 
the number density $n_e$ and the bulk velocity ${\bf v}$ are given by
\begin{equation}
\label{defne}
n_e=\int \frac{d^3 q}{(2\pi)^3}\, g \, , 
\end{equation} 
and 
\begin{equation}
\label{defvg}
v^i= \frac{1}{n_e} \int \frac{d^3q}{(2\pi)^3}\, g \, \frac{q n^i}{E}\, ,
\end{equation}
where one can set $E\simeq m_e$ since we are considering non-relativistic 
particles. 
We will also make use of the following relations when integrating over the 
solid angle $d\Omega$
\begin{equation}
\label{relOmega}
\int d\Omega\, n^i=\int d\Omega\, n^in^jn^k=0\, ,\quad \int 
\frac{d\Omega}{4\pi}\, n^in^j =  \frac{1}{3} \delta^{ij}\, . 
\end{equation}
Finally notice that $dE/dq=q/E$ and $\partial g/\partial q= (q/E) 
\partial g/\partial E$.

Thus the first two integrals just brings $n'_e$ and $(n_e v^i)_{,i}$. 
Notice that all the terms proportional to the second-order 
vector and tensor perturbations of the metric give a vanishing contribution 
at second order since in this case we can take the 
zero-order distribution functions which depends only on $\eta$ and $E$, 
integrate over the direction 
and use the fact that $\delta^{ij} \chi_{ij}=0$. The trick to solve 
the remaining integrals is an integration  by parts over $q^i$.  
We have an integral like (the one multiplying $( \Psi'-{\cal H})$)
\begin{equation}
\label{r1}
\int \frac{d^3q}{(2 \pi)^3} q^i \frac{\partial g}{\partial q^i} = 
-3 \int \frac{d^3q}{(2 \pi)^3} g =-3 n_e \, ,
\end{equation}
after an integration by parts over $q^i$. The remaining integrals can 
be solved still by integrating by parts over $q^i$.  
The integral proportional to $\Phi^{,i}$ in Eq.~(\ref{Dg}) gives  
\begin{equation}
\int \frac{d^3q}{(2 \pi)^3} E =- v_i \, n_{e}\, ,
\end{equation}
where we have used the fact that $dE/dq^i=q^i/E$. For the integral 
\begin{equation}
\int \frac{d^3q}{(2 \pi)^3} \frac{q^iq^k}{E} 
\frac{\partial g}{\partial q^i}\, , 
\end{equation}
the integration by parts brings two pieces, one from the derivation 
of $q^iq^k$ and one from the derivation of the energy $E$
\begin{equation}
\label{exint}
- 4 \int \frac{d^3q}{(2 \pi)^3} g\frac{q^k}{E} +     
\int \frac{d^3q}{(2 \pi)^3} g \frac{q^2}{E} \frac{q^k}{E} 
=-4 v^k\, n_e + 
\int \frac{d^3q}{(2 \pi)^3} g \frac{q^2}{E^2} \frac{q^k}{E} \, .
\end{equation}
The last integral in Eq.~(\ref{exint}) can indeed be neglected. To check 
this one makes use of the explicit  expression~(\ref{gel}) for the 
distribution function $g$ to derive 
\begin{equation}
\frac{\partial g}{\partial v^i}=g\frac{q_i}{T_e}-\frac{m_e}{T_e}v_ig\, ,
\end{equation} 
and 
\begin{equation}
\label{gpp}
\int \frac{d^3q}{(2 \pi)^3} g q^i q^j =\delta^{ij} n_e m_e T_e+ n_e m_e^2 
v^i v^j\, .
\end{equation}
Thus it is easy to compute
\begin{equation}
\label{finaleener}
\frac{\Psi_{,k}}{m_e^3} \int \frac{d^3q}{(2 \pi)^3} g q^2q^k
= - \Psi_{,k}v^2\frac{T_e}{m_e} 
+3 \Psi_{,k}v_kn_e\frac{T_e}{m_e}+
\Psi_{,k}v_kv^2 \, , 
\end{equation}
which is negligible taking into account that $T_e/m_e$ is of the 
order of the thermal velocity squared.  

With these results we are now able to compute the left-hand side  of the 
Boltzmann equation~(\ref{boltzep}) integrated over 
$d^3q/(2\pi)^3$. The same operation must be done for the collision terms 
on the right hand side. For example for the 
first of the equations in~(\ref{boltzep}) this brings to the integrals 
$ \langle c_{ep} \rangle_{QQ'qq'} +\langle c_{e\gamma} \rangle_{pp'qq'}$. 
However looking at Eq.~(\ref{collisionterm}) one realizes 
that $\langle c_{e\gamma} \rangle_{pp'qq'}$ vanishes because the 
integrand is antisymmetric under the change ${\bf q} 
\leftrightarrow {\bf q'}$ and ${\bf p} 
\leftrightarrow {\bf p}'$. In fact this is simply a consequence of the 
fact that the electron number is conserved for this process. 
The same argument holds for the other term $\langle c_{ep} \rangle_{QQ'qq'}$. 
Therefore the right-hand side of 
Eq.~(\ref{boltzep}) integrated over $d^3q/(2\pi)^3$ vanishes and we can give 
the evolution equation for $n_e$. Collecting 
the results of Eq.~(\ref{r1}) to~(\ref{finaleener}) we find 
\begin{equation}
\label{cont_bar}
\frac{\partial n_e}{\partial \eta}+e^{\Phi+\Psi} 
\frac{\partial(v^i n_e)}{\partial x^i}+3(
{\cal H}-\Psi')n_e + e^{\Phi+\Psi} v^k n_e \left(\Phi_{,k} -2 \Psi_{,k}\right) 
= 0\, . 
\end{equation}
Similarly, for CDM particles, we find
\begin{eqnarray}
\frac{\partial n_{\rm CDM}}{\partial \eta}&+&e^{\Phi+\Psi} \frac{\partial(v^i 
n_{\rm CDM})}{\partial x^i}+3(
{\cal H}-\Psi')n_{\rm CDM} \nonumber \\
&+& e^{\Phi+\Psi} v_{\rm CDM}^k\,
n_{\rm CDM} \biggl( \Phi_{,k} - 2 \Psi_{,k} \biggr) = 0 \, .
\end{eqnarray}

\subsubsection{Momentum continuity equations}
 
Let us now multiply Eq.~(\ref{Dg}) by $(q^i/E) /(2 \pi)^3$ and 
integrate over $d^3q$. In this way we will find the continuity 
equation for the momentum of baryons. The first term just gives 
$(n_e v^i)'$. The second integral is of the type 
\begin{equation}
\frac{\partial}{\partial x^j} \int \frac{d^3q}{(2 \pi)^3} g\, 
\frac{q n^j}{E} \frac{qn^i}{E} =
\frac{\partial}{\partial x^j}\left( n_e \frac{T_e}{m_e} \delta^{ij}+n_e 
v^i v^j \right)\, ,
\end{equation} 
where we have used Eq.~(\ref{gpp}) and $E=m_e$. The third term proportional 
to $({\cal H}-\Psi')$ is 
\begin{eqnarray}
\label{FirstI}
\int \frac{d^3q}{(2 \pi)^3} q^k \frac{\partial g}{\partial q_k} \frac{q^i}{E}=
4 n_e+\int \frac{d^3q}{(2 \pi)^3} g \frac{q^2}{E^2}\frac{q^i}{E}\, ,
\end{eqnarray}
where we have integrated by parts over $q^i$. Notice that the last 
term in Eq.~(\ref{FirstI}) is negligible being the same integral we 
discussewd above in Eq.~(\ref{finaleener}). By the same arguments that 
lead to neglect the term of Eq.~(\ref{finaleener}) it is easy to 
check that all the remaining integrals proportional to the gravitational
 potentials are negligible except for 
\begin{equation}
- e^{\Phi+\Psi} \Phi_{,k}\int \frac{d^3q}{(2 \pi)^3} 
\frac{\partial g}{\partial q_k} q^i=n_ee^{\Phi+\Psi}\Phi^{,i}\, . 
\end{equation}
The integrals proportional to the second-order vector and tensor 
perturbations vanish as 
vector and tensor perturbations are traceless and divergence-free. The 
only one which survives is the term proportional to $\omega^{i\prime}
+{\cal H}\omega^{i}$ in Eq.~(\ref{Dg}). 

Therefore  for the integral over $d^3q q^i/E$ 
of the left-hand side of the Boltzmann equation~(\ref{Dg})
for a massive particle with mass $m_e$ ($m_p$) and distribution 
function~(\ref{gel}) we find 
\begin{eqnarray}   
\label{dgetau}   
&&\int \frac{d^3q}{(2\pi)^3} \frac{q^i}{E} 
\frac{d g_e}{d\eta} =  \frac{\partial (n_e v^i)}{\partial \eta}+
4 ({\cal H}-\Psi') n_e v^i +\Phi^{,i} e^{\Phi+\Psi} 
n_e \nonumber \\
&&+ e^{\Phi+\Psi} \left( n_e \frac{T_e}{m_e} \right)^{,i}+
 e^{\Phi+\Psi} \frac{\partial}{\partial x^j}(n_e v^j v^i) +
\frac{\partial \omega^i}{\partial \eta} n_e + {\cal H} \omega^i n_e \, . 
\nonumber \\
\end{eqnarray}

Now, in order to derive the momentum conservation equation for baryons, we 
take the first moment of both Eq.~(\ref{boltzep}) 
and~(\ref{boltzep2}) multiplying them by ${\bf q}$ and ${\bf Q}$ 
respectively and integrating over the momenta. Since previously we 
integrated the left-hand side of these equations over $d^3q q^i/E$, 
we just need to multiply 
the previous integrals by $m_e$ for the electrons and for $m_p$ for 
the protons. Therefore if we sum the 
first moment of Eqs.~(\ref{boltzep}) and~(\ref{boltzep2}) the dominant 
contribution on the left-hand side will be that of the protons
\begin{eqnarray}
\int \frac{d^3 Q}{(2 \pi)^3} Q^i\, \frac{dg_p}{d\eta}=\langle c_{ep} 
(q^i+Q^i)\rangle_{QQ'qq'} +\langle c_{e\gamma} q^i\rangle_{pp'qq'}\, .
\end{eqnarray}  
Notice that the integral of the Coulomb collision term $c_{ep} (q^i+Q^i)$ 
over all momenta vanishes simply because of momentum conservation 
(due to the Dirac function $\delta^4(q+Q-q'-Q')$). As far as the Compton 
scattering is concerned we have that, following 
Ref.~\cite{Dodelsonbook}, 
\begin{equation}
\langle c_{e\gamma} q^i \rangle_{pp'qq'} =- \langle c_{e\gamma} p^i 
\rangle_{pp'qq'} \, ,  
\end{equation}   
still because of the total momentum conservation. Therefore what we 
can compute now is the integral over all momenta of 
$c_{e\gamma} p^i$. Notice however that this is equivalent just to 
multiply the Compton collision term $C({\bf p})$ 
of Eq.~(\ref{collisionterm}) by $p^i$ and integrate over $d^3p/(2\pi^3)$ 
\begin{equation}
\label{Ci}
\langle c_{e\gamma} p^i \rangle_{pp'qq'} = a e^{\Phi} \int 
\frac{d^3p}{(2\pi)^3} p^i C({\bf p})\, .
\end{equation}
where $C({\bf p})$ has been already computed in Eqs.~(\ref{C1p}) 
and~(\ref{C2p}). 

We will do the integral~(\ref{Ci}) in the following. First let us 
introduce the definition of the velocity of photons in terms of 
the distribution function 
\begin{equation}
\label{vp}
(\rho_\gamma+p_\gamma) v^i_\gamma = \int \frac{d^3 p}{(2 \pi)^3} f p^i\, ,
\end{equation}  
where $p_\gamma = \rho_\gamma/3$ is the photon pressure and 
$\rho_\gamma$ the energy density. At first order we get
\begin{equation}
\label{vp1}
\frac{4}{3} v^{(1) i}_\gamma= \int \frac{d\Omega}{4\pi} 
\Delta^{(1)}\, n^i \, ,
\end{equation}
where $\Delta$ is the photon distribution anisotropies defined 
in Eq.~(\ref{Delta2}). At second order we instead find
\begin{equation}
\label{vp2}
\frac{4}{3} \frac{v^{(2) i}_\gamma}{2}= \frac{1}{2} \int 
\frac{d\Omega}{4\pi} \Delta^{(2)}\, n^i-\frac{4}{3} \delta^{(1)}_\gamma 
v^{(1)i}_\gamma \, .
\end{equation}
Therefore the terms in Eqs.~(\ref{C1p}) and~(\ref{C2p}) 
proportional to $f^{(1)}({\bf p})$ and $f^{(2)}({\bf p})$ 
will give rise to terms containing the velocity of the photons. 
On the other hand 
the terms proportional to $f^{(1)}_0(p)$ and $f^{(2)}_{00}(p)$, 
once integrated, vanish because of the integral over the momentum direction
$n^i$, $\int d\Omega n^i=0$. Also the integrals involving 
$P_2({\bf {\hat v}}\cdot {\bf n})=[3 ({\bf {\hat v}}\cdot 
{\bf n})^2-1]/2$ 
in the first line of Eq.~(\ref{C1p}) and~(\ref{C2p}) 
vanish since 
\begin{equation}
\int d\Omega P_2({\bf {\hat v}}\cdot {\bf n})\,  n^i= {\hat v}^k 
{\hat v}^j  \int d\Omega n_kn_jn^i=0\, ,
\end{equation}
where we are using the relations~(\ref{relOmega}). Similarly all 
the terms proportional to $v$, 
$({\bf v} \cdot {\bf n})^2$ and $v^2$ do not give any contribution 
to Eq.~(\ref{Ci}) and, 
in the second-order collision term, one can check that 
$\int d\Omega Y_2({\bf n}) n^i =0$. Then there are terms proportional to 
$({\bf v}\cdot {\bf n}) f^{(0)}(p)$, $({\bf v}\cdot {\bf n}) p 
\partial f^{(0)}/\partial p$ and 
$({\bf v}\cdot {\bf n}) p \partial f^{(1)}_0/\partial p$ for which 
we can use the rules~(\ref{rules}) when 
integrating over $p$ while the integration over the momentum
 direction is   
\begin{equation}
\int \frac{d\Omega}{4\pi} ({\bf v}\cdot {\bf n}) n^i =v_k  \int 
\frac{d\Omega}{4\pi}  n^kn^i= \frac{1}{3} v^i \, .
\end{equation}
Finally from the second line of Eq.~(\ref{C2p}) we get three integrals. One is 
\begin{equation}
\int \frac{d^3p}{(2\pi)^3} p^i \, ({\bf v}\cdot {\bf n}) 
f^{(1)}({\bf p})= \bar{\rho}_\gamma 
\int \frac{d\Omega}{4 \pi} \Delta^{(1)} ({\bf v}\cdot {\bf n})  n^i\, , 
\end{equation} 
where $\bar{\rho}_\gamma$ is the background energy density of the photons. 
The second comes from  
\begin{eqnarray}
&&\frac{1}{2} \int \frac{d^3p}{(2\pi)^3} p^i\,   ({\bf v} 
\cdot {\bf n}) P_2({\bf {\hat v}}\cdot {\bf n}) \left(f^{(1)}_2(p)-
p \frac{\partial f^{(1)}_2(p) }{\partial p} \right) \\
&&= \frac{5}{4} 
\bar{\rho}_\gamma \Delta^{(1)}_2 
\left[ 3 
v_j  {\hat v}_k {\hat v}_l \int \frac{d\Omega}{4 \pi} n^in^jn^kn^l-v_j 
\int \frac{d\Omega}{4 \pi} n^in^j 
\right] = \frac{1}{3} \bar{\rho}_\gamma \Delta^{(1)}_2 {\hat v}^i\, ,
\nonumber 
\end{eqnarray}
where we have used the rules~(\ref{rules}), Eq.~(\ref{relOmega}) and 
$\int (d\Omega/4 \pi)\, n^in^jn^kn^l = (\delta^{ij} \delta^{kl} 
+\delta^{ik} \delta^{lj}+\delta^{il} \delta^{jk})/15$. In fact the 
third integral 
\begin{equation}
- \int \frac{d^3p}{(2\pi)^3} p^i ({\bf v} \cdot {\bf n}) f^{(1)}_2(p)\, , 
\end{equation}
exactly cancels the previous one. Summing the various integrals we find  
\begin{eqnarray}
\label{Ciint}
&&\int\frac{d{\bf p}}{(2\pi)^3} C({\bf p}) {\bf p}= 
n_e \sigma_T {\bar \rho_\gamma} \Bigg[ \frac{4}{3} 
({\bf v}^{(1)}-{\bf v}_\gamma^{(1)})
-\int \frac{d\Omega}{4\pi} \frac{\Delta^{(2)}}{2} {\bf n} \nonumber\\
&&+ \frac{4}{3} 
\frac{{\bf v}^{(2)}}{2} 
+\frac{4}{3} \delta^{(1)}_e ({\bf v}^{(1)}-{\bf v}_\gamma^{(1)})+ 
\int \frac{d\Omega}{4\pi} \Delta^{(1)} ({\bf v} \cdot {\bf n}) {\bf n} 
+ \Delta_0^{(1)} {\bf v} \Bigg] \;.
\end{eqnarray}
Eq.~(\ref{Ciint}) can be further simplified. Recalling that 
$\delta^{(1)}_\gamma = \Delta^{(1)}_0$ we use Eq.~(\ref{vp2}) and notice that 
\begin{equation}
\int \frac{d\Omega}{4\pi} \Delta^{(1)}\, ({\bf v} \cdot {\bf n}) 
n^i = v_j^{(1)} \Pi^{ji}_\gamma+\frac{1}{3} v^i \Delta^{(1)}_0 \, ,
\end{equation}
where the photon quadrupole $\Pi^{ij}_\gamma$ is defined as 
\begin{equation}
\label{quadrupole}
\Pi^{ij}_{\gamma}=\int\frac{d\Omega}{4\pi}\,\left(n^i n^j-\frac{1}{3}
\delta^{ij}\right)\left(\Delta^{(1)}+\frac{\Delta^{(2)}}{2}\right)\, . 
\end{equation}

Thus, our final expression for the integrated collision term~(\ref{Ci}) reads
\begin{eqnarray}
\label{Cifinal}
&&\int \frac{d^3 p}{(2\pi)^3} C({\bf p}) p^i =
n_e \sigma_T {\bar \rho_\gamma}
\left[ \frac{4}{3} (v^{(1)i}-v_\gamma^{(1)i}) + \frac{4}{3} 
\left( \frac{v^{(2)i}}{2}-\frac{v_\gamma^{(2)i}}{2} \right) 
\right. \nonumber \\ && 
+ \left. \frac{4}{3} \left(\delta^{(1)}_e +\Delta^{(1)}_0\right) (v^{(1)i}-
v_\gamma^{(1)i})+ v^{(1)}_j \Pi^{ji}_\gamma \right ]\, .
\end{eqnarray}

We are now able to give the momentum continuity equation for baryons
by combining $m_p dg_p/d\eta$ from Eq.~(\ref{dgetau}) with the 
collision term~(\ref{Ci})
\begin{eqnarray}
\label{mcc}
\frac{\partial (\rho_b v^i)}{\partial \eta} &+&4 ({\cal H}-\Psi') 
\rho_b v^i +\Phi^{,i} e^{\Phi+\Psi} 
\rho_b+ e^{\Phi+\Psi} \left( \rho_b \frac{T_b}{m_p}\right)^{,i}
\nonumber \\
& + & e^{\Phi+\Psi} \frac{\partial}{\partial x^j}(\rho_b v^j v^i) +
\frac{\partial \omega^i}{\partial \eta} \rho_b+ {\cal H} \omega^i 
\rho_b  \nonumber \\
&= & - n_e \sigma_T a\, {\bar \rho_\gamma}
\left[ \frac{4}{3} (v^{(1)i}-v_\gamma^{(1)i}) + \frac{4}{3} 
\left( \frac{v^{(2)i}}{2}-\frac{v_\gamma^{(2)i}}{2} \right) 
\right. \nonumber \\
&+& \left. 
\frac{4}{3} \left( \delta^{(1)}_b+\Delta^{(1)}_0+\Phi^{(1)} 
\right) (v^{(1)i}-v_\gamma^{(1)i})+v^{(1)}_j \Pi^{ji}_\gamma \right ] \, ,
\end{eqnarray}
where $\rho_b$ is the baryon energy density and, as we previously explained, 
we took into account that to a good 
approximation the electrons do not contribute to the mass of baryons. 
In the following we will expand explicitly at first and second-order 
Eq.~(\ref{mcc}).

\subsubsection{Second-order momentum continuity equation for baryons}

At first order we find
\begin{equation}
\label{mcc1}
\frac{\partial v^{(1)i}}{\partial \eta} +{\cal H} v^{(1)i}+\Phi^{(1),i}=
\frac{4}{3} \tau' \frac{{\bar \rho}_\gamma}{{\bar \rho}_b} 
\left(  v^{(1)i}-v^{(1)i}_\gamma \right)\, .
\end{equation} 
At second order there are various simplifications. In particular notice 
that the term on the right-hand side of Eq.~(\ref{mcc}) which is proportional 
to $\delta_b$ vanishes when matched to expansion of the left-hand side by 
virtue of the first-order equation~(\ref{mcc1}). Thus, at the end 
we find a very simple equation
\begin{eqnarray}
& & \frac{1}{2} \left( \frac{\partial v^{(2)i}}{\partial \eta} 
+{\cal H} v^{(2)i} + 
2 \frac{\partial \omega^i}{\partial \eta} +2 {\cal H} \omega_i  + 
\Phi^{(2),i}\right) - \frac{\partial \Psi^{(1)}}{\partial \eta} 
v^{(1)i} \nonumber \\
&& + v^{(1)j}\partial_jv^{(1)i}+(\Phi^{(1)}+\Psi^{(1)}) \Phi^{(1),i} 
+\left( \frac{T_b}{m_p} \right)^{,i} = \frac{4}{3} \tau' 
\frac{{\bar \rho}_\gamma}{{\bar \rho}_b} \\
&& \times 
\left[ \left(  \frac{v^{(2)i}}{2}-\frac{v^{(2)i}_\gamma}{2}  \right) +
\left(\Delta^{(1)}_0+\Phi^{(1)}  \right) 
\left(  v^{(1)i}-v^{(1)i}_\gamma  \right)
+\frac{3}{4} v^{(1)}_j \Pi^{ji}_\gamma 
\right] \, \nonumber ,
\end{eqnarray}   
with $\tau'=- {\bar n}_e \sigma_T a$.

\subsubsection{Second-order momentum continuity equation for CDM}

Since CDM particles are collisionless, at first order we find
\begin{equation}
\label{mcc1CDM}
\frac{\partial v_{\rm CDM}^{(1)i}}{\partial \eta} +{\cal H} 
v_{\rm CDM}^{(1)i}+\Phi^{(1),i}=0\, .
\end{equation}
At second order we find 
\begin{eqnarray}
\label{mcc2CDM}
&&\frac{1}{2} \left( \frac{\partial v_{\rm CDM}^{(2)i}}{\partial \eta} 
+{\cal H} v_{\rm CDM}^{(2)i} + 
2 \frac{\partial \omega^i}{\partial \eta} +2 {\cal H} \omega_i  + 
\Phi^{(2),i}\right) - \frac{\partial \Psi^{(1)}}{\partial \eta} 
v_{\rm CDM}^{(1)i} \nonumber \\
&&+ v_{\rm CDM}^{(1)j}\,  \partial_j v_{\rm CDM}^{(1)i}+
(\Phi^{(1)}+\Psi^{(1)}) \Phi^{(1),i} 
+\left( \frac{T_{\rm CDM}}{m_{\rm CDM}} \right)^{,i} 
=0\, .
\end{eqnarray}   

\section{CMB anisotropies at second-order at all scales: analytical approach}
\label{Oscill}
\subsection{Towards a second-order CMB radiation transfer function} 
\label{To}
As pointed out in Sec.~\ref{allscales} various non-primordial sources of non-Gaussianity for the CMB anisotropies can arise from the 
non-linear evolution of the cosmological perturbations, including the Sunyaev-Z'eldovich, ISW (Rees-Sciama) effects, and the gravitational lensing, with 
possible correlations between these contributions. Here we will focus on another relevant source of non-Gaussianity: the non-linear effects operating at
the recombination epoch. The dynamics at recombination is quite involved because all the non-linearities in the evolution of the baryon-photon fluid at 
recombination and  the ones coming from general relativity should be accounted for. 

The following section can be considered as an application of the Boltzmann equations found previously. Despite they depend on a high numbers of terms, 
because of the apperance within second-order perturbation theory of products of fist-oder perturbations, it is remarkable that an analytical study of the 
complicated recombination dynamics is possible. This allows to account for those effects that at the last scattering surface produce 
a non-Gaussian contribution to the CMB anisotropies that add to the primordial one. Such a contribution is so relevant because it 
represents a major part of the second-order radiation transfer function which must be determined in order to have a 
complete control of both the primordial and non-primordial part of NG in the CMB anisotropies and to gain from the theoretical side 
the same level of precision that could be reached experimentally in the near future~\cite{review,Baumann,SL}.

In order to achieve this goal, we consider the Boltzmann equations derived in Sec.~\ref{allscales} 
at second-order describing the evolution of the photon, baryon and CDM  fluids, and we manipulate them further under the 
assumption of tight coupling beteween photons and baryons. This leads to the generalization at second-order of the equations for the 
photon energy density and velocity perturbations which govern the acoustic oscillations of the photon-baryon fluid for modes that are 
inside the horizon at recombination. The evolution is that of a damped harmonic oscillator, with a source term which is given by the 
gravitational potentials generated by the different species. An interesting result is that, unlike the linear case, 
at second-order the quadrupole moment of the photons is not suppressed in the tight coupling limit and 
it must be taken into account.

The analytical solutions for the acoustic oscillations of the photon-baryon fluid 
at second-order are derived adopting some simplifications which are also standard for an analytical treatment of 
the linear CMB anisotropies, and which nonetheless allow to catch most of the physics at recombination. One of these simplifications 
is to study separately two limiting regimes: intermediate scales which enter the horizon in between the equality epoch 
($\eta_{eq}$) and the recombination epoch ($\eta_r$), with $\eta_r^{-1} \ll k \ll \eta_{eq}^{-1}$, and shortwave perturbations, 
with $k \gg \eta_{eq}^{-1}$, which enter the horizon before the equality epoch. 
Here our main concern is to provide a simple estimate of the quantitative behaviour of the non-linear evolution taking place at recombination, 
offering at the same time all the tools for a more accurate computation.  We find  that the second-order 
CMB anisotropies generated on the last scattering surface do not reduce only to the energy density and 
velocity perturbations of the photons evaluated at recombination, but a number of second-order corrections at last 
scattering arise from the Boltzmann equations of Ref.~\cite{paperI} in the form of products of first-oder perturbations. 

We will see that the dynamics at recombination is indeed dominated on small scales by the non-linear 
evolution of the second-order gravitational potentials feeded by the Cold Dark Matter density perturbations. The gravitational potentials determine the 
energy density fluctuations of the photons at recombination and their effects show up in the CMB anisotropies as 
\begin{equation}
\label{SW}
\frac{\Delta T}{T}=\frac{1}{4} \Delta^{(2)}_{00}+\Phi^{(2)}\, ,
\end{equation}
which is the usual term due to  the intrinsic fractional temperature fluctuation  $\Delta^{(2)}_{00}/4$ on the last scattering surface ($\Delta^{(2)}_{00} $ is the monople of the photon distribution function) and the gravitational redshift due to the gravitational potential.  
However the analysis of the remaining contributions that come in the form of products of first-oder perturbations is equally important. The reason is that 
one of the central quantities we are interested in is the \emph{contamination} to the primordial non-Gaussianity that is produced by the secondary effects. In that respect the reasonable question to ask is which kind of primordial non-Gaussianity a given secondary effect can contaminate most, wether it is of the so called ``local type'' or of the ``equilateral type'', for example. Therefore it might well be the case that, even if some secondary effects appear to be the dominant ones, they might give a high contamination to a given type of primordial non-Gaussianity, but a low contamination to a different kind of primordial non-Gaussianity,  for which ``subdominant '' terms, on the other hand,  represent the strongest contaminant. This section and the following two provide a clear example: we anticipate here that while the term in Eq.~(\ref{SW}) will mainly mimic an equilateral type of primordial non-Gaussianity, second-order effects that come as products of first-order times first-order perturbations actually mainly contaminate local primordial non-Gaussianity.

Notice that the case $k \gg \eta_{eq}^{-1}$  has been treated in two steps. First we 
just assume a radiation dominated universe, and then we provide a much more complete analysis by solving 
the evolution of the perturbations from the equality epoch onwards taking into account that the dark matter perturbations 
around the equality epoch tend to dominate the second-order gravitational potentials. As a byproduct, this last step  
provides the Meszaros effect at second-order. In deriving the analytical solutions we have accurately accounted for the initial 
conditions set on superhorizon scales by the primordial non-Gaussianity. In fact the primordial contribution is always transferred 
linearly, while the real new contribution to the radiation transfer function is given by all the additional terms provided in the 
source functions of the equations. Let us stress here that the analysis of the CMB bispectrum performed so far, as for example in 
Ref.~\cite{KomatsuWMAP}, adopt just the linear radiation transfer function (unless the bispectrum originated by specific secondary 
effects, such as Rees-Sciama or Sunyaev-Zel'dovich effects, are considered, see on this Sec.~\ref{cont}). \\
The formalism to provide in a systematic way the second-order CMB radiation transfer function will be reviewd in Sec.~\ref{NumAn}.






\subsection{The Boltzmann equations in the tightly coupled limit}
\label{one}
We now derive the moments of the Boltzmann equations for photons in the limit when the 
photons are tightly coupled to the the baryons (the electron-proton system) due to  Compton scattering. This leads to the governing equations for the acoustic 
oscillations of the photon-baryon fluid. The well-known computation at linear order is briefly reviewed in ~\ref{ApplinearBoltz} under some standard simplifying 
assumptions. Here we will focus on the derivation    
of  the equations at second-order in the perturbations, pointing out some interesting differences with respect to the linear case. 
In particular note that ,while we already know that the L.H.S. of the equations at second-order will have the same form as for the linear case, 
the source term on the R.H.S. of the moments of the Boltzmann equations will also consist of first-order squared terms.

\subsubsection{Energy continuity equation}
Let us start by integrating Eq.~(\ref{B2}) over $d  \Omega_{\bf n}/4\pi$ to get the evolution equation for the second-order 
photon energy density perturbations $\Delta^{(2)}_{00}$
\begin{eqnarray}
\label{delta200}
\Delta^{(2 )\prime}_{00}&+&\frac{4}{3} \partial_i v^{(2)i}_\gamma +\frac{8}{3} \partial_i 
\left( \Delta^{(1)}_{00} v^{(1)i}_\gamma \right) -4 \Psi^{(2) \prime} + \frac{8}{3} (\Phi^{(1)}+\Psi^{(1)} )
\partial_i v^{(1)i}_\gamma \nonumber \\
&+&2  \int \frac{d\Omega_{\bf n}}{4 \pi} \left[(\Phi^{(1)}_{,j}+\Psi^{(1)}_{,j})n^in^j 
-(\Phi^{,i}+\Psi^{,i})\right]
\frac{\partial \Delta^{(1)}}{\partial n^i}\nonumber \\
&-&8 \Psi^{(1)\prime} \Delta^{(1)}_{00}
+\frac{32}{3} \Phi^{(1)}_{,i}v^{(1)i}_\gamma=- \frac{8}{3} \tau' v^{(1)}_i \left( v^{(1)i}-v^{(1)i}_\gamma \right)\, ,
\end{eqnarray} 
where we have used the explicit definition for the second-order velocity of the photons~\cite{paperI} 
\begin{equation}
\label{vp2}
\frac{4}{3} \frac{v^{(2) i}_\gamma}{2}= \frac{1}{2} \int \frac{d\Omega}{4\pi} \Delta^{(2)}\, 
n^i-\frac{4}{3} \delta^{(1)}_\gamma 
v^{(1)i}_\gamma \, .
\end{equation}
We can now make use of the tight coupling expansion to simplify Eq.~(\ref{delta200}). In the L.H.S. we use 
$\partial_i v^{(1)i}_\gamma=\partial_i v^{(1)i}=3 \Psi^{(1)\prime}-\delta^{(1)\prime}_b=
3 \Psi^{(1)\prime}-3\Delta^{(1)\prime}_{00}/4$ 
obtained in the tightly coupled limit from Eqs.~(\ref{LH2bcont}) and~(\ref{LH2D100d1b}). On the other 
hand in the R.H.S. of Eq.~(\ref{delta200}) 
\begin{eqnarray}
\label{rel}
\left( v^{(1)i}-v^{(1)i}_\gamma \right)&=&\frac{R}{\tau'}
\left(v^{(1)i \prime }_\gamma+{\cal H} v^{(1)i}_\gamma+\Phi^{(1),i}  \right)
\nonumber \\
&=&
\frac{R}{\tau'} \left( 
\frac{\cal H}{1+R} v^{(1)i}_\gamma-\frac{1}{4}\frac{\Delta^{(1),i}_{00}}{1+R} \right)\, , 
\end{eqnarray}
by using Eq.~(\ref{LH2vv}) and the evolution equation for the photon veleocity~(\ref{LH2vphotontight}). 
Here we introduce the baryon-photon ratio
\begin{equation}
R=\frac{3}{4} \frac{\rho_b}{\rho_\gamma}\, .
\end{equation}
We thus 
arrive at the following equation
\begin{equation}
\label{D2eq}
\Delta^{(2)'}_{00}+\frac{4}{3} \partial_i v^{(2)i}_\gamma-4\Psi^{(2)'}={\cal S}_\Delta\, ,
\end{equation}
where the source term is given by
\begin{eqnarray}
\label{SD}
{\cal S}_\Delta&=&\left( \Delta^{(1)2}_{00} \right)^{\prime}-2(\Phi^{(1)}+\Psi^{(1)})(4\Psi^{(1)\prime}
-\Delta^{(1)\prime}_{00})-\frac{8}{3} v^{(1)i}_\gamma (\Delta^{(1)}_{00}+4\Phi^{(1)})_{,i}
\nonumber \\
&+&\frac{16}{3}(\Phi^{(1)}+\Psi^{(1)})^{,i}v_i-\frac{8}{3}R 
\left( \frac{\cal H}{1+R} v^{(1)2}_\gamma-\frac{1}{4}\frac{v^{(1)}_{\gamma i} \Delta^{(1),i}_{00}}{1+R}\right)
\, . \nonumber \\
\end{eqnarray}
Notice that the integral in the second line of Eq.~(\ref{delta200}) has been computed by expanding the linear anisotropies 
as $\Delta^{(1)}=\sum_\ell (2\ell+1) \Delta^{(1)}_{\ell} P_{\ell}(\hat{\bf v} \cdot {\bf n})$ and expressing explicitly the dependence on $n^i$ as 
$P_1(\hat{\bf v} \cdot {\bf n})= \hat{v}^in_i$, $P_2(\hat{\bf v} \cdot {\bf n})=(3(\hat{\bf v} \cdot {\bf n})^2-1)/2$ and so on. This allows to perform the 
derivative  $\partial \Delta^{(1)}/{\partial n^i}$. It turns out that the term proportional to the first-order quadrupole vanishes, and higher order terms can be neglected because of the 
tight coupling approximation. Therefore the only term that is non negligible is the dipole term and using $\Delta^{(1)}_1=4 v/3$ in the tight coupling limit, one 
obatins that the integral is equal to $(16/3)(\Phi^{(1)}+\Psi^{(1)})^{,i}v_i$.

\subsubsection{Velocity continuity equation}
We now derive the second moment of the Boltzmann equation~(\ref{B2}) and then we take its tight coupling limit. The integration of 
Eq.~(\ref{B2}) over $d\Omega_{\bf n} n^i/4 \pi$ yields the continuity equation for the photon velocity
\begin{eqnarray}
\label{v2eq}
\frac{4}{3}\frac{v^{(2)i\prime}_\gamma}{2}&+&\frac{1}{2} 
\partial_j \Pi^{(2)ji}_\gamma+\frac{1}{3} \frac{\Delta^{(2),i}_{00}}{2}
+\frac{2}{3} \Phi^{(2),i}+\frac{4}{3}\omega^{i\prime}=
-\frac{4}{3}\left(\Delta^{(1)}_{00} v^{(1)i}_\gamma \right)^\prime \nonumber \\
&+&\frac{16}{3} \Psi^{(1)\prime} v^{(1)i}_\gamma-
4\Phi^{(1)}_{,j}\Pi^{(1)ji}_\gamma
- \frac{4}{3}\Phi^{(1),i} \Delta^{(1)}_{00}
\nonumber \\
&-&(\Phi^{(1)}+\Psi^{(1)})\partial_j \Pi^{(1)ji}_\gamma
-\frac{1}{3}(\Phi^{(1)}+\Psi^{(1)})\Delta^{(1),i}_{00}
\nonumber \\
&+&\int \frac{d\Omega_{\bf n}}{4 \pi} n^i \left[(\Phi^{(1)}_{,j}+\Psi^{(1)}_{,j})n^in^j 
-(\Phi^{,i}+\Psi^{,i})\right]
\frac{\partial \Delta^{(1)}}{\partial n^i}
\nonumber \\
&-&\frac{\tau'}{2} \left[\frac{4}{3}(v^{(2)i}-v^{(2)i}_\gamma)+\frac{8}{3}(\delta^{(1)}_b+\Phi^{(1)} +\Delta^{(1)}_{00})(v^{(1)i}-
v^{(1)}_\gamma) \right. \nonumber \\
& & \left. + 2v^{(1)}_j\Pi^{(1)ji}_\gamma \right] - \frac{4}{3}\Phi^{(1),i}(\Phi^{(1)}+\Psi^{(1)})\, .
\end{eqnarray}
The difference between the second-order baryon and photon velocities $(v^{(2)i}-v^{(2)i}_\gamma)$ appearing in 
Eq.~(\ref{v2eq}) is obtained from
the baryon continuity equation which can be written as (see Ref.~\cite{paperI})
\begin{eqnarray}
\label{vb2}
v^{(2)i}&=&v^{(2)i}_\gamma+\frac{R}{\tau'}\left[\left(v^{(2)i\prime}+{\cal H} v^{(2)i}+2\omega^{i\prime}
+2{\cal H}\omega^i+\Phi^{(2),i}\right)
\right. \nonumber \\
&-& \left.  2\Psi^{(1)\prime} v^{(1)i} \partial_i v^{(1)2}+2\Phi^{(1),i}
(\Phi^{(1)}+\Psi^{(1)})\right] -\frac{3}{2}v^{(1)}_j\Pi^{(1)ji}_\gamma \nonumber \\
&-& 2 (\Delta^{(1)}_{00}+\Phi^{(1)}) (v^{(1)i}-v^{(1)i}_\gamma)\, . 
\end{eqnarray}
We want now to reduce Eq.~(\ref{v2eq}) in the tightly coupled limit. 
We first insert the expression~(\ref{vb2}) in 
Eq.~(\ref{v2eq}). Notice that the last three terms in Eq.~(\ref{vb2}) will cancel out. On the other hand 
in the tight coupling limit expansion one can set $v^{(1)i}=v^{(1)i}_\gamma$ and $v^{(2)i}=v^{(2)i}_\gamma$
in the remaining terms on the R.H.S. of Eq.~(\ref{vb2}). Thus Eq.~(\ref{v2eq}) becomes
\begin{eqnarray}
\label{intermediate}
& &(v^{(2)i}_\gamma +2\omega^i)^\prime+{\cal H} \frac{R}{1+R}(v^{(2)i}_\gamma+2\omega^i)
+\frac{1}{4} \frac{\Delta^{(2),i}_{00}}{1+R}+\Phi^{(2),i}= \nonumber \\
& & - \frac{3}{4(1+R)} \partial_j \Pi^{(2)ji}_\gamma 
-\frac{2}{1+R}\left(\Delta^{(1)}_{00}v^{(1)i}_\gamma \right)^\prime + 
\frac{8}{1+R} \Psi^{(1)\prime} v^{(1)i}_\gamma \nonumber \\
& & - \frac{2}{1+R}\Phi^{(1),i} \Delta^{(1)}_{00}-\frac{2}{1+R}\Phi^{(1),i}
(\Phi^{(1)}+\Psi^{(1)})
\nonumber \\
& & +2 \frac{R}{1+R} \Psi^{(1)\prime} v^{(1)i}_\gamma 
- \frac{1}{2(1+R)}(\Phi^{(1)}+\Psi^{(1)})\Delta^{(1),i}_{00} 
-\frac{R}{1+R} \partial^i v^{(1)2}_\gamma
\nonumber \\
& & -2\frac{R}{1+R} 
(\Phi^{(1)}+\Psi^{(1)}) \Phi^{(1),i} - \tau' \frac{2}{1+R} \delta^{(1)}_b (v^{(1)i}-v^{(1)i}_\gamma)
\, ,
\end{eqnarray}
where in the tightly coupled limit we are neglecting the first-order quadrupole and (higher-order moments) of the photon  
distribution since it is suppressed by $1/\tau$ with respect to the other terms. Moreover the integral in the fourth line of Eq.~(\ref{v2eq}) has been computed following the 
same steps used for the analogous integral in Eq.~(\ref{delta200}). In this case it turns out that the contribution from the first-order dipole term vanishes, while the one 
from thequadrupole is non zero, but this and higher-order terms can be neglected because of the tight coupling approximation, so that the integral is in fact negligible.  
Next for the 
term like $\tau' \delta^{(1)}_b (v^{(1)i}-v^{(1)i}_\gamma)$ we employ the relation previously derived
in Eq.~(\ref{rel}) with $\delta^{(1)}_b=3\Delta^{(1)}_{00}/4$ 
and we use the first order tight coupling equations~(\ref{LH2B1l}) and~(\ref{LH2vphotontight}) in order 
to further simplify Eq.~(\ref{intermediate}). We finally obtain 
\begin{eqnarray}
\label{v2eqf}
v^{(2)i\prime}_\gamma +{\cal H} \frac{R}{1+R}v^{(2)i}_\gamma
+\frac{1}{4} \frac{\Delta^{(2),i}_{00}}{1+R}+\Phi^{(2),i}={\cal S}^{i}_V\, ,
\end{eqnarray}
where
\begin{eqnarray}
\label{SV}
{\cal S}^{i}_V&=&-\frac{3}{4(1+R)} \partial_j \Pi^{(2)ji}_\gamma -2 \omega'_i-2{\cal H} \frac{R}{1+R} \omega^i
+2 \frac{{\cal H}R}{(1+R)^2} \Delta^{(1)}_{00} v^{(1)i}_\gamma
\nonumber \\
&+&\frac{1}{4(1+R)^2}\left(  \Delta^{(1)2}_{00} \right)^{,i}+\frac{8}{3(1+R)}v^{(1)i}_\gamma \partial_j 
v^{(1)j}_\gamma +2 \frac{R}{1+R}\Psi^{(1)\prime} v^{(1)i}_\gamma
\nonumber \\ 
&-&2(\Phi^{(1)}+\Psi^{(1)}) \Phi^{(1),i} 
-\frac{1}{2(1+R)}(\Phi^{(1)}+\Psi^{(1)}) \Delta^{(1),i}_{00}
\nonumber \\
&-&\frac{R}{1+R} \partial^i v^{(1)2}_\gamma 
-\frac{3}{2}\frac{R}{1+R} \Delta^{(1)}_{00}\left(
\frac{{\cal H}}{1+R} v^{(1)i}_\gamma-\frac{1}{4} \frac{\Delta^{(1),i}_{00}}{1+R}
   \right)\, .
\end{eqnarray}
We have spent some time in  giving the details of the computation 
for the photon Boltzmann equations at second-order in the perturbations. As a summary of the results obtained so far 
we refer the reader to Eqs.~(\ref{D2eq}) and (\ref{v2eqf}) as our master equations
which we will solve in the next sections. In particular Eq.~(\ref{v2eqf}) is the second-order counterpart of 
Eq.~(\ref{LH2vphotontight}) for the photon velocity in the tight coupling regime. Notice that there are two important 
differences with respect to the linear case. One is that, in Eq.~(\ref{v2eqf}), there will be a contribution not only from 
scalar perturbations but also from vector modes which, at second-order, are inevitably generated as non-linear 
combinations of first-order scalar perturbations. In particular we have included the vector metric perturbations 
$\omega^i$ in the source term. Second, and most important, we have also kept in the source term the 
second-order quadrupole of the photon distribution $\Pi^{(2)ij}_\gamma$. At linear order we can neglect it 
together with higher order moments of the photons since they turn out to be suppressed with respect to the first two 
moments in the tight coupling limit by increasing powers of $1/\tau$. However in the next section we will 
show that at second order this does not hold anymore, as the photon quadrupole is no longer suppressed. 

Finally following the same steps leading to Eq.~(\ref{LH2eqoscill}) 
at linear order we can derive a similar equation for the second-order 
photon energy density perturbation $\Delta^{(2)}_{00}$ which now will be characterized by the source terms 
${\cal S}_\Delta$ and ${\cal S}^i_V$
\begin{eqnarray}
\label{eqoscille}
& & \left( \Delta^{(2)\prime \prime}_{00}-4\Psi^{(2)\prime \prime} \right) +{\cal H}\frac{R}{1+R} 
\left( \Delta^{(2)\prime}_{00}-4\Psi^{(2)\prime} \right) -c_s^2 \nabla^2 
\left( \Delta^{(2)}_{00}-4\Psi^{(2)} \right)   
\nonumber \\
& &= \frac{4}{3} \nabla^2 \left( \Phi^{(2)}+\frac{\Psi^{(2)}}{1+R} \right)
+{\cal S}'_\Delta+{\cal H}\frac{R}{1+R} {\cal S}_\Delta -\frac{4}{3}\partial_i {\cal S}^i_V\, , 
\end{eqnarray}
where we have introduced the photon-baryon fluid sound speed $c_s=1/\sqrt{3(1+R)}$.
\subsubsection{Second-order quadrupole moment of the photons in the tight coupling limit}
\label{Pi2}
Let us now consider the quadrupole moment of the photon distribution defined in Eq.~(\ref{quadrupole}) and show 
that at second-order it cannot be neglected in the tightly coupled limit, unlike for the linear case. 
We first integrate the R.H.S. of Eq.~(\ref{B2}) over $d\Omega_{\bf n} (n^in^j-\delta^{ij}/3)/4\pi$ 
and then we set it to be vanishing in the limit of tight coupling. 

The integration involves various pieces to 
compute. For clarity we will consider each of them separately. The term $\Delta^{(2)}_{00}$ does not contribute. 
For the third term we can write, from Eq.~(\ref{Dlm2})
\begin{eqnarray}
\label{3term}
- \sum_{m=-2}^{m=2} \frac{\sqrt{\pi}}{5^{3/2}} \Delta^{(2)}_{2m} Y_{2m} 
&=& \frac{\Delta^{(2)}}{10}-\frac{1}{10} \sum_{\ell \neq 2} 
\sum_{m=-\ell}^{\ell} \Delta^{(2)}_{\ell m} (-i)^\ell \sqrt{\frac{4\pi}{2\ell+1}} Y_{\ell m}, \nonumber \\
\end{eqnarray} 
 so that the integral just brings $\Pi^{(2)ij}_\gamma/10$, since the only contribution in Eq.~(\ref{3term}) comes from
$\Delta^{(2)}/10$ with all the other terms vanishing. The following nontrivial integral is 
\begin{eqnarray}
\int \frac{d\Omega}{4\pi} \left( n^i n^j -\frac{1}{3} \delta^{ij} \right) \Delta^{(1)}_2  
P_2(\hat{\bf v} 
\cdot {\bf n}) &=&
\hat{v}_k \hat{v}_l  \int \frac{d\Omega}{4\pi} \left( n^i n^j -\frac{1}{3} \delta^{ij} \right)  
\Delta^{(1)}_2 \nonumber \\
&\times&
\left(\frac{3}{2}n^k n^l -\frac{1}{2} \right)
\nonumber \\
&=& \frac{\Delta^{(1)}_2}{5} 
\left(\hat{v}^i\hat{v}^j -\frac{1}{3} 
\delta^{ij}  \right)\, ,
\end{eqnarray}
where the baryon velocity appearing in $P_2(\hat{\bf v} \cdot {\bf n})$ is first order and we make use of 
the relations~(\ref{relOmega}) together with 
\begin{equation}
\int \frac{d\Omega}{4 \pi}\, n^in^jn^kn^l = \frac{1}{15} (\delta^{ij} \delta^{kl} 
+\delta^{ik} \delta^{lj}+\delta^{il} \delta^{jk})\, .
\end{equation}
The integrals of $\delta^{(1)}_e \Delta^{(1)}_0$ , 
$\delta^{(1)}_e ({\bf v} \cdot {\bf n})$ and ${\bf v}^{(2)}\cdot {\bf n}$ vanish and 
\begin{eqnarray}
v \Delta^{(1)}_1 
\int \frac{d\Omega}{4\pi} P_2(\hat{{\bf v}} \cdot {\bf n}) \left( n^i n^j -\frac{1}{3} \delta^{ij} \right)
&=&\frac{1}{5} v \Delta^{(1)}_1\left(\hat{v}^i\hat{v}^j -\frac{1}{3} 
\delta^{ij}  \right) 
\nonumber  \\
&=& \frac{4}{15} \left( v^iv^j -\frac{1}{3} \delta^{ij} v^2 \right)\, ,   
\end{eqnarray}
 where in the last step we take $\Delta^{(1)}_1=4v/3$ in the tight coupling limit. Similarly the integral of 
$14 ({\bf v} \cdot {\bf n})^2$ brings 
\begin{equation}
14 v^k v^\ell \int \frac{d\Omega}{4\pi}  n_kn_\ell \left( n^i n^j -\frac{1}{3} \delta^{ij} \right)=
\frac{28}{15} \left(v^iv^j-\frac{1}{3} \delta^{ij} v^2   \right) \, .
\end{equation}
The integral of $2 ({\bf v} \cdot {\bf n}) \Delta^{(1)}$ can be performed by expanding the linear anisotropies 
as $\Delta^{(1)}=\sum_\ell (2\ell+1) \Delta^{(1)}_{\ell} P_{\ell}(\hat{\bf v} \cdot {\bf n})$. We thus find 
\begin{eqnarray}
v^k \hat{v}^m \int \frac{d\Omega}{4\pi} n_k \left( n^i n^j -\frac{1}{3} \delta^{ij} \right)  n_m \Delta^{(1)}_1 +
{\cal O}_{\ell  > 2}&=& \frac{8}{15} \left(v^iv^j-\frac{1}{3} \delta^{ij} v^2   \right), \nonumber \\
\end{eqnarray}
where we have used Eq.~(\ref{relOmega}) and ${\cal O}_{\ell >2}$ indicates 
all the integrals coming from the multipoles $\ell > 2$ in the expansion (for $\ell=0$ and $\ell=2$ they
vanish.) In fact we have dropped the ${\cal O}(\ell >2)$ 
since they are proportional to firts-order photon moments $\ell > 2$ which turn out to be suprressed in the tight 
coupling limit. Finally the term proportional to 
$({ \bf v} \cdot {\bf n}) \Delta^{(1)}_2 (1-P_2(\hat{\bf v} \cdot {\bf n})/5$ gives a vanishing contribution.

Collecting all the various pieces we find that the third moment of the R.H.S. of Eq.~(\ref{B2}) is given by
\begin{eqnarray}
\label{2quadRHS}
&-& \frac{\tau'}{2} \left[ 
-\Pi^{(2)ij}_\gamma+\frac{1}{10} \Pi^{(2)ij}_\gamma+2\delta^{(1)}_e\left(-\Pi^{(1)ij}_\gamma+\frac{1}{10} 
\Delta^{(1)}_2 
   (\hat{v}^i\hat{v}^j-\frac{1}{3} \delta^{ij}) \right)  
\right. \nonumber  \\
&+& \left. \frac{12}{5}\left(v^iv^j-\frac{1}{3} \delta^{ij} 
v^2  \right) 
\right]\, .
\end{eqnarray}
Therefore in the limit of tight coupling, when the interaction rate is very high, the second-order 
quadrupole moment turns out to be 
\begin{equation}
\label{2quad}
\Pi^{(2)ij}_\gamma \simeq \frac{8}{3} \left(v^iv^j-\frac{1}{3} \delta^{ij} v^2  \right)\, ,
\end{equation}
by setting Eq.~(\ref{2quadRHS}) to be vanishing (the term multiplying $\delta^{(1)}_e$ goes to zero in the tight 
coupling limit since it just comes from the first-order collision term). 
At linear order one would simply get the term $9 \tau' \Pi^{(1)ij}_\gamma/10$ 
implying that, in the limit of a high scattering rate $\tau'$,
$\Pi^{(1)ij}_\gamma$ goes to zero. However at second-order the quadrupole is not suppressed in the tight 
coupling limit 
becasue it turns out to be sourced by the linear velocity squared.

\subsection{Second-order CMB anisotropies generated at recombination}
\label{ar}
The previous equations allow us to follow the evolution of 
the monopole and dipole of CMB photons at recombination. As at linear order, they will appear in the expression for 
the CMB anisotropies today $\Delta^{(2)}({\bf k},{\bf n},\eta_0)$ together with various integrated effects.  
Our focus now will be 
to obtain an expression for the second-order  
CMB anisotropies today $\Delta^{(2)}({\bf k},{\bf n},\eta_0)$ 
from which we can extract all those contributions generated specifically 
at recombination due to the non-linear dynamics of the 
photon-baryon fluid. This expression will not only relate the moments $\Delta^{(2)}_{\ell m}$ 
today to the second-order monopole and dipole at recombination as it happens at linear order, but one has to properly
account also for additional first-order squared contributions. Let us see how to achieve this goal in some details. 

As we have seen previously, 
it is possible to write down an integral solution of the photon Boltzmann
equation~(\ref{B2}) in Fourier space. We wrote 
\begin{equation} 
\Delta^{(2) \prime}+ik\mu \Delta^{(2)}-\tau' \Delta^{(2)}=
e^{-ik\mu+\tau} \frac{d}{d \eta}[\Delta^{(2)} e^{ik \mu \eta-\tau}]=S({\bf k},{\bf n},\eta) 
\end{equation} 
in order to derive a solution of the form 
\begin{eqnarray}
\label{IS}
\Delta^{(2)}({\bf k},{\bf n},\eta_0)=\int_0^{\eta_0} d\eta S({\bf k},{\bf n},\eta) e^{ik\mu(\eta-\eta_0)}e^{-\tau}\, .
\end{eqnarray}
Here $\mu=\cos \vartheta=\hat{\bf k} \cdot {\bf n}$ is the polar angle of the photon momentum in a coordinate system 
such that ${\bf e}_3=\hat{\bf k}$.
At second-order the source term has been  computed in Ref.~\cite{paperI} and can be read off Eq.~(\ref{B2}) and 
Eq.~(\ref{D}) to be 
\begin{eqnarray}
\label{sourceS}
S&=&-\tau' \Delta^{(2)}_{00} -4 n^i\Phi^{(2)}_{,i} +4 \Psi^{(2)\prime}-8 \omega'_i n^i-4\chi'_{ij}n^in^j \nonumber \\
&+&8 \Delta^{(1)}(\Psi^{(1)\prime}-n^i \Phi^{(1)}_{,i})
- 2n^i (\Phi^{(1)}+\Psi^{(1)})(\Delta^{(1)}+4\Phi^{(1)})_{,i} 
\nonumber \\
&-&2\left[(\Phi^{(1)}+\Psi^{(1)})_{,j}n^in^j-(\Phi^{(1)}+\Psi^{(1)})^{,i} \right] 
\frac{\partial \Delta^{(1)}}{\partial n^i} 
\nonumber \\
&-& \tau' \Bigg[  
- \frac{1}{2} \sum_{m=-2}^{2} \frac{\sqrt{4 \pi}}{5^{3/2}}\, \Delta^{(2)}_{2m} \, Y_{2m}({\bf n}) 
+4{\bf v}^{(2)} \cdot {\bf n} 
\nonumber \\
&+& 2 \delta^{(1)}_e \left( \Delta^{(1)}_0
-\Delta^{(1)}+4 {\bf v} \cdot {\bf n}+\frac{1}{2} \Delta^{(1)}_2 P_2({\bf {\hat v}} \cdot {\bf n})
\right)\nonumber \\
&+ &  2 ({\bf v} \cdot {\bf n}) \left[ \Delta^{(1)}+3\Delta^{(1)}_0-\Delta^{(1)}_2 \left(1-\frac{5}{2} P_2({\bf {\hat v}} 
\cdot {\bf n})
\right)\right]
\nonumber \\
&-&v\Delta^{(1)}_1 \left(4+2 P_2({\bf {\hat v}} \cdot {\bf n}) \right) 
+14 ({\bf v} \cdot {\bf n})^2-2 v^2  \Bigg]\, .
\end{eqnarray}
The key point here is to isolate all those terms that multiply the differential optical depth $\tau'$. The reason is that 
in this case in the integral~(\ref{IS}) one recognizes the visibility function $g(\eta)=- e^{-\tau} \tau'$ which is sharply 
peaked at the time of recombination and whose integral over time is normalized to unity. Thus for these terms the integral 
just reduces to the remaining integrand (apart from the visibility function) evaluated at recombination. The standard example 
that one encounters also at linear order is given by the first term appearing in the source $S$, Eq.~(\ref{sourceS}), that is 
$-\tau' \Delta^{(2)}_{00}$. The contribution of this term  to the integral~(\ref{IS}) just reduces to 
\begin{eqnarray}
\label{ex1exp0}
\Delta^{(2)}({\bf k},{\bf n}, \eta_0)&=&\int_0^{\eta_0} d\eta e^{ik\mu(\eta-\eta_0)}e^{-\tau} (-\tau')
\Delta^{(2)}_{00} \simeq e^{ik\mu(\eta_*-\eta_0)}\Delta^{(2)}_{00}(\eta_*)\, ,  \nonumber \\
\end{eqnarray}   
where $\eta_*$ is the epoch of recombination and, in the multipole decomposition~(\ref{angular1}),  Eq.~(\ref{ex1exp0}) brings the 
standard result
\begin{equation}
\label{ex1exp}
\Delta^{(2)}_{\ell m}(\eta_0) \propto \Delta^{(2)}_{00}(\eta_*)j_\ell(k(\eta_*-\eta_0))\, , 
\end{equation}
having used the Legendre expansion $e^{i{\bf k} \cdot {\bf x}}=
\sum_\ell (i)^\ell (2\ell+1) j_\ell(kx) P_\ell( {\bf {\hat k}} \cdot {\bf {\hat x}})$. In Eq.~({\ref{ex1exp}) the monopole 
at recombination is found by solving the Boltzmann equations~Eqs.~(\ref{D2eq})-(\ref{v2eqf}) derived in tight coupling limit. 

Looking at Eq.~(\ref{sourceS}) we recognize immediately some terms which multiply explicitly $\tau'$ (the first one 
discussed in 
the example above and the last two lines of Eq.~(\ref{sourceS})). However it is easy to realize from the standard 
procedure 
adopted at the linear-order that such terms are not the only ones. This is clear by focusing, as an example, on the term 
$-4n^i \Phi^{(2)}_{,i}$ in the source $S$ which appears in the same form also at linear order. In Fourier space one can 
replace 
the angle $\mu$ with a time derivative and thus this term gives rise to
\begin{eqnarray}
\label{ex2}
& -& 4 ik \int_0^{\eta_0} d\eta\, e^{ik \mu(\eta-\eta_0)} e^{-\tau} \mu \Phi^{(2)}=
-4 \int_0^{\eta_0} d\eta\,  \Phi^{(2)} e^{-\tau} \frac{d}{d\eta}
\left( e^{ik\mu(\eta-\eta_0)} \right)
\nonumber \\
&=&=4 \int_0^{\eta_0} d\eta\, 
 e^{ik \mu(\eta-\eta_0)} e^{-\tau} \left( \Phi^{(2)\prime}-\tau' \Phi^{(2)} \right)\, ,
\end{eqnarray}  
where, in the last step, we have integrated by parts. 
In Eq.~(\ref{ex2}) the time derivative of the gravitational potential contributes to the Integrated Sachs-Wolfe effect, but also 
also a $\tau'$ results implying that we have also to evaluate $\Phi^{(2)}$ at recombination. 
Thus, in the following, we look for those terms in the source~(\ref{sourceS}) which give rise to a $\tau'$ factor in the 
same way as for $-4n^i \Phi^{(2)}_{,i}$. In particular let us consider the combination in Eq.~(\ref{sourceS})
\begin{eqnarray}
\label{C1}
C &\equiv & 8 \Delta^{(1)}(\Psi^{(1)\prime}-n^i \Phi^{(1)}_{,i})
-2n^i (\Phi^{(1)}+\Psi^{(1)})(\Delta^{(1)}+4\Phi^{(1)})_{,i} 
\nonumber \\
&=& 
8 \Delta^{(1)} \Psi^{(1)\prime} -8 n^i (\Delta^{(1)} \Phi^{(1)})_{,i} +4 \Phi^{(1)} n^i \Delta^{(1)}_{,i} 
- 8n^i(\Phi^{(1)2})_{,i}
\, ,  
\end{eqnarray}
where for simplicity we are setting $\Phi^{(1)} \simeq \Psi^{(1)}$. We already recognize terms of the form $n^i 
\partial_i(\cdot)$. Moreover we can use the Boltzmann equation~(\ref{B1}) to replace $n^i \Delta^{(1)}_{,i}$ in Eq.~(\ref{C1}). 
This brings 
\begin{eqnarray}
\label{Cfinal}
C & =  & 8 \Delta^{(1)} \Psi^{(1)\prime}-4\Psi^{(1)} \Delta^{(1)\prime} -8 (\Psi^{(1)2})'
-8 n^i (\Delta^{(1)} \Phi^{(1)}+2\Phi^{(1)2})_{,i} 
\nonumber \\
&-& 4 \tau' \Psi^{(1)} \left[\Delta^{(1)}_{00}-\Delta^{(1)}
+4{\bf v}^{(1)} \cdot {\bf n} +\frac{1}{2}  \Delta^{(1)}_2 P_2(\hat{\bf v} \cdot {\bf n})\right]\, . \nonumber \\
\end{eqnarray}
In fact we will not be interested for our purposes in the first three terms of Eq.~(\ref{Cfinal}), since they will not 
contribute to the anisotropies generated at recombination. 

Therefore, as a result of Eqs.~(\ref{sourceS}),~(\ref{ex2}) and~(\ref{Cfinal}), we can rewrite the source 
term~(\ref{sourceS}) as  
\begin{eqnarray}
\label{split}
S=S_*+S'
\end{eqnarray}  
where 
\begin{eqnarray}
\label{Sstar}
S_*&=& -\tau' \Bigg[ \Delta^{(2)}_{00}+4\Phi^{(2)}
- \frac{1}{2} \sum_{m=-2}^{2} \frac{\sqrt{4 \pi}}{5^{3/2}}\, \Delta^{(2)}_{2m} \, Y_{2m}({\bf n}) 
+4{\bf v}^{(2)} \cdot {\bf n} 
\nonumber \\
&+&
2 \delta^{(1)}_e \left( \Delta^{(1)}_0
-\Delta^{(1)}+4 {\bf v} \cdot {\bf n}+\frac{1}{2} \Delta^{(1)}_2 P_2({\bf {\hat v}} \cdot {\bf n})
\right)\nonumber \\
&+ &  2 ({\bf v} \cdot {\bf n}) \left[ \Delta^{(1)}+3\Delta^{(1)}_0-\Delta^{(1)}_2 \left(1-\frac{5}{2} P_2({\bf {\hat v}} 
\cdot {\bf n})
\right)\right]
\nonumber \\
&-&v\Delta^{(1)}_1 \left(4+2 P_2({\bf {\hat v}} \cdot {\bf n}) \right) 
+14 ({\bf v} \cdot {\bf n})^2-2 v^2  +8\Delta^{(1)} \Phi^{(1)} \nonumber \\
&+&16 \Phi^{(1)2}
+4 \Psi^{(1)} \left[  \Delta^{(1)}_0-\Delta^{(1)}
+4{\bf v}^{(1)} \cdot {\bf n} +\frac{1}{2}  \Delta^{(1)}_2 P_2(\hat{\bf v} \cdot {\bf n})  \right]
\Bigg]\, , \nonumber \\
\end{eqnarray}
and 
\begin{eqnarray} 
S'&=& 4(\Phi^{(2)}+\Psi^{(2)})^\prime -8 \omega'_i n^i-4\chi'_{ij}n^in^j
-2\left[(\Phi^{(1)}+\Psi^{(1)})_{,j}n^in^j
\right. \nonumber \\
&-& \left. (\Phi^{(1)}+\Psi^{(1)})^{,i} \right] 
\frac{\partial \Delta^{(1)}}{\partial n^i} 
+ 8(\Delta^{(1)} \Phi^{(1)})^\prime+8 \Delta^{(1)} \Psi^{(1)\prime}
-4\Psi^{(1)} \Delta^{(1)} 
\nonumber \\
&+&16 \Psi^{(1)} \Psi^{(1)\prime}\, .
\end{eqnarray}
In Eq.~(\ref{split}) $S_*$ contains the contribution to the second-order 
CMB anisotropies created on the last scattering surface at 
recombination, while $S'$ includes all those effects which are integrated in time from the last scattering surface up to now, 
including the second-order Integrated Sachs-Wolfe effect and the second-order lensing effect. Since the main concern 
of this Section is the CMB anisotropies generated at last scattering, from now on we will focus only on the contribution from the 
last scattering surface $S_*$. In particular notice that, following the same steps that lead to Eq.~(\ref{ex1exp0}}), 
the first two terms of Eq.~(\ref{Sstar}) give rise to the CMB temperature anisotropies as written in Eq.~(\ref{SW}).

\subsection{Tightly coupled solutions for the second-order perturbations}
\label{ac}
In this section we will solve the tightlty coupled limit of the Boltzmann equations~(\ref{D2eq}) and~(\ref{v2eq}) 
at second-order in perturbation theory. We will proceed as for the linear case, focusing on the two limiting cases 
of perturbation modes entering the horizon respectively much before and much after the time of equality. The solution 
of Eq.~(\ref{eqoscille}) can be written as 
\begin{eqnarray}
\label{soltot2}
& & [1+R(\eta)]^{1/4} (\Delta^{(2)}_{00}-4\Psi^{(2)})=
A\, \cos[kr_s(\eta)]+B\,\sin[kr_s(\eta)] \nonumber \\
& &- 4\frac{k}{\sqrt{3}}\int_0^\eta d\eta' [1+R(\eta')]^{3/4} \left(\Phi^{(2)}(\eta')+\frac{\Psi^{(2)}(\eta')}{1+R} \right) 
\sin_k[\eta,\eta'] \nonumber \\
& &+ \frac{\sqrt{3}}{k}\int_0^\eta d\eta' [1+R(\eta')]^{3/4} \left( {\cal S}'_\Delta+\frac{{\cal H} R}{1+R} {\cal S}_\Delta 
-\frac{4}{3} ik_i\, {\cal S}^i_V
\right) 
\sin_k[\eta,\eta']\, , \nonumber \\
\end{eqnarray}
where for simplicity we use the shorthand notation $\sin_k[\eta,\eta']\equiv \sin[k(r_s(\eta)-r_s(\eta'))]$, and $r_s(\eta)$ is the sound horizon 
\begin{equation}
r_s(\eta)=\int_0^\eta d\eta' c_s(\eta') \, .
\end{equation}
The source terms are given in Eq.~(\ref{SD}) and~(\ref{SV}). Notice that we can write 
$ {\cal S}'_\Delta+\frac{{\cal H} R}{1+R} {\cal S}_\Delta=({\cal S}_\Delta(1+R))^\prime/1+R$ so that we can perform 
an integration by parts in Eq.~(\ref{soltot2}) leading to  
\begin{eqnarray}
\label{soltotint}
& & [1+R(\eta)]^{1/4} (\Delta^{(2)}_{00}-4\Psi^{(2)})=
A\, \cos[kr_s(\eta)]+B\,\sin[kr_s(\eta)] \nonumber \\
&&-4\frac{k}{\sqrt{3}}\int_0^\eta d\eta'\, [1+R(\eta')]^{3/4}\, 
\left(\Phi^{(2)}(\eta')+\frac{\Psi^{(2)}(\eta')}{1+R} \right)\, 
\sin_k[\eta,\eta'] \nonumber \\
&&+\int_0^\eta d\eta'\, {\cal S}_\Delta (\eta')\, (1+R(\eta'))^{1/4}\,   
\cos_k[\eta,\eta'] -\frac{\sqrt{3}}{k} {\cal S}_\Delta(0) \sin[kr_s(\eta)]\nonumber \\
&&-\frac{4}{\sqrt{3}} \frac{ik_i}{k} \int_0^\eta d\eta'\, {\cal S}^i_V (\eta')\, (1+R(\eta'))^{3/4}\,  
\sin_k[\eta,\eta'] \nonumber \\
&&+\frac{\sqrt{3}}{4k} \int_0^\eta d\eta'\, {\cal S}_\Delta (\eta')\, (1+R(\eta'))^{-1/4} R'(\eta')\, 
\sin_k[\eta,\eta'] \, .
\end{eqnarray}  
In order to give an analytical solution that catches most of the physics underlying Eq.~(\ref{soltot2}) and which remains 
at the same time very simple to treat, we will make some simplifications which are also usually adopted in linear theory, see e.g. 
Ref.~\cite{Huthesis}. 
In particular we will treat $R=R_*$ as a constant 
evaluated at the time of recombination. In this way, with the presence of $R$ in the varying cosines and sines, we can   
keep track of a damping of the photon velocity amplitude with respect to the case 
$R=0$ which prevents the acoustic peaks in the power-spectrum to disappear.   
Treating $R$ as a constant is justified by the fact that for modes within the horizon the 
time scale of the ocillations is much shorter than the time scale on which $R$ varies. 
If $R$ is a constant the sound speed is just a constant 
\begin{equation}
\label{soundspeed}
c_s=\frac{1}{\sqrt{3(1+R_*)}} \, ,
\end{equation}
and the sound horizon is simply $r_s(\eta)=c_s \eta$.

There is also another good reason not to forget about $R$, even tough $R <1$, 
which is particularly relevant when estimating the second-order contributions to the 
CMB anisotropies in~(\ref{SW}). As we will see in Sec.~\ref{An}, that combination on small scales comes actually as proportional to $R$, so that 
neglecting $R$ would lead to miss one of the most significant CMB contribution at second order. 

Another simplification consists in solving the evolutions of the perturbations 
in two well distinguished 
limiting regimes. One regime is for those perturbations which enter 
the Hubble radius when matter is the dominant 
component, that is at times much bigger than the equality epoch, with 
$k \ll k_{eq} \sim \eta^{-1}_{eq}$, 
where $k_{eq}$ is the wavenumber of the Hubble radius at the equality 
epoch. The other regime is for those perturbations 
with much smaller wavelenghts which enter the Hubble radius when the 
universe is still radiation dominated, that is 
perturbations with wavenumbers $k \gg k_{eq}\sim \eta_{eq}^{-1}$. In 
fact we are interested in  
perturbation modes which are within the horizon by the time of 
recombination $\eta_*$. Therefore 
we will further suppose that $\eta_* \gg \eta_{eq}$ in order to study 
such modes in the first regime. Even though $\eta_* \gg \eta_{eq}$ is 
not the real case, it allows to obtain some  
analytical expressions.

\subsubsection{Setting the initial conditions: primordial non-Gaussianity}
\label{ICCMB}
The integration constants $A$ and $B$ are fixed according to the initial conditons for the second-order cosmological 
perturbations. 
These refer to the values of the perturbations on superhorizon scales deep in the radiation dominated period. We will 
consider the case of initial adiabatic perturbations, for which there exist some useful conserved quantities on 
large scales which as such carry directly the information about the initial conditons.

As explained in Sec.~\ref{PNG} such a conserved quantity is given by the curvature perturbation $\zeta$
and its conserved value allows to  set the initial 
conditions for the metric and matter perturbations accounting for the 
primordial contributions. At linear order during the radiation-dominated epoch 
and on large scales $\zeta^{(1)}=-2\Psi^{(1)}/3$. On the other hand, after some calculations, one can easily compute 
$\Delta \zeta^{(2)}$ for a radiation dominated epoch
\begin{equation}
\Delta \zeta^{(2)}=\frac{7}{2} \left( \Psi^{(1)} \right)^2 \, ,
\end{equation}   
where in Eq.~(\ref{deltaz2}) one uses that on large scales $\delta^{(1)} \rho_\gamma/\rho_\gamma=-2\Psi^{(1)}$ and the
energy continuity equation $\delta^{(1)\prime} \rho_\gamma+4{\cal H} \delta^{(1)} \rho_\gamma-4 \Psi^{(1)\prime} 
\rho_\gamma=0$. Therefore we find 
\begin{equation}
\label{zetar}
\zeta^{(2)}=-\Psi^{(2)}+\frac{\Delta^{(2)}_{00}}{4}+\frac{7}{2} \Psi^{(1)2}(0)\, , 
\end{equation}
where we are evaluating the quantities in the large scale limit for $\eta \rightarrow 0$. Using the 
parametrization~(\ref{param}) at the initial times the quantity $\Delta^{(2)}_{00}-4\Psi^{(2)}$ is 
given by 
\begin{equation}
\label{DPsi20}
\Delta^{(2)}_{00}-4\Psi^{(2)}=2(9 a_{\rm NL}-7) \Psi^{(1)2}(0)\, .
\end{equation}
Since for adiabatic perturbations such a quantity is conserved on superhorizon scales, 
it follows that the constant $B=0$ and $A=2(9 a_{\rm NL}-7) \Psi^{(1)2}(0)$.

Eqs.~(\ref{soltot2}) and~(\ref{soltotint}) are analytical expressions describing 
the acoustic oscillations of the photon-baryon fluid 
induced at second-order for perturbation modes within the horizon at recombination. In the following we will adopt similar 
simplifications already used for the linear case in order to provide some analytical solutions. In particular, if in 
Eq.~(\ref{soltotint}) we treat $R$ as a constant we can write, using the initial conditions determined above, 
\begin{eqnarray}
\label{solsem}
(\Delta^{(2)}_{00}-4\Psi^{(2)})&=&
2(9 a_{\rm NL}-7) \Psi^{(1)2}(0)\, \cos[kr_s(\eta)] \nonumber \\
&-& 4kc_s\int_0^\eta d\eta'\, \, 
\left(\Phi^{(2)}(\eta')+\Psi^{(2)}(\eta') +R \Phi^{(2)}\right)\, 
\sin_k[\eta,\eta'] \nonumber \\
&+&\int_0^\eta d\eta'\, {\cal S}_\Delta (\eta')\, \,   
\cos_k[\eta,\eta'] -\frac{\sqrt{3}}{k} {\cal S}_\Delta(0) \sin[kr_s(\eta)] \nonumber \\
&-&\frac{4}{3} \frac{ik_i}{kc_s} \int_0^\eta d\eta'\, {\cal S}^i_V (\eta')\, \,  
\sin_k[\eta,\eta']\, . 
\end{eqnarray}

\subsection{Perturbation modes with $k \gg k_{eq}$}
\label{kggkeq2}
In order to study the contribution to the second-order CMB anisotropies coming from perturbation modes that 
enter the horizon during the radiation dominated epoch, we will assume that the second-order gravitational potentials 
are the ones of a pure radiation dominated universe throughout the evolution. Though not strictly correct, this approximation 
will give us the basic picture of the acoustic oscillations for the baryon-photon fluid 
occuring for these modes. Also for the second-order case, in Section~\ref{improved} we 
will provide the appropriate corrections accounting for the transition from radiation to matter domination which is indeed 
(almost) achieved by the recombination epoch. Before moving into the details a note of caution is in order here. 
At second order in the perturbations all the relevant quantities are 
expressed as convolutions of linear perturbations, bringing to a mode-mode mixing. In 
some cases in our treatment for a given regime under analysis ($k \gg k_{eq}$ or $k \ll k_{eq}$) we use for the first-order 
perturbations the solutions corresponding to that particular regime, while the mode-mode mixing would require 
to consider in the convolutions (where 
one is integrating over all the wavenumbers) a more general expression for the first-order perturbations (which analytically does not exist 
anyway). For the computation of the CMB bispectrum this would be equivalent to consider just some specific scales, 
{\it i.e.} all the three scales involved in the bispectrum should correspond approximately to wavenumbers $k \gg k_{eq}$ or 
$k \ll k_{eq}$, and not a combination of the two regimes
(a step towards the evaluation of the three-point correlation function has been taken on Ref. \cite{rec} where it was computed in the 
in so-called  squeezed triangle limit, when one mode has a wavelength much larger than the other two and is outside the horizon).

In Appendix C we show how to solve the Boltzmann equations at the linear level in the various regimes. 
In fact for $k \gg k_{eq}$ the acoustic oscillations can be solved in an alternative way (see Ref.~\cite{paperII}) for this procedure at 
linear order). One can start directly from Eq.~(\ref{D2eq}) where we can neglect the gravitational potential term $\Psi^{(2)'}$. The 
reason is that, as it happens at linear order, the second-order gravitational potentials decay at late times as $\eta^{-2}$, while the 
second-order velocity $v^{(2)i}_\gamma$ oscillates in time. Let us now see that in some details. 
     
The evolution equation for the gravitational potential $\Psi^{(2)}$ is given by Eq.~(\ref{P2radeq}) and is characterized by the 
source term $S_\gamma$, Eq.~(\ref{Sgamma}). In particular the source term contains the second-order quadrupole moment of the 
photons $\Pi^{(2)ij}_\gamma$. We saw in Section~\ref{Pi2} that at second-order the quadrupole moment is not suppressed in the 
tight coupling limit, being fed by the non-linear combination of the first-order velocities, Eq.~(\ref{2quad}). For the 
pertubation modes we are considering here the velocity at late times is oscillating being given in Appendix C in 
Fourier space. Since the linear gravitational potential decays in time and for a radiation dominated period 
${\cal H}=1/\eta$, it is easy to check that the dominant 
contribution at late times to the source term $S_\gamma$ simply reduces to
\begin{eqnarray}
\label{Sgammadom}
S_\gamma &\simeq& \frac{3}{2} {\cal H}^2 \frac{\partial_i \partial^j}{\nabla^2} \Pi^{(2)i}_{\gamma~~~j} \equiv
\frac{F({\bf k}_1,{\bf k}_2,{\bf k})}{\eta^2} C \, \Psi^{(1)}_{{\bf k}_1}(0)  \Psi^{(1)}_{{\bf k}_2}(0) \sin(k_1c_s \eta) 
\sin(k_2 c_s \eta)\, ,   
\end{eqnarray}
where  
\begin{equation}
\label{C}
C = - \frac{9}{c_s^2k_1k_2}\, 
\end{equation}
and the sound speed is $c_s=1/\sqrt{3(1+R)}$. Before proceeding further let us explain the notation
that we are using. The equivalence symbol will be used to indicate that we are evaluating the expression in Fourier space. At 
second-order in perturbation theory most of the Fourier transforms reduce to some convolutions. We will not indicate these 
convolutions explicitly but just through their kernel. For example in Eq.~(\ref{Sgammadom}) by 
$F({\bf k}_1,{\bf k}_2,{\bf k})$ we actually indicate the convolution operator  
\begin{equation}
\label{F}
F \equiv \frac{1}{2\pi^3} \int d^3k_1 d^3k_2 \delta^{(1)}({\bf k}_1+{\bf k}_2-{\bf k}) F({\bf k}_1,{\bf k}_2,{\bf k})\, .
\end{equation}  
In the specific case of Eq.~(\ref{Sgammadom}) the kernel is given by 
\begin{equation}
\label{Fkernel}
F({\bf k}_1,{\bf k}_2,{\bf k})= \frac{({\bf k}\cdot {\bf k}_1)({\bf k}\cdot {\bf k}_2 )}{k^2}-\frac{1}{3}{\bf k}_1 \cdot 
{\bf k}_2\, . 
\end{equation}
The choice of these conventions is due not only for simplicity and to keep our expressions shorter, but also because at 
the end we will be interested to the bispectrum of the CMB anisotropies generated at recombination, and the 
relevant expressions entering in the bispectrum are just the kernels of the convolution integrals. 

Having determined the leading contribution to the source term at late times, we can now solve the evolution 
equation~(\ref{P2radeq}). Since the source term scales like $\eta^{-2}$ it is useful to introduce the rescaled variable 
$\chi = \eta^2 \Psi^{(2)}$. Eq.~(\ref{P2radeq}) then reads
\begin{equation}
\chi''+\left(k^2 c_s^2-\frac{2}{\eta^2}  \right) \chi= \eta^2 S_\gamma\, .
\end{equation}
For perturbation modes which are subhorizon with $k \eta \gg 1$ the solution of the homogeneous equation is given by
\begin{equation}
\chi_{\rm hom.}=A \cos(kc_s \eta)+ B \sin(c_s k \eta)\, ,
\end{equation} 
from which we can build the general solution 
\begin{equation}
\chi=\chi_{\rm hom.} + \chi_+\int_0^\eta d\eta'\frac{\chi_{-}(\eta')}{W(\eta')}
S_\gamma(\eta')-\chi_-\int_0^\eta d\eta'
\frac{\chi_+(\eta')}{W(\eta')}
S_\gamma(\eta')\, ,
\end{equation}
where $W=-kc_s$ is the Wronskian, and $\chi_+=\cos(kc_s \eta)$, $\chi_-=\sin(c_s k \eta)$. 
Using Eq.~(\ref{Sgammadom}) the integrals involve products of sines and cosines which can be performed giving
\begin{eqnarray}
\chi&=&\chi_{\rm hom.}-\frac{FC}{c_s^2} \Psi^{(1)}_{{\bf k}_1}(0)  \Psi^{(1)}_{{\bf k}_2}(0) \nonumber\\
&\times&\frac{k\left[2k_1k_2 \cos(k_1c_s\eta)\cos(k_2c_s\eta)-2k_1k_2 \cos(kc_s\eta) +
(k_1^2+k_2^2-k^2) \sin(k_1 c_s \eta) \sin(k_2 c_s \eta) \right] }{k_1^4+k_2^4+k^4-
2k_1^2k_2^2-2k_1^2k^2-2k_2^2k^2}.
\end{eqnarray}
Thus the gravitational potential $\Psi^{(2)}$ at late times is given by 
\begin{eqnarray}
\label{Psi2rf}
\Psi^{(2)}_{\bf k}(\eta)&=& -3\Psi^{(2)}(0) \frac{\cos(kc_s\eta)}{(kc_s\eta)^2} -\frac{FC}{\eta^2 c_s^2} \Psi^{(1)}_{{\bf k}_1}(0)  \Psi^{(1)}_{{\bf k}_2}(0) \nonumber\\
&\times&\frac{\left[2k_1k_2 \cos(k_1c_s\eta)\cos(k_2c_s\eta)-2k_1k_2 \cos(kc_s\eta) +
(k_1^2+k_2^2-k^2) \sin(k_1 c_s \eta) \sin(k_2 c_s \eta) \right] }{k_1^4+k_2^4+k^4-
2k_1^2k_2^2-2k_1^2k^2-2k_2^2k^2}\, , \nonumber \\
\end{eqnarray}
where we have set the integration constant $B=0$ and $A=-3\Psi^{(2)}(0)/(kc_s)^2$ in order to 
match the homogenoeus solution at 
late times. Here $\Psi^{(2)}(0)$ is the intial condition for 
the gravitational potential taken on large scales deep in the radiation dominated era which will be determined in 
Section~\ref{icrd}. 

Eq.~(\ref{Psi2rf}) shows the result that we anticipated: also at second order  the 
gravitational potential varies in time oscillating 
with an amplitude that decays as $\eta^{-2}$. Let us then take the divergence of the $(i-0)$ Einstein 
equation~(\ref{i0}) expanded at second-order 
\begin{eqnarray}
&&\partial_i\left[ \frac{1}{2}\partial^i \Psi^{(2)\prime}+\frac{\cal H}{2} \partial^i 
\Phi^{(2)} +
2\Psi^{(1)} \partial^i \Psi^{(1)\prime}+2{\cal H} \Psi^{(1)} \partial^i\Phi^{(1)}-\Psi^{(1)\prime} 
\partial^i \Phi^{(1)} \right]\nonumber\\
&=&-2 {\cal H}^2 \partial_i \left[ \frac{1}{2} v^{(2)i}_\gamma +(\Phi^{(1)}+\Psi^{(1)}) v^{(1)i}_\gamma  
+  \Delta^{(1)}_{00} v^{(1)i}_\gamma \right]\, ,
\end{eqnarray}   
which, using the first-order $(i-0)$ Einstein equation and $\Phi^{(1)} \simeq \Psi^{(1)}$, reduces 
to 
\begin{eqnarray}
\partial_i\left[ \frac{1}{2}\partial^i \Psi^{(2)\prime}+\frac{\cal H}{2} \partial^i 
\Phi^{(2)} -\Psi^{(1)\prime} \partial^i \Psi^{(1)} \right]&=& 
-2 {\cal H}^2 \partial_i \left[ \frac{1}{2} v^{(2)i}_\gamma 
+ \Delta^{(1)}_{00} v^{(1)i}_\gamma \right]\, .\nonumber\\
&&
\end{eqnarray}
Since $\Psi^{(1)}$ during a radiation dominated period is given by Eq.~(\ref{LH2Phir}) and at late 
times it decays oscillating, it is easy to see that $(\Psi^{(1)\prime} \partial^i \Psi^{(1)})$ will be oscillating and 
decaying as $\eta^{-4}$ and thus   
can be neglected with respect to $\Psi^{(2)\prime}$, which oscillates with an amplitude decaying as 
$\eta^{-2}$. Also ${\cal H} \Phi^{(2)}$ turns out to be subdominant. Recall that    
$\Phi^{(2)}=\Psi^{(2)}- Q^{(2)}$ and  $Q^{(2)}$ is dominated by 
the second-order quadrupole of the photons in Eq.~(\ref{Sgamma}), so that $\Phi^{(2)}$ scales like 
$\Psi^{(2)}$ but there is the additional damping factor of the Hubble rate ${\cal H}=1/\eta$. 
Thus the dominant terms give
\begin{equation}
\label{v2rsol}
\partial_i v^{(2)i}_\gamma \simeq -\frac{1}{2 {\cal H}^2} \nabla^2 \Psi^{(2)\prime} -2 \partial_i 
(\Delta^{(1)}_{00} v^{(1)i}_\gamma )
\end{equation}
Eq.~(\ref{v2rsol})  allows to proceed further in a similar way 
as for the linear case by using the results found so far, Eqs.~(\ref{Psi2rf}) 
and~(\ref{v2rsol}), in the energy 
continuity equation~(\ref{D2eq}). In Eq.~(\ref{D2eq}) the first- and second-order gravitational potentials 
can be neglected with respect to the remaining terms given by $\Delta^{(1)}_{00}$ 
and $v^{(1)i}_\gamma$ which oscillate in time. Thus, replacing the divergence of the second-order velocity 
by the expression~(\ref{v2rsol}), Eq.~(\ref{D2eq}) becomes  
\begin{equation}
\Delta^{(2)\prime}_{00}=\frac{2}{3{\cal H}^2}\nabla^2 \Psi^{(2)\prime}+\frac{8}{3} \partial_i v^{(1)i}_\gamma
\Delta^{(1)}_{00} +\left( \Delta^{(1)2}_{00} \right)^\prime\, ,
\end{equation}
which, using the first-order equation~(\ref{LH2B1l}), further simplifies to 
\begin{equation}
\label{toint}
\Delta^{(2)\prime}_{00} = \frac{2}{3{\cal H}^2} \nabla^2 \Psi^{(2)\prime} \, ,
\end{equation} 
where we have kept only the dominant terms at late times. 

The gravitational potential $\Psi^{(2)}$ is given in Eq.~(\ref{Psi2rf}), so the integration of 
Eq.~(\ref{toint}) gives  
\begin{eqnarray}
\label{D2f}
\Delta^{(2)}_{00}&=&6 \Psi^{(2)}(0) \cos(kc_s\eta) +2 \frac{FC}{3c_s^2} \Psi^{(1)}_{{\bf k}_1}(0)
\Psi^{(1)}_{{\bf k}_2}(0) k^2 \nonumber\\
&\times&
\frac{\left[2k_1k_2 \cos(k_1c_s\eta)\cos(k_2c_s\eta)-2k_1k_2 \cos(kc_s\eta) +
(k_1^2+k_2^2-k^2) \sin(k_1 c_s \eta) \sin(k_2 c_s \eta) \right] }{k_1^4+k_2^4+k^4-
2k_1^2k_2^2-2k_1^2k^2-2k_2^2k^2}\, . \nonumber \\
\end{eqnarray} 
Needless to say, modes for $k \gg k_{D}$, where $k_D^{-1}$ indicates the usual
damping length, are supposed to be multiplied by an exponential 
$e^{-(k/k_D)^2}$ (see, e.g. \cite{Dodelsonbook}).

\subsubsection{Vector perturbations}
So far we have discussed only scalar perturbations. However at second-order in perturbation theory an 
unavoidable prediction is that also vector (and tensor) perturbation modes are produced dynamically as non-linear 
combination of first-order scalar perturbations. In particular notice that the second-order velocity
appearing in Eq.~(\ref{Sstar}), giving rise to a second-order Doppler effect at last scattering, will 
contain a scalar and a vector (divergence free) part. Eq.~(\ref{v2rsol}) provides the scalar component of the 
second-order velocity. We now derive an expression for the velocity that includes also the vector 
contribution.   

The (second-order) vector metric perturbation $\omega^i$ when 
radiation dominates can be obtained from Eq.~(\ref{omegair})  
\begin{equation}
\label{omegasempl}
-\frac{1}{2}\nabla^2 \omega^i+3{\cal H}^2 \omega^i=-4{\cal H}^2 \left(\delta^i_j-\frac{\partial^i\partial_j}
{\nabla^2}   \right) \left(  \frac{v^{(2)j}_\gamma}{2}+\Delta^{(1)}_{00}v^{(1)j}_\gamma  \right)\, ,
\end{equation}
where we have dropped the gravitational potentials
$\Psi^{(1)} \simeq \Phi^{(1)}$ 
which are subdominant at late times. On the other hand from the velocity 
continuity equation~(\ref{v2eqf})  we get 
\begin{equation}
v^{(2)i\prime}_\gamma+\frac{1}{4} \Delta^{(2),i}_{00}= \frac{1}{4} \left(\Delta^{(1)2}_{00}   \right)^{,i}
+\frac{8}{3} v^{(1)i}_\gamma \partial_jv^{(1)j}_\gamma-2\omega^{i\prime}-\frac{3}{4}\partial_k 
\Pi^{(2)ki}_\gamma\, ,
\end{equation}
neglecting the term proportional to $R$ and the decaying gravitational potentials. Using the tight coupling 
equations at first-order and integrating over time one finds
\begin{equation}
\label{combinazione}
v^{(2)i}_\gamma+2(v^{(1)i}_\gamma \Delta^{(1)}_{00})=-2\omega^i-\frac{1}{4} \int d\eta' \Delta^{(2),i}_{00}
-\frac{3}{4} \int d\eta' \partial_k \Pi^{(2)ki}_\gamma\, .
\end{equation}   
We can thus plug Eq.~(\ref{combinazione}) into Eq.~(\ref{omegasempl}) to find that at late times (for $k\eta \gg 1$) 
\begin{equation} 
\label{omegafinal}
\nabla^2 \omega^i=- 3{\cal H}^2\left(\delta^i_j-\frac{\partial^i\partial_j}
{\nabla^2}\right) \int d\eta' \partial_k \Pi^{(2)kj}_\gamma\, . 
\end{equation}
We will come later to the explicit expression for the term on the R.H.S. of Eq.~(\ref{omegafinal}). Here it 
is enough to notice that the second-order 
quadrupole oscillate in time and thus $\omega^i$ will decay in time as ${\cal H}^2=1/\eta^2$. This shows that
$\omega^i$ in Eq.~(\ref{combinazione}) can be in fact neglected with respect to the other terms giving 
\begin{equation}
\label{vgamma}
v^{(2)i}_\gamma=-2 (v^{(1)i}_\gamma \Delta^{(1)}_{00})-\frac{1}{4} \int d\eta' \Delta^{(2),i}_{00}
-\frac{3}{4} \int d\eta' \partial_k \Pi^{(2)ki}_\gamma\, .
\end{equation}
It can be useful to compute the combination on the R.H.S. of 
Eq.~(\ref{omegafinal}) 
$(\delta^i_j-\partial^i\partial_j/\nabla^2) \partial_k \Pi^{(2)kj}_\gamma$. 
The second-order quadrupole moment of the photons in 
the tightly coupled limit is given by Eq.~(\ref{2quad}), and 
\begin{equation}
\partial_k \Pi^{(2)kj}_\gamma=\frac{8}{3}\left[ \partial_k(v^k v^j)-2v^k \partial^j v_k \right]= 
\frac{8}{3} \left[v^j \partial_k v^k -v^k \partial^j v_k \right]\, ,
\end{equation}  
 where in the last step we have used that the linear velocity is the gradient of a scalar perturbation. We thus find
\begin{eqnarray}
\left( \delta^i_{~j}-\frac{\partial^i \partial_j}{\nabla^2} \right) \partial_k  \Pi^{(2)kj}_\gamma&=&
\frac{8}{3}\left(v^i \partial_k v^k -v^k \partial^i v_k \right) \nonumber \\
&-&\frac{8}{3} \frac{\partial^i}{\nabla^2} 
\left[ (\partial_k v^k)^2+v^j \partial_j\partial_k v^k -\partial_jv^k\partial^jv_k-v^k \nabla^2v_k \right] \, . \nonumber \\ 
\end{eqnarray}
Notice that if we split the quadrupole moment into a scalar, vector (divergence-free) and tensor (divergence-free and 
traceless) parts as 
\begin{equation}
\Pi^{(2)kj}_\gamma=\Pi^{(2),kj}_\gamma-\frac{1}{3}\nabla^2\delta^{kj} \Pi^{(2)}_\gamma+ 
\Pi^{(2)k,j}_\gamma+\Pi^{(2)j,k}_\gamma+\Pi^{(2)kj}_{\gamma T}\, ,
\end{equation}
then it turns out that 
\begin{equation}
\left( \delta^i_{~j}-\frac{\partial^i \partial_j}{\nabla^2} \right) \partial_k  \Pi^{(2)kj}_\gamma 
= \nabla^2 \Pi^{(2)i}_\gamma\, ,
\end{equation}
where $\Pi^{(2)i}_\gamma$ is the vector part of the quadrupole moment. Therefore one can rewrite Eq.~(\ref{omegafinal}) as 
\begin{equation}
\omega^i=-3{\cal H}^2 \int d\eta' \Pi^{(2)i}_\gamma\, .
\end{equation}
\subsubsection{Initial conditions for the second-order gravitational potentials}
\label{ICCMB1}
\label{icrd}
In order to complete the study of the CMB anisotropies at second-order for modes $k \gg k_{eq}$ we have to specify 
the initial conditions $\Psi^{(2)}(0)$ appearing in Eq.~(\ref{D2f}). These are set on super-horizon scales deep in the 
standard radiation dominated epoch (for $\eta \rightarrow 0$) 
by exploiting the conservation in time of the curvature perturbation $\zeta$.  
On superhorizon scales $\zeta^{(2)}$ is given 
by Eq.~(\ref{zetar}) during the radiation dominated epoch and, 
using the $(0-0)$-Einstein equation in the large scale limit $\Delta^{(2)}_{00}=-2\Phi^{(2)}+4\Phi^{(1)2}$, we find
\begin{equation}
\label{zrad}
\zeta^{(2)}=-\frac{3}{2} \Psi^{(2)}(0)-\frac{1}{2}\left(   \Phi^{(2)}(0)-\Psi^{(2)}(0) \right)+\frac{9}{2} 
\Psi^{(1)2}(0)\, . 
\end{equation} 
The conserved value of $\zeta^{(2)}$ is parametrized by $\zeta^{(2)}=2a_{\rm NL} \zeta^{(1)2}$, where, as explained in 
Section~\ref{ac} the parameter $a_{\rm NL}$ specifies the level of primordial non-Gaussianity depending on the particular 
scenario for the generation of the cosmological perturbations. On  the other hand ,
during radiation-domination
\begin{equation}
\label{PmP}
\Phi^{(2)}(0)-\Psi^{(2)}(0)=-Q^{(2)}(0)\, ,
\end{equation}  
where 
\begin{eqnarray}
\label{Q20}
Q^{(2)}(0)&=& -2 \nabla^{-2} \partial_k \Phi^{(1)}(0) \partial^k \Phi^{(1)}(0)+6\frac{\partial_i\partial^j}{\nabla^4}
\left( \partial^i \Phi^{(1)}(0) \partial_j \Phi^{(1)}(0)  
\right)+\frac{9}{2} {\cal H}^2\frac{\partial_i\partial^j}{\nabla^4} \Pi^{(2)i}_{\gamma~~j}\, ,\nonumber\\
\end{eqnarray}
where we are evaluating Eq.~(\ref{Q2rad}) in the limit $k\eta \ll 1$. The contribution from the 
second-order quadrupole moment in this limit reads 
\begin{eqnarray}
\frac{9}{2} {\cal H}^2\frac{\partial_i\partial^j}{\nabla^4} \Pi^{(2)i}_{\gamma~~j} &=&  
\frac{9}{2\eta^2}\frac{8}{3}\frac{\partial_i\partial^j}{\nabla^4}\left(v^iv^j-\frac{1}{3}\delta^{i}_{~j} v^2   \right)
\equiv -3 \frac{FC}{k^2\eta^2} \Psi^{(1)}_{{\bf k}_1}(0) \Psi^{(1)}_{{\bf k}_2}(0) \sin(k_1c_s\eta) 
\sin(k_2c_s\eta)\nonumber\\
&\rightarrow& \frac{27}{k^2} F \Psi^{(1)}_{{\bf k}_1}(0) \Psi^{(1)}_{{\bf k}_2}(0)\, ,
 \end{eqnarray} 
where $F$ and $C$ are defined in Eqs.~(\ref{F}) and~(\ref{C}). Therefore we find that in Fourier space 
\begin{equation}
Q^{(2)}(0)=33 \frac{F}{k^2}\Psi^{(1)}_{{\bf k}_1}(0) \Psi^{(1)}_{{\bf k}_2}(0)\, ,
\end{equation}  
and from Eq.~(\ref{zrad}) we read off the intial condition as (convolution products are understood)  
\begin{equation}
\label{Psi20}
\Psi^{(2)}(0)=\left[-3(a_{\rm NL}-1)+11\frac{F({\bf k}_1,{\bf k}_2,{\bf k})}{k^2}  
\right] \Psi^{(1)}_{{\bf k}_1}(0) \Psi^{(1)}_{{\bf k}_2}(0)\, .  
\end{equation}

\subsection{Perturbation modes with $k \ll k_{eq}$}
\label{klkeq}
Let us consider the photon perturbations which enter the horizon between the equality epoch and 
the recombination epoch, with wavelenghts $\eta_*^{-1} < k  < \eta^{-1}_{eq}$. In fact, in order to find 
some analytical solutions, we will assume that by the time of recombination the universe is 
matter dominated $\eta_{eq} \ll \eta_*$. In this case the gravitational potentials are sourced by the 
dark matter component and their evolution is given in Sec~.\ref{LH2Appmatter}. At linear order the gravitational potentials remain 
constant in time, while at second-order they are given by Eq.~(\ref{solmatter}).  In turn the 
gravitational potentials act as an external force on the CMB photons as in the equation~(\ref{eqoscille}) 
describing the CMB energy density evolution in the tightly coupled regime.    

For the regime of interest it proves convenient to use the solution of Eq.~(\ref{eqoscille}) found in~(\ref{solsem}). The source 
functions ${\cal S}_\Delta$ and ${\cal S}^i_V$ are given by Eqs.~(\ref{SD}) and~(\ref{SV}), respectively. In particular  
${\cal S}_\Delta$ at early 
times -- ${\cal S}_\Delta(0)$ appearing in Eq.~(\ref{solsem})-- vanishes. For a matter dominated period     
\begin{eqnarray}
{\cal S}_\Delta(R=0)&=&\left(\Delta^{(1)2}_{00} \right)^\prime-\frac{16}{3} \Psi^{(1)}\partial_i v^{(1)i}_\gamma+\frac{16}{3}
\left( v_\gamma^2 \right)^\prime
\nonumber \\
&+&\frac{32}{3} \partial^i \Psi^{(1)} v_i\, ,
\end{eqnarray}
where we have used the linear evolution equations~(\ref{LH2B1l}) and~(\ref{LH2vphotontight}) with $\Phi^{(1)}=\Psi^{(1)}$, and 
\begin{eqnarray}
{\cal S}^i_V(R=0)&=&\frac{8}{3}v^{(1)i}_\gamma \partial_jv^{(1)j}_\gamma+\frac{1}{4} \partial^i \Delta^{(1)2}_{00}-2 
\partial^i \Psi^{(1)2}-\Psi^{(1)} \partial^i \Delta^{(1)}_{00}
\nonumber \\
&-&2\omega^{i'}-\frac{3}{4} \partial_j \Pi^{(2)ij}\, .
\end{eqnarray}
As at linear order we are 
evaluating these expressions in the limit $R=3 \rho_b/4 \rho_\gamma \rightarrow 0$, while retaining a non-vanishing 
and constant value for $R$ in the expression for the photon-baryon fluid sound speed entering in the sines and cosines. In fact this approximation gives the dominant contributions to the source terms ${\cal S}_\Delta$ and 
${\cal S}^i_V$ made by first-order squared terms.  
Using the linear solutions~(\ref{LH2D001sol}) 
and~(\ref{LH2v1sol}) for the energy density and velocity of photons, the source functions 
in Fourier space read
\begin{eqnarray}
{\cal S}_\Delta(R=0)&=&\left[ -2 \left(  \frac{6}{5}\right)^2 k_2c_s\, \cos(k_1c_s\eta) \sin(k_2c_s\eta)+
\frac{108}{25} k_2c_s \sin(k_2c_s\eta)
\right. \nonumber \\
&-& \left. \frac{32}{3} \left(  \frac{9}{10}\right)^2 \frac{{\bf k}_1}{k_1} \cdot {\bf k}_2 c_s^3\, 
\sin(k_1c_s\eta) \cos(k_2c_s \eta) +\frac{32}{3} \left( \frac{9}{10} \right)^2 \right. \nonumber \\
& \times& \left. \frac{{\bf k}_1}{k_1} \cdot {\bf k}_2 c_s\, 
\sin(k_1c_s\eta) \right] \Psi^{(1)}_{{\bf k}_1}(0) \Psi^{(1)}_{{\bf k}_2}(0)\, , \nonumber \\
\end{eqnarray} 
and 
\begin{eqnarray}
\label{SVF}
{\cal S}^i_V(R=0)&=&\left[-i \frac{2}{3} \left(  \frac{9}{10}\right)^2 c_s^2\, \frac{k_1^i}{k_1}k_2\, \sin(k_1c_s \eta) 
\sin(k_2c_s \eta)
\right. \nonumber \\
&+& \left. \frac{i}{4} k^i \left(\frac{6}{5} 
\cos(k_1c_s \eta)-\frac{18}{5} \right) \left(\frac{6}{5} \cos(k_2c_s \eta)-\frac{18}{5} \right) 
\right. \nonumber \\
&-& \left. 2i k^i \left(  \frac{9}{10}\right)^2 - i \frac{9}{10} k_2^i \left(\frac{6}{5} \cos(k_2c_s \eta)-\frac{18}{5} \right)     
-2\omega^{i'} 
\right. \nonumber \\
&+& \left. i \frac{2}{3} \left(  \frac{9}{10}\right)^2 
c_s^2 \frac{{\bf k}_2}{k_2} \cdot {\bf k}_1 \frac{k_1^i}{k_1} \sin(k_1c_s \eta) 
\sin(k_2c_s \eta) \right]  
\Psi^{(1)}_{{\bf k}_1}(0) \Psi^{(1)}_{{\bf k}_2}(0)\, . \nonumber \\
\end{eqnarray}
In ${\cal S}^i_V$ we have used the expression~(\ref{2quad}) for the second-order quadrupole moment $\Pi^{(2)ij}_\gamma$ of the 
photons in the tight coupling limit, with the velocity $v^{(1)}=v^{(1)}_\gamma$.  
Notice that, for the modes crossing the horizon at $\eta > \eta_{eq}$,
we have expressed the gravitational potential during the matter dominated period in terms of the initial value on 
superhorizon scales deep in the radiation dominated epoch as $\Psi^{(1)}=9 \Psi^{(1)}(0)/10 $.   

As for the second-order gravitational potentials we have to compute the combination $\Phi^{(2)}+\Psi^{(2)}$ appearing in 
Eq.~(\ref{solsem}). The gravitational potential $\Psi^{(2)}$ is given by Eq.~(\ref{solmatter}), while $\Phi^{(2)}$ is given by 
\begin{equation}
\label{PPmatterrel}
\Phi^{(2)}=\Psi^{(2)}-Q^{(2)}\, ,
\end{equation}  
according to the relation~(\ref{Q}), where for a matter-dominated period 
\begin{equation}
\label{PPmatter}
Q^{(2)}=5 \nabla^{-4}\partial_i\partial_j(\partial^i \Psi^{(1)} \partial_j \Psi^{(1)})-\frac{5}{3}\nabla^{-2} (\partial_k 
\Psi^{(1)} \partial^k \Psi^{(1)})\, .
\end{equation}
We thus find 
\begin{eqnarray}
\label{PP}
& & \Phi^{(2)}+\Psi^{(2)}= 
- 5 \nabla^{-4} \partial_i \partial^j ( \partial^i \Psi^{(1)} \partial_j \Psi^{(1)} ) 
+ \frac{5}{3} \nabla^{-2} (\partial_k \Psi^{(1)} \partial^k \Psi^{(1)})
\nonumber \\
& & 
+2\Psi^{(2)}_m(0)-\frac{1}{7} \left(\partial_k\Psi^{(1)} \partial^k \Psi^{(1)}-\frac{10}{3} 
\nabla^{-2} \partial_i \partial^j (\partial^i \Psi^{(1)} \partial_j \Psi^{(1)})   \right)\, \eta^2\, , \nonumber \\ 
\end{eqnarray}
which in Fourier space reads
\begin{eqnarray}
\label{PPF} 
\Phi^{(2)}+\Psi^{(2)}&=&2 \Psi^{(2)}_m(0)+\left[ \frac{1}{7}G({\bf k}_1,{\bf k}_2,{\bf k})\, \eta^2-\frac{5}{k^2} 
F({\bf k}_1,{\bf k}_2,{\bf k}) \right] 
\nonumber \\
&\times& \left( \frac{9}{10}\right)^2 \Psi^{(1)}_{{\bf k}_1}(0)  \Psi^{(1)}_{{\bf k}_2}(0)\, , 
\end{eqnarray}
where the kernels of the convolutions are given by Eq.~(\ref{Fkernel}) and 
\begin{equation}
\label{Gkernel}
G({\bf k}_1,{\bf k}_2,{\bf k})={\bf k}_1 \cdot {\bf k}_2-\frac{10}{3} \frac{({\bf k}\cdot {\bf k}_1)
 ({\bf k}\cdot {\bf k}_2)}{k^2}\, .
\end{equation}
In Eq.~(\ref{PP}) $\Psi^{(2)}_m(0)$ is the initial condition for the gravitational potential fixed at some time 
$\eta_i > \eta_{eq}$. For the regime of interest it corresponds to the value of the gravitational potential on superhorizon scales 
during the matter-dominated epoch. 

Notice a property that will be useful later on. By looking at Eqs.~(\ref{PPmatter}) and the 
explicit solution for $\Psi^{(2)}$~(\ref{solmatter}) which grows as $\eta ^2$, it is easy to realize that on very small scales, for $k \eta \gg 1$,
the two gravitational potentials are equal with 
\begin{equation}
\label{phi2=psi2}
\Phi^{(2)} \simeq \Psi^{(2)} = \Psi^{(2)}_m(0)+\frac{1}{14}G({\bf k}_1,{\bf k}_2,{\bf k})\, \eta^2\, .
\end{equation}

We are now able to compute the integrals entering in the solution~(\ref{solsem}). The one involving the second-order gravitional potentials 
is straightforward to compute 
\begin{eqnarray}
\label{primoint}
& -& 4kc_s \int_0^\eta d\eta'\, \left( \Phi^{(2)}+\Psi^{(2)} +R \Phi^{(2)} \right) \sin[kc_s(\eta-\eta')]=
\nonumber \\
&-& 8 \Psi^{(2)}_m(0) \left(1-\cos(kc_s \eta)\right) 
- 4 \left[ -\frac{5F}{k^2}  \left(1-\cos(kc_s \eta)\right) 
\right. \nonumber \\
& +& \left. \frac{1}{7k^2c_s^2} G \left(-2+(kc_s \eta)^2+2 \cos(kc_s \eta)\right) 
\right] \left( \frac{9}{10}\right)^2 \Psi^{(1)}_{{\bf k}_1}(0)  \Psi^{(1)}_{{\bf k}_2}(0) \nonumber \\ 
&-&4 R \Psi^{(2)}_m(0) \left(1-\cos(kc_s \eta)\right) 
- 4 R \left[ -\frac{5F}{k^2} \left(1-\cos(kc_s \eta)\right) 
\right. \nonumber \\
& +& \left. \frac{1}{14k^2c_s^2} G \left(-2+(kc_s \eta)^2+2 \cos(kc_s \eta)\right) 
\right] \left( \frac{9}{10}\right)^2 \Psi^{(1)}_{{\bf k}_1}(0)  \Psi^{(1)}_{{\bf k}_2}(0) \, , \nonumber \\
\end{eqnarray}  
where for brevity $F$ and $G$ stand for the convolutions~(\ref{Fkernel}) and~(\ref{Gkernel}). Notice also the we have isolated the terms proportional to 
$R$ as it will be useful later on.  

For the two remaining integrals, in the following we will show only the terms that in the final expression for $\Delta^{(2)}_{00}$ 
and the second-order velocity $v^{(2)i}_\gamma$ give the dominant contributions for $k\eta \gg 1$, even though we have 
perfomed a fully computation. The integral over the source function ${\cal S}_\Delta$ yields a sum of oscillating functions 
(cosines) which turn out to be subdominant, so we do not display the full result. 
For the last integral we find ($(1 \leftrightarrow 2)$ stands by an exchange of indices)
\begin{eqnarray}
\label{terzoint}
&-&\frac{4}{3}\frac{i k_i}{kc_s} \int_0^\eta d\eta' S^i_V \sin[kc_s(\eta-\eta')]=
\left[ \frac{27}{25}
\frac{2 {\bf k} \cdot{\bf k}_2+k^2}{k^2c_s^2} +(1 \leftrightarrow 2) \right] 
\Psi^{(1)}_{{\bf k}_1}(0)  \Psi^{(1)}_{{\bf k}_2}(0)\, , \nonumber \\
\end{eqnarray}
where the terms that have been dropped vary in time as sines and cosines. We have written the contribution in Eq.~(\ref{terzoint}) 
because, upon integration over time, it will give a non-negligible contribution to the velocity 
$v^{(2)i}_\gamma$.

From the general solution~(\ref{solsem}) and the expression~(\ref{solmatter}) for the second-order 
gravitational potential $\Psi^{(2)}$ we thus obtain
\begin{eqnarray}
\label{quasigen}
\Delta^{(2)}_{00}&=& -4 \left(1+R \right) \Psi^{(2)}_m(0) \\
&+& \left[
2(9a_{\rm NL}-7) \Psi^{(1)}_{{\bf k}_1}(0) \Psi^{(1)}_{{\bf k}_2}(0)+(8+4 R)  
\Psi^{(2)}_m(0) \right] \cos(kc_s\eta) \nonumber \\
&-&\frac{2}{7}  G({\bf k}_1,{\bf k}_2,{\bf k}) \eta^2  (1+R)  \left(\frac{9}{10} \right)^2 
\Psi^{(1)}_{{\bf k}_1}(0) \Psi^{(1)}_{{\bf k}_2}(0)\, . \nonumber  
\end{eqnarray}  
We warn  the reader that in writing Eq.~(\ref{quasigen}) we have kept all those terms that contain the primordial non-Gaussianity 
parametrized by $a_{\rm NL}$, and the terms which dominate at late times for $k\eta \gg1$.


\subsubsection{Initial conditions for the second-order gravitational potentials}
\label{ICCMB2}
The initial condition $\Psi^{(2)}_m(0)$ for the modes that cross the horizon after the equality epoch is fixed by 
the value of the gravitational potential on superhorizon scales during the matter dominated epoch. To compute this value we use 
the conservation on superhorizon scales of the curvature perturbation $\zeta^{(2)}$ defined in Eq.~(\ref{defz2}). For a 
matter-dominated period the curvature perturbation on large-scales turns out ot be
\begin{equation}
\zeta^{(2)}=-\Psi^{(2)}_m(0)+\frac{1}{3} \frac{\delta^{(2)} \rho_m}{\rho_m}+\frac{38}{9} \Psi^{(1)2}_m(0)\, ,
\end{equation} 
where we used the energy continuity equation $\delta^{(1)} \rho_m^\prime+3{\cal H}\delta^{(1)} \rho_m-3\rho_m \Psi^{(1)'}=0$ and 
the $(0-0)$ Einstein equation $\delta^{(1)}\rho_m/\rho_m=-2\Psi^{(1)}$ in the superhorizon limit.  

From the $(0-0)$ Einstein equation on large scales $\delta^{(2)} \rho_m/\rho_m=-2\Phi^{(2)}+4\Phi^{(1)2}$ bringing
\begin{equation}
\label{zPmatter}
\zeta^{(2)}=-\frac{5}{3} \Psi^{(2)}_m(0)-\frac{2}{3} \left( \Phi^{(2)}_m(0)-\Psi^{(2)}_m(0) \right)
+\frac{50}{9} \Psi^{(1)2}_m(0)\, .
\end{equation}
The conserved value of $\zeta^{(2)}$ is parametrized as in Eq.~(\ref{param}), 
$\zeta^{(2)}=2 a_{\rm NL} \zeta^{(1)2}=(50 a_{\rm NL}/9) \Psi^{(1)2}$, 
with $\zeta^{(1)}=-5 \Psi^{(1)}/3$ on large scales after the equality
epoch. At second-order the two gravitational 
potentials in a matter dominated epoch differ according to Eq.~(\ref{PPmatter}) and using Eq.~(\ref{zPmatter}) we find 
\begin{eqnarray}
\label{imatter}
\Psi^{(2)}_m(0)&=&-\frac{27}{10} (a_{\rm NL}-1) \left( \Psi^{(1)}(0)\right)^2
+ \left( \frac{9}{10}\right)^2 \left[ 
2\nabla^{-4}\partial_i\partial^j(\partial^i\Psi^{(1)}(0)
\right. \nonumber \\
&\times& \left.
\partial_j\Psi^{(1)}(0))
- \frac{2}{3} \nabla^{-2}(\partial_k\Psi^{(1)}(0)
\partial^k \Psi^{(1)}(0))
\right]\, ,
\end{eqnarray}  
we have expressed the gravitational potential during the matter dominated period $\Psi^{(1)}$ in terms of the initial value on 
superhorizon scales after the equality epoch as $\Psi^{(1)}=9 \Psi^{(1)}(0)/10 $. In Fourier space Eq.~(\ref{imatter}) becomes 
\begin{equation}
\label{Psi20m}
\Psi^{(2)}_m(0)=\left[ 
-\frac{27}{10}(a_{\rm NL}-1)+2 \left( \frac{9}{10} \right)^2 \frac{F({\bf k}_1,{\bf k}_2,{\bf k}) }{k^2} 
\right] \Psi^{(1)}_{{\bf k}_1}(0) \Psi^{(1)}_{{\bf k}_2}(0)\, ,
\end{equation}   
where $F$ is the kernel defined in Eq.~(\ref{Fkernel}). 

We can use the explicit expression for $\Psi^{(2)}_m(0)$ in Eq.~(\ref{quasigen}), still keeping only the terms 
 that contain the primordial non-Gaussianity 
parametrized by $a_{\rm NL}$, and the terms which dominate at late times for $k\eta \gg1$ to find 
\begin{eqnarray}
\Delta^{(2)}_{00}&=&\left[\frac{54}{5}(1+R) (a_{\rm NL}-1)
\right. 
\nonumber \\
&-& \left. \frac{2}{5}(9a_{\rm NL}-19) \cos(kc_s\eta) + \frac{54}{5} R (a_{\rm NL}-1)  \cos(kc_s\eta) 
\right. \nonumber \\
&-& \left. \frac{2}{7} \left( \frac{9}{10} \right)^2 G({\bf k}_1,{\bf k}_2,{\bf k}) \eta^2 (1+R)\right] 
 \Psi^{(1)}_{{\bf k}_1}(0) \Psi^{(1)}_{{\bf k}_2}(0)\, .
\end{eqnarray}

\subsubsection{Second-order photon velocity perturbation}
The second-order velocity of the photons can be obtained from Eq.~(\ref{v2eqf})
\begin{equation}
\label{vgammamatter}
v^{(2)i}_\gamma\simeq\int_0^\eta d\eta'\, \left( {\cal S}^i_V-\partial^i\Phi^{(2)}-\frac{1}{4} \partial^i\Delta^{(2)}_{00}\right)\, .
\end{equation}
Notice that for simplicity in writing Eq.~(\ref{vgammamatter}) we have dropped off the dependence on $R$. In fact  
the main conclusions of this Subsection remains unchanged. 
The second-order gravitional potential in matter-dominated universe 
can be obtained from Eqs.~(\ref{PPmatterrel})-(\ref{PPmatter}) and Eq.~(\ref{solmatter}) as  
\begin{eqnarray}
\Phi^{(2)}&=&
\Psi^{(2)}_m(0)-\frac{1}{14} \left(\partial_k\Psi^{(1)} \partial^k \Psi^{(1)}-\frac{10}{3} 
\nabla^{-2} \partial_i \partial^j (\partial^i \Psi^{(1)} \partial_j \Psi^{(1)})   \right)\, \eta^2 
\nonumber \\
&-& 5 \nabla^{-4} \partial_i \partial^j ( \partial^i \Psi^{(1)} \partial_j \Psi^{(1)} )
+ \frac{5}{3} \nabla^{-2} 
(\partial_k \Psi^{(1)} \partial^k \Psi^{(1)})\, .
\end{eqnarray}
In Fourier space this becomes
\begin{eqnarray}
\label{PhiF} 
\Phi^{(2)}&=&\Psi^{(2)}_m(0)+\left[ \frac{1}{14}G({\bf k}_1,{\bf k}_2,{\bf k})\, \eta^2-\frac{5}{k^2} 
F({\bf k}_1,{\bf k}_2,{\bf k}) \right] 
\nonumber \\
&\times& 
\left( \frac{9}{10}\right)^2 \Psi^{(1)}_{{\bf k}_1}(0)  \Psi^{(1)}_{{\bf k}_2}(0)\, ,
\end{eqnarray}
where the kernels of the convolutions are given by Eqs.~(\ref{Fkernel}) and~(\ref{Gkernel}). The integral over $\Phi^{(2)}$ in 
Eq.~(\ref{vgammamatter}) is then easily computed
\begin{eqnarray}
\label{primointv}
-\int_0^\eta d\eta'\, \partial^i \Phi^{(2)}&\equiv& -i k^i\left[ \Psi^{(2)}_m(0) \eta+ 
\left(\frac{1}{42} G({\bf k}_1,{\bf k}_2,{\bf k}) \eta^3
\right. \right.\nonumber \\
&-& \left. \left. \frac{5}{k^2}F({\bf k}_1,{\bf k}_2,{\bf k})
\eta  \right)  \left( \frac{9}{10}\right)^2  
 \Psi^{(1)}_{{\bf k}_1}(0) \Psi^{(1)}_{{\bf k}_2}(0) \right]\, ,
\end{eqnarray}  
where as usual the equivalence symbol means that we are evaluating a given expression in Fourier space.
For the integral over the source function ${\cal S}^i_V$ we use its expression in Fourier space, Eq.~(\ref{SVF}), and the 
dominant terms for $k\eta \gg 1$ are
\begin{eqnarray}
\label{SVint}
\int_0^\eta d\eta' {\cal S}^i_V &\equiv& \left[ - 2ik^i \left( \frac{9}{10}\right)^2+ik^i\frac{81}{25}+i\frac{8}{3} 
\left( \frac{9}{10}\right)^2 \frac{1}{k^4} (k_2^2-k_1^2) {\bf k}\cdot{\bf k}_1 k_2^i 
\right] \eta \, . \nonumber \\  
\end{eqnarray}
Notice that, in order to compute this integral, 
we must know the second-order vector metric petturbation $\omega^i$. This is easily obtained 
for a matter-dominated universe from Eq.~(\ref{omegamatter}). Using Eqs.~(\ref{LH2deltamatter1}) and~(\ref{LH2velocitymatter1}) one 
finds   
\begin{eqnarray}
\omega^{i}&=&-\frac{4}{3}\left( \frac{9}{10}\right)^2  \nabla^{-4} \partial_j 
\left[ \partial^i\nabla^2\Psi^{(1)}(0) \partial^j\Psi^{(1)}(0)-\partial^j\nabla^2\Psi^{(1)}(0)
\right. \nonumber \\
&\times& \left. \partial^i\Psi^{(1)}(0)
\right]\eta \, , 
\end{eqnarray}
giving rise to the third term in Eq.~(\ref{SVint}).

Finally for the integral over $\Delta^{(2)}_{00}$ some caution is nedeed. Since in the final expression for $v^{(2)i}_\gamma$ 
the dominant terms at late times turn out to be proportional $\eta$, one has to use an expression 
for $\Delta^{(2)}_{00}$ that keep track of all those contributions that, upon integration, scale like $\eta$. Thus we must use 
the expression written in Eq.~(\ref{quasigen}), plus Eq.~(\ref{terzoint}), and some terms of 
Eq.~(\ref{primoint}) that have been previously neglected in Eq.~(\ref{quasigen}). Then we find for $k\eta \gg 1$ 
\begin{eqnarray}
\label{terzointv}
&+&\frac{1}{4} \int_0^\eta d\eta' \partial^i \Delta^{(2)}_{00} \equiv 
\frac{ik^i}{4} \left[ -4 \Psi^{(2)}_m(0) \eta+
\left( 2 (9a_{\rm NL}-7)   \Psi^{(1)}_{{\bf k}_1}(0) \Psi^{(1)}_{{\bf k}_2}(0)
\right. \right. \nonumber \\
&+& \left. \left. 8 \Psi^{(2)}_m(0)  
\right) \frac{\sin(kc_s \eta)}{kc_s} \right] 
+\frac{ik^i}{4} \left[- \frac{2}{21} 
\left( \frac{9}{10} \right)^2 G \eta^3  
+ \left( \frac{20}{3c_s^2} 
\left( \frac{9}{10} \right)^2 \frac{F}{k^2}
\right. \right. \nonumber \\
&+& \left. \left. \frac{8}{21c_s^4}\left(\frac{9}{10} \right)^2 \frac{G}{k^2} 
+\frac{54}{25c_s^2} \frac{k^2+{\bf k}\cdot ({\bf k}_1+{\bf k}_2)}{k^2} \right) \eta
\right] 
\Psi^{(1)}_{{\bf k}_1}(0)  \Psi^{(1)}_{{\bf k}_2}(0) \, . \nonumber \\
\end{eqnarray} 
Using Eqs.~(\ref{primointv}),~(\ref{SVint}) and~(\ref{terzointv}) we get  
\begin{eqnarray}
\label{vgm}
&v^{(2)i}_\gamma&=\Bigg[ i\frac{k^i}{k} \frac{1}{10 c_s} (9a_{\rm NL}-19) \sin(kc_s\eta)+ \left( -i 
\frac{2}{21c_s^4} k^i \left( \frac{9}{10} \right)^2 \frac{G}{k^2} 
\right. \nonumber \\
&-& \left. 2 i \left( \frac{9}{10} \right)^2 k^i+ i\frac{81}{50}(k_2^i+k_1^i) 
- i\frac{27}{50c_s^2} \frac{k^2+{\bf k} \cdot({\bf k}_1
+{\bf k}_2)}{k^2} k^i \right. \nonumber \\
&+& \left. i\frac{4}{3} \left( \frac{9}{10} \right)^2 \frac{k_2^2-k_1^2}{k^4}({\bf k} \cdot{\bf k}_1k_2^i-
{\bf k} \cdot{\bf k}_2 k_1^i ) \right) \eta
\Bigg] \Psi^{(1)}_{{\bf k}_1}(0)  \Psi^{(1)}_{{\bf k}_2}(0)\, . \nonumber \\
\end{eqnarray}
To obtain Eq.~(\ref{vgm}) we have also used the explicit expression~(\ref{Psi20m}) for $\Psi^{(2)}_m(0)$ and we have kept the terms 
depending 
on $a_{\rm NL}$ parametrizing the primordial non-Gaussianity and the terms that dominate at late times for $k\eta \gg1$.

\subsection{Perturbation modes with $k \gg k_{eq}$: Cold Dark Matter perturbations at second order 
and improved analytical solutions for CMB anisotropies} 
\label{improved}
In Sec.~\ref{kggkeq2} we have computed the perturbations of the CBM photons at last scattering 
for the modes that cross the horizon at $\eta < \eta_{eq}$ under the approximation that the universe  
is radiation-dominated. However around the equality epoch, through recombination, 
the dark matter component will start to dominate. In this section we will account for its 
contribution to the gravitational potential and for the resulting perturbations of the photons 
from the equality epoch onwards. This leads to a more realistic and accurate abalytical solutions 
for the acoustic oscillations of the photon-baryon fluid for the modes of interest. 

The starting point is to consider the density perturbation in the dark matter component for 
subhorizon modes during the radiation dominated epoch. Its value at the equality epoch will 
fix the magnitude of the gravitational potential at $\eta_{eq}$ and hence the initial conditions 
for the subsequent evolution of the photons fluctuations during the matter dominated period. At linear order the procedure is standard 
(see, {\rm e.g} ~\cite{Dodelsonbook}), and we will use a similar one at 
second-order in the perturbations. Thus this section serves also as a guide through the evolution of the CDM density perturbations at second-order accounting for those modes that enter the horizon during the radiation dominated epoch. 
This allows to determine  the second-order transfer function for the density perturbations with a generalization at second-orde of the 
Meszaros effect.

\subsubsection{Subhorizon evolution of CDM perturbations for $\eta < \eta_{eq}$}

From the energy and velocity continuity equations for CDM it is possible to isolate an evolution 
equation for the density perturbation $\delta_{\d}=
\delta \rho_{\d}/\rho_{\d}$, where the subscript $d$ stands for cold dark matter. In Ref.~\cite{paperI} we have obtained the 
Boltzmann equations up to second-order for CDM. The number density of CDM evolves according to~\cite{paperI}
\begin{eqnarray}
& & \frac{\partial n_{\d}}{\partial \eta}+e^{\Phi+\Psi} \frac{\partial(v^i_{\d} n_{\d})}{\partial x^i}+3(
{\cal H}-\Psi')n_{\d} -2 e^{\Phi+\Psi} \Psi_{,k}v_{\d}^k\,  n_{\d}
\nonumber \\
&& +e^{\Phi+\Psi} \Phi_{,k}v_{\d}^k \, 
n_{\d}=0\, . 
\end{eqnarray} 
At linear order $n_{\d}=\bar{n}_{\d}+\delta^{(1)} n_{\d}$ and one recovers the usual energy continuity equation
\begin{equation}
\label{Dd1}
\delta^{(1)'}_\d+v^{(1)i}_{\d,i}-3\Psi^{(1),i}=0\, ,
\end{equation} 
with $\delta^{(1)}_\d=\delta^{(1)} \rho_{\d}/\bar{\rho}_{\d}=\delta^{(1)} n_\d/\bar{n}_\d$. The CDM velocity  
at the same order of perturbation obeys~\cite{paperI}
\begin{equation}
\label{vd1}
{v^{(1)i}}'_\d+{\cal H}v^{(1)i}_\d=-\Phi^{(1),i}\, .
\end{equation} 

Perturbing $n_\C$ up to second-order 
we find
\begin{eqnarray}
\delta^{(2)'}_\d+v^{(2)i}_{\d ,i}-3\Psi^{(2)'}&=&-2(\Phi^{(1)}+\Psi^{(1)})v^{(1)i}_{\d,i}-2 v^{(1)i}_{\d,i}\delta^{(1)}_\d
-2 v^{(1)i}_\d \delta^{(1)}_{\d,i}
\nonumber \\
&+&6\Psi^{(1)'} \delta^{(1)}_\d+(4\Psi^{(1)}_{,k}-2\Phi^{(1)}_{,k}) v^{(1)k}_\d \, .
\end{eqnarray}
The R.H.S. of this equation can be further manipulated by using the linear equation~(\ref{Dd1}) to replace $v^{(1)i}_{\d~~,i}$ 
yielding
\begin{eqnarray}
\label{Dd2}
\delta^{(2)'}_\d+v^{(2)i}_{\d~~,i}-3\Psi^{(2)'}&=&4\delta^{(1)'}_\d\Psi^{(1)}-6 \left( \Psi^{(1)2} \right)^\prime 
+\left(\delta^{(1)2}_\d \right)^\prime-2v^{(1)i}_\d \delta^{(1)}_{\d,i}
\nonumber \\
&+&2\Psi^{(1)}_{,k} v^{(1)k}_\d\, ,
\end{eqnarray}
where we use $\Phi^{(1)}=\Psi^{(1)}$. 
In Ref.~\cite{paperI} the evolution equation for the second-order CDM velocity perturbation has been already obtained 
\begin{eqnarray}
\label{vd2}
{v^{(2)i}}'_\d+{\cal H}v^{(2)i}_\d+2\omega^{i'}+2{\cal H}\omega^i+\Phi^{(2),i}&=&2\Psi^{(1)'} v^{(1)i}_\d-2v^{(1)j}_\d \partial_j 
v^{(1)i}_\d
\nonumber \\
&-&4\Phi^{(1)}\Phi^{(1),i}\, .
\end{eqnarray}

At linear order we can take the divergence of Eq.~(\ref{vd1}) and, using Eq.~(\ref{Dd1}) to replace the velocity perturbation, we 
obtain a differental equation for the CDM density contrast
\begin{equation}
\label{comb}
\left[a \left(3\Psi^{(1)'}-\delta^{(1)'}_\d \right)     \right]^\prime = -a \nabla^2 \Phi^{(1)}\, ,
\end{equation} 
which can be rewritten as 
\begin{equation}
\label{master1}
\delta^{(1)''}_\d+{\cal H}\delta^{(1)'}_\d=S^{(1)}\, ,
\end{equation}
where 
\begin{equation}
S^{(1)}=3\Psi^{(1)''}+3{\cal H}\Psi^{(1)'}+\nabla^2 \Phi^{(1)}\, .
\end{equation}
When the radiation is dominating the gravitational potential is mainly due to the perturbations in the photons, and $a(\eta) 
\propto \eta$.
For subhorizon scales Eq.~(\ref{master1}) can be solved following the procedure introduced in Ref.~\cite{HSsmallscales}. 
Using the Green method the general solution to Eq.~(\ref{master1})  (in Fourier space) is given by
\begin{equation}
\label{sold}
\delta^{(1)}_\d({\bf k},\eta)=C_1+C_2\ln(\eta)-\int_0^\eta d\eta' S^{(1)}(\eta')\, \eta'(\ln(k\eta')-\ln(k\eta))\, ,
\end{equation}
where the first two terms correspond to the solution of the homogeneous equation. At early times the density contrast is constant 
with 
\begin{equation}
\label{C1f}
\delta^{(1)}_d(0)=\frac{3}{4} \Delta^{(1)}_{00}(0)=-\frac{3}{2} \Phi^{(1)}_{\bf k}(0)\, ,
\end{equation}
having used the adiabaticity condition, and thus we fix the integration constant as 
\begin{equation}
C_1=-3 \Phi^{(1)}_{\bf k}(0)/2\, ,
\end{equation}
and $C_2=0$. The gravitational potential during the radiation-dominated epoch 
 starts to decay as a given mode enters the horizon. Therefore the source term $S^{(1)}$  
behaves in a similar manner and this implies that
the integrals over $\eta'$ reach asymptotically a constant value. Once the mode has crossed the horizon we can thus write the solution as  
\begin{equation}
\label{Mes1}
\delta^{(1)}_\d({\bf k},\eta)=A^{(1)} \Phi^{(1)}(0) \ln[B^{(1)} k\eta]\, ,
\end{equation}   
where the constants $A^{(1)}$ and $B^{(1)}$ are defined as 
\begin{equation}
\label{A1}
A^{(1)} \Phi^{(1)}(0)=\int_0^\infty d\eta' S^{(1)}(\eta')\eta'\, , 
\end{equation}
and 
\begin{equation}
\label{defiB1}
A^{(1)} \Phi^{(0)} \ln(B^{(1)})=-\frac{3}{2}\Phi^{(1)}(0)-\int_0^\infty d\eta' S^{(1)}(\eta')\, \eta' \ln(k\eta')\, .
\end{equation}
The upper limit of the integrals can be taken to infinity because the main contribution comes form when $k\eta \sim 1$ and  
once the mode has entered the horizon the result will change by a very small quantity. Performing the integrals in Eq.~(\ref{A1}) and~(\ref{defiB1}) one finds that $A^{(1)}=-9.6$ and 
$B^{(1)}\simeq 0.44$.

Before moving to the second-order case, a useful quantity to compute is the CDM velocity in a radiation dominated epoch.  From 
Eq.~(\ref{vd1}) it is given by  
\begin{equation}
\label{vd1sol}
v^{(1)i}_d-\frac{1}{a}\int_0^\eta d\eta' \, \partial^i \Phi^{(1)} a(\eta') \equiv -3 (ik^i) \Phi^{(1)}(0) \frac{kc_s\eta
-\sin(kc_s\eta)}{k^3c_s^3\eta^2}\, ,
\end{equation}
where the last equality holds in Fourier space and we have used Eq.~(\ref{LH2Phir}) (and the fact that $a(\eta)\propto \eta$ when 
radiation dominates).

Combining Eq.~(\ref{Dd2}) and~(\ref{vd2}) we get the analogue of Eq.~(\ref{master1}) at second-order in perturbation theory 
\begin{equation}
\label{masterq2}
\delta^{(2)''}_\d+{\cal H} \delta^{(2)'}_\d=S^{(2)}\, ,
\end{equation}  
where the source function is 
\begin{eqnarray}
S^{(2)}&=&3\Psi^{(2)''}+3{\cal H} \Psi^{(2)'} +\nabla^2\Phi^{(2)}-2\partial_i(\Psi^{(1)}v^{(1)i}_\d)+\nabla^2 v^{(1)2}_\d
\nonumber \\
&+&2 \nabla^2\Phi^{(1)2}+\frac{1}{a}\left[ a\left( 
4\delta^{(1)'}_\d \Psi^{(1)}-6 \left( \Psi^{(1)2} \right)^\prime 
+\left(\delta^{(1)2}_\d \right)^\prime \right. \right. \nonumber \\
&-& \left.  \left. 2v^{(1)i}_\d \delta^{(1)}_{\d,i}+2\Psi^{(1)}_{,k} v^{(1)k}_\d
\right) \right]^\prime \, .
\end{eqnarray} 
In fact we write Eq.~(\ref{masterq2}) in a more convenient way as 
\begin{eqnarray}
\label{master2}
\delta^{(2)''}_\d-3\Psi^{(2)''}-s'_1 + {\cal H}( \delta^{(2)'}_\d-3\Psi^{(2)'}-s_1)=s_2\, ,  
\end{eqnarray}
where for simplicity we have introduced the two functions 
\begin{equation}
\label{s1}
s_1=4\delta^{(1)'}_\d \Psi^{(1)}-6 \left( \Psi^{(1)2} \right)^\prime 
+\left(\delta^{(1)2}_\d \right)^\prime -2v^{(1)i}_\d \delta^{(1)}_{\d,i}+2\Psi^{(1)}_{,k} v^{(1)k}_\d\, ,
\end{equation} 
and 
\begin{equation}
\label{s2}
s_2=\nabla^2\Phi^{(2)}-2\partial_i(\Psi^{(1)'}v^{(1)i}_\d)+\nabla^2 v^{(1)2}_\d+2\nabla^2 \Phi^{(1)2}\, .
\end{equation}
In this way we get an equation of the same form as~(\ref{master1}) in the variable 
$[\delta^{(2)}-3\Psi^{(2)}-\int_0^\eta d\eta' s_1(\eta')]$ with source $s_2$ on the R.H.S.. Its solution in Fourier space 
therefore is just as Eq.~(\ref{sold})   
\begin{eqnarray}
\label{sold2}
\delta^{(2)}_\d-3\Psi^{(2)}-\int_0^\eta d\eta' s_1(\eta')&=&
C_1+C_2\ln(\eta) \\
&-&\int_0^\eta d\eta' s_2(\eta')\, \eta'\, [\ln(k\eta')-\ln(k\eta)]\, . \nonumber 
\end{eqnarray}
As we will see, Eq.~(\ref{sold2}) provides the generalization of the Meszaros effect at second-order in 
perturbation theory.   
 
\subsubsection{Initial conditions}
In the next two sections we will compute explicitly the expression~(\ref{sold2}) for the 
second-order CDM density contrast on subhorizon scales during the radiation dominate era. First let 
us fix the constants $C_1$ and $C_2$ by appealing to the initial conditions. At $\eta \rightarrow 0$ the L.H.S. of 
Eq.~(\ref{sold2}) is constant, as one can check by using the results of Sec.~\ref{icrd} and the 
condition of adiabaticity at second-order (see, {\it e.g.}, Ref.~\cite{evolution,review}) which relates the CDM density contrast 
at early times on superhorizon scales to the energy density fluctuations of photons by    
\begin{equation}
\label{adia2nd}
\delta^{(2)}_\d(0)=\frac{3}{4} \Delta^{(2)}_{00}(0)-\frac{1}{3} \left( \delta^{(1)}_d(0) \right)^2=
\frac{3}{4} \Delta^{(2)}_{00}(0)-\frac{3}{4} \left(\Phi^{(1)}(0)\right)^2
\, ,
\end{equation}
where in the last step we have used Eq.~(\ref{C1f}). 
Therefore we can fix $C_2 =0$ and 
\begin{equation}
C_1=\delta^{(2)}_d(0)-3\Psi^{(2)}(0)\, .
\end{equation}    
 
Eq.~(\ref{DPsi20}) gives $\Delta^{(2)}_{00}(0)-4\Psi^{(2)}(0)$ in terms of the primordial non-Gaussianity parametrized 
by $a_{\rm NL}$, and the expression for $\Psi^{(2)}(0)$ have been already computed in Eq.~(\ref{Psi20}). Thus we find (in Fourier space)
\begin{equation}
\Delta^{(2)}_{00}=\left[2(3 a_{\rm NL}-1)+44 \frac{F({\bf k}_1,{\bf k}_2,{\bf k})}{k^2}
\right] \Psi^{(1)}_{{\bf k}_1}(0) \Psi^{(1)}_{{\bf k}_2}(0)\, ,
\end{equation} 
and from Eqs.~(\ref{adia2nd}) we derive the initial density contrast for CDM at second-order
\begin{equation}
\label{Dd20}
\delta^{(2)}_\d(0)=\left[\frac{3}{2}(3 a_{\rm NL}-1)+33 \frac{F({\bf k}_1,{\bf k}_2,{\bf k})}{k^2}
-\frac{3}{4} \right] \Psi^{(1)}_{{\bf k}_1}(0) \Psi^{(1)}_{{\bf k}_2}(0)\, .
\end{equation}
Eq.~(\ref{Dd20}) togheter with Eq.~(\ref{Psi20}) allows to compute the constant $C_1$ as 
\begin{equation}
\label{C12}
C_1=\delta^{(2)}_d(0)-3\Psi^{(2)}(0)=\left[\frac{27}{2}(a_{\rm NL}-1)+\frac{9}{4}  
\right] \Psi^{(1)}_{{\bf k}_1}(0) \Psi^{(1)}_{{\bf k}_2}(0)\, .
\end{equation}

\subsubsection{Meszaros effect at second order}
We now compute the integrals over the functions $s_1$ and $s_2$ appearing in Eq.~(\ref{sold2}). Let us first focus on the integral 
$\int_0^\eta d\eta'\, s_1(\eta')$. 

Notice that, using the linear equations~(\ref{Dd1}) and~(\ref{vd1}) for the CDM density and velocity perturbations, the function 
$s_1(\eta')$ can be written in a more convenient way as 
\begin{equation}
s_1(\eta)=-6\Psi^{(1)} v^{(1)i}_{\d~~,i}+\left(\delta^{(1)2}_d\right)'-2v^{(1)i}_{\d}\delta^{(1)}_{\d,i}+2(\Psi^{(1)}v^{(1)k}_\d)_{,k}\, ,
\end{equation} 
and then 
\begin{eqnarray}
\label{intint}
\int_0^\eta d\eta'\, s_1(\eta')&=&\left( \delta^{(1)}_d(\eta)\right)^2-\left(\delta^{(1)}_d(0)\right)^2
\\
&+&\int_0^\eta d\eta'\, \left[ 
-2v^{(1)i}_{\d}\delta^{(1)}_{\d,i}+2(\Psi^{(1)}v^{(1)k}_\d)_{,k}
-6\Psi^{(1)} v^{(1)i}_{\d~~,i} \right] \, . \nonumber 
\end{eqnarray}
In Eq.~(\ref{intint}) all the quantities are known being first-order perturbations: the linear gravitional potential $\Psi^{(1)}$ 
for a radiation dominated era is given in Eq.~(\ref{LH2Phir}), the CDM velocity perturbation corresponds to Eq.~(\ref{vd1sol}) and the 
CDM density contrast is given by Eq.~(\ref{Mes1}). Thus the integral in Eq.~(\ref{intint}) reads (in Fourier space)
\begin{eqnarray}
\label{intintF}
\int_0^\eta &d\eta'& \, \left[ -3 A^{(1)}\, {\bf k}_1 \cdot{\bf k}_2 \frac{k_1c_s\eta'-\sin(k_1c_s\eta')}{k_1^3c_s^3\eta^{'2}}\, 
\ln(B^{(1)}k_2\eta')   \right.  \nonumber \\
&+& \left.
 (9({\bf k} \cdot {\bf k}_1)-27k_1^2) \frac{k_1c_s\eta'-\sin(k_1c_s\eta')}{k_1^3c_s^3\eta^{'2}} 
\right.
\nonumber \\
&\times& \left.
\frac{\sin(k_sc_s\eta')-k_2c_s\eta'
\cos(k_2c_s\eta')}{k_2^3c_s^3\eta^{'3}} \right] \Psi^{(1)}_{{\bf k}_1}(0) \Psi^{(1)}_{{\bf k}_2}(0) \, .
\end{eqnarray}
Let us recall that we are interested in the evolution of the CDM second-order density contrast  
on subhorizon scales during the radiation dominated epoch. Therefore once we compute the integrals we are interested in the limit of 
their expression for late times $(k\eta \gg 1$). For the first contribution to Eq.~(\ref{intintF}) we find that at late times it is 
well approximated by the expression 
\begin{eqnarray}
\label{int1sol}
& & \int_0^\eta d\eta' \, 3 A^{(1)}({\bf k}_1 \cdot{\bf k}_2) \frac{k_1c_s\eta'-\sin(k_1c_s\eta')}{k_1^3c_s^3\eta^{'2}}\, 
\ln(B^{(1)}k_2\eta') \simeq  
\nonumber \\
& & 3A^{(1)} \frac{{\bf k}_1 \cdot{\bf k}_2}{k_1^2c_s^2} \left[ 
2.2\left(-\frac{1.2}{2}\left[\ln(k_1c_s\eta)\right]^2+\ln(B^{(1)}k_2\eta) \ln(k_1c_s\eta) \right) \right]\, . \nonumber \\
\end{eqnarray}
We have computed also the remaining integral in Eq.~(\ref{intintF}), 
but it turns out to be negligible compared to Eq.~(\ref{int1sol}). The reason is that 
the integrand oscillates with an amplitude decaying in time as $\eta^{-3}$, and thus it leads just to a constant (the argument is the 
same we used at linear order to compute the integrals in Eq.~(\ref{sold})). Thus we can write
\begin{eqnarray}
\label{ints1sol}
& & \int_0^\eta d\eta' s_1(\eta')=\left( \delta^{(1)}_d(\eta)\right)^2-\left(\delta^{(1)}_d(0)\right)^2-3A^{(1)} \frac{{\bf k}_1 \cdot{\bf k}_2}{k_1^2c_s^2}
\\
&&\times \left[ 
2.2\left(-\frac{1.2}{2}\left[\ln(k_1c_s\eta)\right]^2+\ln(B^{(1)}k_2\eta) \ln(k_1c_s\eta) \right) \right] 
\Psi^{(1)}_{{\bf k}_1}(0) \Psi^{(1)}_{{\bf k}_2}(0) 
\, . \nonumber 
\end{eqnarray} 


We now compute the integrals over the function $s_2(\eta)$ given in Eq.~(\ref{s2}). Since at late times $\phi^{(1)2} \sim 1/\eta^4$ 
and $(\Psi^{(1)'} v^{(1)i}_\d)_{,i} \sim 1/\eta^3$ the main contribution to the integral will come from the two remaining terms, 
$\Phi^{(2)}$ and $v^{(1)2}_d$, whose amplitudes scale at late times as $1/\eta^2$   
\begin{equation}
\label{s2c}
s_2 \simeq \nabla^2\Phi^{(2)}+\nabla^2 v^{(1)2}_\d \, .
\end{equation}
Two are the integrals that we have to compute 
\begin{equation}
\label{ints21}
\int_0^\eta d\eta' s_2(\eta') \eta' \ln(k\eta')\, , 
\end{equation}
and the one multiplying
$\ln(k\eta)$ 
\begin{equation}
\label{ints22}
\int_0^\eta d\eta' s_2(\eta') \eta'\, . 
\end{equation} 
Let us first consider the contributions from $\nabla^2 v^{(1)2}_\d$. 
The second integral is easily computed using the expression~(\ref{vd1sol}) for the linear CDM velocity. We find that at late times 
the dominant term is (for $k\eta \gg 1$)
\begin{eqnarray}
\label{cv2}
-\int_0^\eta d\eta'\, \nabla^2v^{(1)2}_\d \eta' &\equiv& -\left[
\frac{9}{c_s^4} k^2 \frac{{\bf k}_1 \cdot {\bf k}_2}{k_1^2k_2^2}\, \ln(k\eta)
\right]  \Psi^{(1)}_{{\bf k}_1}(0) \Psi^{(1)}_{{\bf k}_2}(0) \, .
\end{eqnarray}
The first integral can be computed by making the following approximation: we split the integral between $0< k\eta <1$ and $k\eta >1$ 
and for $0< k\eta <1$ we use the asymptotic expression
\begin{equation}
v^{(1)i}_\d \approx -\frac{1}{2} ik^i\eta\,  \Psi^{(1)}_{\bf k}(0)   \quad\quad (k\eta \ll 1)\, ,
\end{equation}
while for $ k\eta >1$ we use the limit 
\begin{equation}
v^{(1)i}_\d \approx -i \frac{3}{c_s^2} \frac{k^i}{k} \frac{1}{k\eta} \Psi^{(1)}_{\bf k}(0)  \quad\quad (k\eta \gg 1)\, .
\end{equation}
The the integral for $0< k\eta <1$ just gives a constant, while the integral for $ k\eta >1$ brings the dominant contribution at late 
times being proportional to $[\ln(k\eta)]^2$ so that we can write ($k\eta \gg 1$)
\begin{equation}
\label{cv1}
\int_0^\eta d\eta' \nabla^2 v^{(1)2}_d \eta'\,\ln(k\eta') = \frac{9}{2c_s^4}\, k^2 \frac{{\bf k}_1 \cdot {\bf k}_2}{k_1^2 k_2^2}\, 
\left[\ln(k\eta)\right]^2 \Psi^{(1)}_{{\bf k}_1}(0) \, \Psi^{(1)}_{{\bf k}_2}(0)\, .  
\end{equation}   
As far as the contribution to the integrals~(\ref{ints21}) and~(\ref{ints22}) due to $\nabla^2 \Phi^{(2)}$ is concerned 
we have just to keep track of the initial condition provided by the primordial non-Gaussianity. We have verified that all the other 
terms give a negligible contribution. This is easy to understand: 
the integrand function on large scale is a constant while at late times it oscillates with 
decreasing amplitudes as $\eta^{-2}$, and thus the integrals will tend asymptotically to a constant. We find that 
\begin{equation}
\label{int1Phi}
\int_0^\eta d\eta' \nabla^2 \Phi^{(2)}\eta' \simeq -9 \Phi^{(2)}(0)\, ,
\end{equation} 
and 
\begin{equation}
\label{int2Phi}
\int_0^\eta d\eta' \nabla^2 \Phi^{(2)}\eta' \ln(k\eta')\simeq\left( -9+9\gamma-9\frac{\ln3}{2} \right) \Phi^{(2)}(0)\, ,
\end{equation}
where $\gamma=0.577...$ is the Euler constant, and $\Phi^{(2)}(0)$ is given by Eq.~(\ref{PmP}). 

Therefore, from Eqs.~(\ref{cv1}), (\ref{cv2}), and~(\ref{int1Phi})-(\ref{int2Phi}) we find that for $k\eta \gg 1$ 
\begin{eqnarray}
\label{ints2sol}
& & \int_0^\eta d\eta' s_2(\eta') \eta' [\ln(k\eta')-\ln(k\eta)] = -\frac{9}{2c_s^4} k^2 \frac{{\bf k}_1 \cdot {\bf k}_2}{k_1^2k_2^2} \, 
[\ln(k\eta)]^2 
\nonumber \\
&& \times \Psi^{(1)}_{{\bf k}_1}(0) \Psi^{(1)}_{{\bf k}_2}(0) 
+9 \Phi^{(2)}(0) 
\left(-9+9\gamma-9\frac{\ln(3)}{2}\right) \ln(k\eta) \Phi^{(2)}(0)\, . \nonumber \\
\end{eqnarray}

Let us collect the results of Eqs.~(\ref{C12}),~(\ref{ints1sol}) and~(\ref{ints2sol}) into Eq.~(\ref{sold2}). 
We find that for $k\eta \gg 1$ 
\begin{eqnarray}
\label{sold2f}
\delta^{(2)}_\d(k\eta \gg 1)&=&
\Big[-3 (a_{\rm NL}-1) A_1 \ln(B_1k\eta) + A_1^{2} \ln(B_{1}k_1\eta) \ln(B_{1}k_2\eta)
\nonumber \\
&+&
\Big[ -\frac{3}{2} A_{1} \frac{{\bf k}_1 \cdot{\bf k}_2}{k_1^2c_s^2} 
\, 2.2\Big(-\frac{1.2}{2}\left[\ln(k_1c_s\eta)\right]^2 + \ln(B_{1}k_2 \eta) 
 \nonumber \\
&\times&  
\ln(k_1c_s\eta) \Big) 
+ (1 \leftrightarrow 2) \Big]
+\frac{9}{2c_s^4} k^2 \frac{{\bf k}_1 \cdot {\bf k}_2}{k_1^2k_2^2} \, 
\left[\ln(k\eta)\right]^2 \Big]
\nonumber \\
&\times&
\Psi^{(1)}_{{\bf k}_1}(0) \Psi^{(1)}_{{\bf k}_2}(0) \, .
\end{eqnarray}     
Notice that in Eq.~(\ref{sold2}) we have neglected $\Psi^{(2)}$, which decays on subhorizon scales during the radiation dominated 
epoch (see Eq.~(\ref{Psi2rf}), and we have used Eqs.~(\ref{C1f}) and~(\ref{Mes1}). 
Eq.~(\ref{sold2f}) represents the second-order Meszaros effect: the CDM density contrast on small scales (inside the horizon) 
slowly grows starting from the initial conditions that, 
at second-order, are set by the primordial non-Gaussianity parameter $a_{\rm NL}$. As one could have guessed the primordial 
non-Gaussianity is just transferred linearly. The other terms scale in time as a logarithm squared.  We stress that 
the computation of these terms allows one to derive the full transfer function for the matter 
perturbations at second order accounting for the dominant second-order corrections. 
In the next section we will use~(\ref{sold2f}) to fix the initial conditions for the evolution on subhorizon scales of the 
photons density fluctuations $\Delta^{(2)}_{00}$ after the equality epoch.

\subsubsection{Second-order CMB anisotropies for modes crossing the horizon during the radiaton epoch}
\label{Dms}
In this section we derive the energy density perturbations $\Delta^{(2)}_{00}$ of the photons during the matter dominated epoch, for the 
modes that cross the horizon before equality. In Sec.~\ref{klkeq}
we have already solved the problem assuming matter domination for modes crossing the horizon after equality.  
Thus it is sufficient to take the result~(\ref{quasigen}) and replace the initial conditions
\begin{eqnarray}
\label{gen}
\Delta^{(2)}_{00}&=& -4 \left(1+R \right) \Psi^{(2)}_m(0) \\
&+& \left[
A+(8+4 R)  
\Psi^{(2)}_m(0) \right] \cos(kc_s\eta)+B \sin(kc_s\eta) \nonumber \\
&-&\frac{2}{7}  G({\bf k}_1,{\bf k}_2,{\bf k}) \eta^2  (1+R)  \left(\frac{9}{10} \right)^2 
\Psi^{(1)}_{{\bf k}_1}(0) \Psi^{(1)}_{{\bf k}_2}(0)\, . \nonumber  
\end{eqnarray}  
where we have restored the generic integration constants $A$ and $B$, $\Psi^{(1)}$ is the linear gravitational potential 
(which is constant for the matter era) and $\Psi^{(2)}_m(0)$ represents   
the initial condition for the second-order gravitational potential fixed at some time $\eta_i > \eta_{eq}$.
Eq.~(\ref{sold2f}) allows to fix the proper initial conditions for 
the gravitational potentials on subhorizon scales (accounting for the fact that around the equality epoch they are 
mainly determined by the CDM density perturbations). 
At linear order this is achieved by solving the equation for $\delta^{(1)}_\d$ which is obtained 
from Eq.~(\ref{comb}) and the $(0-0)$-Einstein equation which reads (see Eq.~(\ref{00}))
\begin{equation}  
\label{001}
3{\cal H}\Psi^{(1)'}+3{\cal H}^2 \Psi^{(1)}-\nabla^2\Psi^{(1)}=-\frac{3}{2} {\cal H}^2 \left(\frac{\rho_\d}{\rho} \delta^{(1)}_\d + 
\frac{\rho_\gamma}{\rho} \Delta^{(1)}_{00}   \right)\, .
\end{equation}
On small scales one neglects the time derivatives of the gravitational potential in Eqs.~(\ref{comb}) and~(\ref{001}) to obtain 
\begin{equation}
\delta^{(1)''}_\d+{\cal H}\delta^{(1)'}_\d=\frac{3}{2} {\cal H}^2 \delta^{(1)}_\d\, ,
\end{equation}
where we have also dropped the contribution to the gravitational potential from the radiation component. 
The solution of this equation is matched to the value that $\delta^{(1)}_\d$ has during 
the radiation dominated epoch on subhorizon scales, Eq.~(\ref{Mes1}), and one finds that 
for $\eta \gg \eta_{eq}$ on subhorizon scales the gravitational potential remains constant with
\begin{equation}
\label{ex1}
\Psi^{(1)}_{\bf k}(\eta > \eta_{eq}) = \frac{\ln(0.15 k\eta_{eq}) }{(0.27 k\eta_{eq})^2} \Psi^{(1)}_{\bf k}(0)\, .
\end{equation}  
We skip the details of the derivation of Eq.~(\ref{ex1}) since it is a standard computation that the reader can find, for example, in 
Refs.~\cite{Dodelsonbook}. Since around $\eta_{eq}$ the dark matter begins to dominate, an approximation to the result~(\ref{ex1}) 
can be simply achieved by requiring that during matter domination 
the gravitational potential remains constant to a value determined by the density contrast~(\ref{Mes1}) at the equality epoch
\begin{equation}
\nabla^2 \Psi^{(1)}|_{\eta_{eq}} \simeq \frac{3}{2}{\cal H}^2 \delta^{(1)}_\d|_{\eta_{eq}}\, ,
\end{equation}   
from Eq.~(\ref{001}) on small scales, leading to 
\begin{equation}
\Psi^{(1)}_{\bf k}(\eta > \eta_{eq}) \simeq -\frac{6}{(k\eta_{eq})^2} \delta^{(1)}_\d |_{\eta_{eq}}= 
\frac{\ln(B_1k \eta_{eq})}{(0.13k\eta_{eq})^2} 
\Psi^{(1)}_{\bf k}(0)\, ,
\end{equation}
where we used $a(\eta) \propto \eta^2$ during matter domination and Eq.~(\ref{Mes1}) with $A_1=-9.6$ and $B_1=0.44$. 

At second-order we follow a similar approximation. The general solution for the evolution of the the second-order  
gravitational potential $\Psi^{(2)}$ for $\eta > \eta_{eq}$ is given by Eq.~(\ref{solmatter}). We have 
to determine the initial conditions for those modes that
cross the horizon during the radiation epoch. The $(0-0)$-Einstein equation reads
\begin{eqnarray}
\label{002}
&-&\frac{3}{2} {\cal H}^2
\left(\frac{\rho_\d}{\rho} \delta^{(2)}_\d + \frac{\rho_\gamma}{\rho} \Delta^{(2)}_{00} \right)= 
3 {\cal H}\Psi^{(2)'}+3{\cal H}^2 \Phi^{(2)}-\nabla^2 \Psi^{(2)}
\nonumber \\
&-&6{\cal H}^2 \left( \Phi^{(1)} \right)^2
-12{\cal H}\Phi^{(1)}\Psi^{(1)'}-3\left( \Psi^{(1)'}\right)^2
+\partial_i \Psi^{(1)} \partial^i\Psi^{(1)}
\nonumber \\
&-&4\Psi^{(1)}\nabla^2\Psi^{(1)}
\, .
\end{eqnarray}
We fix the initial conditions with the matching at equality (neglecting the radiation component)  
\begin{equation}
\label{matchingequ}
\nabla^2\Psi^{(2)}-\partial_i\Psi^{(1)}\partial^i\Psi^{(1)}+4\Psi^{(1)}\nabla^2\Psi^{(1)}|_{\eta_{eq}} \simeq \frac{3}{2}{\cal H}^2 
\delta^{(2)}_\d|_{\eta_{eq}}\, ,
\end{equation}
where for small scales we neglected the time derivatives in Eq.~(\ref{002}). Using 
Eq.~(\ref{sold2f}) to evaluate $\delta^{(2)}_\d|_{\eta_{eq}}$ and Eq.~(\ref{ex1}) to evaluate $\Psi^{(1)}_{\bf k}(\eta_{eq})$ 
we find in Fourier space  
\begin{eqnarray}
\label{Psieq}
\Psi^{(2)}(\eta_{eq}) &=&
\Bigg[ -3(a_{\rm NL}-1) \frac{\ln(B_1k\eta_{eq})}{(0.13k\eta_{eq})^2}+\left( \frac{{\bf k}_1\cdot{\bf k}_2}{k^2}-4 \right)
\frac{\ln(0.15k_1\eta_{eq})}{(0.27k_1\eta_{eq})^2} 
\nonumber \\
&\times& \frac{\ln(0.15k_2\eta_{eq})}{(0.27k_2\eta_{eq})^2} +
A_1\ln(B_1k_1\eta_{eq}) \frac{\ln(B_1k_2\eta_{eq})}{(0.13k\eta_{eq})^2} -\frac{27}{c_s^4} k^2 
\nonumber \\
&\times& \frac{{\bf k}_1 
\cdot {\bf k}_2}{k_1^2k_2^2} \frac{\left[ \ln(k\eta_{eq})\right]^2}{(k\eta_{eq})^2}+
\frac{3}{2} \frac{{\bf k}_1 \cdot {\bf k}_2}{c_s^2 k_1^2} 
2.2 \Big[\frac{1.2}{2} \frac{\left[ \ln(k_1c_s\eta_{eq})\right]^2}{(0.13k\eta_{eq})^2}
\nonumber \\
&-&
\ln(k_1c_s\eta_{eq}) \frac{\ln(B_1k_2\eta_{eq})}{(0.13 \eta_{eq})^2}+(1\leftrightarrow2)
\Big] \Bigg] \Psi^{(1)}_{{\bf k}_1}(0) \Psi^{(1)}_{{\bf k}_2}(0)\, . \nonumber \\
\end{eqnarray}
In Eq.~(\ref{gen}) the initial condition $\Psi^{(2)}_m(0)$ is given by 
Eq.~(\ref{Psieq}) and $\Psi^{(1)}$ is given by Eq.~(\ref{ex1}).  
The integration constants can be fixed by comparing at $\eta\simeq \eta_{eq}$ the oscillating part of Eq.~(\ref{gen}) to the solution 
$\Delta^{(2)}_{00}$ obtained for modes crossing the horizon before equality and for $\eta <\eta_{eq}$, Eq.~(\ref{D2f}). 
Thus for $\eta \gg \eta_{eq}$ and $k\gg \eta_{eq}^{-1}$ we find that 
\begin{eqnarray}
\Delta^{(2)}_{00}&=&-4(1+R) \Psi^{(2)}(\eta_{eq})+{\bar A} \cos(kc_s\eta)
\nonumber \\
&-&\frac{2}{7} (1+R) G({\bf k}_1,{\bf k}_2,{\bf k}) \eta^2 
\Psi^{(1)}_{{\bf k}_1}(\eta_{eq}) \Psi^{(1)}_{{\bf k}_2}(\eta_{eq})\, ,
\end{eqnarray} 
where
\begin{eqnarray}
{\bar A}&=&6\Psi^{(2)}(0) -\frac{6\, ({\bf k}\cdot{\bf k}_1)({\bf k} \cdot {\bf k}_2)}{c_s^4 \,k_1k_2 \,\cos(kc_s\eta_{eq})} \Psi^{(1)}_{{\bf k}_1}(0) 
\Psi^{(1)}_{{\bf k}_2}(0) \left[2k_1k_2 \cos(k_1c_s\eta_{eq}) 
\right. \nonumber \\
&\times& \left. \cos(k_2c_s\eta_{eq})-2k_1k_2 \cos(kc_s\eta_{eq}) +
(k_1^2+k_2^2-k^2) \sin(k_1 c_s \eta_{eq}) 
\right. \nonumber \\
&\times& \left. \sin(k_2 c_s \eta_{eq}) \right]  
\frac{1}{k_1^4+k_2^4+k^4-2k_1^2k_2^2-2k_1^2k^2-2k_2^2k^2}\, ,
\end{eqnarray}
and $\Psi^{(2)}(0)$ is given in Eq.~(\ref{Psi20}).

\section{Secondary effects and contamination to primordial NG}
\label{cont}
Given that a detection of a sizable primordial non-Gaussianity (and its shape) 
would represent a real breakthrough into the understanding of the dynamics of the universe during its very first stages, it is 
crucial that all sources of contamination to the primordial signal are well understood and kept under control. In fact any non-linearities can 
make initially Gaussian perturbations non-Gaussian. Such non-primordial effects can thus complicate the extraction of the primordial 
non-Gaussianity: we have to be sure we are not ascribing a primordial origin to
a signal that is extracted from the CMB (or LSS) data using estimators of non-Gaussianity when that signal has a different origin. Moreover, as 
stressed in Section.~\ref{To}, we must always specify of which primordial non-Gaussianity we study the contamination from the non-primordial 
sources (e.g. primordial non-Gaussianity of ``local'' , ``equilateral'' or ``folded'' shape). 

Broadly speaking, non-primordial sources of non-Gaussianity can be classified into four categories: instrumental systematic effects; residual 
foregrounds and unresolved point sources; some well known secondary CMB anisotropies, such as Sunyaev-Zel'dovich (SZ) effect, gravitational lensing, 
Rees-Sciama effect, and finally previously unknown effects coming from 
non-linearities in the Boltzmann equations, which are related to the non-linear nature of General Relativity and to the non-linear dynamics of 
the photon-baryon system. This Review focuses on the last category, but in this Section we offer also a summary of some recent results of the 
other types of possible contaminations. Before doing that, however, it is important  to be more precise about how secondary effects may impact the extraction of a primordial non-Gaussian signal. Mainly they act in two ways. They might ``mimic'' a three-point correlation function similar in shape to the primordial one. This would produce a \emph{bias} or a \emph{contamination} to the estimator of primordial non-Gaussianity.  On the other hand secondary effects might increase the variance of the estimator  without contributing to the signal-to-noise ratio. In other words, in this case they degrade the signal-to-noise ratio. Let us define in a quantitative way how to characterize these effects.

\subsection{Signal-to-noise ratio, shapes, and contamination to primordial non-Gaussianity}
\label{SNSC}
The angular bispectrum, $B_{\ell_1 \ell_2 \ell_3}$, the harmonic transform of the angular three-point function, of the CMB anisotropies is often used to 
measure non-Gaussianity (see for example~\cite{review}). From the usual spherical harmonic coefficients of the temperature anisotropies
\begin{equation}
a_{\ell m}=\int d^2{\bf n} Y^*_{\ell m}({\bf n}) \frac{\Delta T({\bf n})}{T}\, ,
\end{equation}  
the angle-averaged bispectrum is given by (for more details see, e.g., Ref.~\cite{ks})
\begin{eqnarray}
\label{AAB}
B_{\ell_1\ell_2\ell_3}\equiv\sum_{{\rm all}\,m}
\left(
\begin{array}{ccc}\ell_1&\ell_2&\ell_3\\m_1&m_2&m_3\\ \end{array}
\right)
\langle a_{\ell_1m_1}a_{\ell_2m_2}a_{\ell_3m_3}\rangle \, ,
\end{eqnarray}
where the brackets indicate ensemble average. 

We shall quantify the degree to which the primordial and the secondary bispectra 
are correlated, as well as their expected signal-to-noise ratio following a method that as now become standard
\cite{ks,Goldberg:xm,CoorayHu}. Namely, in the limit of weak non-Gaussianity, one can introduce the Fisher matrix for the amplitudes 
of the bispectra, $F_{ij}$, given by
\begin{eqnarray}
\label{FMB}
F_{ij}\equiv \sum_{2\le \ell_1\le \ell_2\le \ell_3}\frac{B_{\ell_1\ell_2\ell_3}^{(i)}B_{\ell_1\ell_2\ell_3}^{(j)}}{\sigma_{\ell_1\ell_2\ell_3}^2},
\end{eqnarray}
where the variance of the bispectrum is 
\begin{eqnarray}
\sigma^2_{\ell_1\ell_2\ell_3}\equiv 
\langle B_{\ell_1\ell_2\ell_3}^2\rangle-{\langle B_{\ell_1\ell_2\ell_3}\rangle}^2\approx C_{\ell_1}C_{\ell_2}C_{\ell_3}\Delta_{\ell_1\ell_2\ell_3},
\end{eqnarray}
and $\Delta_{\ell_1\ell_2\ell_3}$ takes values 1, 2, and 6 when all $\ell$'s are different, two of them are equal and all are the same,
respectively. The power spectrum, $C_\ell$, is the sum of the theoretical
CMB and  the detector noise. 

The signal-to-noise ratio is given by
\begin{eqnarray}
\label{SN}
{\left(\frac{S}{N}\right)}_i=\frac{1}{\sqrt{F_{ii}^{-1}}},
\end{eqnarray}
and we define the cross-correlation coefficient between different
shapes $i$ and $j$, $r_{ij}$, as
\begin{eqnarray}
\label{corrij}
r_{ij}\equiv \frac{F^{-1}_{ij}}{\sqrt{F^{-1}_{ii}F^{-1}_{jj}}}\, ,
\end{eqnarray}
which does not depend on the amplitudes of the different bispectra, but on the shapes, thus measuring how similar are two bispectra $i$ and $j$. 
One can also define a degradation parameter $d_i=F_{ii}F^{-1}_{ii}$ for a degraded $(S/N)$ due to $r_{ij}$. 

If a secondary bispectra has some 
correlation with the primordial one, one expects that the signal-to-noise ratio will be degraded: if one does not account for any secondary 
bispectra $(S/N)_0=\sqrt{F_{\rm prim, prim}}$ while, marginalizing over a (single) secondary bispectrum,  
$(S/N)_{\rm prim}$ gets modified from its zero-order value to $(S/N)_{\rm prim}=(S/N)_0 (1-r_{ij}^2)^{1/2}$. This means that the minimum detectable value 
of the primordial non-linearity parameter $f_{\rm NL}$ which is obtained by imposing $(S/N)_{\rm prim}=1$
(or the 1-$\sigma$ uncertainty on $f_{\rm NL}$ given by $\delta f_{\rm NL}=\sqrt{(F)^{-1}_{\rm prim, \rm prim} |_{f_{\rm NL}=1}}$) 
gets shifted by a quantity $\Delta f_{\rm NL}/(f_{\rm NL})_0 =(1-r_{ij}^2)^{-1/2}-1$. 

How much does a given secondary bispectra contaminate the extraction of the primordial bispectrum? 
If, for example, the predicted shape of the secondary bispectra is
sufficiently different from that of the primordial bispectrum, then one
would hope that the contamination would be minimal. We can quantify 
the contamination of the primordial bispectrum as follows: we fit the primordial bispectrum template to
the second-order bispectrum, and find the best-fitting $f^{\rm
con}_{NL}$ (``con'' stands for contamination) by
minimizing the $\chi^2$ given by
\begin{eqnarray}
\chi^2=\sum_{2\le \ell_1\le \ell_2\le \ell_3}\frac{{\left(f_{\rm NL}B_{\ell_1\ell_2\ell_3}^{prim}-B_{\ell_1\ell_2\ell_3}^{2nd}\right)}^2}{\sigma_{\ell_1\ell_2\ell_3}^2},
\end{eqnarray}
with respect to $f_{\rm NL}$. Here,  $B_{\ell_1\ell_2\ell_3}^{prim}$ is the primordial bispectrum with amplitude $f_{\rm NL}=1$ \cite{ks}.  We obtain
\begin{eqnarray}
\label{fNLcont}
f_{\rm NL}^{\rm con}&=& \frac{1}{N}\sum_{2\le \ell_1\le \ell_2\le \ell_3}\frac{B_{\ell_1\ell_2\ell_3}^{2nd}B_{\ell_1\ell_2\ell_3}^{prim}}{\sigma_{\ell_1\ell_2\ell_3}^2},\nonumber\\
N&=&\sum_{2\le \ell_1\le \ell_2\le \ell_3}\frac{\left({B_{\ell_1\ell_2\ell_3}^{prim}}\right)^2}{\sigma_{\ell_1\ell_2\ell_3}^2}\label{eq:fNL}.
\end{eqnarray}
This is the value of $f_{\rm NL}$ one would find, if one did not know that
the primordial bispectrum did not exist but there was only the
secondary bispectrum. In other words the effective non-linearity parameter $f_{\rm NL}^{\rm con}$ gives the value that a bispectrum estimator designed for constraining primordial non-Gaussianity would measure due to 
the presence of the secondary bispectrum $B_{\ell_1\ell_2\ell_3}^{2nd}$. As one can see $f_{\rm NL}^{\rm con}$ turns out to be proportional to the mixed entry of the Fisher matrix, since this 
governs the correlation between different types of bispectra on the basis of their shape dependence. 
\subsection{Some ``well-known'' secondary sources of non-Gaussianity}
Instrumental  systematic effects are one of the principal issues in the search for primordial non-Gaussianity. Since they depend on the specific instrument, here we 
do not intend to go into any detail, but we refer the reader to Ref.~\cite{Donzelli} as an interesting example of this kind of effects for the currently flying mission of the satellite 
\emph{Planck}.  Residual foregrounds and unresolved point sources represent other important possible contaminants and particular attention have been devoted to them 
in the analysis of the WMAP data~\cite{KomatsuWMAP,SZ,leo}.  For an experiment like WMAP secondary effects such as  Sunyaev-Zel'dovich effect, gravitational lensing, 
Integrated-Sachs-Wolfe (or Rees-Sciama) effect should be not so relevant for $\ell<500$ with a bias of just 1.5 to 2~\cite{SerraCooray} for local primordial non-Gaussianity (see also\cite{ks} and~\cite{SZ}). \footnote{In order to understand the relative importance of the bias from a given secondary effect let us recall that, including the information from polarization,  the minimum detectable (local) $f^{\rm loc}_{\rm NL}$ for an experiment like WAMP is of the order of $10$ while for \emph{Planck} it is of the order of 3; the minimum detectable value for an equilateral primordial non-Gaussianity is  $f^{\rm eq.}_{\rm NL} \simeq 30$ for Planck, see, e.g., Refs.~\cite{ks,BZ,Baumann,SL}.
} However for higher resolution and more sensitive experiments like \emph{Planck}, such effects must be taken into account with care. Here we report the results of some recent analyses about the contamination to primordial non-Gaussianity from some secondary effects. Such a brief summary is by no means exhaustive. An interesting case is given by the cross-correlation of the ISW and the lensing effects, see Ref.~\cite{Goldberg:xm}. In the detailed analysis of~\cite{Hanson} it is shown that the ISW-lensing bispectrum produces a bias of $f^{\rm cont}_{\rm NL}\simeq 9$ to local primordial non-Gaussianity (see also~\cite{SerraCooray}), while the bias to equilateral primordial non-Gaussianity is negligible (see also~\cite{SZ,SerraCooray}). The impact is mainly on the local type of primordial non-Gaussianity because the ISW-lensing correlation correlates the large-scale gravitational potential fluctuations sourcing the ISW effect with the small scale lensing effects of the CMB, thus producing a bispectrum which peaks in the squeezed configurations as the local shape.  As far as the increase of the variance of the estimator of primordial non-Gaussianity induced by CMB lensing is concerned it turns out to be of the order of $20\%$ for an experiment like \emph{Planck} 
(see Refs.~\cite{Hanson,BZ}). Notice that, interestingly enough, gravitational lensing can also have another peculiar effect on the extraction of primordial non-Gaussianity: it can modify its shape by smoothing its acoustic features. However, according to the analysis of Ref.~\cite{Hanson} (and contrary to the results of~\cite{CooraySarkar}) such a ``distorsion'' of the shape should be small, of the order of $10\%$ for $l<2000$.  

A similar analysis has been performed looking also at scales where non-linear effects can become relevant. In this case a bispectrum from the correlation of the Rees-Sciama and lensing effects has been studied in Ref.~\cite{MangilliVerde}, and it has been found that in this case the bias to a local primordial non-Gaussianity corresponds to $f^{\rm cont}_{\rm NL}\simeq 10$, agreeing with the results of~\cite{Hanson} where the two analyses can be compared. 

Another secondary source of contamination that has been studied quite recently in Ref.~\cite{PierpaoliBabich} deals with bispectra generated by correlations of number density and lensing magnification of radio and SZ point sources with the ISW effect. Also in this case the large-scale modulation of  small-scale number density due to  fluctuations which source the ISW  generates a bispectrum which peaks on squeezed configurations and it has been found that it corresponds to a contamination of local primordial non-Gaussianity of  $f^{\rm cont}_{\rm NL}\simeq 1.5$, which should be not so relevant for an experiment like \emph{Planck}. The novelty of this effect is that it is different from the usual bispectrum of point sources when treated with a Poisson distribution (see, e.g., ~\cite{ks}).

Other secondary effects include, for example, the SZ-SZ-SZ bispectrum~\cite{CooraysoloSZ} and non-Gaussianities from the kinetic SZ and Ostriker-Vishniac effect
~\cite{Castro,CoorayHu}. However the SZ-SZ-SZ bispectrum can be considered as an extra Poisson contribution to the unresolved point sources for $l<1500$ ~\cite{SerraCooray}. Moreover recently an analysis of  the bispectrum generated from inhomogeneous reionization has been performed in~\cite{WandeltReio}, where it has been shown that it can give rise to a very small contamination to the local primordial non-Gaussianity of $f^{\rm cont}_{\rm NL}\simeq - 0.1$.

Finally let us  also recall some recent studies of a specific effect arising at second-order in perturbation theory, when looking at the fluctuations in the Boltzmann equations for the photon-baryon fluid. In Refs.~\cite{khatri09,khatri09bis,senatore08I,senatore08II} the perturbations to the phase of recombination 
between electrons and baryons has been considered (``inhomogeneous recombination'') . It turns out that the electron density perturbation can be of the order of $5$ times bigger than the one in the density of the baryons. This gives rise to the prospect of a non-Gaussianity which corresponds to $|f^{\rm cont}_{\rm NL}|\simeq 5$. In fact the detailed study of this effect shows that the contamination is much smaller. The bispectrum from the perturbed recombination peaks in the squeezed configuration with 
a contamination to the primordial non-Gaussianity of the local type of $|f^{\rm cont}_{\rm NL}|\simeq 0.7$. 

Let us conclude this section by mentioning that the non-linearities emerging from secondary effects can be very interesting by themeselves, since they carry a lot of information about the evolution of the universe and the growth of structures after recombination till very low redshift. Detailed examples and reviews about this aspect can be found, e.g., in Refs.~\cite{VerdeSpergel,Aghanim,KomProp}.

\section{How to analytically estimate the signal-to-noise ratio from the NG at recombination and its contamination to the 
primordial NG}
\label{An}
\noindent
As we have seen in Sec.~\ref{Oscill}, the 
 dynamics at recombination is quite involved because
all the non-linearities in the evolution of the baryon-photon fluid at 
recombination and  the ones 
coming from general relativity should be accounted for. 
Such a contribution is so relevant because it 
represents a major part of the second-order 
radiation transfer function which must be determined in order to have a 
complete control of both the primordial and non-primordial part of NG in 
the CMB anisotropies. The NG generated at the surface of last scattering comprises 
various effects. In this Section, we devote our attention to one particular relevant contribution, the 
one  coming from the non-linear evolution 
of the second-order gravitational potential  
which grows in time on small scales. The analysis of this contribution offers the opportunity to understand how some secondary effects can contaminate mostly a given type of primordial NG, while leaving almost untouched other different forms of primordial NG. In this case this effect is a causal one, 
developing on small scales, so we expect that the NG it generates 
will be mainly of the equilateral type, rather than of local type. 
Therefore, a reasonable
question is to which extent the NG from recombination alters the possibile detection 
of the primordial NG of the equilateral type.
The goal of this Section is therefore to illustrate how to estimate in a semi-analytical way the contribution to NG from recombination.

\subsection{Signal-to-Noise ratio for the primordial equilateral bispectrum}
In this subsection we wish to recover the estimate for the signal-to-noise ratio $(S/N)$ 
given in  Ref.~\cite{SZ} for the 
primordial bispectra of ``equilateral'' 
type~\cite{CN} by adopting a simple model. In other words,  we test the goodness
of the semi-analytical model we will be using in the next Section to   
estimate the bispectrum from the recombination era.

Our starting point is the primordial equilateral bispectrum~\cite{CN}
\be
\label{equil}
\langle \Phi({\bf k}_1) \Phi({\bf k}_2) \Phi({\bf k}_3) \rangle = (2 \pi)^3 \delta^{(3)}
\big({\bf k}_1 + {\bf k}_2 + {\bf k}_3 \big)
B_{\rm equil}( k_1,  k_2 ,  k_3) \, ,
\ee
where
\begin{eqnarray}
\label{eq:ours}
B_{\rm equil}(k_1,k_2,k_3) &=& f_{\rm NL}^{\rm equil} \cdot 6  
 A^2 \cdot \left(-\frac1{k_1^3 k_2^3} - 
\frac1{k_1^3 k_3^3} - \frac1{k_2^3 k_3^3}  - \frac2{k_1^2 k_2^2 k_3^2}\right. \nonumber\\
&+& \left.
\frac1{k_1 k_2^2 k_3^3}
+ (5 \; {\rm perm.}) \right)\, , 
\end{eqnarray}
and  the permutations act only on the last term in parentheses. The parameter $f_{\rm NL}^{\rm equil}$ 
quantifies the level of NG while
$A=17.46 \times 10^{-9}$ 
is the amplitude of the  primordial gravitional potential 
power spectrum computed at first-order

\be
\langle \Phi^{(1)}({\bf k}_1) \Phi^{(1)}({\bf k}_2) \rangle = (2 \pi)^3 \delta^{(3)}
\big({\bf k}_1 + {\bf k}_2 \big) P(k_1)\, ,
\ee 
with $P(k)=A/k^3$. Since the signal-to-noise ratio $(S/N)$ will be some function of the maximum multipole a given 
experiment can reach, $\ell_{\rm max}\gg 1$, we can use the flat-sky approximation ~\cite{Hulensing,BZ} and write
for the bispectrum 

\be
\langle a(\vec{\ell}_1)a(\vec{\ell}_2)a(\vec{\ell}_3) \rangle 
   = (2\pi)^2\delta^{(2)}(\vec{\ell}_{123}) B(\ell_1,\ell_2,\ell_3)\, ,
\ee 
where $\vec{\ell}_{123}=\vec{\ell}_1+\vec{\ell}_2+\vec{\ell}_3$, 
with~\cite{Shapes}  
\begin{eqnarray} 
\label{bispectflat}
B_{\rm equil}(\ell_1,\ell_2,\ell_3) &=& \frac{(\eta_0 - \eta_r)^2}{(2\pi)^2} \int
dk^z_1dk^z_2dk^z_3 \delta^{(1)}(k^z_{123}) B_{\rm equil}(k'_1,k'_2,k'_3)\nonumber\\  
&\times&\tilde{\Delta}^T(\ell_1,k^z_1) \tilde{\Delta}^T(\ell_2,k^z_2) \tilde{\Delta}^T(\ell_3,k^z_3)\, ,
\end{eqnarray}
where $k'$ means $k$ evaluated such that $\vec{k}^{\parallel} = \vec{\ell}/(\eta_0-\eta_r)$ and 
   \begin{equation}
        \tilde{\Delta}^T(\ell,k^z) = \int_0^{\eta_0} \frac{d\eta}{(\eta_0-\eta)^2} 
          S(\sqrt{(k^z)^2 + \ell^2/(\eta_0-\eta)^2},\eta) 
          e^{ik^z(\eta_r-\eta)} \, , 
   \end{equation}
is the radiation transfer function defined by the CMB source function $S(k,\eta)$. 
In this notation, $\eta_0$ and $\eta_r$  represent the
present-day and the recombination conformal time, respectively and $k^z$ and $\vec{k}^{\parallel}$ are 
the momentum components
perpendicular and parallel respectively to the plane orthogonal to the line-of-sight.

The $(S/N)$ ratio in the flat-sky formalism is~\cite{Hulensing,BZ} 

\begin{equation}
\label{S/N}
\left( \frac{S}{N} \right)^2= 
\frac{f_{\rm sky}}{\pi} \frac{1}{(2 \pi)^2} \int d^2 \ell_1 d^2 \ell_2  d^2 \ell_3 
\,\delta^{(2)}(\vec{\ell}_{123})\, 
\frac{B_{\rm equil}^2(\ell_1,\ell_2,\ell_3)}{6\, C(\ell_1)\,C(\ell_2)\, C(\ell_3)}\, ,
\end{equation}
where $f_{\rm sky}$ stands for  the portion of the observed sky. 
In order to compute the bispectrum $B_{\rm equil}(\ell_1,\ell_2,\ell_3)$ 
and the power spectrum $C(\ell)$ we adopt the following model   
\be
\label{toymodel}
a({\vec \ell})= \int \frac{d k^z}{2\pi} e^{ik^z(\eta_0-\eta_r)} \Phi({\bf k}') \tilde{\Delta}^T(\ell,k^z)
\ee 
where we mimic the effects of the transfer function on small scales as 

\be
\tilde{\Delta}^T(\ell,k^z)=a\,(\eta_0-\eta_r)^{-2}  e^{-1/2(\ell/\ell_*)^{1.2}}
e^{-1/2(|k_z|/ k_*)^{1.2}}\, ,
\ee 
{\it i.e.} a simple exponential and a normalization 
coefficent $a$ to be 
determined to match the amplitude of the angular power spectrum at the characteristic scale 
$\ell \simeq \ell_*=k_* (\eta_0-\eta_r)$.\footnote{We could equally choose a transfer function as 
$\tilde{\Delta}^T(\ell,k^z)=a\, (\eta_0-\eta_r)^{-2}\,e^{-1/2(\ell/\ell_*)^{1.2}}  \theta(k_*-|k^z|)$, 
the relevant approximation 
being that the integral over $k^z$ is cut at the 
scale $k_*$.} It is important to make clear what are the reasons underlying the choice of such a
model. When computing the $(S/N)$, Eq.~(\ref{toymodel}) with $\ell_*=k_*(\eta_0-\eta_r)\simeq750$ and $a\simeq 3$ 
is able to account for the combined effects of ``radiation driving'', 
which occours at $\ell>\ell_{\rm eq}\simeq 160$ and boosts the angular power spectrum with respect to 
the Sachs-Wolfe plateau, 
and the effects of 
Silk damping which tend to suppress the CMB anisotropies for scales $\ell>\ell_D\simeq 1300$. 
The combination of these 
effects produces a decrease in the angular power spectrum from a scale $\ell_*\simeq 750$.\footnote{
The choice of the exponent $1.2$ derives from the study of the diffusion damping envelope in 
Ref. ~\cite{HuWhitedamping}.}  
The power spectrum in 
the flat-sky approximation is given 
by  $\langle a(\vec{l}_1)a(\vec{l}_2) \rangle 
   = (2\pi)^2\delta^{(2)}(\vec{l}_{12}) C(\ell_1)$ with 
\be
C(\ell)=\frac{(\eta_0-\eta_r)^2}{(2 \pi)} \int dk^z |\tilde{\Delta}^T(\ell,k^z)|^2\, P(k)\, .
\ee       
The exponential of the transfer function for Eq.~(\ref{toymodel}) allows to cut off the 
integral for $k\simeq k_*$ and one finds 
(see also Ref.~\cite{BZ})
\be
\label{Cl}
C(\ell)=a^2 \frac{A}{\pi \ell^2} \frac{e^{-(\ell/\ell_*)^{1.2}}}{\sqrt{1+\ell^2/\ell_*^2}}
\simeq a^2 \frac{A}{\pi}\frac{\ell_*}{\ell^3}\,
e^{-(\ell/\ell_*)^{1.2}}\, ,
\ee
where the last equality holds for $\ell\gg\ell_*$. 
To compute the bispectrum we proceed in a similar way. One first uses the Dirac deltas, 
$\delta^{(1)}(k^z_{123})$ and $\delta^{(2)}(\vec{\ell}_{123})$. Then it 
proves to be useful the change of 
variable $k^z_1=x_1 \ell_1/(\eta_0-\eta_r)$, $k^z_2=x_2 \ell_2/(\eta_0-\eta_r)$. 
In this way the transfer functions become $ \tilde{\Delta}^T(\ell_i,k_i^z)\propto 
e^{-1/2(|x_i| \ell_i/\ell_*)^{1.2}}$ which allows to cut the integrals over $x_i$ ($i=1,2$) at $\ell_*/\ell_i$. Now, as a good approximation to see the effects of the transfer functions, we 
can take $\ell \gg \ell_*$ and thus the integral over $x_i$ can be easily computed 
by just evaluating the integrand 
in $x_i=0$ times $4(\ell_*/\ell_1)(\ell_*/\ell_2) $. 
With this approximation the integral in $k^z_i$ is easily obtained and we get for the bispectrum 
\begin{eqnarray}
\label{Bapprox}
B_{\rm equil}(\ell_1,\ell_2,\ell_3)&=& \frac{24 f_1}{(2 \pi)^2}\, f^{\rm equil}_{\rm NL} a^3\, A^2\,  
e^{-(\ell_1^{1.2}+\ell_2^{1.2}+\ell_3^{1.2})/2\ell^{1.2}_*}\,
\ell_*^2 \nonumber\\
&\times&  \left(-\frac1{\ell_1^3 \ell_2^3} - 
\frac1{\ell_1^3 \ell_3^3} - \frac1{\ell_2^3 \ell_3^3}  - \frac2{\ell_1^2 \ell_2^2 \ell_3^2} +
 \frac1{\ell_1 \ell_2^2 \ell_3^3}
+ (5 \; {\rm perm.}) \right)\, , \nonumber\\
&&
\end{eqnarray}
where 
\be
\label{l3}
\ell_3^2=\ell_1^2+\ell_2^2+2\,\vec{\ell}_1\cdot\vec{\ell}_2\, . 
\ee
The coefficient $f_1\simeq 1/1.4=0.7$ is a fudge factor that improves 
the matching between our approximation for the bispectrum 
and numerical results that have been consistenly checked. Notice that, 
according to our approximation, the equilateral structure of 
Eq.~(\ref{eq:ours}) is preserved in $\ell$ 
space.\footnote{The expression~(\ref{Bapprox}) can be also written as 
\newline
$B_{\rm equil}(\ell_1,\ell_2,\ell_3)=  (2 \pi)^{-2} 48 f_1\, f^{\rm equil}_{\rm NL} \,
a^3\,A^2\,  \ell_*^2 e^{-(\ell_1^{1.2}+\ell_2^{1.2}+\ell_3^{1.2})/2\ell^{1.2}_*}
\,(1+\cos\theta) (\ell_1+\ell_2-\ell_3)/\ell_1^2 \ell_2^2\ell_3^3$, 
$\theta$ being the angle between ${\vec \ell}_1$ and ${\vec \ell}_2$.}
$\ell_1 =\ell_2$, 
in correspondence of the maximum 
following expression
In computing the signal-to-noise ratio, consistency with our approximation~(\ref{Bapprox}) 
requires that we integrate over 
$\ell_1,\ell_2$ starting from a minimum $\ell_{\rm min} > \ell_*$ up to $\ell_{\rm max}$ and 
paying attention to the fact that even $\ell_3$ in 
Eq. (\ref{l3}) must be larger than $\ell_{\rm min}$.  The scaling with $\ell_{\rm max}$ with 
respect to the case of a local type bispectrum turns out to be much milder. 
While for the local type $(S/N)^2 \propto \ell_{\rm max}^2$ \cite{BZ}, 
 for the equilateral 
bispectrum~(\ref{equil}) we find\footnote{
These scalings can be 
easily understood by analyzing the expressions $(S/N)^2$ for the 
local and equilateral primordial NG. In the local case,  $(S/N)^2$ is 
proportional to $\int d^2\ell_1 d^2\ell_2 d^2\ell_3 \delta^{(2)}
(\vec{\ell}_{123})(\ell^3_1+\ell^3_2+\ell^3_3)^2/(\ell_1\ell_2\ell_3)^3$~\cite{BZ};
since the squeezed configuration, {\it e.g.} $\ell_1\ll \ell_2,\ell_3$,  
is dominating the local bispecrum, the integral becomes proportional to $\int 
d\ell_1 d\ell_2 (\ell_2/\ell^2_1)\propto \ell^2_{\rm max}$. In the equilateral
case, however, $(S/N)^2$ receives contributions from the configuration which is 
peaked at $\ell_1\sim\ell_2\sim\ell_3$ and therefore 
it can be written as  $\int d^2\ell_1
d^2 \ell_2 \delta(\ell_1-\ell_2)/\ell_1^2\propto \ell_{\rm max}$.}  
$(S/N)^2 \propto \ell_{\rm max}$ and, setting $\ell_*=750$ and $\ell_{\rm min}\simeq 1200$, 
\begin{eqnarray}
\label{SNprim}
\left( \frac{S}{N} \right)_{\rm equil}^2 &=& 0.48 \times 10^5\, \frac{f_{\rm sky}}{2^5 \pi^3 6} 
\,A\, (f^{\rm equil}_{\rm NL})^2\, \, 
\ell_{\rm max} \simeq  
8 \, f_{\rm sky} \,A\,   (f^{\rm equil}_{\rm NL} )^2\,  
\ell_{\rm max} \, . \nonumber\\
&&
\end{eqnarray}
By choosing $f_{\rm sky}=0.8$ and $\ell_{\rm max}=2000$ we find a minimum detectable 
\be
\label{fnleq}
f^{\rm equil}_{\rm NL}\simeq 66\, ,
\ee
obtained imposing $(S/N)_{\rm equil} =1$. Both the estimate of the minimum value of 
$f^{\rm equil}_{\rm NL}$ and the scaling $(S/N)^2 \propto \ell_{\rm max}$ are 
 in remarkable agreement with the result obtained in Ref.~\cite{SZ}  
where the full transfer function is used and a value of $f^{\rm equil}_{\rm NL}= 67$ is 
obtained.\footnote{We thank M. Liguori for discussions about 
the minimum value of $f^{\rm equil}_{\rm NL}$ detectable by Planck
and for its scaling with $\ell_{\rm max}$.} Notice that our estimate is independent 
from the coefficient $a$ and the exponetial $e^{-1/2(\ell/\ell_*)^{1.2}}$ 
introduced below Eq.~(\ref{toymodel}) to mimic the full transfer function.  This is 
because  there is an equal number of transfer 
functions in the
numerator and denominator of the expression (\ref{S/N}) for the signal-to-noise ratio 
and their effect tend to cancel despite they 
are not simple
multiplicative factors (see discussion in Ref.~\cite{BZ}).

\subsection{Non-Gaussianity from recombination}
Comforted by the goodness of our model, in this Section we wish to estimate 
the level of NG generated at the recombination era. One can check that 
on small scales the second-order anisotropies are dominated 
by the second-order gravitational potential $\Phi^{(2)}$ which grows as $\eta^2$, as 
first pointed out in Ref.~\cite{fr}.
Let us first briefly show how this result can be obtained using the calculations of Sec.~\ref{Oscill}. Consider the tight coupling regime, and 
small scales, well inside the diffusion Silk-damping scale $\lambda_D$. Then look at Eq.~(\ref{Sstar}). One realizes that, apart from the 
combination $\Delta^{(2)}_{00}+4\Phi^{(2)}$, all the other terms scale at most as $\eta$ or they are suppressed either because
in the tight coupling regime  or because of the damping diffusion on small scales. On the other hand  
one finds that the  main contribution to the bispectrum generated at recombination comes 
from   
\begin{eqnarray}
\label{monopole}
\Theta^{(2)}_{\rm SW}=\frac{1}{4} \Delta^{(2)}_{00}+\Phi^{(2)}\, ,
\end{eqnarray}
which is the usual term appearing in the CMB anisotropies due to the intrinsic photon 
energy density fluctuations 
$\Delta^{(2)}_{00}$ and the gravitational redshift due to the potential. 
In fact such a term on large scales reduces to the 
Sachs-Wolfe effect, while on small scales at recombination, using Eq.~(\ref{phi2=psi2}) and~(\ref{quasigen}), we find  

\begin{eqnarray}
\label{dominant}
\frac{1}{4} \Delta^{(2)}_{00}+\Phi^{(2)} &\simeq& -R\,  \Phi^{(2)}
\nonumber \\
&=& -\frac{R\eta_r^2}{14} G({\bf k_1}, 
{\bf k_2},{\bf k})T(k_1) \Phi^{(1)}({\bf k}_1)
 T(k_2)\Phi^{(1)}({\bf k}_2)\, ,\nonumber\\
&&
\end{eqnarray}
where we have evaluated the expression at the recombination time $\eta_r$ and $R=3\rho_{\rm b}/4
\rho_\gamma$ is the baryon-to-photon energy density ratio. In writing Eq.~(\ref{dominant}) we have dropped off the contribution from primordial 
non-Gaussianity (which, anyway, always propagates linerarly) since here we are focusing on the relevant secondary effects that can constitute 
a source of contamintation to it (see Eq.(\ref{SW}) and the discussion below). 
Notice also that this expression has been obtained assuming all the momenta much larger than 
$k_{\rm eq}$. 

Eq.~(\ref{dominant}) has been confirmed numerically in Ref.~\cite{fr}.  
There is also another way to obtain the analytical expression in the first line of Eq.~(\ref{dominant}), 
as discussed in details in Ref.~\cite{fr}, which exploits some well-known results 
at linear order  extending them at second-order. At second-order in the perturbations the solution for the acoustic oscillations 
will have a form very similar to the linear solution written in Eq.(\ref{LB}), except, as usual, of some source terms $S$ made by products of fistr-order 
perturbations. 
\begin{equation}
\label{LBII}
\frac{\Delta^{(2)}_{00}}{4}=\Big[\frac{\Delta^{(2)}_{00}}{4}(0)+(1+R)\Phi^{(2)} \Big] \cos(kr_s) -(1+R)\Phi^{(2)}+S\, .
\end{equation}
Now on sufficiently small scales the products of first-order terms are indeed suppressed, due either to Silk damping, or 
because gravitational potentials are suppressed as $(k/k_{\rm equ})^2$. Recall that only acoustic oscillations are damped by Silk damping diffusion, so that 
the cosine in Eq.(\ref{LBII}) is multiplied by $e^{-(k/k_D)^2}$  $k \gg k_{D}$, where $k_D^{-1}=\lambda_D $ indicates 
damping length, but the baryon induced shift $(1+R)\Phi^{(1)}$ is left untouched. Therefore on small scales the combination of the damping effects and the 
growth of the second-order gravitational potential $\Phi^{(2)}$ as $\eta^2$ single out the dominant effect in Eq.(\ref{LBII}) to be 
$\Theta^{(2)}_{SW}\simeq -R \Phi^{(2)}$.

\subsubsection{Bispectrum from recombination due to the non-linear growth of the gravitational potential}
Let us start to go into some details of the form of the bispectrum induced by the contribution~(\ref{dominant}). 
In Eq.~(\ref{dominant}) the kernel $G$ is given by
\be
\label{kernelG}
G({\bf k_1}, {\bf k_2},{\bf k})={\bf k_1} \cdot {\bf k_2}-\frac{10}{3} \frac{({\bf k}\cdot {\bf k_1})
({\bf k}\cdot {\bf k_2})}{k^2}.
\ee  
From the form of the kernel we see that the NG at recombination is dominated by an 
equilateral configuration, as expected 
from the fact that its origin is gravitational. 
Here and in the following we are implicitly assuming that a convolution is acting on 
the kernel as  
$$
\frac{1}{(2 \pi)^3} \int d{\bf k}_1 d{\bf k}_2\, \delta^{(3)}({\bf k}_1+{\bf k}_2+{\bf k}_3) 
G({\bf k_1}, {\bf k_2},{\bf k})  
T(k_1)\Phi^{(1)}({\bf k}_1)
T(k_2) \Phi^{(1)}({\bf k}_2)\, .
$$
The reader should remember that, at first-order in perturbation theory,  
the  combination
$\Theta^{(1)}_{\rm SW}+R\,\Phi^{(1)}$ is exponentially suppressed by the Silk damping, but still greater than the 
term $R\Phi^{(1)}$ (which does not suffer the damping) for the maximum multipole of interest, 
$\ell_{\rm max}\sim 2000$. This is meanly due to the fact that the first-order gravitational 
potential rapidly decays on small scales. On the contrary, at second-order in perturbation theory, the
gravitational potential grows like the scale factor on small scales and it turns out that the 
$R\Phi^{(2)}$ dominates on small scales 
(see Ref.~\cite{fr}).

The gravitational potential at linear order can be expressed as usual in terms of the transfer function  $T(k)$
\be
T(k) \approx 12 \left( \frac{k_{\rm eq}}{k} \right)^2 \ln[k/8 k_{\rm eq}]\, ,  
\ee
where the last step is an approximation valid on  scales smaller the the equivalence scale, $k\gg k_{\rm eq}$.
In the following we will account for the logarithmic growth just with a 
coeffcient $T_0(k)=12 \ln[k/8 k_{\rm eq}] \approx 11$ for the scales of interest. 

In the flat-sky approximation one arrives at an expression similar to~(\ref{bispectflat}), 
where now one of the linear 
transfer functions must replaced by a transfer function at second-order. Specifically one finds
\begin{eqnarray} 
\label{bis}
B_{\rm rec}(\ell_1,\ell_2,\ell_3)&=& \frac{(\eta_0 - \eta_r)^2}{(2\pi)^2} \int
dk^z_1dk^z_2dk^z_3 \delta^{(1)}(k^z_{123}) \Big[ G({\bf k}'_1,{\bf k}'_2,{\bf k}'_3)T(k'_1)
 T(k'_2)P(k'_1) P(k'_2)  
 \nonumber\\
& \times &\tilde{\Delta}^{T}(\ell_1,k^z_1) \tilde{\Delta}^T(\ell_2,k^z_2) \tilde{\Delta}^{T(2)}(\ell_3,k^z_3) +
{\rm cyclic} \Big]\, .\nonumber\\
&&
\end{eqnarray}
By using our model~(\ref{toymodel}) and 
\be
\tilde{\Delta}^{T(2)}(\ell,k^z)=- \frac{R}{14} \frac{\tau^2_r}{(\eta_0-\eta_r)^2}\, ,   
\ee
for the second-order radiation transfer function, we find
\bea
&&B_{\rm rec}(\ell_1,\ell_2,\ell_3)=-\frac{R}{14} \frac{(\eta_0-\eta_r)^{-4}}{(2 \pi)^2}k^4_{\rm eq}\,
 \eta_r^2A^2 a^2 T_0^2 
e^{-1/2(\ell_1/\ell_*)^{1.2}} e^{-1/2(\ell_2/\ell_*)^{1.2}}\nonumber\\
&\times&  \int
dk^z_1dk^z_2dk^z_3 \delta^{(1)}(k^z_{123})  \left[ G({\bf k}'_1,{\bf k}'_2,{\bf k}'_3) \frac{1}{{k'_1}^5 {k'_2}^5}  e^{-1/2(|k_{1z}|/ k_*)^{1.2}}
 e^{-1/2(|k_{2z}|/ k_*)^{1.2}} +{\rm cyclic} \right]\, .\nonumber\\
 &&
\eea
At this point we proceed further by employing the same approximation described after Eq. (\ref{Cl}). We 
use the Dirac delta to replace the variable $k_{3z}$, and the exponential allow us to 
evaluate the integral for $k_{1z}=k_{2z}=0$, for scales $\ell_i \gg \ell_*$. This leads to 
\bea
\label{Bisprec}
B_{\rm rec}(\ell_1,\ell_2,\ell_3)&=&-\frac{4 f_2}{(2 \pi)^2}\frac{R}{14} A^2 a^2 T_0^2 (k_{\rm eq} \eta_r)^2
 \ell_{\rm eq}^2 \ell_*^2 
e^{-1/2(\ell_1/\ell_*)^{1.2}} e^{-1/2(\ell_2/\ell_*)^{1.2}}\nonumber\\
&\times& \frac{1}{\ell_1^5 \ell_2^5} \Big[{\vec\ell}_1 \cdot {\vec\ell}_2-\frac{10}{3} 
\frac{({\vec\ell}_3 \cdot {\vec\ell}_1)({\vec\ell}_3 \cdot {\vec\ell}_2)}{\ell_3^2} \Big]+ 
 {\rm cyclic}\, .
\eea
Again here $f_2$ is a coefficient to better calibrate our approximations with numerical results that we 
have performed in order to 
test the validity of our approach. Not surprisingly, it turns out that $f_2\simeq f_1 \simeq 1/1.4$.  
\subsection{Contamination to primordial non-Gaussianity from recombination: Fisher matrices}
Our goal now is to quantify the level of NG coming from the recombination era and to estimate the level of 
degradation it causes
on the possible measurement of the equilateral primordial bispectrum. The reader should keep in mind that, 
given the form 
of the kernel function (\ref{kernelG}), the NG from recombination is expected to be of the equilateral type. 
Following Sec.~\ref{SNSC}, a rigorous procedure is to define the Fisher matrix (in flat-sky approximation) as 
\be 
F_{ij}=\int d^2 \ell_1 d^2 \ell_2  d^2 \ell_3 
\,\delta^{(2)}(\vec{\ell}_{123})\,\frac{
B^{i}(\ell_1,\ell_2,\ell_3)\, B^{j}(\ell_1,\ell_2,\ell_3)}{6\, C(\ell_1)\,C(\ell_2)\, C(\ell_3)}\, ,
\ee
where $i$ (or $j$)$=({\rm rec},{\rm equil})$, and to define the signal-to-noise ratio for a component $i$, 
$(S/N)_i=1/\sqrt{F^{-1}_{ii}}$,  and the degradation parameter $d_i=F_{ii} F^{-1}_{ii}$ due to the correlation 
bewteen the 
different components $r_{ij}=F^{-1}_{ij}/\sqrt{F^{-1}_{ii}F^{-1}_{jj}}$. The first entry $F_{\rm equil,equil}$ 
of the Fisher matrix corresponds to 
the $(S/N)^2$ ratio computed in Eq.~(\ref{SNprim}) which does not account for any kind of cross-correlation. 
Due to the equilateral form of the 
NG generated at recombination we expect that the minimum value detectable for $f_{\rm NL}^{\rm equil}$ will 
be higher that the one reported in Eq.~(\ref{fnleq}). For the mixed entry we find

\bea
\label{S/Nij}
F_{\rm rec,equil}&=& 
\frac{f_{\rm sky}}{\pi} \frac{1}{(2 \pi)^2} \int d^2 \ell_1 d^2 \ell_2  d^2 \ell_3 
\,\delta^{(2)}(\vec{\ell}_{123})\, 
\frac{B_{\rm rec}(\ell_1,\ell_2,\ell_3)  B_{\rm equil}(\ell_1,\ell_2,\ell_3)}
{6\, C(\ell_1)\,C(\ell_2)\, C(\ell_3)}\, \nonumber\\
&=&- 3 f_1 f_2 \frac{f_{\rm sky}}{\pi^3} \frac{4 R}{14} \frac{48}{2^5 6}\frac{T^2_0}{a} 
\,(k_{\rm eq} \eta_r)^2 \ell^2_{\rm eq} \ell_*\, A f_{\rm NL}^{\rm equil}\nonumber\\
&\times&\int d\ell_1 d\ell_2 (1+\vec{\ell}_1\cdot\vec{\ell}_2/\ell_1 \ell_2)\, 
e^{1/2(\ell_3/\ell_*)^{1.2}} \nonumber\\
&\times& \frac{1}{\ell_1^3 \ell_2^3}\left(\ell_1+\ell_2 -\ell_3 \right) \left[{\vec\ell}_1 
\cdot {\vec\ell}_2-\frac{10}{3} 
\frac{({\vec\ell}_3 \cdot {\vec\ell}_1)({\vec\ell}_3 \cdot {\vec\ell}_2)}{\ell_3^2} \right]\, ,
\eea 
where $\ell_3$ is given by Eq.~(\ref{l3}). The factor 3 in front of this expression comes from cyclic 
permutations. 
The integral can be performed numerically and, 
integrating from a minimum $\ell_{\rm min}\simeq 1200$ up to $\ell_{\rm max}=2000$, and 
by taking $R\simeq 0.3$ when evaluated at recombination, $a \simeq 3$, $T_0 \simeq 11$, 
$(k_{\rm eq} \eta_r)^2\simeq 26$, $\ell_{\rm eq}=150$, $\ell_*=750$, we find $F_{\rm rec,equil} \simeq 
9.4 \times 10^{-4}$. Finally for the entry $F_{\rm rec,rec}$ we get

\bea
\label{S}
F_{\rm rec,rec}&=&  
\frac{f_{\rm sky}}{\pi} \frac{1}{(2 \pi)^2} \int d^2 \ell_1 d^2 \ell_2  d^2 \ell_3 
\,\delta^{(2)}(\vec{\ell}_{123})\, 
\frac{B^2_{\rm rec}(\ell_1,\ell_2,\ell_3)}
{6\, C(\ell_1)\,C(\ell_2)\, C(\ell_3)}\, \nonumber\\
&=& f_2^2 \frac{f_{\rm sky}}{2^5 \pi^3 6} \left( \frac{4 R}{14} \right)^2 \left( \frac{T^2_0}{a} \right)^2 
(k_{\rm eq} \eta_r)^4\, \ell^4_{\rm eq} \ell_*  A\nonumber\\
&\times&\Bigg[
3 \int d\ell_1 d\ell_2 e^{(\ell_3/\ell_*)^{1.2}} \frac{\ell_3^3}{\ell_1^6 \ell_2^6} \left({\vec\ell}_1 \cdot {\vec\ell}_2-\frac{10}{3} 
\frac{({\vec\ell}_3 \cdot {\vec\ell}_1)({\vec\ell}_3 \cdot {\vec\ell}_2)}{\ell_3^2} \right)^2\nonumber \\   
&+& 6 \int d\ell_1 d\ell_2  e^{1/2(\ell_3/\ell_*)^{1.2}}  e^{1/2(\ell_2/\ell_*)^{1.2}} \frac{1}{\ell_1^6 \ell_2 \ell_3^2} 
 \left({\vec\ell}_1 \cdot {\vec\ell}_2-\frac{10}{3} 
\frac{({\vec\ell}_3 \cdot {\vec\ell}_1)({\vec\ell}_3 \cdot {\vec\ell}_2)}{\ell_3^2} \right)\nonumber \\
&\times& \left({\vec\ell}_1 \cdot {\vec\ell}_3-\frac{10}{3} 
\frac{({\vec\ell}_2 \cdot {\vec\ell}_3)({\vec\ell}_1 \cdot {\vec\ell}_2)}{\ell_2^2} \right)
\Bigg]\, , 
\eea
and we find a value $F_{\rm rec, rec}\simeq 0.014$. We are now able to compute the entries of inverse 
of the Fisher matrix, $F^{-1}_{ij}$. In 
the following we report our results for the signal-to-noise ratios and the degradation parameters
\bea
\label{SN1}
\left( \frac{S}{N} \right)_{\rm equil}&=& \frac{1}{\sqrt{F^{-1}_{\rm equil,equil}}} \simeq
12.6 \times 10^{-3}  f_{\rm NL}^{\rm equil}
\, ,\\
\left( \frac{S}{N} \right)_{\rm rec}&=& \frac{1}{\sqrt{F^{-1}_{\rm rec,rec}}}\simeq 0.1\, , \\ 
r_{\rm rec,equil}&=&\frac{F^{-1}_{\rm rec,equil}}{\sqrt{F^{-1}_{\rm equil,equil}F^{-1}_{\rm rec,rec}}}
\simeq -0.53\, , \\
d_{\rm rec}&=&F_{\rm rec,rec}F^{-1}_{\rm rec,rec}\simeq 1.4\, , \\
d_{\rm equil}&=&F_{\rm equil,equil}F^{-1}_{\rm equil,equil}\simeq 1.4\, .
\eea
As a confirmation of our expectations, we find that the NG of the type given by Eq.~(\ref{kernelG}) 
has a quite high correlation with an 
equilateral primordial bispectrum.
This translates into a degradation in the mimimum detectable value for 
$f_{\rm NL}^{\rm equil}$ with respect to the value given in~(\ref{fnleq}). In fact from the signal-to-noise
 ratio~(\ref{SN1}) we find a minimum detectable value of 
\be
f_{\rm NL}^{\rm equil} \simeq 79\, ,
\ee  
imposing that $(S/N)_{\rm equil}=1$ with an an increase of $\Delta f^{\rm equil}_{\rm NL}={\cal O}(10)$. 
This corresponds to an increase of the 1-$\sigma$ uncertainty on $f^{\rm equil}_{\rm NL}$ of $20\%$ (see Sec.~\ref{SNSC}).\footnote{Due to a non-vanishing correlation, 
$r_{ij}$, $(S/N)$ gets modified from its zero-order value to $(S/N)=(S/N)_0 (1-r_{ij}^2)^{1/2}$, so that the minimum detectable value 
of $f^{\rm equil}_{\rm NL}$ gets shifted by a quantity $\Delta f^{\rm equil}_{\rm NL}/(f^{\rm equil}_{\rm NL})_0 =(1-r_{ij}^2)^{-1/2}-1$.} 

Following Sec.~\ref{SNSC}, in order to measure the contamination to the primordial bispectra 
one can define that effective non-linearity parameter $f^{\rm con}_{\rm NL}$ which minimizes the $\chi^2$ defined as 
\begin{eqnarray}
\chi^2=\int d^2 \ell_1 d^2 \ell_2  d^2 \ell_3 
\,\delta^{(2)}(\vec{\ell}_{123})\,
\frac{\left(f^{\rm equil}_{\rm NL}\,B_{\rm eq}(\ell_1,\ell_2,\ell_3;f_{\rm NL}^{\rm equil}=1 )
- B_{\rm rec}(\ell_1,\ell_2,\ell_3)\right)^2}{6\, C(\ell_1)\,C(\ell_2)\, C(\ell_3)}\, , \nonumber \\
\end{eqnarray}
to find 
\begin{equation}
\label{effr}
f^{\rm con}_{\rm NL}= \frac{F_{\rm rec,equil}}{F_{\rm equil ,equil}} \Big|_{f_{\rm NL}^{\rm equil}=1}\, ,
 \nonumber
\end{equation}
and an analogous expression to compute the contamination to the local primordial bispectrum. 
In this case we find an effective  non-linearity parameter $f^{\rm cont}_{\rm NL} \simeq 5$.\footnote{Notice that 
sometimes in the literature one can find also an effective non-linearity parameter  $f^{\rm eff.}_{\rm NL}$ defined in such a way that 
the equilateral (or the local) bispectrum has the same Fisher matrix errors as the recombination bispectrum 
$$ 
f^{\rm eff.}_{\rm NL}=\frac{\sqrt{F_{\rm rec,rec}}}{\sqrt{F_{\rm equil,equil}}}\Big|_{f^{\rm equil}_{\rm NL}=1}\, .
$$
However this is not the proper quantity to define the contamination level to primordial NG; here we are just comparing 
signal-to-noise ratios, while the contamination $f^{\rm con}_{\rm NL}$ defined in Eq.~(\ref{effr}) contains a somewhat richer information, since we 
are asking what is 
the value of equilateral (local) $f_{\rm NL}$ which best mimics the bispectrum from recombination.}

Similarly we can compute the Fisher matrix accounting for the NG generated at recombination and the primordial 
NG of the local type 
\be
\label{equilloc}
\langle \Phi({\bf k}_1) \Phi({\bf k}_2) \Phi({\bf k}_3) \rangle = (2 \pi)^3 \delta^{(3)}
\big({\bf k}_1 + {\bf k}_2 + {\bf k}_3 \big)
B_{\rm loc}( k_1,  k_2 ,  k_3) \, ,
\ee
where
\be
\label{eq:loc}
B_{\rm loc}(k_1,k_2,k_3) = f_{\rm NL}^{\rm loc} \cdot 2  
 A^2 \cdot \left(\frac1{k_1^3 k_2^3} +
\frac1{k_1^3 k_3^3} + \frac1{k_2^3 k_3^3}  \right) \, .
\ee
The bispectrum and the signal-to-noise ratio as defined in Eq.~(\ref{S/N}) have already been computed in the flat-sky approximation in 
Ref.~\cite{BZ}. The result is that $(S/N)_{\rm loc}^2=4 \pi^{-2} 
 f_{\rm sky} (\ell_*/\ell_{\rm min}) (f^{\rm loc}_{\rm NL})^2 
A\, \ell^2_{\rm max}$, corresponding to 
a minimum 
detectable value of $f^{\rm loc}_{\rm NL}={\cal O}(7)$ for $\ell_{\rm max}=2000$ 
(when other possible sources of NG are ignored). We can compute the 
off-diagonal entry of the Fisher matrix in a similar way to what we have described in this section, and we get 
$F_{\rm rec,loc}\simeq 8 \times 10^{-3} f_{\rm NL}^{\rm loc}$. Finally the entry $F_{\rm rec,rec}\simeq 0.014$ 
has already been computed above. 
From inverting the Fisher matrix, we get the following signal-to-noise ratios and the degradation parameters
\bea
\label{SN2}
\left( \frac{S}{N} \right)_{\rm loc}&=& \frac{1}{\sqrt{F^{-1}_{\rm loc,loc}}} \simeq 14 \times 10^{-2}  
f_{\rm NL}^{\rm loc}
\, ,\\
\left( \frac{S}{N} \right)_{\rm rec}&=& \frac{1}{\sqrt{F^{-1}_{\rm rec,erec}}}\simeq 0.1\, , \\ 
r_{\rm rec,loc}&=&\frac{F^{-1}_{\rm rec,loc}}{\sqrt{F^{-1}_{\rm loc,loc}F^{-1}_{\rm rec,rec}}}\simeq -0.44\, , \\
d_{\rm rec}&=&F_{\rm rec,rec}F^{-1}_{\rm rec,rec}\simeq 1.2\, , \\
d_{\rm loc}&=&F_{\rm equil,equil}F^{-1}_{\rm equil,equil}\simeq 1.2\, .
\eea
In particular,  from Eq.~(\ref{SN2}) we see that now the minimum detectable value of $f^{\rm loc}_{\rm NL}$ 
remains basically
 unchanged in the presence of the recombination signal. 
Similarly the effective $f^{\rm con}_{\rm NL}$ reads 
\begin{equation}
\label{effr}
f^{\rm con}_{\rm NL}= \frac{F_{\rm rec,loc}}{F_{\rm loc ,loc}} \Big|_{f_{\rm NL}^{\rm loc}=1}\simeq 0.3\, ,
 \nonumber
\end{equation}
which is much smaller than the effective non-linearity parameter~(\ref{effr}) for the equilateral case. 
We have also checked the cross-correlation between 
the primordial local and equilateral bispectra finding a value of $r_{\rm loc,equil}\simeq0.23$, which is in 
agreement with the 
value reported in Ref.~\cite{SZ}. This reflects the fact that the primordial local and equilateral signals are 
not fully uncorrelated. The reason  is due to the fact that the equilateral and 
local bispectrum~(\ref{eq:loc}) and~(\ref{eq:ours}) 
approach the same shape in the equilateral configuration. This is also the reason why the 
cross-correlation between the primordial local and recombination bispectra is not so small.

\section{How to perform a numerical analysis of the CMB bispectrum produced by second-order perturbations}
\label{NumAn}

In the previous section, through a specific example, we have learned some basic principles to determine a reasonable and quite fast analytic estimate for the contamination to the primordial NG.  
In this section we provide all the tools necessary to a full numerical implementation of the second-order Boltzmann equations which allow to obtain a systematic numerical evaluation of the CMB angular bispectrum produced by second-order cosmological perturbations. This section is mainly based on the results of Ref.~\cite{prep}.

In particular we will apply this formalism to numerically evaluate the contamination to primordial NG from the second-order 
fluctuations of the Boltzmann equations that come as products of the first-order perturbations,
and ignore the intrinsically second-order terms, or the
effects of the perturbed recombination \cite{khatri09,senatore08I,senatore08II}.
\footnote{At the moment of writing this review 
the calculations that also include the intrinsically second-order terms
and the perturbed recombination are in progress and will be presented in Ref.~\cite{intrin}.}
 Here therefore we come to the discussion of the second example mentioned in Sec.~\ref{To}. As anticipated there we will see that, unlike the intrisically second-order contribution considered in the previous section, the products of the first-order perturbations give a CMB bispectrum that peaks in the squeezed configuration. Therefore  we shall study the contamination of the primordial 
 NG of the local type. 
 
\subsection{CMB Bispectrum from second-order perturbations}
\subsubsection{Definitions}
Again, we expand the temperature fluctuation into the linear (first-order)
part and the second-order part as
\begin{equation}
\frac{\Delta T(\hat{\bf n})}{T}=\frac{\Delta T^{(1)}(\hat{\bf n})}{T}+\frac{\Delta T^{(2)}(\hat{\bf n})}{T}+\ldots.
\end{equation}
The spherical harmonic coefficients of temperature anisotropy, \break
$a_{\ell m}=T^{-1}\int d^2\hat{\bf n}Y^{*}_{\ell m}(\hat{\bf n})\Delta  T(\hat{\bf n})$,
are therefore expanded as
\begin{equation}
a_{\ell m}=a_{\ell m}^{(1)}+a_{\ell m}^{(2)}+\ldots.
\end{equation}
Recall that to expand the Boltzmann equation up to the second order in
perturbations, we had to  expand the distribution function,
\begin{eqnarray}
f({\bf x},p,{\hat{\bf n}},\eta)=2\left[\exp\bigg\{\frac{p}{T(\eta)e^{\Theta({\bf x},{\hat{\bf n}},\eta)}}\bigg\}-1\right]^{-1},
\end{eqnarray}
up to the second order in perturbations: $\Theta=\Theta^{(1)}+\Theta^{(2)}/2+\ldots,$ and accordingly $f=f^{(0)}+f^{(1)}+f^{(2)}/2+\ldots$.
We then had to compute the fractional perturbation in photon's energy density
at the $i$-th order in perturbations, $\Delta^{(i)}$, by multiplying $f^{(i)}$
by $p$, and integrating over $p^2dp$: 
\begin{equation}
\Delta^{(i)}\equiv \frac{\int dpp^3f^{(i)}}{\int dpp^3f^{(0)}}.
\end{equation}
At the linear order, we recovered the usual relation between the
linear fractional temperature fluctuation, $\Theta^{(1)}=\Delta T^{(1)}/T$, and the
linear fractional
energy density perturbation, $\Delta^{(1)}=\delta\rho_\gamma^{(1)}/\rho_\gamma$, i.e.,
$\Delta^{(1)}=4\Theta^{(1)}$. 
At the second order we have 
\begin{eqnarray}
\Delta^{(2)}=4\Theta^{(2)}+16[\Theta^{(1)}]^2,
\end{eqnarray}
which is related to the second-order temperature fluctuation as
\begin{eqnarray}
{\frac{\Delta
T}{T}}^{(2)}&=&\frac{1}{8}\left(\Delta^{(2)}-\langle\Delta^{(2)}\rangle\right)
-\frac{3}{2}\left([\Theta^{(1)}]^2-\langle[\Theta^{(1)}]^2\rangle\right)\nonumber\\
&=&\frac{1}{2}\left(\Theta^{(2)}-\langle\Theta^{(2)}\rangle
+[\Theta^{(1)}]^2-\langle[\Theta^{(1)}]^2\rangle\right),
\end{eqnarray}
where we have subtracted the average of the temperature fluctuation so
that the average of $\Delta T^{(2)}/T$ vanishes. 

We compute $a_{\ell m}^{(2)}$ from $\Delta T^{(2)}/T$ using 
\begin{eqnarray}
a^{(2)}_{\ell m}&=&\int d^2\hat{\bf n}Y^{*}_{\ell m}(\hat{{\bf n}})
{\frac{\Delta T}{T}}^{(2)}\\
&=&{\tilde a}^{(2)}_{\ell m}-\frac{3}{2}\sum_{\ell' m'}\sum_{\ell'' m''}(-1)^m{\cal G}_{\ell \ell' \ell''}^{-mm'm''}
(a^{(1)}_{\ell' m'}a^{(1)}_{\ell'' m''}-\langle a^{(1)}_{\ell' m'}a^{(1)}_{\ell'' m''}\rangle) \nonumber,
\end{eqnarray}
where we define 
\begin{equation}
{\tilde a}^{(2)}_{\ell m}\equiv\frac{1}{8}\int d^2\hat{\bf n}Y^{*}_{\ell m}(\hat{\bf n})(\Delta^{(2)}(\hat{\bf n})
-\langle\Delta^{(2)}(\hat{\bf n})\rangle),
\end{equation}
\begin{eqnarray}
& & {\cal G}_{\ell_1\ell_2\ell_3}^{m_1m_2m_3}\equiv\int d^2\hat{\bf n}Y_{\ell_1m_1}(\hat{\bf n})Y_{\ell_2m_2}(\hat{\bf n})Y_{\ell_3m_3}(\hat{\bf n})
\nonumber \\
& & =\sqrt{\frac{(2\ell_1+1)(2\ell_2+1)(2\ell_3+1)}{4\pi}}
\left(
\begin{array}{ccc}\ell_1&\ell_2&\ell_3\\0&0&0\\ \end{array}
\right)
\left(
\begin{array}{ccc}\ell_1&\ell_2&\ell_3\\m_1&m_2&m_3\\ \end{array}
\right) \nonumber \, .
\end{eqnarray}
Here the matrix is the Wigner 3$j$ symbol.

The CMB angular-averaged bispectrum,
$B_{\ell_1\ell_2\ell_3}$, is related to the ensemble average of
$a_{\ell_1m_1}a_{\ell_2m_2}a_{\ell_3m_3}$ as
\begin{eqnarray}
B_{\ell_1\ell_2\ell_3}\equiv\sum_{{\rm all}\,m}
\left(
\begin{array}{ccc}\ell_1&\ell_2&\ell_3\\m_1&m_2&m_3\\ \end{array}
\right)
\langle a_{\ell_1m_1}a_{\ell_2m_2}a_{\ell_3m_3}\rangle.
\end{eqnarray}
This definition guarantees rotational invariance for the bispectrum, and
the Wigner $3j$ symbol ensures that the bispectrum must satisfy triangle
conditions: 
$\ell_i-\ell_j|\le \ell_k\le \ell_i+\ell_j$ for all permutations of indices, and
selection rules: $m_1+m_2+m_3=0$.

The ensemble average is given by
\begin{eqnarray}
& & \langle a_{\ell_1m_1}a_{\ell_2m_2}a_{\ell_3m_3}\rangle=\langle a^{(1)}_{\ell_1m_1}a^{(1)}_{\ell_2m_2}a^{(2)}_{\ell_3m_3}\rangle+cyclic
= \nonumber\\
&&=\langle a^{(1)}_{\ell_1m_1}a^{(1)}_{\ell_2m_2}{\tilde a}^{(2)}_{\ell_3m_3}\rangle
-\frac{3}{2}\sum_{\ell_3'm_3'}\sum_{\ell_3''m_3''}(-1)^{m_3}{\cal G}_{\ell_3\ell_3'\ell_3''}^{-m_3m_3'm_3''}
\nonumber\\
&&\times
(\langle a^{(1)}_{\ell_1m_1}a^{(1)}_{\ell_2m_2}a^{(1)}_{\ell_3'm_3'}a^{(1)}_{\ell_3''m_3''}\rangle-
\langle a^{(1)}_{\ell_1m_1}a^{(1)}_{\ell_2m_2}\rangle\langle a^{(1)}_{\ell_3'm_3'}a^{(1)}_{\ell_3''m_3''}\rangle)
+{\rm cyclic} \nonumber \, , \\ 
\label{eq:one}
\end{eqnarray}
where cyclic means that we have to sum the cyclic permutations of
Eq.~(\ref{eq:one}) for indices $(1,2,3) \to(3,1,2) \to(2,3,1)$. 

As we assume that $a_{\ell m}^{(1)}$'s are Gaussian random variables, 
the four-point function of $a_{\ell m}^{(1)}$'s in Eq.~(\ref{eq:one}) 
is given by the sum of products of all possible pairs. 
Each pair gives the angular power spectrum, $C_\ell$:
\begin{eqnarray}
\langle a^{(1)}_{\ell m}a^{(1)}_{\ell' m'}\rangle=(-1)^mC_\ell\delta_{\ell \ell'}\delta_{-mm'}.
\end{eqnarray}
We obtain
\begin{eqnarray}
\langle a^{(1)}_{\ell_1m_1}a^{(1)}_{\ell_2m_2}a^{(1)}_{\ell_3'm_3'}a^{(1)}_{\ell_3''m_3''}\rangle
-\langle a^{(1)}_{\ell_1m_1}a^{(1)}_{\ell_2m_2}\rangle\langle a^{(1)}_{\ell_3'm_3'}a^{(1)}_{\ell_3''m_3''}\rangle\nonumber\\
=(-1)^{m_1+m_2}C_{\ell_1}C_{\ell_2}[\delta_{\ell_1\ell_3'}\delta_{-m_1m_3'}\delta_{\ell_2\ell_3''}\delta_{-m_2m_3''}
+(1\leftrightarrow 2)]\label{eq:two}.
\end{eqnarray}
Substituting the right hand side of equation (\ref{eq:two}) for the
second term of equation (\ref{eq:one}), and using $\ell_1+\ell_2+\ell_3=$ even,
we obtain the angular averaged bispectrum,  
\begin{eqnarray}
B_{\ell_1\ell_2\ell_3}
=\tilde{B}_{\ell_1\ell_2\ell_3}-3I_{\ell_1\ell_2\ell_3}(C_{\ell_1}C_{\ell_2}+{\rm cyclic}),
\label{eq:16}
\end{eqnarray}
where we have defined the quantities,
\begin{eqnarray}
I_{\ell_1\ell_2\ell_3}\equiv
\sqrt{\frac{(2\ell_1+1)(2\ell_2+1)(2\ell_3+1)}{4\pi}}
\left(
\begin{array}{ccc}\ell_1&\ell_2&\ell_3\\0&0&0\\ \end{array}
\right),\nonumber\\
\end{eqnarray}
and
\begin{eqnarray}
\tilde{B}_{\ell_1\ell_2\ell_3}=\sum_{{\rm all}\,m}
\left(
\begin{array}{ccc}\ell_1&\ell_2&\ell_3\\m_1&m_2&m_3\\ \end{array}
\right)
\langle a^{(1)}_{\ell_1m_1}a^{(1)}_{\ell_2m_2}{\tilde a}^{(2)}_{\ell_3m_3}\rangle
+{\rm cyclic}.\label{eq:18}
\end{eqnarray}
\subsubsection{The second-order CMB radiation transfer function}
In this section we show how it is possible to define in a rigorous way the radiation transfer function for CMB anisotropies 
at second-order in the perturbations. It is a generalization of the well known quantity used at linear-order and as such it allows us to develop a systematic 
numerical analysis of the angular bispectrum produced by second-order perturbations 
in the very same way as at linear-order various numerical codes are nowadays available for the computation of the CMB power spectrum, such as  {\sf CMBFAST} or CAMB.

The starting point is the Boltzmann equation that governs the evolution of
$\Delta^{(1)}(k,\mu,\eta)$ and $\Delta^{(2)}({\mathbf k},\hat{\mathbf{n}},\eta)$, 
where $\mu=\hat{k}\cdot\hat{n}$ and $\mathbf{n}$ is the direction of
propagation of photons. Note that for the linear perturbation there is
azimuthal symmetry such that $\Delta^{(1)}$ depends only on the angle
between $\mathbf{k}$ and $\mathbf{n}$; however, for the second-order
perturbation there is no such symmetry.
We write again the Boltzmann equations in Fourier space  
\begin{eqnarray}
{\Delta^{(1)}}'+ik\mu\Delta^{(1)}-\tau'\Delta^{(1)}=S^{(1)}(k,\mu,\eta),\label{eq:1stBol}\\
{\Delta^{(2)}}'+ik\mu\Delta^{(2)}-\tau'\Delta^{(2)}=S^{(2)}(\bf{k,\hat{n}},\eta),\label{eq:2ndBol}
\end{eqnarray}
where the primes denote derivatives with respect to the conformal
 time $\partial/\partial\eta$, 
 $S^{(1)}$ and $S^{(2)}$ are the source functions at the first and
 the second orders, respectively, and $\tau'$ is the differential
 optical depth which is defined by using the mean electron number
 density, $\bar{n}_e$, the Thomson scattering cross-section, $\sigma_T$,
 and the scale factor, $a$, as
\begin{eqnarray}
\tau'=-\bar{n}_e\sigma_Ta. 
\end{eqnarray} 
We expand again the angular dependence of $\Delta^{(i)}$ as
\begin{eqnarray}
\Delta_{\ell m}^{(i)}({\bf k},\eta)=i^{\ell}\sqrt{\frac{2\ell+1}{4\pi}}\int
 d^2\hat{\bf n}Y_{\ell m}^{*}(\hat{\bf n})\Delta^{(i)}({\bf k},\hat{\bf
 n},\eta),
\label{eq:delta_lm}
\end{eqnarray}
and that of the source terms as
\begin{eqnarray}
\label{DecSo}
S_{\ell m}^{(i)}({\bf k},\eta)=i^{\ell}\sqrt{\frac{2\ell+1}{4\pi}}\int d^2\hat{\bf n}Y_{\ell m}^{*}(\hat{\bf n})S^{(i)}({\bf k},\hat{\bf n},\eta),
\end{eqnarray}
where $i=1,2$.

The source functions relate the observed $a_{\ell m}$'s to 
the primordial curvature perturbations in comoving gauge,
$\zeta({\mathbf k})$. The relations contain
the linear radiation transfer function, $g_\ell(k)$, and the second-order
radiation transfer function, $F_{\ell m}^{\ell' m'}(k)$, and are given by
\begin{eqnarray}
&&a^{(1)}_{\ell m}=4\pi(-i)^\ell\int\frac{d^3k}{(2\pi)^3}g_\ell(k)Y^*_{\ell m}(\hat{\bf k})\zeta({\bf k}),
\label{eq:19}\\
&&\tilde{a}^{(2)}_{\ell m}=\frac{4\pi}{8}(-i)^l\int\frac{d^3k}{(2\pi)^3}\int\frac{d^3k'}{(2\pi)^3}\int d^3k''
\delta^{(3)}({\bf k'+k''-k})
\nonumber\\&&\quad\quad\quad\times
\sum_{\ell' m'}F^{\ell' m'}_{\ell m}({\bf k',k'',k})Y^*_{\ell' m'}(\hat{\bf k})
\zeta({\bf k'})\zeta({\bf k''})\label{eq:20}.
\end{eqnarray}
The linear transfer function is given by
\begin{eqnarray}
g_\ell(k)&=&\int_0^{\eta_0} d\eta e^{-\tau}
\left[S_{00}^{(1)}(k,\eta)+S_{10}^{(1)}(k,\eta)\frac{d}{du}+S_{20}^{(1)}(k,\eta) 
\right. \nonumber \\
&\times & \left. \left(\frac{3}{2}\frac{d^2}{du^2}+ \frac{1}{2}\right)\right]j_\ell(u), 
\end{eqnarray}
where $u\equiv k(\eta_0-\eta)$ and $S_{\ell m}^{(1)}$ is the standard linear
source function
\begin{eqnarray}
 S_{00}^{(1)}(k,\eta)&=& 4{\Psi^{(1)}}'(k,\eta)-\tau'\Delta_0^{(1)}(k,\eta) ,\label{eq:s00}\\
 S_{10}^{(1)}(k,\eta)&=& 4k\Phi^{(1)}(k,\eta)-4\tau'v_0^{(1)}(k,\eta),\label{eq:s10}\\
 S_{20}^{(1)}(k,\eta)&=& \frac{\tau'}{2}\Delta_2^{(1)}(k,\eta),\label{eq:s20}
\end{eqnarray}
where $\Phi^{(1)}(k,\eta)$ and $\Psi^{(1)}(k,\eta)$ are the metric perturbations at the linear order in the longitudinal gauge and $\Delta_0^{(1)}(k,\eta)$, $\Delta_1^{(1)}(k,\eta)$, and
$\Delta_2^{(1)}(k,\eta)$ are the coefficients of the expansion
 in Legendre polynomials of $\Delta^{(1)}(k,\mu,\eta)$. These coefficients $\Delta_\ell^{(1)}(k,\eta)$
are related to $\Delta_{\ell m}^{(1)}$ (Eq.~(\ref{eq:delta_lm})) via
\begin{equation}
\label{expc}
\Delta^{(1)}_{\ell m}({\bf k})=i^\ell \sqrt{\frac{4 \pi}{2\ell+1}}Y^*_{\ell m}(\hat{\bf k}) 
\Delta^{(1)}_\ell({\bf k}) (2\ell+1)\, .
\end{equation}
The first-order velocity perturbation, $v_0^{(1)}(k,\eta)$, is the
irrotational part of the baryon velocity defined by ${\bf v}({\bf
k})=-iv_0(k)\hat{\bf k}$. 

The new piece, \emph{the second-order radiation transfer
function}, is the line-of-sight integral of the second-order source terms
in the Boltzmann equation:
\begin{eqnarray}
F^{\ell' m'}_{\ell m}({\bf k',k'',k})&=&i^\ell\sum_{\lambda\mu}(-1)^m(-i)^{\lambda-\ell'}
{\cal G}_{\ell\ell'\lambda}^{-mm'\mu}\sqrt{\frac{4\pi}{2\lambda+1}}
\nonumber\\&\times&
\int_0^{\eta_0}d\eta e^{-\tau}{\cal S}^{(2)}_{\lambda\mu}({\bf k',k'',k},\eta)
j_{\ell'}[k(\eta-\eta_0)].\nonumber\\
\label{eq:Fdefinition}
\end{eqnarray}
Here, we have introduced a new function, ${\cal S}_{\ell m}^{(2)}({\bf
 k}',{\mathbf k}'', \mathbf{k},\eta)$, which is defined by the following equation:
\begin{eqnarray}
\label{kernelS}
S_{\ell m}^{(2)}({\bf k},\eta)&=&\int\frac{d{\bf k'}}{(2\pi )^3} d{\bf k''}\delta^{(3)}({\bf k'+k''-k})
{\cal S}_{\ell m}^{(2)}({\bf k',k'', k},\eta)\zeta({\bf k'})\zeta({\bf k''}). \nonumber \\
\end{eqnarray}
Basically, ${\cal S}_{\ell m}^{(2)}({\bf k',k'', k},\eta)$ is the second-order
source function divided by \break $\zeta({\bf k'})\zeta({\bf  k''})$. 
The explicit expression of the second-order source function can be  be read from Eq.~(\ref{B2}). 

Using equation (\ref{eq:19}) and (\ref{eq:20}), we calculate the first
term in Eq.~(\ref{eq:16}), $\tilde{B}_{\ell_1\ell_2\ell_3}$, as follows:
\begin{eqnarray}
\langle a_{\ell_1m_1}^{(1)}a_{\ell_2m_2}^{(1)}\tilde{a}_{\ell_3m_3}^{(2)}\rangle
&=&
\frac{(-i)^{\ell_1+\ell_2+\ell_3}}{(2\pi)^3}\sum_{\ell_3M_3}\prod_i\int d^3k_i\delta^{(3)}({{\bf k}_{123}})
Y_{\ell_1m_1}^*({\bf\hat{k}_1}) \nonumber \\
& \times & 
Y_{\ell_2m_2}^*({\bf\hat{k}_2})Y_{\ell_3M_3}^*({\bf\hat{k}_3}) g_{\ell_1}(k_1)g_{\ell_2}(k_2)P_{\zeta}(k_1)P_{\zeta}(k_2)
\nonumber \\
&\times & \{F_{\ell_3m_3}^{\ell_3M_3}({\bf k_1,k_2,k_3})+F_{\ell_3m_3}^{\ell_3M_3}({\bf
k_2,k_1,k_3})\} \, , \nonumber \\
\label{eq:a}
\end{eqnarray}
where, as in the previous section we will sometimes use the notation ${\bf k}_{123}={\bf k}_1+{\bf k}_2+{\bf k}_3$. 
$P_{\zeta}(k)$ is the power spectrum of $\zeta$ given by the usual definition:
\begin{eqnarray}
\langle\zeta({\bf k_1})\rangle=0,\quad \langle\zeta({\bf k_1})\zeta({\bf k_2})\rangle=(2\pi)^3
\delta^{(3)}({\bf k_1+k_2})P_{\zeta}(k_1).
\end{eqnarray}
In order to perform the integral over angles, $\hat{{\bf k}}$, we expand the three-dimensional $\delta$-function using 
Rayleigh's formula,
\begin{eqnarray}
\delta^{(3)}({\bf k_1+k_2+k_3})&=&8\sum_{{\rm all}\,l'm'}i^{\ell_1'+\ell_2'+\ell_3'}{\cal G}_{\ell_1'\ell_2'\ell_3'}^{m_1'm_2'm_3'}
Y_{\ell_1'm_1'}({\bf\hat{k}_1})Y_{\ell_2'm_2'}({\bf\hat{k}_2})\nonumber\\
&\times& Y_{\ell_3'm_3'}({\bf\hat{k}_3}) \int drr^2j_{\ell_1'}(rk_1)j_{\ell_2'}(rk_2)j_{\ell_3'}(rk_3)\, ,\nonumber\\
&&
\end{eqnarray}
and also expand the angular dependence of ${\cal S}_{\ell m}^{(2)}({\bf
 k_1,k_2,k_3},\eta)$ by introducing the transformed source function,
${\cal S}_{\lambda_1\lambda_2\lambda_3}^{\mu_1\mu_2\mu_3}(k_1,k_2,k_3,\eta)$, as
\begin{eqnarray}
{\cal S}^{(2)}_{\lambda_3\mu_3}({\bf k_1},{\bf k_2},{\bf k_3},\eta)&=&\sum_{\lambda_1,\mu_1}\sum_{\lambda_2,\mu_2}(-i)^{\lambda_1+\lambda_2}
\sqrt{\frac{4\pi}{2\lambda_1+1}}\sqrt{\frac{4\pi}{2\lambda_2+1}}\nonumber\\
&\times&{\cal S}_{\lambda_1\lambda_2\lambda_3}^{\mu_1\mu_2\mu_3}(k_1,k_2,k_3,\eta)
Y_{\lambda_1\mu_1}({\bf\hat{k}_1})Y_{\lambda_2\mu_2}({\bf\hat{k}_2})\, . \nonumber \\
\label{eq:28}
\end{eqnarray}
This result shows that ${\cal S}^{(2)}_{\lambda_3\mu_3}({\bf k_1},{\bf
 k_2},{\bf k_3},\eta) = {\cal S}^{(2)}_{\lambda_3\mu_3}({\bf k_1},{\bf
 k_2},k_3,\eta)$, and thus $F_{\ell m}^{\ell' m'}({\bf k_1},{\bf k_2},k_3)$
 follows (see Eq.~(\ref{eq:Fdefinition})).

Now we can perform the angular integration of Eq.~(\ref{eq:a}) to obtain
\begin{eqnarray}
& & \tilde{B}_{\ell_1\ell_2\ell_3}=
\frac{4}{\pi^2}(-i)^{\ell_1+\ell_2+\ell_3}\!\sum_{{\rm all}\,m}\sum_{{\rm all}\,\ell'm'}\sum_{{\rm all}\,\lambda\mu}
i^{\ell_1'+\ell_2'+\ell_3'-\lambda_1-\lambda_2-\lambda_3}\nonumber\\
& & \times \sqrt{\frac{4\pi}{(2\lambda_1+1)(2\lambda_2+1)(2\lambda_3+1)}} \left(
\begin{array}{ccc}\ell_1&\ell_2&\ell_3\\m_1&m_2&m_3\\ \end{array}
\right) \nonumber \\
& & \times 
{\cal G}_{\ell'_1\ell'_2\ell'_3}^{m'_1m'_2m'_3}
{\cal G}_{\ell'_1\ell_1\lambda_1}^{m'_1-m_1\mu_1}
{\cal G}_{\ell'_2\ell_2\lambda_2}^{m'_2-m_2\mu_2}
{\cal G}_{\ell'_3\ell_3\lambda_3}^{m'_3-m_3\mu_3}\nonumber\\
&&\times
\int drr^2 \prod_{i=1}^{3}\int k_i^2dk_i j_{\ell_i'}(rk_i) 
g_{\ell_1}(k_1)g_{\ell_2}(k_2)P_{\zeta}(k_1))P_{\zeta}(k_2)\nonumber\\
&&\times i^{\ell_3+\ell_3'}\int d\eta e^{-\tau}\{ {\cal S}_{\lambda_1\lambda_2\lambda_3}^{\mu_1\mu_2\mu_3}(k_1,k_2,k_3,\eta)+
1 \leftrightarrow 2 \}j_{\ell_3'}[k_3(\eta-\eta_0)]\nonumber\\
&&+ {\rm cyclic}, 
\label{eq:tildeBcyclic}
\end{eqnarray}
where for brevity we indicate $P_{\lambda_1 \lambda_2 \lambda_3}=\sqrt{4\pi/[(2\lambda_1+1)(2\lambda_2+1)(2\lambda_3+1)]}$, 
and we have used the following relation of the Wigner $9j$ symbol,
\begin{eqnarray}
& & (-1)^{\ell_1'+\ell_2'+\ell_3'}\sum_{{\rm all}\,mm'}
\left(
\begin{array}{ccc}\ell_1&\ell_2&\ell_3\\m_1&m_2&m_3\\ \end{array}
\right)
{\cal G}_{\ell'_1\ell'_2\ell'_3}^{m'_1m'_2m_3'}
{\cal G}_{\ell'_1\ell_1\lambda_1}^{m'_1-m_1\mu_1}
\\
& & \times {\cal G}_{\ell'_2\ell_2\lambda_2}^{m'_2-m_2\mu_2}
{\cal G}_{\ell'_3\ell_3\lambda_3}^{m'_3-m_3\mu_3}= (-1)^{R}I_{\ell'_1\ell'_2\ell'_3}I_{\ell_1\ell_1'\lambda_1}I_{\ell_2\ell_2'\lambda_2}I_{\ell_3\ell_3'\lambda_3}
\nonumber \\
& & \times 
\Bigg\{
\begin{array}{ccc}\ell_1&\ell_2&\ell_3\\\ell_1'&\ell_2'&\ell_3'\\\lambda_1&\lambda_2&\lambda_3\\ \end{array} 
\Bigg\}
\left(
\begin{array}{ccc}\lambda_1&\lambda_2&\lambda_3\\ \mu_1&\mu_2&\mu_3\\ \end{array}
\right)\, , \nonumber 
\end{eqnarray}
with $R\equiv \ell_1+\ell_2+\ell_3+\ell_1'+\ell_2'+\ell_3'+\lambda_1+\lambda_2+\lambda_3$.
The Wigner 9$j$ symbols have the permutation symmetry:
\begin{eqnarray}
(-1)^{R}
\Bigg\{
\begin{array}{ccc}\ell_1&\ell_2&\ell_3\\\ell_1'&\ell_2'&\ell_3'\\\lambda_1&\lambda_2&\lambda_3\\ \end{array}
\Bigg\}
&=&
\Bigg\{
\begin{array}{ccc}\ell_2&\ell_1&\ell_3\\\ell_2'&\ell_1'&\ell_3'\\\lambda_2&\lambda_1&\lambda_3\\ \end{array}
\Bigg\}
=
\Bigg\{
\begin{array}{ccc}\ell_1&\ell_3&\ell_2\\\ell_1'&\ell_3'&\ell_2'\\\lambda_1&\lambda_3&\lambda_2\\ \end{array}
\Bigg\}
\nonumber\\
&=&
\Bigg\{
\begin{array}{ccc}\ell_1'&\ell_2'&\ell_3'\\\ell_1&\ell_2&\ell_3\\\lambda_1&\lambda_2&\lambda_3\\ \end{array}
\Bigg\}
=
\Bigg\{
\begin{array}{ccc}\ell_1&\ell_2&\ell_3\\\lambda_1&\lambda_2&\lambda_3\\\ell_1'&\ell_2'&\ell_3'\\ 
\end{array}
\Bigg\} \, , \nonumber \\
\label{eq:9j}
\end{eqnarray}
and the coefficients $I_{\ell'_1\ell'_2\ell'_3}$, $I_{\ell_1\ell_1'\lambda_1}$,
$I_{\ell_2\ell_2'\lambda_2}$, and $I_{\ell_3\ell_3'\lambda_3}$, ensure
$\ell_1'+\ell_2'+\ell_3'={\rm even}$, $\ell_1+\ell_1'+\lambda_1={\rm even}$,
$\ell_2+\ell_2'+\lambda_2={\rm even}$, and $\ell_3+\ell_3'+\lambda_3={\rm even}$,
respectively, which gives $R={\rm even}$. Hence the Wigner $9j$ coefficients
are invariant under the permutations. 

Finally, we obtain the angular averaged bispectrum, 
\begin{eqnarray}
\tilde{B}_{\ell_1\ell_2\ell_3}&=&\frac{4}{\pi^2}\sum_{{\rm all}\,\ell'\lambda}
\sqrt{\frac{4\pi}{(2\lambda_1+1)(2\lambda_2+1)(2\lambda_3+1)}} i^{\ell_3-\ell_3'+R}
\nonumber \\
& \times & I_{\ell'_1\ell'_2\ell'_3}I_{\ell_1\ell_1'\lambda_1} 
I_{\ell_2\ell_2'\lambda_2}I_{\ell_3\ell_3'\lambda_3}
\Bigg\{
\begin{array}{ccc}\ell_1&\ell_2&\ell_3\\\ell_1'&\ell_2'&\ell_3'\\\lambda_1&\lambda_2&\lambda_3\\ \end{array} 
\Bigg\}\nonumber\\
&\times&\int drr^2\prod_{i=1}^{2}\int dk_ik_i^2P_{\zeta}(k_i)g_{\ell_i}(k_i)j_{\ell_i'}(rk_i)
\int dk_3k_3^2j_{\ell_3'}(rk_3)\nonumber\\
&\times& \int dr' e^{-\tau(r')}j_{\ell_3'}(r'k_3)
{\cal S}_{\lambda_1\lambda_2\lambda_3}(k_1,k_2,k_3,r')
+{\rm perm}.\label{eq:bis}\, ,
\end{eqnarray}
where $r'\equiv (\eta_0-\eta)$ and we have used the relation of the spherical Bessel function,
$j_\ell(-x)=(-1)^\ell j_\ell(x)$. We also define the ``angular-averaged source
function,''
\begin{eqnarray}
{\cal S}_{\lambda_1\lambda_2\lambda_3}(k_1,k_2,k_3,r)&\equiv&\sum_{{\rm all}\mu}
\left(
\begin{array}{ccc}\lambda_1&\lambda_2&\lambda_3\\ \mu_1&\mu_2&\mu_3\\ \end{array}
\right)
{\cal
S}_{\lambda_1\lambda_2\lambda_3}^{\mu_1\mu_2\mu_3}(k_1,k_2,k_3,r)
\label{aas}
\nonumber \\
&=&i^{\lambda_1+\lambda_2}\sqrt{\frac{2\lambda_1+1}{4\pi}}\sqrt{\frac{2\lambda_2+1}{4\pi}}\sum_{{\rm all}\mu}
\left(
\begin{array}{ccc}\lambda_1&\lambda_2&\lambda_3\\ \mu_1&\mu_2&\mu_3\\ \end{array}
\right)\nonumber\\
&\times&\int d^2\hat{\bf k}_1 d^2\hat{\bf k}_2
Y_{\lambda_1\mu_1}^{*}(\hat{\bf k}_1)Y_{\lambda_2\mu_2}^{*}(\hat{\bf k}_2)
{\cal S}_{\lambda_3\mu_3}({\bf k_1},{\bf k_2},r)\, ,
\nonumber\\ 
\end{eqnarray}
where we have used the inverse relation of Eq.~(\ref{eq:28}).
Note that cyclic terms in
Eq.~(\ref{eq:tildeBcyclic}) have become permutations
 because of invariance of the Wigner $9j$ coefficients under the
permutations. 
 
The final analytic formula (\ref{eq:bis}) we have obtained is a general
formula which can be applied to any second-order perturbations. 
The information about the specific second-order terms is contained in 
the angular-averaged source term, ${\cal S}_{\lambda_1\lambda_2\lambda_3}$
(see Eqs.~(\ref{aas}) and (\ref{eq:28}) for the definition).

For terms that are products of the first-order perturbations, there is indeed a further simplification. One can show that  
${\cal S}_{\lambda_1\lambda_2\lambda_3}(k_1,k_2,k_3,\eta)$ does not depend on
$k_3$, i.e., ${\cal S}_{\lambda_1\lambda_2\lambda_3}(k_1,k_2,k_3,\eta)
={\cal
S}_{\lambda_1\lambda_2\lambda_3}(k_1,k_2,\eta)$. This property enables us to
integrate Eq.~(\ref{eq:bis}) over $k_3$. We obtain
\begin{eqnarray}
& & \tilde{B}_{\ell_1\ell_2\ell_3}=\frac{2}{\pi}\sum_{{\rm all}\,l'\lambda}
\sqrt{\frac{4\pi}{(2\lambda_1+1)(2\lambda_2+1)(2\lambda_3+1)}} i^{\ell_3-\ell_3'+R}
\nonumber \\
& & \times I_{\ell'_1\ell'_2\ell'_3}I_{\ell_1\ell_1'\lambda_1}I_{\ell_2\ell_2'\lambda_2}I_{\ell_3\ell_3'\lambda_3}
\Bigg\{
\begin{array}{ccc}\ell_1&\ell_2&\ell_3\\\ell_1'&\ell_2'&\ell_3'\\\lambda_1&\lambda_2&\lambda_3\\ \end{array} 
\Bigg\}
\nonumber\\
& & \times\int dr e^{-\tau}\prod_{i=1}^{2}\int dk_ik_i^2P_{\zeta}(k_i)j_{l_i'}(rk_i)g_{\ell_i}(k_i)
{\cal S}_{\lambda_1\lambda_2\lambda_3}(k_1,k_2,r)+{\rm perm.}\, ,
\nonumber \\
\label{eq:34}
\end{eqnarray}
where $r\equiv (\eta_0-\eta)$, $R=
\ell_1+\ell_2+\ell_3+\ell_1'+\ell_2'+\ell_3'+\lambda_1+\lambda_2+\lambda_3$, and 
we have used
\begin{eqnarray}
\int dk_3k_3^2j_{\ell_3'}(rk_3)j_{\ell_3'}(r'k_3)=\frac{\pi}{2r^2}\delta(r-r').
\end{eqnarray}
Finally, by adding the remaining term in the full bispectrum,
Eq.~(\ref{eq:16}), we obtain
\begin{eqnarray}
&&B_{\ell_1\ell_2\ell_3}=\frac{2}{\pi}\sum_{{\rm all}\,\ell'\lambda}
\sqrt{\frac{4\pi}{(2\lambda_1+1)(2\lambda_2+1)(2\lambda_3+1)}}i^{\ell_3-\ell_3'+R}
\nonumber \\
& & \times I_{\ell'_1\ell'_2\ell'_3}I_{\ell_1\ell_1'\lambda_1}I_{\ell_2\ell_2'\lambda_2}I_{\ell_3\ell_3'\lambda_3}
\Bigg\{
\begin{array}{ccc}\ell_1&\ell_2&\ell_3\\\ell_1'&\ell_2'&\ell_3'\\\lambda_1&\lambda_2&\lambda_3\\ \end{array}
\Bigg\}\nonumber\\
&&\times\int dr e^{-\tau}\prod_{i=1}^{2}\int dk_ik_i^2P_{\zeta}(k_i)j_{\ell_i'}(rk_i)g_{\ell_i}(k_i)
{\cal S}_{\lambda_1\lambda_2\lambda_3}(k_1,k_2,r)
\nonumber \\
& &
-\frac{3}{2}I_{\ell_1\ell_2\ell_3}C_{\ell_1}C_{\ell_2}+{\rm perm}.
\label{eq:fullbis}
\end{eqnarray}
The remaining task is to calculate the angular-averaged source term,
\break 
${\cal S}_{\lambda_1\lambda_2\lambda_3}(k_1,k_2,k_3,r)$ according to Eq.~({\ref{aas}). Up to now our scope has been to offer the reader all the general tools that allow a systematic treatment of the CMB angular bispectrum from second-order perturbations. In the following we will give just some examples of how such tools can be implemented 
in order to obtain numerical results about signal-to-noise ratios and contamination to primordial non-Gaussianity.

\subsubsection{Source Term: an example}
The explicit expression for $S^{(2)}_{\ell m}({\bf k}, \eta)$ can be obtained from  Eq.~(\ref{B2}). Here we do not want to report the complete expression $S^{(2)}_{\ell m}({\bf k}, \eta)$ of the source function at second-order, rather, for the goal of this Review, we think it is more instructive to show explicitly how the calculation of 
${\cal S}_{\lambda_1\lambda_2\lambda_3}$ proceeds focusing on just one simple example.  

Therefore consider, for example, the term of the second-order source term from Eq.~(\ref{B2}) given by  
\begin{equation}
2 \, \tau' \delta^{(1)}_e \times \Delta^{(1)}\, .
\end{equation}

First we compute the  multipole coefficients of the source term $S^{(2)}_{\ell m}({\bf k})$ from this contribution, 
as defined in Eq.~(\ref{DecSo}). 
They are given by the convolution 
\begin{equation}
S^{(2)}_{\ell m}({\bf k})=\int \frac{d^3k'}{(2 \pi)^3} 2 \tau'\,\delta^{(1)}_e({\bf k}-{\bf k'}) \Delta^{(1)}_{\ell m}({\bf k'})\, .
\end{equation}
 Now for $\Delta^{(1)}_{\ell m}({\bf k})$ we use 
\begin{equation}
\label{expc}
\Delta^{(1)}_{\ell m}({\bf k})=i^l \sqrt{\frac{4 \pi}{2l+1}}Y^*_{\ell m}(\hat{\bf k}) 
\Delta^{(1)}_\ell({\bf k}) (2\ell+1)\, ,
\end{equation}
so that 
\begin{eqnarray}
\label{Sc}
S^{(2)}_{\ell m}({\bf k})&=&\int \frac{d^3k_1}{(2 \pi)^3}\int d^3k_2 \delta^{(3)}({\bf k}_1+{\bf k}_2-{\bf k})\,\, 
i^\ell \sqrt{\frac{4 \pi}{2\ell+1}} (2\ell+1) \nonumber 
\\
&\times&
2 \tau' \, \delta_e({\bf k}_2) \Delta^{(1)}_\ell({\bf k}_1) Y^*_{\ell m}(\hat{\bf k}_1)\, .  
\end{eqnarray}
We now compute the  corresponding ``angular-averaged source function'' coefficients ${\cal S}_{\lambda_1 \lambda_2 \lambda_3}(k_1,k_2,r)$ 
defined by Eq.~(\ref{aas}).
From Eq.~(\ref{Sc}) you read the kernel defined in Eq.~(\ref{kernelS})
\begin{equation}
{\cal S}^{(2)}_{\ell m}({\bf k}_1,{\bf k}_2, \eta)=
i^\ell \sqrt{\frac{4 \pi}{2\ell+1}} (2\ell+1) 2 \tau' \, \delta_e({\bf k}_2) \Delta^{(1)}_\ell({\bf k}_1)Y^*_{\ell m}(\hat{\bf k}_1)\, .
\end{equation}
From here 
\begin{eqnarray}
{\cal S}^{(2)}_{\lambda_1\lambda_2\lambda_3}(k_1,k_2,r)
&=&i^{\lambda_1+\lambda_2}\sqrt{\frac{2\lambda_1+1}{4\pi}}\sqrt{\frac{2\lambda_2+1}{4\pi}}\sum_{{\rm all}\mu}
\left(
\begin{array}{ccc}\lambda_1&\lambda_2&\lambda_3\\ \mu_1&\mu_2&\mu_3\\ \end{array}
\right)
\nonumber \\
&\times& 
 \int d^2\hat{\bf k}_1\int d^2\hat{\bf k}_2
Y_{\lambda_1\mu_1}^{*}(\hat{\bf k}_1)Y_{\lambda_2\mu_2}^{*}(\hat{\bf k}_2)
{\cal S}^{(2)}_{\lambda_3\mu_3}({\bf k_1},{\bf k_2},r),
\nonumber \\
&=& i^{\lambda_1+\lambda_2+\lambda_3}
\sqrt{\frac{2\lambda_1+1}{4\pi}}\sqrt{\frac{2\lambda_2+1}{4\pi}} \sqrt{\frac{4\pi}{2\lambda_3+1}} (2\lambda_3+1) 
\nonumber \\
&\times& \sum_{{\rm all}\mu}
\left(
\begin{array}{ccc}\lambda_1&\lambda_2&\lambda_3\\ \mu_1&\mu_2&\mu_3\\ \end{array}
\right) 
\int d^2\hat{\bf k}_1 \int d^2\hat{\bf k}_2
Y_{\lambda_1\mu_1}^{*}(\hat{\bf k}_1)
\nonumber \\
&\times&Y_{\lambda_2\mu_2}^{*}(\hat{\bf k}_2)  Y_{\lambda_3\mu_3}^{*}(\hat{\bf k}_2) 2 \tau' \, \delta_e(k_1) 
\Delta^{(1)}_{\lambda_3}(k_2) 
\nonumber \\
&=& i^{\lambda_2+\lambda_3}\, \delta_{\lambda_10} \delta_{\lambda_2 \lambda_3}\,
\sqrt{\frac{2\lambda_2+1}{4\pi}} \sqrt{\frac{4\pi}{2\lambda_3+1}} (2\lambda_3+1) 
\nonumber \\
&\times& 2 \tau' 
\delta_e(k_1) \Delta^{(1)}_{\lambda_3}(k_2) 
\sum_{\mu_2} (-1)^{\mu_2}
\left(
\begin{array}{ccc}0&\lambda_2&\lambda_2\\ 0&\mu_2&-\mu_2\\ \end{array}
\right) \, , \nonumber \\
\end{eqnarray}
where we have used $Y^*_{\ell m}=(-1)^{-m}Y_{\ell-m}$, and the orthonormality of the spherical harmonics.
Using  the property of the Wigner 3-j symbols 
\begin{equation}
\sum_{\mu_2} (-1)^{\mu_2}
\left(
\begin{array}{ccc}0&\lambda_2&\lambda_2\\ 0&\mu_2&-\mu_2\\ \end{array}
\right)=
\sum_{\mu_2} \frac{(-1)^{\lambda_2}}{\sqrt{2\lambda_2+1}}
=(-1)^{\lambda_2} \sqrt{2\lambda_2+1}\, ,
\end{equation}
we  find
\begin{equation}
\label{Sr}
{\cal S}^{(2)}_{\lambda_1\lambda_2\lambda_3}(k_1,k_2,r)=2 \tau' \, (2\lambda_2+1)^{(3/2)}\,  
\delta_e(k_1) \Delta^{(1)}_{\lambda_3}(k_2,\eta)\delta_{\lambda_10} \delta_{\lambda_2\lambda_3}\, .
\end{equation}
This term therefore corresponds to coefficients ${\cal S}_{0\lambda_2\lambda_2}(k_1,k_2,r)$.

In general the perturbation variables of the source term can be split into two
parts (see, for example, Eq.(\ref{B2})). A part containing perturbations that are intrinsically second-order (these perturbations have superscripts $(2)$,
and $\omega_m$ and $\chi_m$ are also intrinsically second-order). 
Solving for these terms requires solving the full second-order Boltzmann
equations coupled with the Einstein equations.

Another part contains terms that are products of two linear
variables, as the example that we have just considered. Evaluation of these terms is much easier than that of the
intrinsically second-order terms, as the first-order variables have
already been calculated using the standard linearized Boltzmann code
such as {\sf CMBFAST}.

\subsection{Second-order bispectrum from products of the first-order terms}

\subsubsection{A worked example}
 We shall now focus only on the products of the
first-order perturbations and we choose to analyze just  some of these contributions to offer the reader an example 
of the analysis one can perform. We warn the reader that the full analysis is under progress and that the full results will be presented in 
Ref.~\cite{intrin}, including the intrinsically second-order perturbations which 
are equally important, and the contribution from
perturbing the recombination history
\cite{khatri09,senatore08I,senatore08II}.

For the products of the first-order perturbations, 
from now on we will consider the following non-zero four cases for 
the source terms, ${\cal S}_{\lambda_1\lambda_2\lambda_3}$, which have been analyzed in Ref.~\cite{prep}
(for notational simplicity we shall omit the superscripts (1)): \footnote{
The following terms are in fact the only ones that have been fully analyzed numerically; they correspond to the source term $S^{(2)}_{\ell m}({\bf k}, \eta)$ as given in Ref.~\cite{prep}, Eq.~(3.2). However notice that that expression does not include all the products of first-order perturbations; the complete expression for the source term  
will be given in Ref.~\cite{intrin}.
}

\begin{eqnarray}
\label{4comb}
{\cal S}_{000}
&=&4i\tau'v_0(k_1)\Delta_1(k_2)+[2\, \tau' (\delta_e+\Phi)(k_1)+8\Psi'(k_1)]\Delta_0(k_2),\nonumber\\
{\cal S}_{110}
&=&\frac{4}{\sqrt{3}}\big\{-5\tau' v_0(k_1)v_0(k_2)
+2k_1(\Psi+\Phi)(k_1)\sum_{\ell={\rm odd}}(2\ell+1)\Delta_{l}(k_2)\big\},\nonumber\\
{\cal S}_{101}
&=&2i\sqrt{3}\big\{\tau' v_0(k_1)(4\delta_e+4\Phi+2\Delta_0-\Delta_2)(k_2)\nonumber\\
&&+4k_1\Phi(k_1)(\Delta_0-\Psi)(k_2)+k_1\Delta_0(k_1)(\Psi+\Phi)(k_2)\big\},\nonumber\\
{\cal S}_{112}
&=&2\sqrt{\frac{10}{3}}\big\{7\tau' v_0(k_1)v_0(k_2)
-k_1(\Psi+\Phi)(k_1)\sum_{\ell={\rm odd}}(2\ell+1)\Delta_{l}(k_2)\big\}.
\nonumber \\
\end{eqnarray}
In particular it is recognizable the contribution to ${\cal S}_{000}$ from the example just discussed.  
From these results we find that ${\cal S}_{\lambda_1\lambda_2\lambda_3}$
does not depend on $k_3$, i.e., 
${\cal S}_{\lambda_1\lambda_2\lambda_3}={\cal
S}_{\lambda_1\lambda_2\lambda_3}(k_1,k_2,r)$. 
Note also that ${\cal S}_{011}(k_1,k_2,r)={\cal S}_{101}(k_2,k_1,r)$.

\subsubsection{Bispectrum from products of the first-order terms}
 For the four non-vanishing combinations of $\lambda_1$, $\lambda_2$, and
$\lambda_3$ in Eq.~(\ref{4comb}), we rewrite the expression for the bispectrum,
Eq.~(\ref{eq:fullbis}), as 
\begin{eqnarray}
B_{\ell_1\ell_2\ell_3}&=&\sum_{\lambda_1\lambda_2\lambda_3}
B_{\ell_1\ell_2\ell_3}^{(\lambda_1,\lambda_2,\lambda_3)}+B_{\ell_1\ell_2\ell_3}^{Cl}
\nonumber \\
&=&B_{\ell_1\ell_2\ell_3}^{(0,0,0)}+B_{\ell_1\ell_2\ell_3}^{(1,1,0)}
+2B_{\ell_1\ell_2\ell_3}^{(1,0,1)}+B_{\ell_1\ell_2\ell_3}^{(1,1,2)}+B_{\ell_1\ell_2\ell_3}^{Cl}\label{eq:44},
\end{eqnarray}
where we have used $B_{\ell_1\ell_2\ell_3}^{(0,1,1)}=B_{\ell_1\ell_2\ell_3}^{(1,0,1)}$, 
and defined 
\begin{eqnarray}
B_{\ell_1\ell_2\ell_3}^{Cl}\equiv-3I_{\ell_1\ell_2\ell_3}C_{\ell_1}C_{\ell_2}+cyclic,
\end{eqnarray}
and
\begin{eqnarray}
B_{\ell_1\ell_2\ell_3}^{(\lambda_1,\lambda_2,\lambda_3)}&\equiv&\frac{2}{\pi}\sum_{{\rm all}\,l'}
\sqrt{\frac{4\pi}{(2\lambda_1+1)(2\lambda_2+1)(2\lambda_3+1)}}
i^{\ell_3-\ell_3'+R} 
\nonumber \\
& \times&  
I_{\ell'_1\ell'_2\ell'_3}I_{\ell_1\ell_1'\lambda_1}I_{\ell_2\ell_2'\lambda_2}I_{\ell_3\ell_3'\lambda_3}
\Bigg\{
\begin{array}{ccc}\ell_1&\ell_2&\ell_3\\\ell_1'&\ell_2'&\ell_3'\\\lambda_1&\lambda_2&\lambda_3\\ \end{array}
\Bigg\}\nonumber\\
&\times&\int dr e^{-\tau}\prod_{i=1}^{2}\int dk_ik_i^2P_{\zeta}(k_i)j_{l_i'}(rk_i)g_{l_i}(k_i)
{\cal S}_{\lambda_1\lambda_2\lambda_3}(k_1,k_2,r)
\nonumber \\
&+&{\rm perm}.
\label{eq:lambdabis}
\end{eqnarray}
To proceed further, it turns out to be useful to simplify the expression by introducing the
following notation for the integral over $k$ that appears many times:
\begin{equation}
[x]_{\ell\ell'}^{(n)}(r)\equiv\frac{2}{\pi}\int dkk^{2+n}P_{\zeta}(k)j_{\ell'}(rk)g_{\ell}(k)x(k,r).\label{eq:66}
\end{equation}
This function corresponds to the existing functions in the literature in
the appropriate limits. For example, for $x(k,r)=\pi/2$, this function
is the same as $\beta_{\ell\ell'}^{(n)}(r)$ introduced in \cite{lig06}.
In fact, we find that an order-of-magnitude estimate of
$[x]_{\ell\ll'}^{(n)}(r)$ is given by $[x]_{\ell\ell'}^{(n)}(r)\sim
2\beta_{\ell\ell'}^{(n)}(r)/\pi\times x(k=\ell'/r,r)$ for a smooth function of
$x(k,r)$. 
As $\beta_{\ell\ell'}^{(n)}(r)$ is a sharply peaked function at the decoupling
epoch, $r=r_*$, we find that $[x]_{\ell\ell'}^{(n)}(r)$ is also sharply peaked
at $r=r_*$.
 With these tools in hand,  one can calculate 
$B_{\ell_1\ell_2\ell_3}^{(0,0,0)}$,  $B_{\ell_1\ell_2\ell_3}^{(1,1,0)}$, 
$B_{\ell_1\ell_2\ell_3}^{(1,0,1)}$, and $B_{\ell_1\ell_2\ell_3}^{(1,1,2)}$. We do not bother here the reader with the details of this computation whose details  can be found 
in Ref.~\cite{prep}. Rather here we prefer to go straight to the results of the analysis of these bispectra which can be particularly instructive as far as their shape and their 
contamination to primordial non-Gaussianity are concerned.

\subsection{Shape and signal-to-noise of the second-order bispectrum from products of the
 first-order terms}
One of the motivations for calculating the second-order bispectrum is to
see how much the second-order effects in gravity and the photon-baryon fluid
contaminate the extraction of the primordial bispectrum. If, for
example, the predicted shape of the second-order bispectrum is
sufficiently different from that of the primordial bispectrum, then one
would hope that the contamination would be minimal. 
To investigate this, we shall compare the numerical results of the
second-order bispectrum with the so-called ``local'' model of the
primordial bispectrum.

We extract the first-order perturbations from the {\sf CMBFAST} code.
 We use the following cosmological parameters: 
$\Omega_{\Lambda}=0.72,\,\Omega_{m}=0.23,\,\Omega_{b}=0.046,\,h=0.70$,  
and assume a power law spectrum, $P_{\zeta}\propto k^{n-4}$, with
$n=1$. We determine the decoupling time, $\eta_*$, from the peak of
the visibility function. In this model we have $c\eta_0=14.9$ Gpc and
$c\eta_*=288$~Mpc. While the most of the signal is generated in 
the region of the decoupling epoch, in the low-$\ell$ regime we must also
take into account the late time contribution due to the late integrated
Sachs-Wolfe effect; thus, we integrate over the line-of-sight, $r$, in
the following regions: $c(\eta_0-5\eta_*)<r<c(\eta_0-0.7\eta_*)$ for
$\ell>100$, and $0<r<c(\eta_0-0.7\eta_*)$ for $l\le 100$. 
The step size  is $\Delta r=0.1\eta_*$ around the decoupling epoch, and
we use the same time steps used by {\sf CMBFAST} after the decoupling
epoch.

The local primordial bispectrum is given by \cite{ks} 
\begin{eqnarray}
B_{\ell_1\ell_2\ell_3}=2I_{\ell_1\ell_2\ell_3}\int_{0}^{\infty}r^2drb_{\ell_1}^{L}(r)b_{\ell_2}^{L}(r)b_{\ell_3}^{NL}(r)
+{\rm cyclic}\,,
\nonumber\\
\end{eqnarray}
where
\begin{eqnarray}
b_{\ell}^{L}(r)\equiv \frac{2}{\pi}\int_{0}^{\infty}k^2dkP_{\Phi}(k)g^{\rm KS}_{T\ell}(k)j_\ell(kr),\nonumber\\
b_{\ell}^{NL}(r)\equiv \frac{2}{\pi}\int_{0}^{\infty}k^2dkf_{\rm NL}g^{\rm KS}_{T\ell}(k)j_\ell(kr).
\end{eqnarray}
Note that our linear transfer function, $g_\ell(k)$, is related to 
that of \cite{ks}, $g^{\rm KS}_{T\ell}(k)$, by $g_\ell(k)=\frac35g^{\rm KS}_{T\ell}(k)$. 

Figure~\ref{fig:shape1} shows the shape of the bispectrum generated by the
products of the first-order terms f selected in Eq.~(\ref{4comb}), and compares it to the primordial
local bispectrum, for $\ell_3=200$. 
Both shapes (second-order and primordial) have the largest
signals in the squeezed triangles, $\ell_1\ll \ell_2\approx
 \ell_3$. This is an expected result: both the local primordial bispectrum
 and the second-order bispectrum that we have computed here
arises from the products of the first-order terms, also products in
position space.  
However, these two shapes are slightly different when $\ell_1/\ell_3$ is not
so small ($\ell_1/\ell_3={\cal O}(0.1)$): the ways in which the radiation transfer
function (which gives the acoustic oscillations) enters into the
bispectrum are different for the products of the first-order terms and
the primordial bispectrum. The primordial bispectrum contains
$j_\ell(kr_*)g_\ell(k)$, whereas the second-order bispectrum contains
$j_\ell(kr_*)g_\ell(k)x(k,r_*)$ where $x=\Delta_0$, $v_0$, etc., also has the
oscillations. Therefore, the second-order bispectrum has more
interferences between multiple radiation transfer functions. 
Moreover, the second-order effects contain derivatives that the local
primordial effects do not have, which also makes the details of the two
shapes different.

Notice, in particular, that most of these gradients in the source term,
Eq.~(\ref{B2}), are contracted with the direction vector, $\hat{\mathbf n}$.
There is only one term that has a scalar product of two wave-vectors,
${\mathbf k}_1\cdot {\mathbf k}_2$, which vanishes in the squeezed limit.
The resulting bispectrum, Eq.~(\ref{eq:44}), resembles that of a local form,
except for the extra powers of $k$ coming from the derivatives. 
These extra powers of $k$ will affect the scale-dependence of the
bispectrum, i.e., the second-order bispectrum is no longer
scale-invariant. Nevertheless,  the largest signal of the bispectrum
still comes from the squeezed configurations, as the number of extra powers of
$k$ from the derivatives in the source term is not large enough to
change the fact that we have the largest contribution when one of $k_1$,
$k_2$, and $k_3$ is very small. In other words, schematically the
bispectrum looks like $B(k_1,k_2,k_3)\sim
(k_1^{m_1}k_2^{m_1})/(k_1^3k_2^3)+{\rm cyclic}$, where $m_1$ and $m_2$ are the
extra powers of $k$ from the derivatives. Therefore, the largest
contribution is in the squeezed configurations as long as $m_i<3$.
Figure~\ref{fig:shape2} shows the same for $\ell_3=1000$.
The results are similar to those for $\ell_3=200$, but the acoustic oscillations
are more clearly visible.

\begin{figure}
\begin{center}
\includegraphics[width=0.495\textwidth]{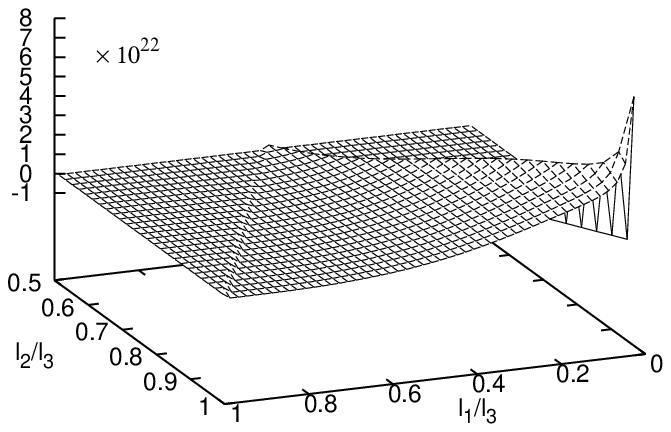}
\includegraphics[width=0.495\textwidth]{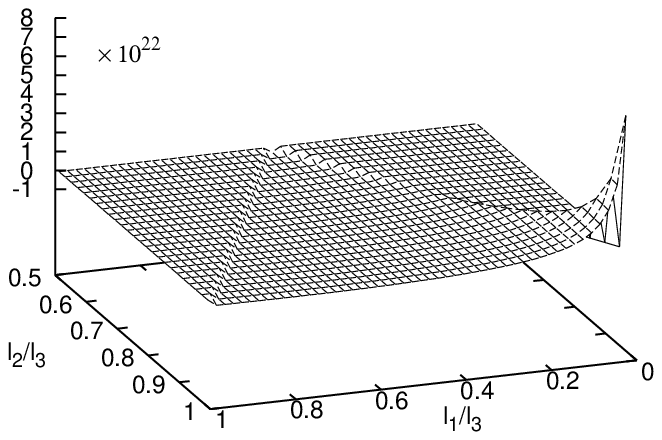}
\end{center}
\caption{
\label{fig:shape1}
Shape dependence of the second-order bispectrum from products of the
 first-order terms (top) and that of the  local primordial bispectrum
 (bottom). We show $\ell_1\ell_2\langle
 a_{\ell_1m_1}^{(1)}a_{\ell_2m_2}^{(1)}a_{\ell_3m_3}^{(2)}\rangle
 {({\cal{G}}_{\ell_1\ell_2\ell_3}^{m_1m_2m_3})}^{-1}/(2\pi)^2\times 10^{22}$ as a
 function of $\ell_1/\ell_3$ and $\ell_2/\ell_3$ where $\ell_3=200$. Both shapes have
 the largest signals in the squeezed triangles, $\ell_1\ll \ell_2\approx
 \ell_3$. 
}
\end{figure}

\begin{figure}
\begin{center}
\includegraphics[width=0.495\textwidth]{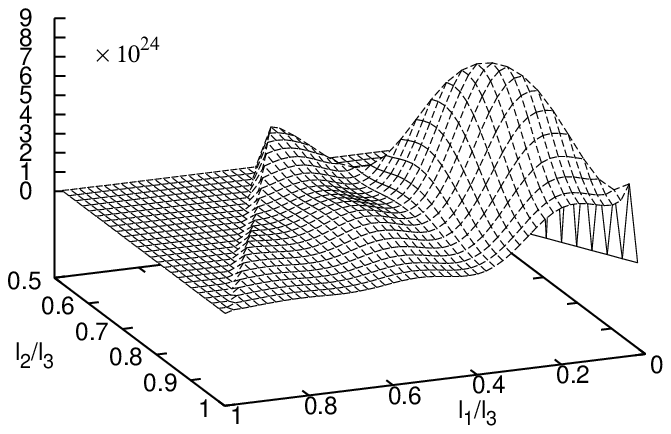}
\includegraphics[width=0.495\textwidth]{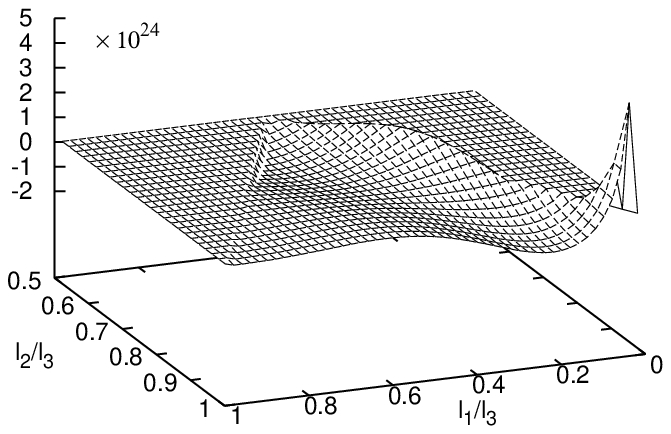}
\end{center}
\caption{
\label{fig:shape2}
Same as Fig.~\ref{fig:shape1} for $\ell_3=1000$. The acoustic oscillations
 are clearly seen.
}
\end{figure}

We can quantify the degree to which the second-order and the primordial
bispectra are correlated, as well as the expected signal-to-noise ratio of the
second-order bispectrum, following the general definitions summarized in Sec.~\ref{cont}. Notice that  we are ignoring the
noise contribution. In other words, we shall only consider ideal
cosmic-variance limited experiments with full sky coverage, which however, at least for the multipole maximum multipole of $l_{max} \sim 2000$ we will consider, is a good reference for an experiment like \emph{Planck} (see, e.g.,~\cite{ks}).

The signal-to-noise ratio is given by Eq.~(\ref{SN}). In Fig.~\ref{fig:SN} we show the cumulative signal-to-noise ratio,
summed up to a maximum multipole of $\ell_{\rm max}$, of the
primordial 
bispectrum, assuming $f_{\rm NL}=1$ and ignoring the second-order
bispectrum, i.e., $(S/N)_{prim}=(F_{prim,prim})^{1/2}$, as well as that of the
second-order bispectrum, ignoring the primordial bispectrum, i.e., 
$(S/N)_{2nd}=(F_{2nd,2nd})^{1/2}$. In both cases $S/N$ increases roughly
as $S/N\propto \ell_{\rm max}$ (or $\propto \sqrt{N_{pix}}$ where $N_{pix}$ is
the number of independent pixels in the map). A larger contribution to the second-order bispectrum at
$\ell \leq 50$ comes from the terms involving the Integrated Sachs-Wolfe
effect. 
The signal-to-noise ratio of the second-order bispectrum reaches $\sim
0.4$ at $\ell_{\rm max}=2000$; thus, this signal is undetectable. While our
calculation includes the temperature anisotropy only, 
including polarization would increase the signal-to-noise by a factor of
two at most, which would not be enough to push the signal-to-noise above
unity. 

\begin{figure}
\begin{center}
\includegraphics[width=0.6\textwidth]{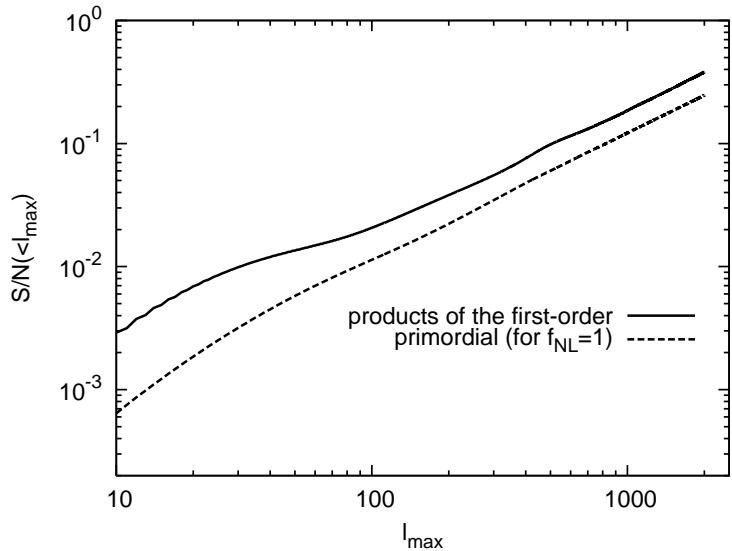}
\end{center}
\caption{\label{fig:SN}
Signal-to-noise ratios for the local primordial bispectrum for $f_{\rm NL}=1$
 (dashed), and the second-order bispectrum from the products of the
 first-order terms (solid), for an ideal full-sky and
 cosmic-variance-limited (noiseless) experiment.
}
\end{figure}

While the total signal-to-noise does not exceed unity, it may still be
instructive to show which terms of
$B_{\ell_1\ell_2\ell_3}^{(\lambda_1,\lambda_2,\lambda_3)}$ and
$B_{\ell_1\ell_2\ell_3}^{C_\ell}$ are more important
than the others.  
To do this we show the following quantity:
\begin{eqnarray}
\left(\frac{S}{N}\right)_{ab}\equiv
\left|\sum_{2\le \ell_1\le \ell_2\le \ell_3}\frac{B_{\ell_1\ell_2\ell_3}^{a}B_{\ell_1\ell_2\ell_3}^{b}}
{\sigma_{\ell_1\ell_2\ell_3}^2}\right|^{1/2},
\label{eq:sncomp}
\end{eqnarray}
where $a,b=1$, 2, 3, 4, and 0 correspond to  $(0,0,0)$, $(1,1,0)$,
$(1,0,1)$, $(1,1,2)$, and $C_\ell$, respectively.

\begin{figure}
\begin{center}
\includegraphics[width=0.6\textwidth]{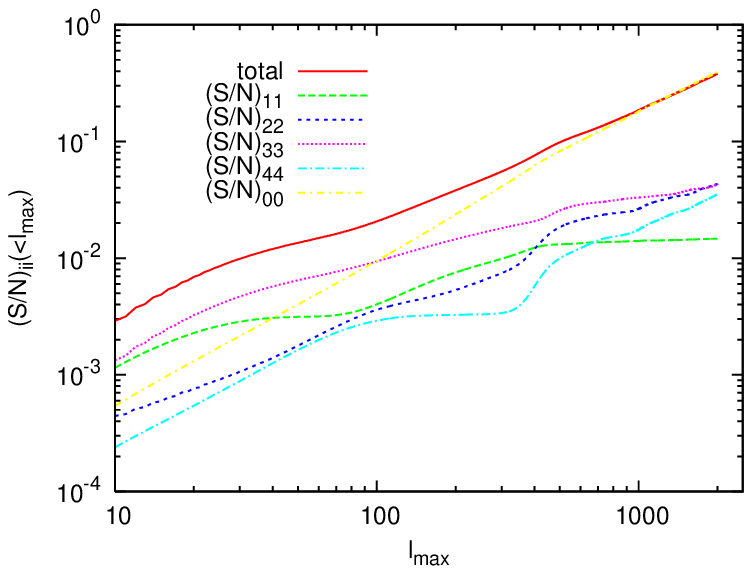}
\includegraphics[width=0.6\textwidth]{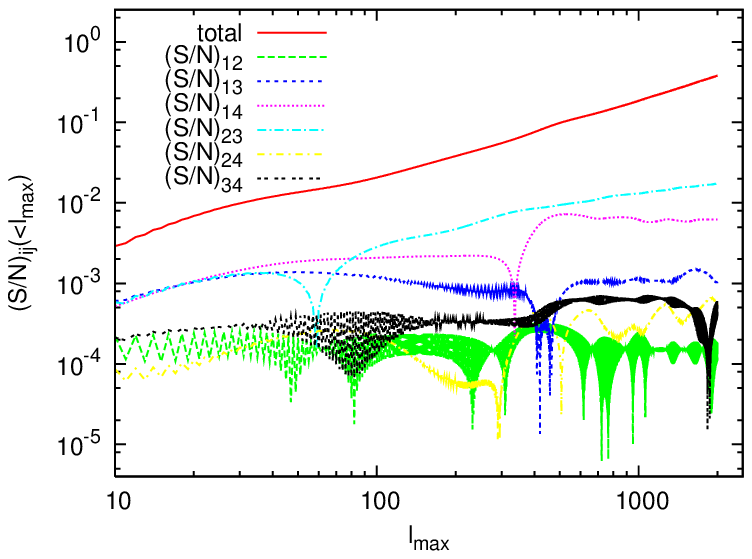}
\includegraphics[width=0.6\textwidth]{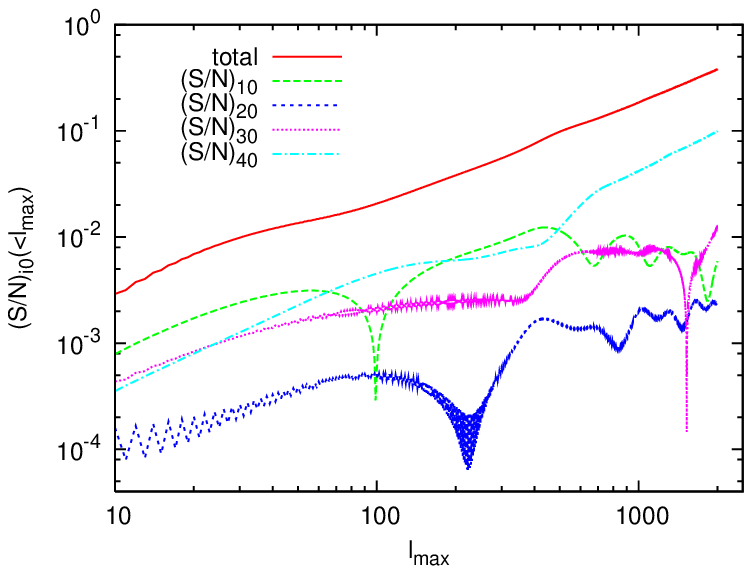}
\end{center}
\caption{\label{fig:SNab}
Absolute values of the contributions to the signal-to-noise ratio from
 each component, $(S/N)_{ab}$, as defined by Eq.~(\ref{eq:sncomp}).
}
\end{figure}

The results are shown in Fig.~\ref{fig:SNab}. We find that 
$(S/N)_{{\rm 2nd}}$ is dominated by
$B_{\ell_1\ell_2\ell_3}^{(\lambda_1,\lambda_2,\lambda_3)}$ for $\ell\leq 100$,
whereas it is 
dominated by $B_{\ell_1\ell_2\ell_3}^{C_\ell}$ for $\ell> 100$ (see the top panel).
Among $B_{\ell_1\ell_2\ell_3}^{(\lambda_1,\lambda_2,\lambda_3)}$, 
the most dominant term is $(1,0,1)$ (the bispectrum from the
second-order dipole created by the first-order dipole and monopole).
The second most dominant is $(0,0,0)$ (from the second-order monopole
created by the first-order monopole) for $l\leq400$ and $(1,1,0)$
(from the second-order monopole 
created by the first-order dipole) for $l> 400$.
The cross terms (middle and bottom panels) are sub-dominant compared to
the auto terms (top panel) at all multipoles.

\begin{figure}
\begin{center}
\includegraphics[width=0.6\textwidth]{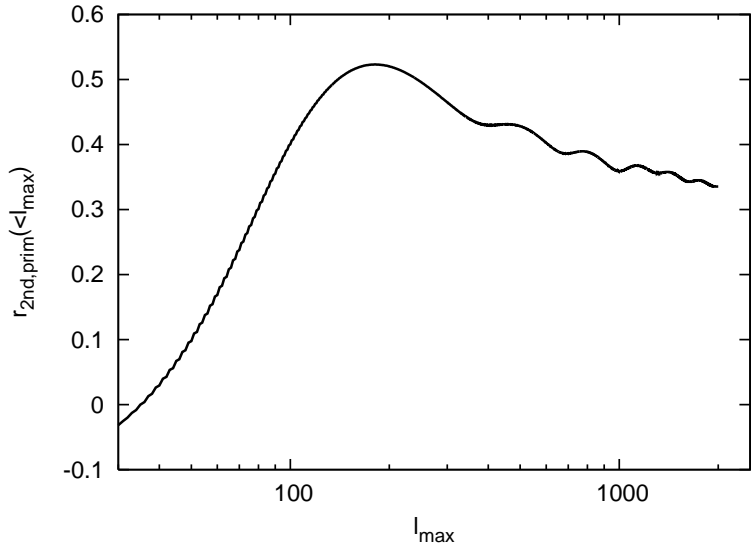}
\end{center}
\caption{\label{fig:r12}
The cross-correlation coefficient between the second-order bispectrum from
 the products of the first-order terms and the local primordial bispectrum. 
}
\end{figure}
\begin{figure}[t]
\begin{center}
\includegraphics[width=0.6\textwidth]{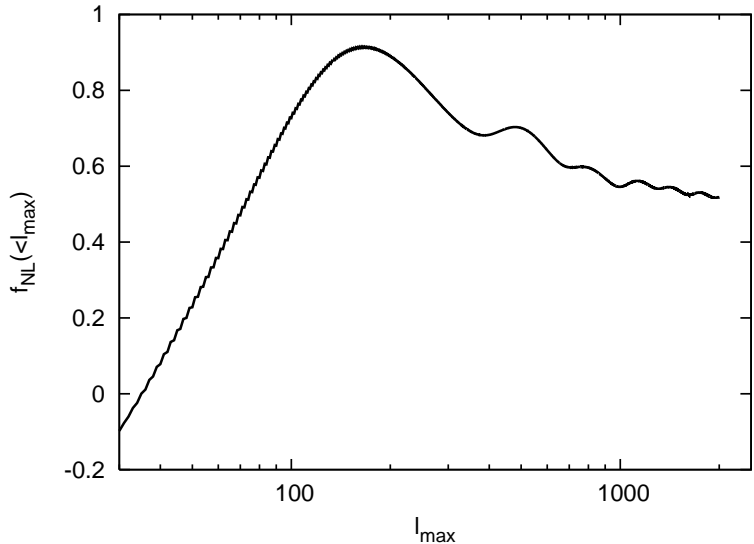}
\end{center}
\caption{\label{fig:fNL}
Contamination of the local primordial bispectrum as measured by
 $f_{\rm NL}^{\rm con}$ (Eq~(\ref{eq:fNL})).
}
\end{figure}
\begin{figure}
\begin{center}
\includegraphics[width=0.6\textwidth]{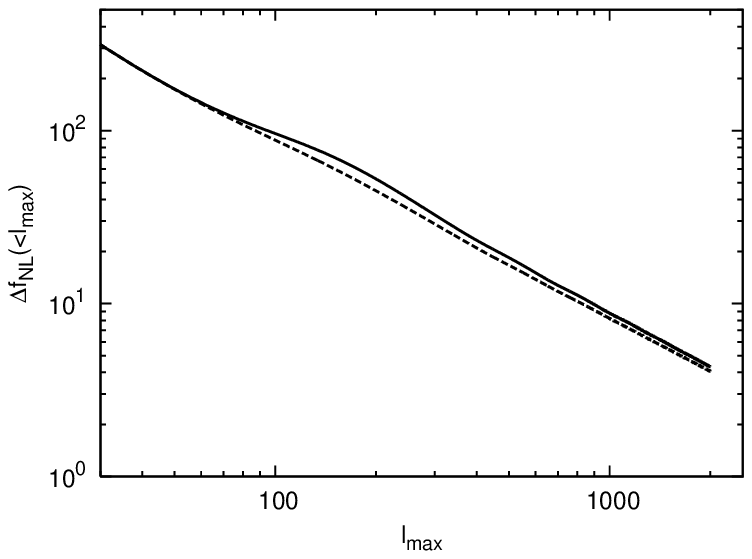}
\end{center}
\caption{\label{fig:DeltafNL}
Projected uncertainty of $f_{\rm NL}$ with (dashed) and without (solid) the
 second-order bispectrum  marginalized over.
}
\end{figure}

How similar are the second-order and the primordial bispectra?
In Fig.~\ref{fig:r12} we show the cross-correlation coefficient, between
the local bispectrum and second-order bispectrum from the products of the first-order terms given in Eq.~(\ref{4comb}). 
The cross-correlation coefficient (see Eq.~(\ref{corrij}), 
reaches $\sim 0.5$ for $\ell_{\rm max}=200$, and the shapes for $\ell_3=200$ are 
shown in Fig.~\ref{fig:shape1}. After $\ell_{\rm max}=200$ the correlation
weakens, and reaches $\sim 0.35$ at $\ell_{\rm max}=1000$, and 
the shapes for $\ell_3=1000$ are shown in Fig.~\ref{fig:shape2}.
These results show that the second-order bispectrum from the products of
the first-order perturbations and the local primordial bispectrum are
fairly similar, with a sizable correlation coefficient. 

How large is the contamination of the primordial
bispectrum? The level of contamination is measured by the effective non-linearity parameter given by Eq.~(\ref{fNLcont}).

In Fig.~\ref{fig:fNL} we show $f_{\rm NL}^{\rm con}$ as a function of the
maximum multipoles, $\ell_{max}$. We find that $f_{\rm NL}^{\rm con}$ reaches
the maximum value, $\sim 0.9$, when the correlation coefficient reaches
the maximum at $\ell_{\rm max}\sim 200$, but then decreases to $\sim 0.5$ at
$\ell_{\rm max}\sim 2000$.  Therefore, we conclude that the contamination of the
primordial bispectrum due to the second-order bispectrum from the terms in Eq.~(\ref{4comb}) is negligible
for CMB experiments.

Finally, one can also calculate the 1-$\sigma$ uncertainty of $f_{\rm NL}$, $\delta
f_{\rm NL}$, with the second-order bispectrum marginalized over. This is
given by $\delta f_{\rm NL}= \sqrt{(F)^{-1}_{\rm prim, \rm prim} |_{f_{\rm NL}=1}}$. Fig.~\ref{fig:DeltafNL} shows that
an increase in the uncertainty of $f_{\rm NL}$ due to marginalization is
totally negligible. 

\section{Conclusions}

In these review we have addressed a basic question in cosmology: how a primordial NG propagates
into an observable like the CMB anisotropy. Answering this question is fundamental as it will help us
in getting some knowledge about the way the primordial cosmological perturbation was generated
at the very early stages of the evolution of the universe.  In the first sections we have
shown how to set the initial conditions at second-order for the (gauge-invariant) CMB anisotropy when some source of primordial NG is present. This was 
more or less straightforward because on large angular scales is basically gravity which dictates the non-linear dynamics. On small angular  scales
the computation of the second-order effects in the CMB anisotropy is far more difficult, as 
so many sources of non-linearities appear. In this review, we have focussed ourselves on the
study of the second-order effects appearing at the recombination era when the CMB anisotropy
is left imprinted. We have shown how to derive the equations which allow to evaluate 
CMB anisotropies, 
by computing the Boltzmann equations describing the evolution of the 
baryon-photon fluid up to second order. This allows to follow the 
time evolution of CMB anisotropies (up to second order) on all scales, 
from the early epoch, when the cosmological perturbations were generated,
to the present time, through the recombination era. 
Through some analytical and simplified example we have also shown how estimate the contamination of the recombination secondary effects 
onto the detection of primordial NG. More refined numerical results confirm our estimates and are also
reported here.  It goes without saying  that  this line of research should be pursued until  the level of accuracy in the theoretical prediction is reached and  is comparable to the one provided by the current or future satellite experiments.
\vspace{2cm}

\noindent
{\bf Acknowledgments}
A.R. is on leave of absence from INFN, Sezione di Padova. This research has been partially supported by ASI contract I/016/07/0 ``COFIS''. 
AR acknowledges support by the EU Marie Curie Network UniverseNet (HPRNCT2006035863). 
We would like to thank Michele Liguori for stimulating discussions, and Daisuke Nitta and Eiichiro Komatsu for past and ongoing enjoyable collaborations.

  
\appendix
\setcounter{equation}{0}
\def\theequation{A.\arabic{equation}}

\section{Energy-momentum tensor}
\subsection{Energy-momentum tensor for photons}

The energy-momentum tensor for photons is defined as

\begin{equation}
T^\mu_{\gamma ~\nu}=\frac{2}{\sqrt{-g}}\int \frac{d^3 P}{(2\pi)^3}\, 
\frac{P^\mu P_\nu}{P^0}\, f\, ,
\end{equation}
where $g$ is the determinant of the metric (\ref{metric}) and $f$
is the distribution function. We thus obtain
\begin{eqnarray}
\label{LH2T00photons}
T^0_{\gamma ~0}&=&-\bar{\rho}_\gamma
\left(1+\Delta_{00}^{(1)}+\frac{\Delta_{00}^{(2)}}{2}\right)\, ,\\
\label{LH2Ti0photons}
T^i_{\gamma ~0}&=&-\frac{4}{3}e^{\Psi+\Phi}\bar{\rho}_\gamma\left(
v_\gamma^{(1)i} +
\frac{1}{2}v_\gamma^{(2)i}+\Delta^{(1)}_{00} v_\gamma^{(1)i}\right)+\frac{1}{3}
\bar{\rho}_\gamma e^{\Psi-\Phi}\omega^i \\
\label{LH2Tijphotons}
T^i_{\gamma ~j}&=& \bar{\rho}_\gamma\left(\Pi^i_{\gamma ~j}+
\frac{1}{3}\delta^i_{~j}\left(1+\Delta_{00}^{(1)}+
\frac{\Delta_{00}^{(2)}}{2}\right)
\right)
\, ,
\end{eqnarray}
where $\bar{\rho}_\gamma$ is the background energy density of photons
and 
\begin{equation}
\Pi^{ij}_{\gamma}=\int\frac{d\Omega}{4\pi}\,\left(n^i n^j-\frac{1}{3}
\delta^{ij}\right)\left(\Delta^{(1)}+\frac{\Delta^{(2)}}{2}\right)\, ,
\end{equation}
is the quadrupole moment of the photons.

\subsection{Energy-momentum tensor for massive particles}

The energy-momentum tensor for massive particles of mass $m$, 
number density $n$  and degrees of freedom $g_d$
\begin{equation}
T^\mu_{m ~\nu}=\frac{g_d}{\sqrt{-g}}\int \frac{d^3 Q}{(2\pi)^3}\, 
\frac{Q^\mu Q_\nu}{Q^0}\, g_m\, ,
\end{equation}
where $g_m$ is the distribution function. We obtain
\begin{eqnarray}
\label{LH2T00massive}
T^0_{m~0}&=&-\rho_m=-\bar{\rho}_m\left(1+\delta^{(1)}_m+
\frac{1}{2}\delta^{(2)}_m
\right) \;, \\
\label{LH2Ti0massive}
T^i_{m~0}&=&-e^{\Psi+\Phi}\rho_m v_m^{i}=
-e^{\Phi+\Psi}\bar{\rho}_m\left(v_m^{(1)i}+
\frac{1}{2}v_m^{(2)i}+\delta^{(1)}_m v_m^{(1)i}\right) \\
\label{LH2Tijmassive}
T^i_{m~j}&=& \rho_m\, \left(\delta^i_{~ j} 
\frac{T_m}{m}+v_m^{i} v_{m~j}\right)=
\bar{\rho}_m\left(\delta^i_{~ j} \frac{T_m}{m}+v_m^{(1)i} v^{(1)}_{m~j}\right)
\, , 
\end{eqnarray}
where $\bar{\rho}_m$ is the background energy density of massive
particles and we have included the equilibrium temperature $T_m$.

\setcounter{equation}{0}
\def\theequation{B.\arabic{equation}}
\section{Solutions of Einstein's equations in various eras}
\label{LH2B}

\subsection{Matter-dominated era}
\label{LH2Appmatter}

During the phase in which the CDM is dominating the energy density
of the Universe, $a\sim \eta^2$ and we may use Eq.~(\ref{LH2trace}) to
obtain an equation for the gravitational potential at first order
in perturbation theory (for which $\Phi^{(1)}=\Psi^{(1)}$)

\begin{equation}
\label{LH2PhiCDM}
\Phi^{(1)''}+3{\mathcal H}\Phi^{(1)'}=0\, ,
\end{equation}
which has two solutions $\Phi^{(1)}_+=$ constant and 
$\Phi^{(1)}_{-}={\mathcal H}/a^2$.
At the same order of perturbation theory, the CDM velocity can be read off
from Eq.~(\ref{i0})

\begin{equation}
\label{LH2velocitymatter1}
v^{(1)i}=-\frac{2}{3{\mathcal H}}\partial^i\Phi^{(1)}\, .
\end{equation}

The matter density contrast $\delta^{(1)}$ satisfies the first-order
continuity equation

\begin{equation}
\delta^{(1)'}=-\frac{\partial v^{(1)i}}{\partial x^i}=-
\frac{2}{3{\mathcal H}}\nabla^2\Phi^{(1)}\, .
\end{equation}
Going to Fourier space, this implies that

\begin{equation}
\label{LH2deltamatter1}
\delta^{(1)}_k=\delta^{(1)}_k(0)+\frac{k^2\eta^2}{6}\Phi^{(1)}_k\, ,
\end{equation}
where $\delta^{(1)}_k(0)$ is the initial condition
in the matter-dominated period.

At second-order, using Eqs.~(\ref{LH2trace}) and~(\ref{Q}) 
and the fact
that the first-order gravitational potential is constant, we find
and equation for the gravitational potential at second-order $\Psi^{(2)}$

\begin{eqnarray}
\Psi^{(2)''}+3{\mathcal H}\Psi^{(2)'}&=&S_m\, ,\\
S_m=-\partial_k\Phi^{(1)}\partial^k\Phi^{(1)}+N&=&
-\partial_k\Phi^{(1)}\partial^k\Phi^{(1)}+\frac{10}{3}
\frac{\partial_i\partial^j}{\nabla^2}\left(\partial_i\Phi^{(1)}
\partial_j\Phi^{(1)}\right)\nonumber \, ,
\end{eqnarray}
whose solution is

\begin{eqnarray}
\label{solmatter}
\Psi^{(2)}&=&\Psi^{(2)}_m(0)+
\int_0^\eta d\eta'\frac{\Phi^{(1)}_+(\eta)\Phi_{-}^{(1)}(\eta')-\Phi^{(1)}_{-}(\eta)\Phi_{+}^{(1)}(\eta')}{W(\eta')}
S_m(\eta')
\nonumber \\
&=&\Psi^{(2)}_m(0)-\frac{1}{14} \left(\partial_k\Phi^{(1)}\partial^k\Phi^{(1)}-\frac{10}{3}
\frac{\partial_i\partial^j}{\nabla^2}\left(\partial_i\Phi^{(1)}
\partial_j\Phi^{(1)}\right)\right)\eta^2
\nonumber \, , \\ 
\end{eqnarray}
with $W(\eta)=W_0/a^3$ ($a_0=1$) the Wronskian obtained from the corresponding
homogeneous equation. In Eq.~(\ref{solmatter}) $\Psi^{(2)}_m(0)$ represents the initial condition
(taken conventionally at $\eta \rightarrow 0$) deep in the matter-dominated phase.

From Eq. (\ref{LH2omegai}), we may compute the vector perturbation in the metric

\begin{equation}
\label{omegamatter}
-\frac{1}{2}\nabla^2\omega^i=3{\mathcal H}^2\frac{1}{\nabla^2}\partial_j\left(
 \partial^i\delta^{(1)} v^{(1)j}-\partial^j\delta^{(1)} v^{(1)i}\right)\, ,
\end{equation}
where we have made use of the fact that the vector part of the CDM velocity
satisfies the relation
$\left(\delta^i_j-\frac{\partial^i\partial_j}{\nabla^2}\right)v^{(2)i}=-
\omega^i$.

\subsection{Radiation-dominated era}

We consider a universe dominated by photons and massless neutrinos. 
The energy-momentum tensor 
for massless neutrinos has the same form as that for photons. 
During the phase in which radiation  is dominating the energy density
of the Universe, $a\sim \eta$ and 
we may combine Eqs. (\ref{00}) and (\ref{LH2trace}) to
obtain an equation for the gravitational potential $\Psi^{(1)}$ at first order
in perturbation theory

\begin{eqnarray}
\label{LH2firstrad}
\Psi^{(1)''}+4{\mathcal H}\Psi^{(1)'}-\frac{1}{3}\nabla^2\Psi^{(1)}
&=& {\mathcal H}Q^{(1)'}+\frac{1}{3}\nabla^2Q^{(1)}
\, ,\nonumber\\
\nabla^2Q^{(1)}&=& \frac{9}{2}{\mathcal H}^2
\frac{\partial_i\partial^j}{\nabla^2}\Pi^{(1)i}_{T~~~j}\, ,
\end{eqnarray}
where the total anisotropic stress tensor is   
\begin{equation}
\Pi^{i}_{T~j}=\frac{\bar{\rho}_\gamma}{\bar{\rho}_T}\, \Pi^{i}_{\gamma~j}
+\frac{\bar{\rho}_\nu}{\bar{\rho}_T}\, \Pi^{i}_{\nu~j}\, .
\end{equation}

We may safely neglect the quadrupole and solve Eq.~(\ref{LH2firstrad}) 
setting $u_\pm=\Phi_\pm^{(1)}\eta$. 
Then Eq.~(\ref{LH2firstrad}), in Fourier space, becomes
\begin{equation}
u''+\frac{2}{\eta}u'+\left(\frac{k^2}{3}-\frac{2}{\eta^2}\right)u=0\, .
\end{equation}
This equation has as independent solutions
$u_+=j_1(k\eta/\sqrt{3})$, the spherical Bessel function of order 1, and 
$u_{-}=n_1(k\eta/\sqrt{3})$, 
the spherical Neumann function of order 1. The latter blows up
as $\eta$ gets small and we discard it on the basis of initial 
conditions. The final solution is therefore 
\begin{equation}
\label{LH2Phir}
\Phi_k^{(1)}=3\Phi^{(1)}(0)\frac{\sin(k\eta/\sqrt{3})-
(k\eta/\sqrt{3})\cos(k\eta/\sqrt{3})}{(k\eta/\sqrt{3})^3}\, 
\end{equation}
where $\Phi^{(1)}(0)$ represents the initial condition deep in the
radiation era.

At the same order in perturbation theory, the radiation
velocity can be read off from Eq.~(\ref{i0})
\begin{equation}
\label{LH20ir}
v^{(1)i}_{\gamma}=-\frac{1}{2{\mathcal H}^2}\frac{\left(a\partial^i
\Phi^{(1)}\right)'}{a}\, .
\end{equation}

At second order, combining Eqs. (\ref{00}), (\ref{LH2trace}), we find

\begin{eqnarray}
\label{P2radeq}
\Psi^{(2)''}+4{\mathcal H}\Psi^{(2)'}-\frac{1}{3}\nabla^2\Psi^{(2)}
= S_\gamma\, ,
\end{eqnarray}
\begin{eqnarray}
\label{Sgamma}
S_\gamma&=&4\left(\Psi^{(1)'}\right)^2+2 \Phi^{(1)'}\Psi^{(1)'}+\frac{4}{3}(\Phi^{(1)}+
\Psi^{(1)}) \nabla^2\Psi^{(1)}
\nonumber \\
&-&\frac{2}{3}(\partial_k\Phi^{(1)} \partial^k\Phi^{(1)}
+\partial_k\Psi^{(1)} \partial^k\Psi^{(1)}-\partial_k\Phi^{(1)} \partial^k\Psi^{(1)}) 
\nonumber \\
&+&{\mathcal H}Q^{(2)'}+\frac{1}{3}\nabla^2Q^{(2)}+\frac{4}{3}(
\Phi^{(1)}+\Psi^{(1)})\nabla^2Q^{(1)}\nonumber \, ,
\end{eqnarray}
\begin{eqnarray}
\label{Q2rad}
\frac{1}{2}\nabla^2Q^{(2)}&=&-\partial_k\Phi^{(1)}\partial^k\Psi^{(1)}-\frac{1}{2}
(\partial_k \Phi^{(1)} \partial^k \Phi^{(1)}-\partial_k \Psi^{(1)} \partial^k \Psi^{(1)}) 
\nonumber \\
&+& 3\frac{\partial_i\partial^j}{\nabla^2}\left[\partial^i\Phi^{(1)}
\partial_j\Psi^{(1)}+\frac{1}{2}(\partial^i\Phi^{(1)}
\partial_j\Phi^{(1)}-\partial^i\Psi^{(1)}
\partial_j\Psi^{(1)})\right] \nonumber \\
&+& \frac{9}{2}{\mathcal H}^2
\frac{\partial_i\partial^j}{\nabla^2}\frac{\Pi^{(2)i}_{T~~~j}}{2}
-9{\cal H}^2\frac{\partial_i\partial^j}{\nabla^2} \left( \Psi^{(1)} \Pi^{(1)i}_{T~~~j} \right)\, ,
\end{eqnarray}

whose solution is

\begin{eqnarray}
\Psi^{(2)}=\Psi^{(2)}_{\rm hom.}+
\int_0^\eta d\eta'\frac{\Phi^{(1)}_+(\eta)\Phi_{-}^{(1)}(\eta')-\Phi^{(1)}_{-}(\eta)\Phi_{+}^{(1)}(\eta')}{W(\eta')}
S_\gamma(\eta')\, , \nonumber \\
\end{eqnarray}
where $W(\eta)=(a(0)/a)^4$ is the Wronskian, and $\Psi^{(2)}_{\rm hom.}$ is the solution of the 
homogeneous equation.

The equation of motion for the
vector metric perturbations reads

\begin{eqnarray}
\label{omegair}
&-&\frac{1}{2}\nabla^2\omega^i
+4{\mathcal H}^2\omega^i=
\left(\delta^i_j-\frac{\partial^i\partial_j}{\nabla^2}\right)
\left[2\Psi^{(1)'}\partial^j\Phi^{(1)}+{\cal H}^2 \frac{\bar{\rho}_\gamma+\bar{\rho}_\nu}{\bar{\rho}_T} \omega^j
\right. \nonumber \\
&-& \left. 2
{\mathcal H}^2 \left( \frac{\bar{\rho}_\gamma}{\bar{\rho}_T} v_\gamma^{(2)j}+
\frac{\bar{\rho}_\nu}{\bar{\rho}_T} v_\nu^{(2)i}
+2 \frac{\bar{\rho}_\gamma}{\bar{\rho}_T} \Delta^{(1)\gamma}_{00} v_\gamma^{(1)j}
+2\frac{\bar{\rho}_\nu}{\bar{\rho}_T} \Delta^{(1)\nu}_{00} v_\nu^{(1)j} 
\right. \right. \nonumber \\
&+& \left. \left. 2 (\Phi^{(1)}-\Psi^{(1)})\frac{\bar{\rho}_\gamma}{\bar{\rho}_T} v_\gamma^{(1)j}
+ 2 (\Phi^{(1)}-\Psi^{(1)})\frac{\bar{\rho}_\nu}{\bar{\rho}_T} v_\nu^{(1)j}
\right) \right]
\, ,
\end{eqnarray}
where ${\bar{\rho}_T}$ is the total background energy density. The Einstein equations for a universe filled by CDM and a relativisitc component 
can be foun in Ref.~\cite{paperII}.

\setcounter{equation}{0}
\def\theequation{C.\arabic{equation}}
\section{Linear solution of the Boltzmann equations}      
\label{LH2LBE}

In this section we will solve the Boltzmann 
equations at first order in perturbation theory. The interested reader 
will find the extension of these formulae to second order in 
Ref.~\cite{paperII}.  
The first two moments of the photon Boltzmann equation 
are obtained by integrating Eq.~(\ref{B1}) 
over $d\Omega_{\bf n}/4\pi$ and 
$d\Omega_{\bf n} n^i /4\pi$ respectively and they lead to 
the density and velocity continuity equations 
\begin{equation}
\label{LH2B1l}
\Delta^{(1)'}_{00}+\frac{4}{3} \partial_i 
v^{(1)i}_\gamma-4\Psi^{(1)'}=0\, ,
\end{equation}
\begin{equation}
\label{LH2B2l}
v^{(1)i\prime}_\gamma+\frac{3}{4} \partial_j 
\Pi^{(1)ji}_\gamma+\frac{1}{4} \Delta^{(1),i}_{00}
+\Phi^{(1),i}=-\tau' \left( v^{(1)i}-v^{(1)}_\gamma \right)\, , 
\end{equation}
where $\Pi^{ij}$ is the photon quadrupole moment, defined in 
Eq.~(\ref{quadrupole}).

Let us recall here that $\delta^{(1)}_\gamma=\Delta^{(1)}_{00}= 
\int d\Omega \Delta^{(1)}/4\pi$ and that 
the photon velocity is given by Eq.~(\ref{vp1}).

The two equations above are complemented by the momentum 
continuity equation for baryons, which can be conveniently written 
as  
\begin{equation}
\label{LH2bv1}
v^{(1)i}=v^{(1)i}_\gamma+\frac{R}{\tau'} \left[v^{(1)i\prime}+
{\cal H} v^{(1)i} +\Phi^{(1),i}   \right]\, ,
\end{equation}
where we have introduced the baryon-photon ratio 
$R \equiv 3 \rho_b/(4\rho_\gamma)$.

Eq.~(\ref{LH2bv1}) is in a form ready for a consistent expansion 
in the small quantity $\tau^{-1}$ which can be performed 
in the tight-coupling limit. By first taking $v^{(1)i}=v^{(1)i}_\gamma$ 
at zero order and then using this relation in the 
left-hand side of Eq.~(\ref{LH2bv1}) one obtains
\begin{equation}
\label{LH2vv}
v^{(1)i}-v^{(1)i}_\gamma=\frac{R}{\tau'} \left[v^{(1)i\prime}_\gamma
+{\cal H} v^{(1)i}_\gamma +\Phi^{(1),i}   \right]\, .
\end{equation}
Such an expression for the difference of velocities can be used in 
Eq.~(\ref{LH2B2l}) to give the evolution equation for 
the photon velocity in the limit of tight coupling
\begin{equation}
\label{LH2vphotontight}
v^{(1)i\prime}_\gamma+{\cal H}\frac{R}{1+R}v^{(1)i}_\gamma 
+\frac{1}{4} \frac{\Delta^{(1),i}_{00}}{1+R}+\Phi^{(1),i} =0\, .
\end{equation}
Notice that in Eq.~(\ref{LH2vphotontight}) 
we are neglecting the quadrupole of the photon distribution $\Pi^{(1) ij}$ 
(and all the higher moments) since it is well known that at linear 
order such moment(s) are suppressed in the tight-coupling limit 
by (successive powers of) $1/\tau$ with respect to the first two 
moments, the photon energy density and velocity. 
Eqs.~(\ref{LH2B1l}) and (\ref{LH2vphotontight}) are the master equations 
which govern the photon-baryon fluid acoustic 
oscillations before the epoch of recombination when photons and baryons 
are tightly coupled by Compton scattering. 

In fact one can combine these two equations to get a single second-order 
differential equation for the photon energy 
density perturbations $\Delta^{(1)}_{00}$. 
Deriving Eq.~(\ref{LH2B1l}) with respect to conformal time and using
Eq.~(\ref{LH2vphotontight}) to replace 
$\partial_i v^{(1)i}_\gamma$ yields
\begin{eqnarray}
\label{LH2eqoscill}
&&\left( \Delta^{(1)\prime \prime}_{00}-4\Psi^{(1)\prime \prime} \right) 
+{\cal H}\frac{R}{1+R} 
\left( \Delta^{(1)\prime}_{00}-4\Psi^{(1)\prime} \right) \nonumber \\
&&-c_s^2 \nabla^2 
\left( \Delta^{(1)}_{00}-4\Psi^{(1)} \right) 
= \frac{4}{3} \nabla^2 
\left( \Phi^{(1)}+\frac{\Psi^{(1)}}{1+R} \right)\, ,
\end{eqnarray}     
where $c_s=1/\sqrt{3(1+R)}$ is the speed of sound of the photon-baryon 
fluid. Indeed, in order to solve Eq.~(\ref{LH2eqoscill}) one needs to know 
the evolution of the gravitational potentials. 
We will come back later to the discussion of the solution of 
Eq.~(\ref{LH2eqoscill}). 

A useful relation is obtained by 
considering the continuity equation for the 
baryon density perturbation. By perturbing at first order 
Eq.~(\ref{cont_bar}) we obtain
\begin{equation}
\label{LH2bcont}
\delta^{(1)\prime}_b+v^i_{,i}-3\Psi^{(1)\prime}=0\, .
\end{equation}
Subtracting Eq.~(\ref{LH2bcont}) form Eq.~(\ref{LH2B1l}) brings 
\begin{equation}
\Delta^{(1)\prime}_{00}-\frac{4}{3} \delta^{(1)\prime}_b+\frac{4}{3} 
(v^{(1)i}_\gamma-v^{(1)i})_{,i}=0\, ,
\end{equation} 
which implies that at lowest order in the tight-coupling approximation 
\begin{equation}
\label{LH2D100d1b}
\Delta^{(1)}_{00}=\frac{4}{3}\delta^{(1)}_b \, ,
\end{equation}
for adiabatic perturbations.
\subsection{Linear solutions in the limit of tight coupling}
\label{ApplinearBoltz}
\label{LH2Tsol1}
In this section we briefly recall how to obtain at linear order the 
solutions of the Boltzmann equations~(\ref{LH2eqoscill}).  
These correspond to the acoustic oscillations of the photon-baryon fluid 
for modes which are within 
the horizon at the time of recombination.   
It is well known that, in the variable $(\Delta^{(1)}_{00}-4\Psi^{(1)})$, 
the solution can be written as~\cite{Huthesis,Husug}
\begin{eqnarray}
\label{LH2soltot}
&&[1+R(\eta)]^{1/4} (\Delta^{(1)}_{00}-4\Psi^{(1)})= 
A\, \cos[kr_s(\eta)]+B\,\sin[kr_s(\eta)] \nonumber \\
&&-4\frac{k}{\sqrt{3}} \int_0^\eta d\eta' [1+R(\eta')]^{3/4} 
\left(\Phi^{(1)}(\eta')+\frac{\Psi^{(1)}(\eta')}{1+R(\eta')} 
\right) \nonumber \\
&&\times \sin[k(r_s(\eta)-r_s(\eta'))] \; ,
\end{eqnarray}
where the sound horizon is given by
$r_s(\eta)=\int_0^\eta d\eta' c_s(\eta')$,
with $R= 3 \rho_b/(4\rho_\gamma)$. 
The constants $A$ and 
$B$ in Eq.~(\ref{LH2soltot}) are fixed by the choice of initial conditions. 

In order to give an analytical we will use some simplifications 
following Ref.~\cite{HZ,Dodelsonbook}. 
First, for simplicity, we are going to neglect the ratio $R$ wherever 
it appears, 
{\it except} in the arguments of the varying cosines and sines, where
 we will treat $R=R_*$ as a constant 
evaluated at the time of recombination. In this way we  
keep track of a damping of the photon velocity amplitude with respect 
to the case 
$R=0$ which prevents the acoustic peaks in the power-spectrum to disappear.   
Treating $R$ as a constant is justified by the fact that for modes 
within the horizon the 
time scale of the ocillations is much shorter than the time scale 
on which $R$ varies. 
If $R$ is a constant the sound speed is just a constant 
$c_s=1/\sqrt{3(1+R_*)}$, 
and the sound horizon is simply $r_s(\eta)=c_s \eta$. 

Second, we are going to solve for the evolutions of the perturbations 
in two well distinguished 
limiting regimes. One regime is for those perturbations which enter 
the Hubble radius when matter is the dominant 
component, that is at times much bigger than the equality epoch, with 
$k \ll k_{eq} \sim \eta^{-1}_{eq}$, 
where $k_{eq}$ is the wavenumber of the Hubble radius at the equality 
epoch. The other regime is for those perturbations 
with much smaller wavelenghts which enter the Hubble radius when the 
universe is still radiation dominated, that is 
perturbations with wavenumbers $k \gg k_{eq}\sim \eta_{eq}^{-1}$. In 
fact we are interested in  
perturbation modes which are within the horizon by the time of 
recombination $\eta_*$. Therefore 
we will further suppose that $\eta_* \gg \eta_{eq}$ in order to study 
such modes in the first regime. Even though $\eta_* \gg \eta_{eq}$ is 
not the real case, it allows to obtain some  
analytical expressions. 

Before solving for these two regimes let us fix our initial conditions,  
which are taken on large scales deep in the radiation dominated era 
(for $\eta \rightarrow 0$). 
During this epoch, for adiabatic perturbations, the gravitational 
potentials remain constant on large scales 
(we are neglecting anisotropic stresses so that $\Phi^{(1)} \simeq 
\Psi^{(1)}$) and from the $(0-0)$-component of Einstein 
equations 
\begin{equation}
\label{LH2initcond1}
\Phi^{(1)}(0)=-\frac{1}{2} \Delta^{(1)}_{00}(0)\, .
\end{equation}
On the other hand, from the energy continuity equation~(\ref{LH2B1l}) 
on large scales 
\begin{equation}
\label{LH2initcond2}
\Delta^{(1)}_{00}-4\Psi^{(1)}={\rm const.}\, ;
\end{equation}
from Eq.~(\ref{LH2initcond1}) the constant on the right-hand side 
of Eq.~(\ref{LH2initcond2}) is 
fixed to be $-6 \Phi^{(1)}(0)$; thus we find $B=0$ and $A=-6 \Phi^{(1)}(0)$. 

With our simplifications Eq.~(\ref{LH2soltot}) then reads 
\begin{eqnarray}
\label{LH2solsempl}
\Delta^{(1)}_{00}-4\Psi^{(1)}=& 
-&6\Phi^{(1)}(0) \cos(\omega_0 \eta) \nonumber \\
&-& \frac{8k}{\sqrt{3}}\int_0^\eta d\eta' \Phi^{(1)}(\eta') 
\sin[\omega_0 (\eta-\eta')]\, ,
\end{eqnarray}
where $\omega_0=kc_s$. 

\subsection{Perturbation modes with $k \ll k_{eq}$}
This regime corresponds to perturbation modes which enter the Hubble 
radius when the universe is matter dominated at times 
$\eta \gg \eta_{eq}$. During matter domination the gravitational potential 
remains constant (both on super-horizon and 
sub-horizon scales), as one can see for example from 
Eq.~(\ref{LH2PhiCDM}), and its value is fixed to $\Phi^{(1)}({\bf k}, 
\eta)=\frac{9}{10} \Phi^{(1)}(0)$, where $\Phi^{(1)}(0)$ 
corresponds to the gravitational potential on large scales during the 
radiation dominated epoch. Since we are interested in 
the photon anisotropies around the time of recombination, when matter 
is dominating, we can perform the integral appearing 
in Eq.~(\ref{LH2soltot}) by taking the gravitational potential equal to 
its value during matter domination so that it is 
easily computed
\begin{equation}
\label{LH2DPhimatter}
2 \int_0^\eta d\eta' \Phi^{(1)}(\eta') 
\sin[\omega_0(\eta-\eta')]=\frac{18}{10} \frac{\Phi^{(1)}(0)}{\omega_0} 
\left( 1-\cos(\omega_0 \eta) 
\right)\, .
\end{equation} 
Thus Eq.~(\ref{LH2solsempl}) gives  
\begin{equation}
\label{LH2D001sol}
\Delta^{(1)}_{00} -4\Psi^{(1)}=\frac{6}{5} \Phi^{(1)}(0)\, 
\cos(\omega_0\eta)-\frac{36}{5} \Phi^{(1)}(0)\, .
\end{equation}
The baryon-photon fluid velocity can then be obtained as 
$\partial_i v^{(1)i}_\gamma=- 3 (\Delta^{(1)}_{00}-4\Psi^{(1)})'/4$
from Eq.~(\ref{LH2B1l}). In Fourier space 
\begin{equation}
ik_i\, v^{(1)i}_\gamma=\frac{9}{10} \Phi^{(1)}(0) \sin(\omega_0 \eta) 
\omega_0\, ,
\end{equation}  
where, going to Fourier space, $\partial_i v^{(1)i}_\gamma \rightarrow i k_i 
\, v^{(1)i}_\gamma({\bf k})$ and
\begin{equation}
\label{LH2v1sol}
v^{(1)i}_\gamma=-i \frac{k^i}{k}\frac{9}{10} \Phi^{(1)}(0) \sin(\omega_0 
\eta) c_s\, ,
\end{equation}
since the linear velocity is irrotational. 
 
Notice that, under the approximations that $R={\rm const.}$ and $\Phi^{(1)}=\Psi^{(1)}={\rm const.}$, it is easy to find 
a more accurate solution to Eq.(\ref{LH2eqoscill}) which better accounts for the presence of the baryons 
(giving rise to the so called baryon-drag effect, as clearly explained in Refs.~\cite{Huthesis,Husug,HuWhitedamping}) 
\begin{equation}
\label{LB}
\frac{\Delta^{(1)}_{00}}{4}=\Big[\frac{\Delta^{(1)}_{00}}{4}(0)+(1+R)\Phi^{(1)} \Big] \cos(kr_s) -(1+R)\Phi^{(1)}\, .
\end{equation}

\subsection{Perturbation modes with $k \gg k_{eq}$}
This regime corresponds to perturbation modes which enter the Hubble 
radius when the universe is still radiation dominated 
at times $\eta \ll \eta_{eq}$. In this case an approximate analytical
 solution for 
the evolution of the perturbations can be obtained by considering the 
gravitational potential for a pure 
radiation dominated epoch, given by Eq.~(\ref{LH2Phir}). For the integral 
in Eq.~(\ref{LH2solsempl}) we thus find 
\begin{equation}
\int_0^\eta \Phi^{(1)}(\eta') \sin[\omega_0(\eta-\eta')] = 
-\frac{3}{2\omega_0} \cos(\omega_0 \eta)\, , 
\end{equation}
where we have kept only the dominant contribution oscillating in 
time, while neglecting terms which decay in time. The 
solution~(\ref{LH2solsempl}) becomes
\begin{equation}
\label{LH2DPsrd}
\Delta^{(1)}_{00} -4\Psi^{(1)}=6 \Phi^{(1)}(0)\, \cos(\omega_0\eta)\, ,
\end{equation} 
and the velocity is given by
\begin{equation}
\label{LH2vsrd}
v^{(1)i}_\gamma=-i \frac{k^i}{k}\frac{9}{2} \Phi^{(1)}(0) 
\sin(\omega_0 \eta) c_s\, ,
\end{equation}

Notice that the solutions~(\ref{LH2DPsrd})--(\ref{LH2vsrd}) 
are actually correct only when radiation dominates.
Indeed, between the epoch of equality and recombination, matter 
starts to dominate. 
Full account of such a period is given e.g. in Section 7.3 of 
Ref.~\cite{Dodelsonbook}, while its consequences for the CMB anisotropy 
evolution can be found e.g. in Ref.~\cite{Muk}. We refer to~\cite{paperII}, where an alternative way to solve for the acoustic 
oscillations in this regime is displayed which turns out to be useful for the corresponding computation at second order in 
Sec.~\ref{kggkeq2}.

\begin{table}[h]
\centering
\begin{tabular}{c c c c c c}
\hline\hline 
\vspace{0.2cm}
Symbol & Definition & Equation \\
\hline  
$\Phi,\Psi$ & Gravitational potentials in Poisson gauge & 
(\ref{metric}) \\
$\omega_i$ & $2$nd-order vector perturbation in Poisson gauge & 
(\ref{metric}) \\
$\chi_{ij}$ & $2$nd-order tensor perturbation in Poisson gauge & 
(\ref{metric}) \\
$\eta$ & Conformal time & (\ref{metric})  \\
$f$ & Photon distribution function & (\ref{Df}) \\
$g$ & Distribution function for massive particles & (\ref{gel}) 
\& (\ref{DG}) \\
$f^{(i)}$ & $i$-th order perturbation of the photon distribution 
function  & (\ref{expf}) \\
$f^{(i)}_{\ell m}$ & Moments of the photon distribution function 
& (\ref{angular}) \\
$C({\bf p})$ & Collision term & (\ref{collisionterm}) \& 
(\ref{Integralcolli}) \\ 
$p$ & Magnitude of photon momentum (${\bf p}=pn^i$) &  
(\ref{defp}) \\
$n^i$ & Propagation direction & (\ref{Pi}) \\
$\Delta^{(1)}(x^i,n^i,\eta)$ & First-order fractional energy 
photon fluctuations & (\ref{Delta1})  \\
$\Delta^{(2)}(x^i,n^i,\eta)$ & Second-order fractional energy 
photon fluctuations & (\ref{Delta2}) \\
$n_e$ & Electron number density & (\ref{defne}) \\
$\delta_e (\delta_b)$  & Electron (baryon) density perturbation 
& (\ref{deltac1}) \\
${\bf k}$ & Wavenumber & (\ref{BF}) \\
$v_m$ & Baryon velocity perturbation & (\ref{vzero}) \& (\ref{vdec}) \\ 
$v^{(2)i}_{\rm CDM}$ & Cold dark matter velocity & (\ref{mcc2CDM}) \\
$v^{(2)i}_\gamma$ & Second-order photon velocity & (\ref{vp2}) \\
$S_{\ell m}$ & Temperature source term & (\ref{H}) \\  
$\tau$ & Optical depth & (\ref{deftau}) \\
${\bar \rho}_\gamma ({\bar \rho}_b)$ & Background photon (baryon) 
energy density & (\ref{mcc1}) \\ 
$B_{\ell_1,\ell_2,\ell_3}$ & CMB angular-averaged bispectrum & (\ref{AAB})\\
$F_{ij}$ & Fisher matrix for the amplitudes of the bispectra & (\ref{FMB})\\
$f^{\rm con}_{\rm NL}$ & contamination to primordial non-Gaussianity & (\ref{fNLcont}) \\
\hline\hline
\end{tabular}
\end{table}


\end{document}